\documentclass[11pt,crest,nopardent,msfonts,fancychap,nomencl]{edengths}

\graphicspath{{chapter1/}{chapter2/}}

\author{Yuanchao Li}

\title{Speech Emotion Recognition with ASR Integration}

\date{2025}

\begin{document}






\abstract{
Speech Emotion Recognition (SER) plays a pivotal role in understanding human communication, enabling emotionally intelligent systems, and serving as a fundamental component in the development of Artificial General Intelligence (AGI). However, deploying SER in real-world, spontaneous, and low-resource scenarios remains a significant challenge due to the complexity of emotional expression and the limitations of current speech and language technologies. This thesis investigates the integration of Automatic Speech Recognition (ASR) into SER, with the goal of enhancing the robustness, scalability, and practical applicability of emotion recognition from spoken language.

As a starting point, we explore the interplay between ASR and SER by conducting an in-depth analysis of speech foundation models on emotionally expressive speech, from both acoustic and linguistic perspectives. Our findings uncover inherent limitations of these models: while they achieve strong performance in ASR, they often fail to capture paralinguistic cues essential for emotion recognition, and exhibit emotion-dependent biases. Additionally, we examine how linguistic properties, such as part-of-speech distributions, affective word ratings, and utterance length, interact with ASR performance across different emotion categories. Extensive experiments reveal that ASR errors are not random but follow systematic patterns associated with specific word types and emotional expressions across diverse speaking conditions.

To overcome paralinguistic limitations while utilizing the hierarchical encoding capabilities of speech foundation models, we propose a joint training framework that integrates ASR hidden representations and lexical outputs in a hierarchical manner for SER. To mitigate transcription quality issues, we introduce two complementary ASR error correction strategies that require only a small amount of emotional speech data: a Large Language Model (LLM)-based method that refines N-best hypotheses using emotion-specific prompts, and a Sequence-to-Sequence (S2S) model that leverages discrete acoustic units to correct 1-best hypothesis. Both approaches substantially improve the emotional accuracy of ASR transcriptions with low-resource speech data and enhance downstream SER performance.

Beyond transcription correction, this thesis introduces robust acoustic-lexical fusion frameworks to handle emotion mismatch across modalities and reduce ASR error propagation. We conduct a comprehensive investigation into cross-modal incongruity and propose incongruity-aware fusion strategies that dynamically adapt to modality conflicts, extending solutions beyond SER to include sarcasm and humor detection. Building on this, we develop an ASR-aware, modality-gated fusion mechanism that integrates ASR error correction with dynamic modality selection in a sequential manner. These fusion strategies achieve strong performance on both controlled and real-world datasets, even under partially erroneous conditions (i.e. ASR transcriptions).

Finally, to reduce reliance on costly labeled emotion data, we present a novel semi-supervised learning framework based on multi-view pseudo-labeling. By leveraging both acoustic similarity metrics and LLM-based confidence estimation, this approach selects high-quality unlabeled samples to augment training. The proposed framework not only outperforms traditional pseudo-labeling methods for SER but also demonstrates generalizability to speech-based Alzheimer's dementia detection.
}

\summary{
Human speech carries more than just words, it reflects our thoughts, intentions, and emotions. The ability to recognize emotions from speech, known as Speech Emotion Recognition (SER), is increasingly important in areas such as mental health monitoring, elderly care, human-computer interaction, and voice-based assistants. A system that can understand how someone feels, simply by listening to them speak, holds great promise for making technology more empathetic and responsive.

However, recognizing emotions in everyday, natural speech is a complex task. In real-life scenarios, speech can vary greatly depending on the speaker, the environment, and the emotional state. This makes it challenging for machines to accurately detect emotions, especially when the speech is spontaneous, noisy, or emotionally intense.

To address this, the research explores how Automatic Speech Recognition (ASR), technology that converts speech into text, can be used to help improve emotion recognition. But there’s a catch: while ASR focuses on what is said, it often loses the emotional tone in how it is said. This creates a gap between recognizing content and capturing emotion.

This project develops new methods to combine both the acoustic (audio-based) and linguistic (text-based) features of speech to better detect emotion. It also uses advanced AI tools, such as large language models, to recover emotional information that might be lost or misinterpreted during speech-to-text conversion, especially when the speaker is upset, excited, or speaking in a non-standard way.

Furthermore, the research introduces techniques that help machines learn from unlabeled or uncertain data, using strategies like confidence scoring and self-training. This makes emotion recognition more robust, flexible, and suitable for real-world applications where clean, annotated data is limited.

The findings are not merely theoretical. They have been tested in a variety of scenarios, including practical applications, and have even been extended to the early detection of conditions such as Alzheimer’s disease.

Overall, this research aims to make machines more emotionally intelligent by helping them understand not just what people say, but how they feel when they say it. It represents a meaningful step toward more natural and compassionate human-computer interactions.
}


\acknowledgements{%
I would like to express my sincere gratitude to my primary supervisor, Dr. Catherine Lai, and my second supervisor, Prof. Peter Bell, for their invaluable mentorship throughout my PhD. Their guidance and expertise have been instrumental in shaping my research at the intersection of SER and ASR. With their support, I was able to pursue my work freely and without pressure.

I am also deeply grateful to Dr. Korin Richmond for his insightful annual reviews and valuable collaboration during my MSc supervision, and to Prof. Simon King and Prof. Carlos Busso for serving as examiners of this thesis and offering their professional feedback.

My heartfelt thanks go to my master’s supervisor, Prof. Tatsuya Kawahara, a lifelong
\begin{CJK}{UTF8}{min}先生\end{CJK} in my life. His continuous support has been invaluable not only in my academic pursuits but also in my personal growth. I am also grateful to Prof. Nigel Ward and Dr. Carlos Ishi for introducing me to the fascinating domain of prosody and dialogue and for guiding my first research project.

Furthermore, I greatly appreciate the warm, collegial, and professional environment at the Centre for Speech Technology Research. I have learned deeply from the weekly reading groups and talks, and I am thankful to all the members of the group for their generosity and kindness.

I would also like to thank my mentors, Dr. Dimitra Emmanouilidou and Dr. Hannes Gamper, at the Microsoft Research Audio and Acoustics Research Group for their support during and beyond my internship. I am grateful to all the collaborators I have worked with throughout this PhD journey.

Foremost, I thank my family for their constant love and support.
}
\standarddeclaration{Yuanchao Li}


\tableofcontents




\printnomenclature



\startbody

\chapter{Introduction}
\label{chap1}

\section{Motivation}
In recent years, the rapid advancement of machine learning in acoustic and language modeling has significantly contributed to the development of conversational agents and voice-driven robotic systems. At the core of these systems lies a dialogue engine, primarily composed of Automatic Speech Recognition (ASR) and Natural Language Processing (NLP) systems, which facilitates grammatical and syntactic analysis of speech to derive semantic meaning and respond to user queries effectively. However, despite their impressive capabilities in processing spoken content, these systems remain fundamentally limited in their ability to interpret non-verbal communication cues.

In other words, while these systems excel at understanding \textit{what is said}, they lack the ability to comprehend \textit{how it is said}, including vocal tones, emotional expressions, underlying psychological states, and personal traits, collectively known as paralinguistic information. Human speech carries rich emotional and social cues that are crucial for natural interactions, and failing to incorporate these aspects restricts the effectiveness of human-computer communication.

To bridge this gap and enable emotionally intelligent interactions, research fields such as Affective Computing, Computational Paralinguistics, and Social Signal Processing have emerged as key areas of study over the past two decades. Among these, Speech Emotion Recognition (SER) has attracted significant attention due to its potential to equip machines with emotional awareness, fostering more natural, engaging, and empathetic human-computer interactions.

SER plays a crucial role in various applications, including:

\begin{itemize}
    \item Human-computer interaction, where emotionally aware systems can provide more engaging and adaptive responses \citep{kwon2003emotion,wang2015speech}.
    \item Mental health monitoring, assisting in early diagnosis and assessment of psychological conditions through speech analysis \citep{elsayed2022speech,cummins2018speech}.
    \item Customer service automation, where emotion-aware AI can enhance user satisfaction by tailoring responses based on emotional cues \citep{li2019acoustic,han2020ordinal}.
    \item Personalized user experiences, enabling virtual assistants, social robots, and entertainment systems to adapt based on a user’s emotional state \citep{lakomkin2018robustness,li2019expressing}.
\end{itemize}





By analyzing vocal attributes such as pitch, intensity, rhythm, spectral features, and temporal dynamics, SER systems aim to infer the speaker's emotional state, enabling intuitive and context-aware interactions.

Despite the remarkable progress in SER over the past decade, several critical challenges hinder its transition from controlled laboratory settings to real-world applications:


\textbf{1. Limited understanding of how emotional speech is represented in cutting-edge speech models}

While speech models have undergone significant improvements, particularly with deep learning and transformer-based architectures, their role in emotionally expressive speech remains underexplored. Most speech models are optimized for word recognition accuracy, often disregarding prosodic and paralinguistic features that carry emotional cues. Additionally, emotion-rich speech is often more spontaneous, containing disfluencies, hesitations, and expressive variations that challenge traditional speech systems. Therefore, it is essential to investigate:

\begin{itemize}
    \item Whether and to what extent speech representations contain emotion-related information.
    \item How to utilize representations from cutting-edge models.
\end{itemize}

A deeper understanding of how state-of-the-art speech models process and interact with emotional speech is essential to bridge the gap between speech recognition (i.e., automatic transcription) and emotion recognition, ultimately improving the effectiveness of SER frameworks.

\textbf{2. Difficult integration of linguistic information from error-prone automatic transcripts}

Advances in ASR offer a promising avenue for enhancing SER by leveraging textual and phonetic information extracted from speech. Integrating ASR with SER enables the fusion of acoustic and linguistic features, enriching SER models with linguistic context. This multimodal approach enhances robustness in noisy environments, multilingual scenarios, and spontaneous speech \citep{li2018towards,lian2024mer,scotti2021combining}. Additionally, ASR-generated text facilitates linguistically-informed feature extraction, complementing prosodic and spectral cues.

However, despite these benefits, existing ASR-enhanced SER methods face several challenges, including:

\begin{itemize}
    \item Error propagation: ASR results are not perfect, and transcription errors can negatively impact emotion recognition \citep{amiriparian2021impact,munot2019emotion}.
    \item Alignment issues: Synchronizing acoustic and textual representations is nontrivial, as speech and text operate on different time scales \citep{xu2019learning,wu2021emotion}.
    \item Computational complexity: Processing both speech and textual features increases computational costs, making real-time SER deployment challenging \citep{peng2021efficient}.
\end{itemize}

Addressing these issues requires novel fusion strategies that effectively integrate linguistic and acoustic characteristics while minimizing the impact of ASR errors on emotion prediction.

\textbf{3. High cost of data labeling}

Emotion annotation in speech is highly subjective and labor-intensive \citep{chou2024embracing}. Unlike speech-to-text transcription, which has objective ground truth, emotion perception varies significantly across individuals, leading to inconsistencies in annotation. As a result, multiple annotators are typically required per speech sample to reach a reliable consensus, making the process time-consuming and expensive.


The high cost of data labeling impedes large-scale data collection, slowing down the development and generalization of SER models. To this end, leveraging a small amount of labeled data to guide learning on a larger pool of unlabeled data is required to:

\begin{itemize}
    \item Reduce reliance on human annotations while maintaining model performance.
    \item Enhance model generalization by incorporating diverse, real-world speech data.
    \item Accelerate training and deployment of SER systems with limited labeled datasets.
\end{itemize}

\section{Research Questions and Objectives}
\label{chap1:questions}
The above-mentioned technical challenges, including the understanding of ASR models\footnote{When this research project began, concepts such as large pre-trained models and speech foundation models had not yet emerged. At that time, the mainstream understanding of the earliest large models, such as wav2vec 2.0, was still as ASR models. Therefore, in this paper, we retain the definition we originally used when starting this work and continue to refer to them as ASR models.} on emotional speech, the fusion of acoustic features and ASR linguistic features, and the high cost of emotion data labeling, need be addressed to bridge the gap between research and practical applications. This research aims to tackle these challenges with the aim of answering the following questions:

\textbf{1. What is the interplay between ASR models and SER?}

While ASR and SER both rely on speech signals as input, their fundamental objectives and processing mechanisms differ significantly. The acoustic model in ASR typically operates on short temporal windows, focusing on word recognition for accurate transcription. In contrast, SER systems analyze longer temporal contexts to capture emotional patterns embedded in speech dynamics.

Previous studies have highlighted a negative impact of emotional expressiveness on ASR performance\footnote{In this thesis, ASR performance refers primarily to the word error rate, unless otherwise specified.} \citep{fernandez2004computational}, as speech affected by strong emotions often exhibits articulatory distortions, prosodic variations, and non-standard intonations. On the other hand, integrating lexical information into SER models has been shown to enhance emotion classification accuracy \citep{lee2005toward}. However, given that ASR performance deteriorates in emotional speech, the full potential of ASR-enhanced SER, particularly in scenarios involving unlabeled or spontaneous emotional speech, remains underexplored.

Objectives include:

\begin{itemize}
    \item Analyzing how ASR systems process emotionally expressive speech and quantify their limitations.
    \item Examining the impact of ASR errors on downstream SER models, particularly in real-world, spontaneous speech settings.
\end{itemize}

\textbf{2. How to improve ASR transcription quality of small-scale emotional speech for SER?}

ASR error correction has been extensively explored as a post-processing strategy to enhance transcription quality. However, its application in the context of SER remains relatively underexplored. One key reason is the dependency of most ASR error correction methods on large-scale training data, which is often unavailable in emotional speech scenarios \citep{mani2020asr, lin2023multi, zhang2021end, tanaka2018neural}. Furthermore, emotional speech introduces unique and systematic ASR errors due to variations in prosody, pitch, and articulation, which existing models, typically trained on neutral or read speech, fail to handle effectively.

This challenge is intensified by the scarcity of labeled emotional speech corpora, hindering the adaptation of general-purpose ASR error correction models to emotion-rich, small-scale datasets.

Objectives include:
\begin{itemize}
    \item Adapting existing ASR error correction methods to low-resource emotional speech domains, with a focus on learning to correct emotion-specific transcription errors.
    \item Improving SER performance\footnote{In this thesis, SER performance refers primarily to the accuracy, unless otherwise specified.} by utilizing corrected ASR transcriptions in both text-only and speech-text fusion approaches.
\end{itemize}

\textbf{3. Can we develop robust speech-text fusion frameworks for ASR-enhanced SER to mitigate error propagation?}

Multimodal approaches that integrate acoustic and lexical features have consistently outperformed unimodal emotion recognition systems \citep{yoon2018multimodal,kumar2021towards,ezzameli2023emotion}. Two primary fusion strategies: feature-level fusion (early fusion) and decision-level fusion (late fusion), have been widely explored. Recent advancements, such as attention-based intermediate-level fusion techniques, have demonstrated further performance gains. For instance, state-of-the-art fusion techniques employ enhanced Transformer models or hybrid architectures \cite{wang2021learning,hasan2025emoformer}.

Most existing fusion methods, however, assume flawless textual transcriptions, an assumption that fails in real-world applications (where ASR errors are common due to noisy environments, varied accents, and imperfect transcriptions), unlike in controlled lab settings where speech quality is clean or transcription can be manually processed. Such errors cause distortions in meaning, which can mislead emotion classifiers based on linguistic features and degrade SER performance.

Objectives include:
\begin{itemize}
    \item Designing robust ASR-enhanced SER frameworks that effectively integrate speech and text cues while mitigating the impact of ASR errors.
    \item Investigating novel fusion techniques, such as error-aware fusion and adaptive attention mechanisms to enhance SER robustness in real-world scenarios.
\end{itemize}

\textbf{4. Can effort-saving strategies be developed for scalable SER to reduce annotation costs?}

Semi-Supervised Learning (SemiSL), such as self-training \citep{zhang2014cooperative} and co-training \citep{liu2007speech}, is a primary approach for SER on limited data. For self-training, the classifier uses its own prediction to teach itself. Whereas co-training tries to exploit the mutual information between two models (views or feature domains), each of which uses its predictions to teach not only itself but also the other.

While existing SemiSL approaches predominantly focus on selecting high-confidence samples through the trained SER model using the labeled data, relatively little work has explored the robustness of the unlabled data for augmenting the training set. Furthermore, although this issue has been extensively studied in ASR \citep{wallington2021learning,zhu2023alternative}, it remains underexplored in SER. In a word, previous SemiSL models in SER use predicted labels from unlabeled data without considering their inherent reliability, which could result in error propagation and bias accumulation with model training.

Objectives include:
\begin{itemize}
    \item Developing novel SemiSL strategies for SER, incorporating both confidence-based and reliability-aware data selection\footnote{In this context, 'confidence' refers to the model's ability to select data based on a predefined threshold, where data points with a prediction confidence higher than the threshold are selected for training. On the other hand, 'reliability' refers to selecting data based on the consensus or agreement among multiple models, where data points that are consistently selected across models are considered more reliable and chosen for further training.}.
    \item Exploring co-training methodologies that exploit multi-view learning across various feature encoders.
    \item Reducing dependence on large-scale annotated emotion datasets and leverage ASR transcription, making SER models more cost-efficient and effort-saving.
\end{itemize}

\section{Thesis Outline and Contributions}
\label{chap1:outline}
This thesis work, initiated in early 2021, represents one of the first efforts in the exploration, integration, and application of ASR within SER.
Given that publicly available emotional speech datasets remain significantly smaller than those for ASR, and considering the aforementioned challenges, practical deployment of SER continues to lag behind ASR despite years of research. Nevertheless, the advent of powerful pre-trained models has significantly improved the potential for utilizing ASR in SER, as the learned representations are now far richer than those produced by traditional ASR systems. At the core of this thesis is the seamless integration of ASR into SER, aiming to bridge the gap between theoretical advancements and real-world applications. This research delves into multiple dimensions, including pre-trained models, Large Language Models (LLMs), ASR error correction, multimodal fusion, and semi-supervised learning. Beyond technical aspects, the study takes a comprehensive approach by examining SER through interdisciplinary perspectives, incorporating insights from acoustics, linguistics, psychology, and cognitive science.

Furthermore, the findings and methodologies presented in this work have been extended to the domain of Alzheimer's dementia detection, demonstrating the broader applicability of our insights in critical speech-related healthcare tasks. This highlights the generalizability of the proposed approaches and underscores their potential impact on real-world speech processing applications.

This chapter serves as an introduction to the thesis. The remainder of this thesis is structured as follows.

\textbf{Chapter~\ref{chap2}} provides an overview of SER and background of integration of ASR in SER, covering fundamental aspects such as feature extraction, machine learning models, commonly used datasets, and persistent challenges in the field.

\textbf{Chapter~\ref{chap3}} first presents a layer-wise analysis of a series of wav2vec 2.0 models, investigating the characteristics of each ASR layer on emotional speech to better understand the strengths and limitations of advanced ASR representations in SER. Subsequently, this chapter introduces a joint ASR-SER training framework with hierarchical fusion, leveraging ASR outputs to improve the robustness of SER systems against ASR errors. Through an exploration of different ASR outputs and fusion strategies, our experiments demonstrate that incorporating both ASR hidden states and text outputs via a hierarchical co-attention fusion method yields the best SER performance, approaching the baseline results obtained using ground-truth transcripts. The research work in this chapter has been presented in the following publications:

\begin{itemize}
    \item \textbf{Li, Y.}, Bell, P., \& Lai, C. (2022). Fusing ASR outputs in joint training for speech emotion recognition. In \textit{ICASSP 2022-2022 IEEE International Conference on Acoustics, Speech and Signal Processing (ICASSP 2022).} \citep{li2022fusing}
    \item \textbf{Li, Y.}, Mohamied, Y., Bell, P., \& Lai, C. (2023). Exploration of a self-supervised speech model: A study on emotional corpora. In \textit{2022 IEEE Spoken Language Technology Workshop (SLT 2022).} \citep{li2023exploration}
\end{itemize}

In \textbf{Chapter~\ref{chap4}}, a detailed word-level analysis is performed on ASR transcriptions of emotional speech. The frequencies of various PoS tags, affective scores, and utterance lengths significantly influence ASR performance. Certain word classes are easier to recognize for one ASR model but may be more difficult for another. Furthermore, SER accuracy can deteriorate substantially with high WER. To further investigate this, two benchmarking experiments are conducted to evaluate SER performance (in both classification and regression tasks) across different WER levels in text-only and bimodal fusion (text+speech) settings. The research work in this chapter has been presented in the following publications:

\begin{itemize}
    \item \textbf{Li, Y.}, Zhao, Z., Klejch, O., Bell, P., \& Lai, C. (2023). ASR and Emotional Speech: A Word-Level Investigation of the Mutual Impact of Speech and Emotion Recognition. In \textit{Proceedings of Interspeech 2023.} \citep{li2023asr}
    \item \textbf{Li, Y.}, Bell, P., \& Lai, C. (2024). Speech emotion recognition with ASR transcripts: A comprehensive study on word error rate and fusion techniques. In \textit{2024 IEEE Spoken Language Technology Workshop (SLT 2024).} \citep{li2024speech}
\end{itemize}

Building upon this, \textbf{Chapter~\ref{chap5}} introduces two innovative methods for joint ASR error correction and emotion recognition: a Sequence-to-Sequence (S2S) model-based approach and an LLM-based method. Their respective advantages are demonstrated, with the S2S model-based approach performing better on 1-best ASR hypotheses, while the LLM-based method proves more effective on N-best ASR hypotheses. The research work in this chapter has been presented in the following publications:

\begin{itemize}
    \item \textbf{Li, Y.}, Chen, P., Bell, P., \& Lai, C. (2024). Crossmodal ASR error correction with discrete speech units. In \textit{2024 IEEE Spoken Language Technology Workshop (SLT 2024).} \citep{li2024crossmodal}
    \item \textbf{Li, Y.}, Gong, Y., Yang, C. H. H., Bell, P., \& Lai, C. (2024). Revise, Reason, and Recognize: LLM-Based Emotion Recognition via Emotion-Specific Prompts and ASR Error Correction. In \textit{ICASSP 2025-2025 IEEE International Conference on Acoustics, Speech and Signal Processing (ICASSP 2025).} \citep{li2024revise}
\end{itemize}

\textbf{Chapter~\ref{chap6}} investigates cross-modal incongruity in emotion recognition, revealing how different modalities (including speech, text, and vision) can convey conflicting affective cues. Furthermore, ASR errors have been found to introduce inconsistencies between the acoustic and linguistic modalities. To address this challenge, an incongruity-aware modality-gated fusion mechanism is proposed, minimizing the negative impact of ASR errors and enhance multimodal SER. The research work in this chapter has been presented in the following publications:

\begin{itemize}
    \item Wang*, Y., \textbf{Li*, Y.}, Liang, P. P., Morency, L. P., Bell, P., \& Lai, C. (2023). Cross-attention is not enough: Incongruity-aware dynamic hierarchical fusion for multimodal affect recognition. \textit{arXiv preprint arXiv:2305.13583.}\footnote{This paper is co-first-authored. Yuanchao proposed the research topic, designed and guided the experiments, rewrote the manuscript, and coordinated discussions with collaborators,} \citep{wang2023cross}
    \item \textbf{Li, Y.}, Bell, P., \& Lai, C. (2024). Speech emotion recognition with ASR transcripts: A comprehensive study on word error rate and fusion techniques. In \textit{2024 IEEE Spoken Language Technology Workshop (SLT 2024).} \citep{li2024speech}
\end{itemize}

In \textbf{Chapter~\ref{chap7}}, a semi-supervised learning framework is introduced, leveraging pre-trained speech models and LLMs for automatic labeling. This approach significantly reduces reliance on manual annotation, making SER more scalable and efficient. The framework is also validated on Alzheimer's dementia detection, further confirming its effectiveness. The research work in this chapter has been presented in the following publications:

\begin{itemize}
    \item \textbf{Li, Y.}, Zhang, Z., Han, J., Bell, P., \& Lai, C. (2024). Semi-Supervised Cognitive State Classification from Speech with Multi-View Pseudo-Labeling. In \textit{ICASSP 2025-2025 IEEE International Conference on Acoustics, Speech and Signal Processing (ICASSP 2025).} \citep{li2025semi}
\end{itemize}

Finally, Chapter~\ref{chap8} concludes the thesis by summarizing key findings and discussing future directions for building more reliable and practical SER systems.

\chapter{Overview of SER and Background of ASR Integration}
\label{chap2}

\section{Introduction}
\label{chap2:intro}
Speech Emotion Recognition (SER) refers to the task of automatically identifying a speaker’s emotional state based on their speech signal. As an interdisciplinary research area, SER lies at the intersection of signal processing \citep{kwon2003emotion,wang2015speech,el2011survey}, machine learning \citep{fayek2017evaluating,han2014speech,schuller2013computational}, linguistics \citep{triantafyllopoulos2022probing,polzehl2011anger}, and psychology \citep{sobin1999emotion,coutinho2013psychoacoustic}. It plays a vital role in affective computing and has a growing number of applications, including human-computer interaction \citep{cowie2001emotion,devillers2015inference}, virtual agents and assistants \citep{li2018towards,hu2022acoustically}, health monitoring \citep{jiang2017investigation,cummins2018speech}, and customer service \citep{li2019acoustic,han2020ordinal}.

In practice, SER systems are typically designed to perform either categorical classification (assigning speech to predefined emotion labels) or dimensional regression (predicting continuous emotional attributes). The choice of approach depends on data availability and application requirements, with some studies investigating hybrid models that combine both perspectives to improve performance \citep{tompkins2023multi}. This chapter presents an overview of SER, covering emotion models, task definitions, features, model architectures, datasets, evaluation metrics, related work, and major research challenges. In addition, this chapter presents the background of the integration of ASR into SER, which leads to this thesis work.

\section{Emotion Theories}
\label{chap2:theory}
A well-defined emotion representation is essential for computational modeling. Two dominant paradigms: \textbf{categorical emotion theory} (also known as discrete emotion theory) \citep{ekman1999basic} and \textbf{dimensional emotion theory} (also referred to as continuous emotion theory) \citep{russell1980circumplex}, form the foundation for emotion annotation and recognition strategies.

\subsection{Categorical Emotion Theory}
Categorical models, such as Ekman's basic emotions framework, classify emotions into discrete categories like happiness, sadness, anger, fear, disgust, and surprise. These are commonly used in classification-based SER systems. We employ categorical annotations for datasets like IEMOCAP and RAVDESS in this thesis (datasets will be introduced in Section~\ref{chap2:datasets}).

\subsection{Dimensional Emotion Theory}
Dimensional models represent emotions along continuous scales. There are four dimensions, which capture affective states via:

\begin{itemize}
    \item \textbf{Valence}: positivity or negativity
    \item \textbf{Arousal}: intensity or activation
    \item \textbf{Dominance}: control or submission
    \item \textbf{Social Rejection}: feeling included or isolated
\end{itemize}

Among the four, the most common is the Valence-Arousal (VA) model, which we use for datasets such as MSP-Podcast and CMU-MOSEI in this thesis (datasets will be introduced in Section~\ref{chap2:datasets}).

Both categorical and dimensional models have been widely adopted in SER research, with the choice often depending on the nature of the available data and the specific requirements of the application \citep{el2011survey}. The categorical approach provides a straightforward method for labeling emotions into discrete classes such as happiness, anger, sadness, or fear. This model is particularly advantageous in applications where distinct emotional states are needed to drive system responses.

In contrast, the dimensional model characterizes emotions along continuous affective dimensions, typically valence (positive–negative), arousal (high–low), and sometimes dominance and social rejection, enabling more nuanced modeling of emotional states \citep{russell1980circumplex,fontaine2007world}. One of the key motivations behind the dimensional approach is its capacity to capture not only broad emotion categories but also subtle intra-category variations \citep{acosta2011achieving}. For instance, it allows distinctions between ``hot anger'' (high arousal, low valence, high dominance) and ``cold anger'' (low arousal, low valence, high dominance), which would otherwise be grouped under a single label ``anger'' in a categorical framework \citep{cowie2003describing}.

Importantly, these two representations are not mutually exclusive. Emotions defined categorically can often be located within a dimensional affective space; for example, ``anger'' is typically associated with high arousal and negative valence, whereas ``calm'' corresponds to low arousal and positive valence \citep{russell1980circumplex}. Conversely, dimensional data can often be clustered into meaningful emotion categories, suggesting a degree of conceptual overlap between the two frameworks.

This complementarity has inspired the development of hybrid models that aim to integrate categorical and dimensional perspectives \citep{park2021dimensional,wu2011automatic}. Such models seek to combine the interpretability and clarity of discrete emotion labels with the expressive flexibility of dimensional features, thereby enhancing the robustness and generalizability of SER systems, particularly in real-world and cross-cultural contexts where emotional expression is often context-dependent \citep{morgan2019categorical}.

\subsection{Alternative Theories}
In addition to the widely used categorical and dimensional theories, alternatives have been proposed to bridge the gap between these two conventional approaches. For example, the Hourglass model \citep{cambria2012hourglass} is organized around four independent yet interrelated affective dimensions: Pleasantness (ranging from extreme sadness to extreme joy), Attention (from boredom to heightened interest), Sensitivity (from calmness to rage), and Aptitude (from disgust to trust). Each dimension comprises six levels of activation, capturing the intensity of affective states (e.g., contentment $\rightarrow$ joy $\rightarrow$ ecstasy within the Pleasantness dimension). This structure preserves the interpretability and categorical clarity of discrete models, while incorporating the continuous gradation of emotional intensity characteristic of dimensional models. Nevertheless, the practical application of such alternative emotion models remains relatively limited, and further empirical validation is required to assess their robustness and utility in real-world scenarios.

\section{SER as A Machine Learning Task}
\label{chap2:machinelearning}
SER is framed as a supervised machine learning task, detecting emotions conveyed in speech signals. It basically consists of two necessary steps: feature extraction and emotion classification/regression \citep{li2019improved}.

\subsection{Feature Extraction}
\label{chap2:features}
Features for SER can be divided into handcrafted and model-learned categories. Handcrafted features are manually designed using domain-specific prior knowledge, whereas model-learned features are automatically extracted from speech signals using deep learning models.

\subsubsection{Handcrafted Features}
Handcrafted features, such as spectral features, prosodic features, format features, and voice quality features, have been intensively explored and used for SER.

\begin{itemize}
    \item \textbf{Spectral features} represent characteristics of the vocal tract. Commonly used spectral features include Mel Frequency Cepstral Coefficients (MFCC) \citep{wang2015speech,daneshfar2020speech}, mel log-mel spectrograms \citep{li2019improved,satt2017efficient}, mel frequency filterbank energies \citep{busso2007using,ancilin2021improved}, and Linear Prediction Cepstral Coefficients (LPCC) \citep{nwe2003speech,pohjalainen2016spectral}.

    \item \textbf{Prosodic features} capture variations in speech intonation, rhythm, and stress patterns, which are crucial for conveying emotional information. Key prosodic features include pitch (fundamental frequency, F0), energy, and speech duration. Since prosody reflects emotional state and speaker variability, it plays an essential role in SER \citep{luengo2005automatic,rao2013emotion,koolagudi2012emotion}.

    \item \textbf{Voice quality features} focus on the vibratory behavior of the vocal cords and the resulting acoustic characteristics. These features help analyze phonation attributes such as breathiness, roughness, and tremor. Commonly used voice quality features include jitter (frequency variations in the voice signal), shimmer (amplitude fluctuations), harmonic-to-noise ratio, and cepstral peak prominence (the periodicity strength of the speech signal), and glottal closure measures (how completely the vocal folds close) \citep{jacob2016speech,lugger2007relevance,li2007stress}.
\end{itemize}

Given the extensive variety of handcrafted features, researchers have developed standardized acoustic feature sets for SER. Some of the most widely recognized handcrafted feature sets include INTERSPEECH 2009 Emotion Challenge (IS09) \citep{schuller2009interspeech}, INTERSPEECH 2013 Computational Paralinguistics Challenge (ComParE 2013) \citep{schuller2013interspeech}, and Extended Geneva Minimalistic Acoustic Parameter Set (eGeMAPS) \citep{eyben2015geneva}.

These feature sets form a broad range of acoustic descriptors, such as spectral, prosodic, and voice-quality related features, along with their statistical functionals. They serve as benchmark feature representations for conventional machine learning approaches to SER. In this thesis, we mainly use the following handcrafted paralinguistic features:

\begin{itemize}
    \item \textbf{Energy} is the perceptual correlate of sound intensity and depends on both amplitude and frequency sensitivity of the ear \citep{moore2012hearing}.

    \item \textbf{Harmonics-to-noise ratio (HNR)} evaluates the proportion of harmonic components to aperiodic noise in voice signals. It is a key indicator of hoarseness and vocal quality \citep{yumoto1982hnr}.

    \item \textbf{Pitch}, associated with the fundamental frequency (F0), reflects the perceived ``highness'' or ``lowness'' of a sound \citep{boersma1993pitch}.

    \item \textbf{Formants (F1, F2, F3)} are vocal tract resonances that characterize different speech sounds. F1 corresponds to vowel height, F2 to tongue advancement, and F3 to more fine-grained articulatory features \citep{peterson1952vowels}.

    \item \textbf{Alpha ratio} represents the energy balance between low (50–1 000 Hz) and high (1–5 000 Hz) frequency bands, often used to assess vocal brightness \citep{barsties2013alpha}.

    \item \textbf{Hammarberg Index} compares the strongest energy peak in the 0–2 000 Hz band to the highest above 2 000 Hz, used to assess vocal effort \citep{hammarberg1980abnormal}.

    \item \textbf{Formant 1, 2, and 3 relative energy} measures the intensity at F1, F2, and F3, providing insight into articulation and voice projection \citep{stevens1998acoustic}.

    \item \textbf{Spectral slope (0–500 Hz, 500–1 500 Hz)} describes how quickly energy decreases with frequency. Steep slopes typically indicate breathiness, while flatter slopes indicate tension \citep{titze1994voice}.

    \item \textbf{Harmonic differences like H1–H2 and H1–A3} are used to infer vocal fold behavior, such as breathy or tense phonation \citep{hanson1997glottal}.

    \item \textbf{Jitter} reflects frequency perturbation from cycle to cycle, revealing irregularity in vocal fold vibration \citep{titze1994voice,ferrand2002aging}.

    \item \textbf{Shimmer} measures amplitude variation between cycles, and is often used in pathology detection \citep{baken2000clinical}.
\end{itemize}

In recent years, however, handcrafted features have been gradually replaced by deep learning model-learned approaches which allow automatic representation learning and high-level data abstraction. Nonetheless, it remains uncertain whether the automatic representations learned by the model can entirely replace the handcrafted features, which is an issue that will be addressed in this thesis.

\subsubsection{Model-Learned Features}
Unlike handcrafted features, direct learning of speech representations through deep learning approaches reduces the requirements for explicit feature engineering. These features provide high-level representations of the speech signal and are typically learned automatically in an end-to-end mannaer. Deep learning models, such as Convolutional Neural Networks (CNNs), Recurrent Neural Networks (RNNs), and Transformers are widely used to directly process waveforms to extract meaningful features for SER. More recently, pre-trained speech foundation models, such as wav2vec 2.0, HuBERT, and Whisper, have emerged as powerful tools for learning universal representation of speech signals, benefiting SER. We will introduce the most widely used speech foundation models in Section~\ref{chap2:foundationmodels}.

\subsubsection{Lexical Features}
As lexical features are also employed in this thesis, we provide a brief overview here. In SER, lexical features are extracted from manually transcribed speech or ASR outputs and are designed to capture affective cues embedded in language use. Common lexical features include \textit{n-grams} (e.g., unigrams and bigrams), \textit{Part-of-Speech (POS) tags} \citep{bandhakavi2016lexicon}, \textit{bag-of-words} \citep{rozgic2012emotion}, and \textit{Term Frequency-Inverse Document Frequency (TF-IDF) weights} \citep{chatterjee2020seminar}. These features capture syntactic and statistical patterns that may correlate with emotional content. Additionally, \textit{affective scores} \citep{warriner2013norms}, which assign values of arousal, valence, and dominance to each word based on an affective score dictionary, have also been utilized to determine the emotion conveyed in transcribed speech.

With the advent of deep learning, more sophisticated lexical representations such as \textit{word embeddings} (e.g., Word2Vec \citep{church2017word2vec}, GloVe \citep{pennington2014glove}) \citep{rezaie2022speech,tzirakis2018end} and \textit{contextual word embeddings} (e.g., BERT, RoBERTa) \citep{padi2022bert,kim2021emoberta} have also been used to encode semantic and contextual information in transcriptions. These representations often complement conventional lexical features by capturing nuanced emotional signals that are context-dependent. Overall, the integration of lexical features, either as standalone inputs or in combination with acoustic and prosodic cues, has proven to significantly enhance the performance of SER systems.

\subsection{Machine Learning Models}
\label{chap2:models}
SER employs a variety of machine learning models, which are typically categorized into conventional machine learning approaches and Deep Neural Networks (DNNs).

\subsubsection{Conventional Machine Learning Models}
Conventional machine learning models for SER typically rely on handcrafted speech features, which serve as inputs to classical machine learning algorithms such as Gaussian mixture models \citep{lanjewar2015implementation,hu2007gmm}, logistic regression \citep{jacob2017modelling,kim2015speech}, support vector machines \citep{lin2005speech,shen2011automatic,hu2007gmm}, random forests \citep{zheng2018speech,noroozi2017vocal}, and K-nearest neighbors \citep{mote2024unsu,venkata2021speech,lanjewar2015implementation}.

Conventional machine learning models were widely used in the early stages of SER research due to their simplicity, interpretability, and relatively low computational requirements. However, they often struggle to capture complex patterns in speech data, requiring extensive feature engineering. Since these models are not used in this thesis, we omit their detailed descriptions.

\subsubsection{Deep Learning Models}
With advancements in deep learning, DNN-based approaches have become the dominant paradigm in SER due to their ability to automatically learn relevant feature representations from speech signals. The models used in this thesis are outlined below.

\begin{itemize}
    \item \textbf{Feed-Forward Neural Networks (FFNNs)} are a fundamental architecture in artificial neural networks, characterized by the unidirectional flow of information from the input layer through one or more hidden layers to the output layer, without any recurrent connections.

Given an input vector $\mathbf{x} \in \mathbb{R}^d$, the transformation at each layer $l$ of an $L$-layer FFNN is defined as follows:

\begin{equation}
\mathbf{z}^{(l)} = \mathbf{W}^{(l)} \mathbf{a}^{(l-1)} + \mathbf{b}^{(l)}, \quad l = 1, 2, ..., L
\end{equation}
\begin{equation}
\mathbf{a}^{(l)} = \phi^{(l)}(\mathbf{z}^{(l)})
\end{equation}

where $\mathbf{a}^{(0)} = \mathbf{x}$ is the input, $\mathbf{W}^{(l)}$ and $\mathbf{b}^{(l)}$ are the weight matrix and bias vector for the $l$-th layer, and $\phi^{(l)}$ is the activation function (e.g., ReLU, sigmoid, or tanh). The intermediate variable ${z}^{(l)}$ represents the pre-activation (linear) output at layer $l$, i.e., the result of the affine transformation before applying the activation function.

For multi-class classification tasks, the output layer uses the \textit{Softmax} activation function to generate a probability distribution over $K$ classes:

\begin{equation}
\hat{y}_i = \frac{e^{z_i^{(L)}}}{\sum_{j=1}^{K} e^{z_j^{(L)}}}, \quad i = 1, 2, ..., K
\end{equation}

The network is trained by minimizing the categorical cross-entropy loss:

\begin{equation}
\mathcal{L}(\hat{\mathbf{y}}, \mathbf{y}) = -\sum_{i=1}^{K} y_i \log(\hat{y}_i)
\end{equation}

where $\mathbf{y} \in \mathbb{R}^K$ is the one-hot encoded true label and $\hat{\mathbf{y}}$ is the predicted probability vector.

For regression tasks, the output layer typically does not include an activation function, allowing it to output continuous values directly:

\begin{equation}
\hat{\mathbf{y}} = \mathbf{z}^{(L)}
\end{equation}

The most commonly used loss function for regression is the Mean Squared Error (MSE):

\begin{equation}
\mathcal{L}(\hat{\mathbf{y}}, \mathbf{y}) = \frac{1}{n} \sum_{i=1}^{n} (\hat{y}_i - y_i)^2
\end{equation}

where $\hat{y}_i$ and $y_i$ denote the predicted and ground truth values, respectively.

The use of FFNNs for SER has been studied at least the last decade \citep{fayek2015towards,fayek2017evaluating}.

    \item \textbf{Convolutional Neural Networks (CNNs)} \citep{lecun1995convolutional} are a specialized class of DNNs designed to exploit the spatial and/or temporal structure of input data. Unlike standard fully connected layers where each neuron is connected to all neurons in the previous layer via matrix multiplication, CNNs replace this operation with convolution: a localized, parameter-shared linear transformation. This makes CNNs particularly efficient and effective for tasks involving grid-like data structures, such as images or spectrograms.

A typical CNN is composed of multiple \textit{convolutional blocks}, where each block generally performs a sequence of operations: convolution, non-linear activation, and pooling. In the convolutional layer, instead of a full weight matrix, a set of learnable kernels (also called filters) $\mathbf{K}$ is used. Each kernel has a small receptive field (e.g., $3 \times 3$ or $5 \times 5$), which significantly reduces the number of parameters and computational cost. The number of output channels (or feature maps) produced by a convolutional layer is denoted as $C^{(l)}$, where $l$ is the layer index.

The output of a convolutional layer can be formally expressed as:

\begin{equation}
h^{(l)}_j = \sigma \left( \sum_{i=1}^{C^{(l-1)}} h^{(l-1)}_i * K^{(l)}_{ij} + b^{(l)}_j \right), \quad \forall j \in \{1, 2, \dots, C^{(l)}\}
\end{equation}

where $*$ denotes the 2D convolution operation, $h^{(l-1)}_i$ is the $i$-th input channel of the previous layer, $K^{(l)}_{ij}$ is the kernel connecting input channel $i$ to output channel $j$, $b^{(l)}_j$ is the bias term, and $\sigma(\cdot)$ is a non-linear activation function such as ReLU.

Following the convolution and activation steps, a \textit{pooling layer} is often applied to down-sample the feature maps. Pooling is a form of non-linear aggregation that improves translation invariance and reduces spatial dimensions. The most common types of pooling are max pooling and average pooling \citep{zhou1988computation}, defined over local regions. Max pooling selects the maximum value, whereas average pooling computes the mean or weighted average of the values within a pooling window.

The hierarchical structure of CNNs allows the network to learn increasingly abstract features at deeper layers. The use of CNNs for SER have proven useful \citep{huang2014speech,peng2021efficient,li2019improved}.

    \item \textbf{Recurrent Neural Networks (RNNs)} \citep{hopfield1982neural,rumelhart1986learning} are designed to process sequential data by allowing information to be propagated through time. RNNs have connections that allow information to persist from one time step to another, making them ideal for tasks where the temporal order of data is crucial, such as time series forecasting, language modeling, and speech recognition.

At each time step $t$, an RNN updates its hidden state $\mathbf{h}_t$ based on the input $\mathbf{x}_t$ and the previous hidden state $\mathbf{h}_{t-1}$:

\begin{equation}
\mathbf{h}_t = \sigma \left( \mathbf{W}_h \mathbf{h}_{t-1} + \mathbf{W}_x \mathbf{x}_t + \mathbf{b}_h \right)
\end{equation}

where $\sigma(\cdot)$ is a non-linear activation function (often $\tanh$ or ReLU), $\mathbf{W}_h$ and $\mathbf{W}_x$ are learnable weight matrices, and $\mathbf{b}_h$ is a bias term. The output of the network at time $t$ is typically a function of the hidden state, $\mathbf{y}_t = \mathbf{W}_y \mathbf{h}_t + \mathbf{b}_y$, where $\mathbf{W}_y$ and $\mathbf{b}_y$ are the weight matrix and bias of the output layer.

However, standard RNNs suffer from the vanishing gradient problem, which makes them difficult to train over long sequences due to the diminishing impact of earlier time steps during backpropagation. This issue is mitigated by the Long Short-Term Memory (LSTM) network \citep{hochreiter1997long}, a specialized type of RNN designed to capture long-range dependencies in sequential data.

LSTMs address the vanishing gradient problem by introducing a more sophisticated architecture that uses gates to control the flow of information. Specifically, LSTM units consist of three gates: the forget gate ($f_t$), the input gate ($i_t$), and the output gate ($o_t$). These gates regulate the information retained or discarded at each time step.

The core update equations for an LSTM unit at time step $t$ are as follows:

\begin{equation}
f_t = \sigma \left( \mathbf{W}_f \mathbf{h}_{t-1} + \mathbf{W}_x \mathbf{x}_t + \mathbf{b}_f \right)
\end{equation}

\begin{equation}
i_t = \sigma \left( \mathbf{W}_i \mathbf{h}_{t-1} + \mathbf{W}_x \mathbf{x}_t + \mathbf{b}_i \right)
\end{equation}

\begin{equation}
o_t = \sigma \left( \mathbf{W}_o \mathbf{h}_{t-1} + \mathbf{W}_x \mathbf{x}_t + \mathbf{b}_o \right)
\end{equation}

\begin{equation}
\mathbf{C}_t = f_t \odot \mathbf{C}_{t-1} + i_t \odot \tanh \left( \mathbf{W}_C \mathbf{h}_{t-1} + \mathbf{W}_x \mathbf{x}_t + \mathbf{b}_C \right)
\end{equation}

\begin{equation}
\mathbf{h}_t = o_t \odot \tanh (\mathbf{C}_t)
\end{equation}

In these equations, $f_t$ is the forget gate, determining how much of the previous memory $\mathbf{C}_{t-1}$ should be forgotten. $i_t$ is the input gate, controlling how much new information should be added to the cell state. $o_t$ is the output gate, dictating the output of the LSTM unit. $\mathbf{C}_t$ represents the cell state, which carries the long-term memory of the network. $\odot$ denotes element-wise multiplication, and $\tanh(\cdot)$ is the hyperbolic tangent activation function.

The LSTM architecture allows the model to learn long-range dependencies effectively, as the cell state $\mathbf{C}_t$ is capable of preserving information over many time steps without suffering from vanishing gradients.

RNNs have proven to be effective and complementary to CNNs in SER \citep{lee2015high,mirsamadi2017automatic,li2019improved}.

    \item \textbf{Transformer} networks \citep{vaswani2017attention} represent a paradigm shift in the field of sequence modeling. Unlike RNNs and LSTM models, which process sequential data in an inherently temporal and recursive manner, Transformers rely entirely on self-attention mechanisms and dispense with recurrence altogether. This architecture enables efficient parallel computation and superior performance on a wide range of sequence tasks, including machine translation, language modeling, and speech recognition.

The core innovation of the Transformer is the self-attention mechanism, which allows each position in the input sequence to attend to all other positions directly. Given an input sequence $\mathbf{X} \in \mathbb{R}^{T \times d_{\text{model}}}$, the self-attention mechanism computes three matrices: queries $\mathbf{Q}$, keys $\mathbf{K}$, and values $\mathbf{V}$, via learned linear transformations:

\begin{equation}
\mathbf{Q} = \mathbf{X}\mathbf{W}_Q, \quad \mathbf{K} = \mathbf{X}\mathbf{W}_K, \quad \mathbf{V} = \mathbf{X}\mathbf{W}_V
\end{equation}

where $\mathbf{W}_Q, \mathbf{W}_K, \mathbf{W}_V \in \mathbb{R}^{d_{\text{model}} \times d_k}$ are learnable weight matrices. The output of scaled dot-product attention is computed as:

\begin{equation}
\text{Attention}(\mathbf{Q}, \mathbf{K}, \mathbf{V}) = \text{softmax} \left( \frac{\mathbf{Q} \mathbf{K}^\top}{\sqrt{d_k}} \right) \mathbf{V}
\end{equation}

This mechanism enables the model to dynamically weigh the relevance of each position in the sequence for the current computation.

To capture multiple types of relationships in parallel, the Transformer employs \textit{multi-head attention}, which consists of several self-attention operations run in parallel, each with its own set of parameters. The outputs of the heads are then concatenated and linearly transformed:

\begin{equation}
\text{MultiHead}(\mathbf{Q}, \mathbf{K}, \mathbf{V}) = \text{Concat}(\text{head}_1, \dots, \text{head}_h) \mathbf{W}^O
\end{equation}

Each attention head is defined as:
\begin{equation}
\text{head}_i = \text{Attention}(\mathbf{Q} \mathbf{W}_i^Q, \mathbf{K} \mathbf{W}_i^K, \mathbf{V} \mathbf{W}_i^V)
\end{equation}

Since the Transformer does not have any recurrence or convolution, it requires \textit{positional encodings} to incorporate the order of the sequence. These are summed to the input embeddings and can be either learned or computed using sinusoidal functions.

The original Transformer is composed of an encoder-decoder structure. The encoder consists of a stack of $N$ identical layers, each comprising two sub-layers: a multi-head self-attention mechanism and a position-wise FFNN. Each sub-layer is followed by residual connections \citep{he2016deep} and layer normalization \citep{ba2016layer}.

The decoder also consists of $N$ layers and includes a third sub-layer that performs multi-head attention over the encoder's output. The decoder is autoregressive, meaning it generates tokens one at a time and uses previously generated tokens as additional input.

Transformers have shown substantial superiority in SER compared to CNNs and RNNs \citep{wang2021novel,wang2021learning,gao2023adversarial}.

\end{itemize}

\subsection{Fusion Techniques}
\label{chap2:fusion}

With the integration of speech features, language features, and machine learning models, SER has greatly benefited from multimodal fusion techniques. These techniques can be broadly categorized into feature-level fusion, intermediate-level fusion, and decision-level fusion, although many approaches inherently involve elements of more than one fusion strategy.

\subsubsection{Feature-Level and Decision-Level Fusion}
Feature-level fusion, also known as early fusion, involves concatenating or combining features extracted from different modalities (e.g., prosodic, spectral, and linguistic features) into a single feature vector before inputting them into a classification model. This approach enables the model to learn joint representations and modality correlations at an early stage. In contrast, decision-level fusion, also referred to as late fusion, processes each modality independently to produce decisions (e.g., predicted emotion labels or confidence scores), which are subsequently combined using ensemble methods such as majority voting, weighted averaging, or rule-based strategies.

For example, \cite{sebastian2019fusion} conducted a comprehensive analysis comparing the emotion classification accuracy and F1 score of a CNN-based audio model, an LSTM-based text model, an early fusion model, and three late fusion models. Specifically, the three late fusion methods employed were sum combination, weighted sum combination, and product fusion. \cite{pepino2020fusion} explored three training strategies, cold-start, pre-trained, and warm-start, for early and late fusion models. The cold-start strategy trained the fusion model from scratch. In the pre-trained strategy, speech and text models were trained separately, and their weights were used to initialize the corresponding layers in the fusion model. The warm-start approach initialized all layers as in the pre-trained strategy, but trained the entire model, not just the post-fusion layers. Additionally, early and late fusion can be combined to complement each other \citep{chuang2004multi}.

However, feature-level fusion methods often struggle when input features from multiple modalities differ in temporal characteristics due to strict synchrony requirements. Furthermore, they are limited in modeling intra-modality dynamics effectively. On the other hand, decision-level fusion cannot capture inter-modality dynamics because the dependencies among modalities are not learned \citep{li2020attention}. To address these limitations, intermediate-level fusion has become the predominant approach in deep learning-based SER.

\subsubsection{Intermediate-Level Fusion}
Intermediate-level fusion, also known as representation-level or hidden-level fusion, integrates modalities at latent or hidden representation layers, typically within a deep neural network. Instead of directly combining features, each modality is first encoded independently, and their resulting embeddings are fused at one or more intermediate layers of the network.

This strategy allows for more flexible and hierarchical modeling of modality-specific patterns. For instance, \citep{zadeh2016multimodal} introduced the memory fusion network, which captures both unimodal and cross-modal interactions through a gated memory mechanism. With the rising popularity of attention mechanisms \citep{vaswani2017attention}, attention-based fusion has become a key trend at the intermediate level. Such methods can dynamically model inter-modality interactions and capture dependencies between modalities. \cite{xu2019learning} used bi-directional LSTMs to extract speech and text features, using text representations as the ``query'' to perform attention-based fusion on speech features. Similarly, \cite{yoon2018multimodal} used attention with speech as the ``query''. Building on this, the authors later proposed a multi-hop attention mechanism that automatically infers inter-modal correlations \citep{yoon2019speech}. In their model, relevant textual segments corresponding to the speech signal were identified via multi-hop attention, and these segments were then used to attend to specific parts of the speech signal.

Although intermediate-level fusion often outperforms feature-level and decision-level fusion, it typically requires careful architectural design and greater computational cost.

\subsection{Speech Foundation Models}
\label{chap2:foundationmodels}
With advancements in deep learning algorithms and computational resources, pre-trained speech models have emerged as powerful tools for learning universal representations of speech signals. These models, trained on large-scale speech corpora, composing the functions of both a feature extractor and a deep learning model for SER. Speech foundation models usually consist of multiple layers of Transformers, reducing the requirements of manually stacking neural networks such as CNNs. To this end, conducting SER using speech foundation models usually requires only simple FFNNs for final prediction output. Below are the speech foundation models that used in this thesis.

\subsubsection{Wav2vec 2.0}

\begin{figure}
    \centering
    \includegraphics[width=0.65\textwidth]{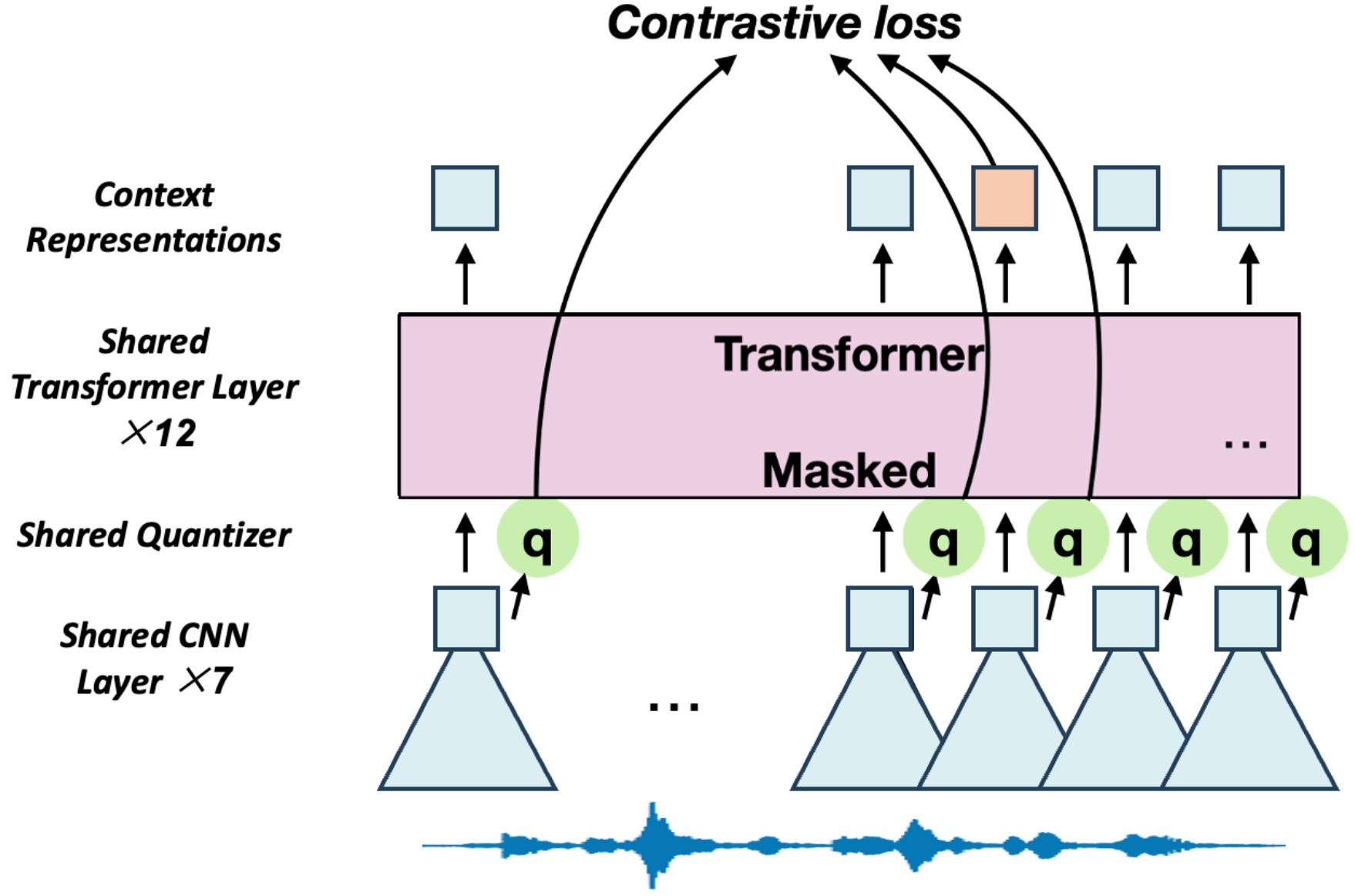}
    \caption{Wav2vec 2.0 model.}
    \label{chap2/fig:w2v-model}
\end{figure}

Wav2vec 2.0 \citep{baevski2020wav2vec} is a self-supervised learning model designed to learn contextualized representations from speech waveforms. As shown in Figure~\ref{chap2/fig:w2v-model}, the model architecture consists of three main components: a CNN-based local encoder that extracts a sequence of embeddings from audio as latent representation; a Transformer network for context representation; and a shared quantizer for discretization across multiple languages.

Given an audio input \( \mathbf{x} \in \mathbb{R}^T \), the CNN-based local encoder maps the waveform into a sequence of latent feature representations \( \mathbf{z} = (z_1, z_2, \ldots, z_{T'}) \). A context network, based on a Transformer, processes these latent features to produce contextualized embeddings \( \mathbf{c} = (c_1, c_2, \ldots, c_{T'}) \).

During pretraining, a subset of latent feature representations \( \mathbf{z} \) is masked, and the model is trained to correctly identify the true quantized targets \( q(\mathbf{z}) \) among a set of distractors. The objective is a contrastive loss, defined as:

\begin{equation}
\mathcal{L}_{\text{contrastive}} = - \sum_{t \in \mathcal{M}} \log \frac{\exp(\text{sim}(c_t, q(z_t)) / \kappa)}{\sum_{\tilde{q} \in \mathcal{Q}_t} \exp(\text{sim}(c_t, \tilde{q}) / \kappa)}
\end{equation}

where \( \mathcal{M} \) is the set of masked time steps, \( \text{sim}(\cdot, \cdot) \) denotes the cosine similarity, \( \mathcal{Q}_t \) includes the true target and negative samples at time \( t \), and \( \kappa \) is a temperature hyperparameter.

Wav2vec 2.0 significantly reduces the amount of labeled data required for competitive performance and has set new benchmarks for semi-supervised and low-resource speech tasks. Due to its ability to encode rich speech features, wav2vec 2.0 has been extensively utilized in SER \citep{pepino2021emotion,chen2023exploring,gao2023two}

\subsubsection{HuBERT}
HuBERT \citep{hsu2021hubert} introduces a self-supervised learning framework for speech representation learning based on masked prediction of discrete hidden units. Inspired by the masked language modeling objective in BERT, HuBERT masks portions of the input speech features and trains the model to predict their corresponding cluster assignments, which serve as pseudo-labels generated through offline clustering.

Same as wav2vec 2.0, the model extracts latent representations \( \mathbf{z} \) from waveform \( \mathbf{x} \) via a convolutional feature encoder. These features are then processed by a Transformer to produce contextualized embeddings \( \mathbf{c} \). Initially, discrete targets are obtained by applying k-means clustering over the extracted features. During training, the model is optimized with a cross-entropy loss over the masked positions:

\begin{equation}
\mathcal{L}_{\text{CE}} = - \sum_{t \in \mathcal{M}} \log p_{\theta}(q(z_t) \mid c_t)
\end{equation}

where \( \mathcal{M} \) denotes the set of masked time steps, \( p_{\theta} \) represents the predicted probability distribution over cluster IDs, and \( q(z_t) \) is the pseudo-label corresponding to frame \( t \).

The overall architecture is similar to wav2vec 2.0. However, unlike wav2vec 2.0, which directly learns to predict quantized latent representations through a contrastive loss, HuBERT relies on a multi-stage training procedure. After the first training phase, hidden representations from the Transformer are reclustered to produce new targets, enabling iterative refinement. This design allows HuBERT to progressively improve the quality of learned representations without requiring a contrastive objective. As a result, HuBERT often outperforms wav2vec 2.0 across various speech processing applications, including SER \citep{pastor2022cross,saliba2024layer}.

\subsubsection{WavLM}
WavLM \citep{chen2022wavlm} extends the line of self-supervised speech representation learning by integrating masked prediction with a denoising pre-training task, aiming to learn universal speech representations that are effective across both low-level and high-level speech processing tasks.

Building upon the masked prediction strategy of HuBERT, WavLM introduces additional noise robustness by augmenting the input signals with speech-like disturbances, such as overlapped speech and additive background noise, during pre-training. The model is then trained to predict the correct masked units despite the presence of noise, encouraging the learning of more generalizable and noise-invariant features.

The general architecture and training loss are the same as HuBERT. However, WavLM further incorporates a gated relative position bias mechanism within the Transformer layers, enhancing its capacity to model long-range dependencies and speaker overlaps, which is crucial for tasks such as speaker diarization, separation, and speech recognition in noisy environments.

Thanks to these innovations, WavLM achieves state-of-the-art performance across a wide range of speech tasks, and serves as a highly effective pre-trained model for SER \citep{diatlova2024adapting,upadhyay2024layer,yang2024single}.

\subsubsection{Whisper}
Whisper \citep{radford2023robust} is a fully supervised, end-to-end ASR system trained on a large-scale, weakly labeled dataset comprising approximately $6.8 \times 10^{5}\ $ hours of multilingual and multitask speech data collected from the web. Unlike previous self-supervised models such as wav2vec 2.0, HuBERT, and WavLM, Whisper does not rely on unsupervised pre-training, but instead utilizes massive supervised training to achieve robustness across diverse domains, languages, and acoustic conditions.

The model architecture follows a standard encoder-decoder Transformer framework. The encoder processes log-mel spectrograms \( \mathbf{x} \in \mathbb{R}^{F \times T} \) into a sequence of hidden states, where \( F \) denotes the number of frequency bins and \( T \) is the number of time frames. The decoder is autoregressive and predicts the corresponding text tokens sequentially, conditioned on the encoder outputs.

Formally, given an input spectrogram \( \mathbf{x} \), the model estimates the probability of the output token sequence \( \mathbf{y} = (y_1, y_2, \ldots, y_N) \) as:

\begin{equation}
P(\mathbf{y} \mid \mathbf{x}) = \prod_{n=1}^{N} P(y_n \mid y_{<n}, \mathbf{x})
\end{equation}

where \( y_{<n} \) denotes the previous tokens, and each conditional probability is computed via the decoder network.

Whisper is trained with a standard cross-entropy loss between the predicted and ground-truth text tokens. Additionally, the model is designed for multitask learning: it can perform language identification, phrase-level timestamps prediction, and multilingual speech recognition within the same framework. Thanks to its scale, diversity of training data, and multitask formulation, Whisper demonstrates strong zero-shot transfer capabilities, achieving competitive or superior performance on many publicly available ASR benchmarks without additional fine-tuning and also demonstrate significant SER performance \citep{goron2024improving,fukuda2025speech,chou2024tiny}.

A brief comparison of the models above is presented in Table~\ref{chap2/tab:model_comparison}.

\begin{sidewaystable*}
\centering
\caption{Comparison of Wav2vec 2.0, HuBERT, WavLM, and Whisper.}
\label{chap2/tab:model_comparison}
\begin{tabular}{lcccc}
\toprule
\textbf{Aspect} & \textbf{Wav2vec 2.0} & \textbf{HuBERT} & \textbf{WavLM} & \textbf{Whisper} \\
\midrule
Training & Self-sup. (contrastive) & Self-sup. (masked pred.) & Self-sup. (denoising) & Supervised \\
Architecture & Encoder (Transformer) & Encoder (Transformer) & Encoder (Relative bias) & Encoder-Decoder \\
Input & Waveform & Waveform & Waveform & Log-mel spec \\
Target & Quantized feats. & K-means labels & K-means + noise & Text tokens \\
Noise Robustness & No & No & Yes & Yes \\
Multilingual & No & No & No & Yes \\
Zero-shot & Limited & Limited & Moderate & Strong \\
\bottomrule
\end{tabular}
\end{sidewaystable*}

\section{Datasets}
\label{chap2:datasets}
High-quality datasets serve as the cornerstone for SER research. This section provides an overview of several widely used SER datasets, with a particular focus on those utilized in this thesis. These datasets play a crucial role in benchmarking SER models and assessing their performance across different experimental settings.

\subsubsection{IEMOCAP}
The \textit{Interactive Emotional Dyadic Motion Capture} (IEMOCAP) dataset~\citep{busso2008iemocap} is one of the most extensively used English-language corpora for SER. It contains approximately 12 hours of multimodal data, including speech recordings, textual transcriptions, and facial motion capture. The dataset consists of five dyadic conversational sessions, each featuring a male-female pair of professional actors, resulting in a total of ten speakers. Within each session, the participants engage in both pre-scripted dialogues and improvised interactions based on predefined emotional scenarios.

IEMOCAP comprises 151 dialogues, segmented into 10039 utterances, with an average duration of 4.5 s per utterance. Each utterance has been manually annotated by three raters for categorical emotions, including \textit{neutral}, \textit{happy}, \textit{sad}, \textit{angry}, \textit{excited}, \textit{disappointed}, \textit{fearful}, \textit{surprised}, and \textit{frustrated}. Additionally, a dimensional annotation scheme is employed, where utterances are rated on a five-point Likert scale for \textit{valence} (1 = negative, 5 = positive), \textit{activation} (1 = calm, 5 = excited), and \textit{dominance} (1 = weak, 5 = strong). If an utterance conveys mixed emotions, multiple labels may be assigned. Final categorical labels are determined by majority voting, while dimensional scores are obtained by averaging the raters' annotations.

\subsubsection{RAVDESS}
The \textit{Ryerson Audio-Visual Database of Emotional Speech and Song} (RAVDESS)~\citep{livingstone2018ryerson} is a multimodal dataset designed for SER and affective computing research. It contains 7,356 recordings of both spoken and sung expressions, performed by 24 professional actors (12 male, 12 female). The speech component features 16 distinct statements, each delivered in eight emotional states: \textit{neutral}, \textit{calm}, \textit{happy}, \textit{sad}, \textit{angry}, \textit{fearful}, \textit{disgusted}, and \textit{surprised}. All recordings are available in both audio-only and audio-visual formats, enabling cross-modal emotion analysis.

However, as the emotional expressions are acted in this dataset, they may not fully capture the complexity and spontaneity of real-world emotional speech. Nevertheless, RAVDESS remains a valuable benchmark for SER models, particularly when controlled and high-fidelity emotional speech data are required.

\subsubsection{MSP-Podcast}
The \textit{MSP-Podcast} corpus~\citep{lotfian2017building} is a large-scale, naturalistic English emotional speech dataset constructed from publicly available podcast recordings. Developed by the Multimodal Signal Processing (MSP) Lab at the University of Texas at Dallas, this dataset is designed to capture authentic emotional expressions in real-world speech. As of Release 1.11, it comprises 151 654 utterances from more than 1500 speakers, totaling over 237 hours of speech.

Annotations are obtained via crowdsourcing, where each utterance is labeled by at least five annotators (with an average of 6.7 annotations per utterance). The labeling process includes both categorical emotion annotations (with a primary emotion and optional secondary emotions) of \textit{anger}, \textit{sadness}, \textit{happiness}, \textit{surprise}, \textit{fear}, \textit{disgust}, \textit{contempt}, \textit{neutral}, and dimensional ratings for \textit{arousal}, \textit{valence}, and \textit{dominance} on a seven-point Likert scale. Majority voting is used to finalize categorical emotion labels, while dimensional ratings are determined by averaging annotators' scores.

To facilitate research and model evaluation, the dataset is partitioned into predefined training (84,030 utterances), development (19,815 utterances), test set 1 (30,647 utterances), and test set 2 (14,815 utterances) subsets. Given its diversity in speaking styles, topics, and recording conditions, MSP-Podcast serves as a crucial resource for developing SER models capable of handling natural, spontaneous speech.

\subsubsection{CMU-MOSI}
The \textit{CMU Multimodal Opinion Sentiment Intensity} (CMU-MOSI) dataset~\citep{zadeh2016multimodal} is a multimodal corpus designed for sentiment and emotion analysis in spoken language. It consists of monologues extracted from YouTube videos, where speech is accompanied by text transcriptions and video. The dataset comprises 2199 utterances from 93 speakers, covering a variety of topics and speaking styles.

Each utterance is annotated for sentiment polarity on a seven-point Likert scale (-3 = highly negative, +3 = highly positive). Due to its combination of audio, textual, and visual modalities, CMU-MOSI is widely used in multimodal emotion recognition research. However, since the speech samples originate from online videos, they often contain background noise and variations in recording quality, posing additional challenges for SER models.

\subsubsection{CMU-MOSEI}
An extension of CMU-MOSI, the \textit{CMU Multimodal Opinion Sentiment and Emotion Intensity Extended} (CMU-MOSEI) dataset~\citep{Zadeh2018} significantly increases the dataset’s scale and speaker diversity. It consists of 23,500 video segments from over 1,000 speakers, capturing a broad range of opinions and emotional expressions.

Similar to CMU-MOSI, CMU-MOSEI includes speech, transcriptions, and video, with each utterance annotated for sentiment on a continuous scale (-3 to +3) and six primary emotions (\textit{happy}, \textit{sad}, \textit{angry}, \textit{surprised}, \textit{fearful}, and \textit{disgusted}). Dimensional emotion ratings for \textit{valence}, \textit{arousal}, and \textit{dominance} are also provided. Compared to CMU-MOSI, CMU-MOSEI offers increased linguistic and emotional diversity, making it a more comprehensive resource for training SER models in real-world settings. However, like its predecessor, it includes variations in background noise and recording quality due to its reliance on online videos.



Current studies on these datasets typically use speech foundation models with FFNNs or Transformers-based architectures on waveforms directly, providing state-of-the-art performance \citep{xu2021head,pepino2021emotion,siriwardhana2020jointly,ando2021speech,leem2023selective}.

\section{Evaluation Metrics}
\label{chap2:metrics}

Evaluation metrics are used to assess the performance of machine learning models and are selected according to the nature of the task: classification or regression. Below, we describe the commonly used metrics in each category, along with their mathematical definitions.

\subsection{Classification Tasks}
\subsubsection{Accuracy}
Accuracy measures the proportion of correctly predicted samples among all predictions.
\begin{equation}
    \text{Accuracy} = \frac{TP + TN}{TP + TN + FP + FN}
\end{equation}
where $TP$, $TN$, $FP$, and $FN$ denote true positives, true negatives, false positives, and false negatives, respectively.

\begin{itemize}
    \item \textbf{Unweighted Accuracy (UA)}: Also known as the overall accuracy, this metric calculates the fraction of instances predicted correctly, without considering class distribution:
\begin{equation}
    \text{Unweighted Accuracy} = \frac{\text{Total correct predictions}}{\text{Total number of samples}}
\end{equation}

    \item \textbf{Weighted Accuracy (WA)}: Weighted accuracy is computed by averaging the accuracy within each class, weighted by the number of samples in that class:
\begin{equation}
    \text{Weighted Accuracy} = \sum_{c=1}^{C} \frac{n_c}{n} \cdot \frac{TP_c}{n_c} = \frac{1}{n} \sum_{c=1}^{C} TP_c
\end{equation}
where $C$ is the number of classes, $n_c$ is the number of samples in class $c$, $n$ is the total sample number, and $TP_c$ is the number of correctly predicted samples in class $c$.
\end{itemize}

The distinction between weighted and unweighted accuracy is especially important in the presence of class imbalance. While unweighted accuracy gives the same weight to every instance regardless of class, weighted accuracy ensures that each class contributes proportionally to its size in the dataset.

\subsubsection{Precision, Recall, and F1 Score} 
\begin{align}
    \text{Precision} &= \frac{TP}{TP + FP} \\
    \text{Recall} &= \frac{TP}{TP + FN} \\
    \text{F1 Score} &= 2 \cdot \frac{\text{Precision} \cdot \text{Recall}}{\text{Precision} + \text{Recall}}
\end{align}

These metrics can be averaged across classes using either \textit{macro} (unweighted average) or \textit{weighted} (weighted by support) averaging to handle class imbalance.


\subsection{Regression Tasks}
\subsubsection{Mean Squared Error (MSE)}
MSE measures the average of the squared differences between the predicted and true values.
\begin{equation}
    \text{MSE} = \frac{1}{n} \sum_{i=1}^{n} (y_i - \hat{y}_i)^2
\end{equation}
where $y_i$ and $\hat{y}_i$ denote the ground truth and predicted value for the $i$-th sample, respectively.

\subsubsection{Pearson Correlation Coefficient (PCC)} PCC measures the linear correlation between predicted and ground truth values.
\begin{equation}
    \text{PCC} = \frac{\sum_{i=1}^{n} (y_i - \bar{y})(\hat{y}_i - \bar{\hat{y}})}{\sqrt{\sum_{i=1}^{n} (y_i - \bar{y})^2} \sqrt{\sum_{i=1}^{n} (\hat{y}_i - \bar{\hat{y}})^2}}
\end{equation}
where $\bar{y}$ and $\bar{\hat{y}}$ are the means of the ground truth and predicted values, respectively.

\subsubsection{Concordance Correlation Coefficient (CCC)} CCC assesses the degree to which pairs of observations fall on the 45-degree line through the origin. It combines measures of both precision and accuracy.
\begin{equation}
    \text{CCC} = \frac{2 \rho \sigma_y \sigma_{\hat{y}}}{\sigma_y^2 + \sigma_{\hat{y}}^2 + (\mu_y - \mu_{\hat{y}})^2}
\end{equation}
where $\rho$ is the Pearson correlation coefficient, $\sigma_y$ and $\sigma_{\hat{y}}$ are standard deviations, and $\mu_y$, $\mu_{\hat{y}}$ are means of the ground truth and predicted values, respectively.

The choice of evaluation metrics depends on the nature of the learning task, the data distribution, and the annotation format of the labels. In this thesis, we primarily focus on \textbf{unweighted accuracy} for classification tasks, as it is the most commonly used metric in SER. For regression tasks, we mainly adopt \textbf{mean squared error} and \textbf{concordance correlation coefficient} when modeling continuous emotional dimensions. Several other metrics will occasionally be included when necessary.

\section{Prerequisites for Integrating ASR into SER}
\label{chap2:SER-ASR}
As the prerequisite of integrating ASR into SER, the relationship between SER and ASR needs to be understood. This section describes the ASR task definition and ASR models used in this thesis, as well as introduces the background of SER-ASR relationship and the use of ASR for SER, leading to the thesis work.

\subsection{ASR Task Definition}

ASR aims to convert spoken language into written text by estimating the most probable word sequence $\hat{W}$ given an acoustic input $X$. This is commonly formulated as a probabilistic inference problem:
\begin{equation}
    \hat{W} = \arg\max_W P(W | X)
\end{equation}
where $W$ denotes a sequence of words and $P(W | X)$ is the posterior probability of the word sequence given the observed acoustic features. 

However, directly modeling the posterior distribution is often challenging. A more tractable approach is to apply Bayes’ theorem to decompose the posterior as follows:
\begin{equation}
    P(W | X) = \frac{P(X | W) P(W)}{P(X)}
\end{equation}
Since $P(X)$ is independent of $W$, it can be omitted during maximization. Therefore, the problem reduces to:
\begin{equation}
    \hat{W} = \arg\max_W P(X | W) P(W)
    \label{formula:asr}
\end{equation}

In this Bayesian formulation:
\begin{itemize}
    \item $P(X | W)$ is the \textit{acoustic model}, which estimates the likelihood of the acoustic input given a word sequence.
    \item $P(W)$ is the \textit{language model}, which provides a prior over plausible word sequences.
\end{itemize}

This decomposition enables separate modeling of acoustic and linguistic knowledge and forms the theoretical basis for conventional ASR systems.

In SER, ASR systems can provide two types of representations:

\textbf{Acoustic representation:} The internal hidden representations of ASR models (e.g., encoder outputs) can be used as high-level acoustic features that are often emotion-sensitive.

\textbf{Lexical representation:} The transcriptions generated by ASR systems provide textual input for text-based SER models, enabling semantic-level emotion inference.

\subsection{Evaluation Metrics}
The performance of ASR systems is commonly evaluated using error rates that measure the mismatch between the predicted transcription and the reference transcription. The most widely used metrics include:

\textbf{Word Error Rate (WER)} measures the proportion of words that are incorrectly predicted. It is calculated based on the minimum number of insertions ($I$), deletions ($D$), and substitutions ($S$) needed to convert the predicted sequence into the reference:
\begin{equation}
    \text{WER} = \frac{S + D + I}{N}
\end{equation}
where $N$ is the total number of words in the reference.

\textbf{Character Error Rate (CER)} is similar to WER, but computed at the character level, which is particularly useful for languages with no explicit word boundaries or for low-resource scenarios:
\begin{equation}
    \text{CER} = \frac{S + D + I}{N}
\end{equation}
where $N$ here refers to the number of characters in the reference.

To evaluate the quality of ASR transcriptions for downstream tasks such as SER, metrics from the machine translation domain that emphasize $n$-gram matching are also commonly applied:

\textbf{BLEU (Bilingual Evaluation Understudy)} is a metric used to evaluate the quality of machine-generated translations by comparing $n$-grams (contiguous sequences of $n$ words) between the hypothesis (machine output) and reference translations \citep{papineni2002bleu}. It primarily measures \textit{precision} but incorporates a \textit{brevity penalty} to avoid overly short translations:

\begin{equation}
    \text{BLEU} = \text{BP} \cdot \exp \left( \sum_{n=1}^{N} w_n \cdot \log p_n \right)
\end{equation}

Where:
\begin{itemize}
    \item $N$ is the highest $n$-gram level considered (typically $N=4$).
    \item $w_n$ is the weight for each $n$-gram level, usually set as $w_n = \frac{1}{N}$.
    \item $p_n$ is the precision for the $n$-gram level, defined as the ratio of the number of matching $n$-grams between the hypothesis and reference to the total $n$-grams in the hypothesis.
    \item BP is the brevity penalty, given by:
    \[
    \text{BP} =
    \begin{cases}
    1, & \text{if } c > r \\
    e^{(1 - \frac{r}{c})}, & \text{if } c \leq r
    \end{cases}
    \]
    where $c$ is the length of the candidate translation and $r$ is the length of the reference translation.
\end{itemize}
 
The BLEU score ranges from 0 to 1, with a score of 1 indicating perfect matching of $n$-grams between the hypothesis and reference(s). The brevity penalty helps to prevent systems from generating very short translations just to match a few words.

\textbf{GLEU (Google BLEU)} is a variant of BLEU that is designed for sentence-level evaluation, making it particularly useful for tasks such as text correction, grammar correction, and short-form translation \citep{wu2016google}. Unlike BLEU, which primarily focuses on precision, GLEU takes both \textit{precision} and \textit{recall} into account:

\begin{equation}
    \text{GLEU} = \min(\text{Precision}, \text{Recall})
\end{equation}

Where:
\begin{itemize}
    \item \textit{Precision} is the ratio of the number of matching $n$-grams in the hypothesis to the total number of $n$-grams in the hypothesis:
    \[
    \text{Precision} = \frac{\text{Number of matching n-grams in hypothesis}}{\text{Total number of n-grams in hypothesis}}
    \]
    \item \textit{Recall} is the ratio of the number of matching $n$-grams in the hypothesis to the total number of $n$-grams in the reference:
    \[
    \text{Recall} = \frac{\text{Number of matching n-grams in hypothesis}}{\text{Total number of n-grams in reference}}
    \]
\end{itemize}

GLEU evaluates both precision and recall, ensuring that the hypothesis captures as many correct n-grams from the reference as possible while maintaining fluency. The final score is the minimum of the precision and recall values.

BLEU is more suited for evaluating larger corpora or multi-sentence outputs, as it primarily measures precision and applies a brevity penalty for shorter translations. GLEU, on the other hand, is optimized for sentence-level tasks and provides a more balanced evaluation by considering both precision and recall, making it ideal for tasks like grammatical error correction or speech recognition.

\subsection{Relationship between SER and ASR}
The relationship between ASR and SER is not yet clearly defined. Although both of them take speech as input, the acoustic model in ASR focuses on the phone and word level, while SER focuses on the utterance level and beyond. Previous studies showed that the presence of emotion in speech degrades the accuracy of ASR \citep{fernandez2004computational}. On the other hand, prior work has reported improvement in SER accuracy when lexical input is added to the acoustic input \citep{lee2005toward}.

\subsubsection{Transferring ASR for SER}
Computer systems do not have the complexity of the human brain and auditory system to perfectly handle multiple tasks simultaneously. However, most speech features can be shared for multiple speech tasks, such as MFCC and F0. Therefore, the same speech features can be used as input for different speech tasks where they may be then further tuned with distinct task characteristics by the model for specific tasks. For this reason, intermediate representations from a trained ASR model can be transferred for SER. These representations can then be used as-is or fine-tuned on given SER-labeled data.  

In 2016, \cite{fayek2016correlation} demonstrated the possibility of knowledge transfer from a convolutional neural acoustic model trained on the TIMIT corpus \citep{garofolo1993darpa} for the emotion recognition task. The authors provided many fine-tuning variants, gradually holding some layers constant and fine-tuning the remaining ones. When compared to an end-to-end CNN model, employing the ASR network as a feature extractor and training only the output softmax layer resulted in a considerable reduction in performance. Allowing more ASR layers to be updated during back-propagation resulted in gradual improvements, but overall, employing ASR for feature extraction had a negative impact on SER performance.

In 2018, \cite{lakomkin2018reusing} compared two options of tuning ASR representations: simply feeding them into a softmax classifier or adding a new Gated Recurrent Units (GRU) layer trained on top of the ASR features. The latter allows the model to train a SER network parallel to the ASR network to capture additional emotion-specific information. The ASR branch of the network remains frozen in both options. In the same year, \cite{tits2018asr} showed that features learned by a trained ASR network can be used for SER and outperform the eGeMAPS feature set \citep{eyben2015geneva}. They also examined the relationship of the emotional dimensions arousal and valence with the ASR layers of audio and text modalities. They showed that for some speakers, arousal is more correlated to hidden features extracted from the first three layers (closer to audio) and valence to the ones closer to the last three layers (closer to text). Then in 2020, \cite{zhou2020transfer} indicated potential feature overlap between ASR and SER, by showing that using representations of the 12th layer of a pre-trained Time-Delay Neural Network (TDNN)-based ASR model for SER can achieve a performance comparable to state-of-the-art SER models. They also noticed a performance degradation in the layers closer to the final output layer since they are more specialized in speech recognition. Another work in the same year \citep{yeh2020speech} first fine-tuned the Listen-Attend-Spell (LAS) model \citep{chan2016listen} on IEMOCAP and then adopted a Singular Value Decomposition (SVD)-based domain adaptation to remove redundant parameters of the pre-trained ASR model to adapt it to SER. Their method improved both the speech recognition rate and the emotion recognition accuracy on IEMOCAP. Furthermore, they also found representations from a deeper layer degrade the emotion recognition accuracy, suggesting that deeper layers of ASR may contain less emotional information.

There are also several work directly using ASR outputs as features for SER and compared them with ground-truth transcripts \citep{schuller2009emotion,sahu2019multi}, which also demonstrated the feasibility of incorporating ASR for SER. All these prior studies have provided solid experimental results and discussion, yet this direction still calls for more attention and detailed analysis, especially at the moment when speech foundation models are becoming mainstream, thus eliciting one of the topics of this thesis work: evaluating and utilizing speech foundation models for SER.

\subsubsection{Joint Training ASR and SER}
The other pathway to incorporate ASR in SER is multitask learning, also known as joint training. To note that, however, due to differences in corpora, data sizes, and modeling approaches, it is not currently feasible to train two tasks simultaneously from scratch. Therefore, joint training here means training the SER model and fine-tuning the pre-trained ASR model simultaneously. Past ASR systems hindered joint training with SER partly due to the need for additional modeling of the language model (as Formula~\ref{formula:asr}). Thanks to the development of end-to-end ASR modeling in recent years \citep{graves2014towards,amodei2016deep}, joint ASR-SER training has become possible and as far as we know, it was firstly attempted in \cite{feng2020end}.

In this work, an end-to-end SER model combined with an acoustic-to-word ASR model \citep{soltau2016neural} was proposed. The authors proposed an ASR feature, which is the hidden state of the decoder in the ASR model, to replace the textual output for the combination in the SER model. As a result, this is a speech-only ASR-SER model but has a performance that is close to that of the models using speech and text. By using multitask learning, the ASR model can be fine-tuned, and the SER model can get more accurate ASR features to improve the recognition performance. In 2021, \cite{cai2021speech} leveraged the pre-trained wav2vec 2.0 as the feature extractor backbone for joint ASR-SER training. They added a Connectionist Temporal Classification (CTC) head and a classification head on top of the backbone, to simultaneously obtain predictions for transcripts and emotion categories. Moreover, they also discussed how ASR affects SER by adjusting the ASR weight in multitask loss. The evaluation results indicated that the ASR task is very important to SER, and a better trained ASR model (lower error rate) yields better SER performance.

\section{Challenges in SER and ASR Integration}
\label{chap2:challenges}
Despite the aforementioned research findings and significant advancements in feature engineering, model development, and dataset collection, SER continues to face several critical challenges that impede its progress. Addressing these challenges is crucial for developing more robust and generalizable SER systems.

\subsection{Limited Availability of Emotional Speech Data}
One of the most pressing challenges in SER research is the scarcity of large-scale emotional speech datasets. Unlike ASR, which benefits from extensive corpora such as LibriSpeech \citep{panayotov2015librispeech} ($\approx 1,000$ hours) and Common Voice \citep{ardila2019common} ($\approx 33,535$ hours), the size of publicly available emotional speech datasets remains significantly smaller (e.g., IEMOCAP \citep{busso2008iemocap} $\approx 12$ hours, MSP-Podcast \citep{lotfian2017building} $\approx 237$ hours). The collection of emotional speech data is inherently challenging due to the time-consuming and costly nature of the process. As a result, existing datasets often suffer from limited speaker diversity, constrained emotional variability, and a lack of varied speaking styles. This data limitation can lead to overfitting in SER models, reducing their generalizability, particularly in real-world scenarios where emotional expressions are highly diverse and context-dependent. To overcome this issue, there is a pressing need for large-scale, diverse, and publicly accessible emotional speech datasets that encompass a wide range of emotions, speaking styles, and recording environments.

\subsection{Mismatch Between Speech Models and Emotional Speech}
A consequence of data scarcity is the lack of speech models specifically designed for emotional speech processing.\footnote{While a pre-trained model tailored for SER, emotion2vec, was introduced in 2023, it emerged after the commencement of this thesis project and is limited to SER tasks without ASR capabilities.}

Most conventional speech models, such as GMM-HMM models in Kaldi \citep{povey2011kaldi}, Deep Speech \citep{hannun2014deep}, and Jasper \citep{li2019jasper}, are optimized for ASR tasks, primarily focusing on phone and word recognition, rather than capturing emotional cues. As a result, adapting pre-trained ASR models for SER has proven to be challenging \citep{lakomkin2017reusing,zhou2020transfer,yeh2020speech}.

Recent speech foundation models, including wav2vec 2.0, HuBERT, and Whisper, have demonstrated impressive performance across various speech processing tasks, including SER. Unlike conventional handcrafted features or speech waveforms that generally work with sophisticated neural networks, the speech features produced from speech foundation models require only simple FFNNs to achieve state-of-the-art SER performance \citep{pepino2021emotion,yang2021superb}.

However, these models are not inherently designed to capture emotional characteristics in speech. First, their training are primarily conducted on non-emotional speech, lacking variations in prosody, voice quality, and speaking style. These aspects may not be fully captured by general-purpose speech models, potentially leading to distortions and suboptimal performance in SER. Second, the training targets of most self-supervised speech foundation models are divided into generative or discriminative approaches based on masked speech segmentation. These targets are mainly for reconstructing content from context \citep{chung2019unsupervised,chung2020vector}, pushing the model to generalize and potentially discard fine-grained paralinguistic cues.

\subsection{Challenges with Lexical Information}
Given the limitations of both conventional speech models and modern foundation models in handling emotional speech, another critical challenge arises: the reliance on imperfect lexical information.

In SER, lexical features complement acoustic cues to enhance emotion recognition. Traditionally, many SER systems have relied on human-annotated transcripts for lexical analysis. However, in real-world applications, such perfectly transcribed text is unavailable due to the absence of human annotators for immediate correction. Instead, ASR systems must be used to generate lexical information. Unfortunately, ASR models often struggle with emotional speech, as they are not explicitly trained to handle variations in prosody and expressiveness \cite{fernandez2004computational}. For example, \cite{goldwater2010words} found that words with extreme prosodic characteristics usually produce higher word error rates in speech recognition. These phenomena result in transcription errors that can negatively impact the accuracy of SER systems. Consequently, improving the robustness of SER in the presence of ASR errors remains a major research challenge.

\subsection{Challenges with Integration of Audio and Text Modalities}
Building on the previous challenge, the integration of audio and text modalities presents an additional hurdle for SER due to inherent discrepancies between these two modalities. While emotions in speech are conveyed through variations in acoustic properties, textual representations primarily rely on emotionally salient words. This fundamental difference makes it difficult to achieve seamless integration between the two sources of information.

Furthermore, ASR-generated transcriptions often contain errors, leading to misalignment between audio and text representations. Such inconsistencies can hinder multimodal fusion strategies and degrade SER performance. For example, \cite{sahu19_interspeech} used two commercial ASR systems to generate transcripts for bimodal SER (audio + text), resulting in a relative loss of 4\% and 5.3\% in unweighted accuracy compared to ground-truth transcripts, respectively. Additionally, certain emotions may be conveyed exclusively through either audio or text, or may even exhibit contradictory expressions across modalities. These complexities further complicate the effective fusion of acoustic and lexical features, posing a significant challenge for multimodal SER research.

\section{Chapter Summary}
\label{chap2:summary}
This chapter introduced the theoretical and practical foundations of SER and the background of integrating ASR into SER, covering task definitions, features, models, datasets, evaluation protocols, related literature, and key challenges. These insights lay the groundwork for the experiments and model proposals in subsequent chapters.

\chapter{Utilizing Self-Supervised Acoustic Representations for SER}
\label{chap3}

\section{Introduction}
The progression from traditional statistical ASR systems to advanced deep learning architectures and large-scale foundation models has greatly transformed the field of speech processing. In particular, these foundation models, while originally trained for ASR, have shown transferable potential across various speech-related downstream tasks, including SER. However, despite their success, the internal mechanisms by which these models capture, encode, and leverage emotional information are still not understood.

Unlike traditional speech modeling approaches that have been extensively researched, self-supervised learning (Self-SL) models have just started to be explored in very recent years, with wav2vec 2.0 (W2V2) attracting the most attention for its wide application potential. For example, \cite{pasad2021layer} conducted layer-wise analysis of W2V2 using a suite of tools and found 1) acoustic and linguistic properties are encoded in different layers; 2) the pre-trained model follows an autoencoder-style behavior; 3) the model encodes some non-trivial word meaning information. \cite{fan2020exploring} showed that W2V2 has the ability to discriminate between speakers and also languages, and this distinction is more obvious in lower layers. They hence proposed multi-task learning of speaker verification and language identification, and verified its feasibility. \cite{yang2021superb} set up benchmark performance using self-supervised speech models on a range of tasks.

Nevertheless, these Self-SL speech models are still understudied and the above-mentioned works have limitations. For example, \cite{pasad2021layer} did not extend their exploration to emotional speech and SER tasks. In \cite{fan2020exploring}, only a portion of the layer difference was shown, so misses a thorough layer-wise analysis. In \cite{yang2021superb}, they presented downstream task performance without further explanation. Furthermore, none of those studies investigated paralinguistic characteristics in W2V2 representations. As such, in the thesis work, we build on previous work while adding new perspectives from detailed quantitative analysis on emotional corpora.

In this chapter, we aim to shed light on how and what speech foundation models contribute to SER, examining their behavior from the acoustic perspective. We perform a series of in-depth experiments to analyze the types of emotional cues embedded in the representations learned by these models. Specifically, we investigate the paralinguistic characteristics, hierarchical structures, and emotion biases, with the goal of understanding how to probe foundation models that are pre-trained upstream for downstream speech tasks like SER. Furthermore, we propose a hierarchical joint ASR-SER training model, integrating W2V2 outputs in SER, aiming to increase the SER accuracy and the reliability of ASR models on emotional speech.

The remainder of this chapter is structured as follows. Section~\ref{chap3:acoustic} details the probing experiments and results examining W2V2 from an acoustic perspective on emotional speech. Section~\ref{chap3:hierarchical} describes a hierarchical attention-based fusion model that incorporates W2V2 outputs into joint ASR-SER training. Finally, Section~\ref{chap3:summary} provides a summary of the chapter.

\section{Exploring wav2vec 2.0 on Emotional Speech}
\label{chap3:acoustic}
Current speech foundation models often operate as black boxes when processing emotional speech, leaving the mechanisms by which they capture and represent emotional cues largely unexplored. Since there are various speech foundation models in terms of training schemes, number of layers, training data size, and more, in this section, we focus on W2V2 series, considering that W2V2 is the most popular speech foundation model and the less well-known black-box characteristics of Self-SL models, to frame our probing experiments and pave the way for probing other speech foundation models.

We begin by evaluating the SER performance of W2V2, aiming to reveal the extent to which it learned representations contribute to effective emotion classification. Following this, we conduct a detailed analysis of their paralinguistic abilities to determine whether it preserves key paralinguistic features known to correlate with human emotional perception, such as pitch, energy, and spectral characteristics.

In addition, we explore the hierarchical nature of W2V2 by probing different layers to identify which levels of their learned representations are most informative for SER tasks. This layer-wise analysis sheds light on whether emotional cues are captured at early (closer to the input), mid-level, or deeper (closer to the output) layers.

Finally, we examine the presence of emotion bias in W2V2, investigating whether their internal representations treat emotional content in a neutral, emotion-agnostic manner, or whether they inherently encode emotion-aware features.

Together, these analyses offer valuable insights into the interpretability of speech foundation models and their potential for emotion recognition applications.

\subsection{Experimental Setup}
\subsubsection{Datasets}
The IEMOCAP (IEM) and RAVDESS (RAV) datasets are used in this section. For IEM, following prior research \citep{li2019improved,peng2018auditory}, we combined \textit{Happy} and \textit{Excited}, and removed utterances that do not have transcriptions, bringing the total number of utterances used in this study to 5 500, each with one label from four classes: \textit{Angry, Happy, Neutral, and Sad}. For RAVDESS, we only use the speech set, which has 1 440 utterances from 24 actors (12 female, 12 male) in eight emotions: \textit{Calm, Happy, Sad, Angry, Fearful, Surprised, Disgust, and Neutral}. The major reason that we choose to use RAVDESS is that, even though other corpora may have a larger size, it provides fixed sentences with different emotional expressions. Such a setting excludes the lexical influence by ``forcing'' different emotions to have the same linguistic content, thus helping us to better explore the effects of the acoustic properties of W2V2 by eliminating the effects raised by lexical content (e.g., word pronunciation causing prosody variation).

\subsubsection{Speech Foundation Models}
We use \textit{wav2vec2-base}, \textit{wav2vec2-base-100h}, and \textit{wav2vec2-base-960h} models, which are the pre-trained and fine-tuned models (on 100h and 960h of Librispeech) respectively. We refer to them as \textit{PT}, \textit{FT100}, and \textit{FT960}. As mentioned before, we choose W2V2 because it is the most widely used speech foundation model (particularly one of the black-box Self-SL models), with the expectation that the exploratory approach can be generalized to other models.

\subsection{Probing SER Performance}
We first implement a layer-wise analysis by using the output of every individual layer within the Transformer network to demonstrate how information encoded by W2V2 contributes to SER. Next, as there is no common practice of how to utilize W2V2 representations as input features for downstream tasks, we compare the performance of three commonly used approaches of using W2V2 representations as input features:

1) taking the last layer output \citep{chen2023exploring,sharma2022multi,cai2021speech};

2) taking the average of all layer outputs \citep{boigne2020recognizing};

3) taking the weighted average of all layer outputs (assigning a trainable weight to each layer output) \citep{pepino2021emotion,yang2021superb}.

We also propose a fourth approach which excludes the last two layers from averaging as they generally underperform other layers. We evaluate the performance using Unweighted Accuracy (UA). Like most downstream tasks, we use W2V2 models as frozen feature extractors.

Since our goal is to explore information in W2V2 representations, we build a simple downstream model comprising only two dense layers (128 and 16 neurons, respectively) with \textit{ReLU} activation and one output layer (four neurons for IEM and eight neurons for RAV) with \textit{Softmax} activation. The learning rate is set as $1.0 \times 10^{-4}$ and $2.0 \times 10^{-4}$ for IEM and RAV with the \textit{AdamW} optimizer, respectively, and the weight decay is set as $1.0 \times 10^{-5}$. The batch size is 64, and we train the models until validation loss converges, as different layer outputs converge at different steps. For IEM, we implement 5-fold Cross-Validation (CV) in accordance with prior work \citep{li2019improved}. For RAV, we randomly divide 24 speakers into four groups and implement 4-fold cross validation.

\begin{figure}[t]
  \centering
  \subfigure{\includegraphics[width=0.8\textwidth]{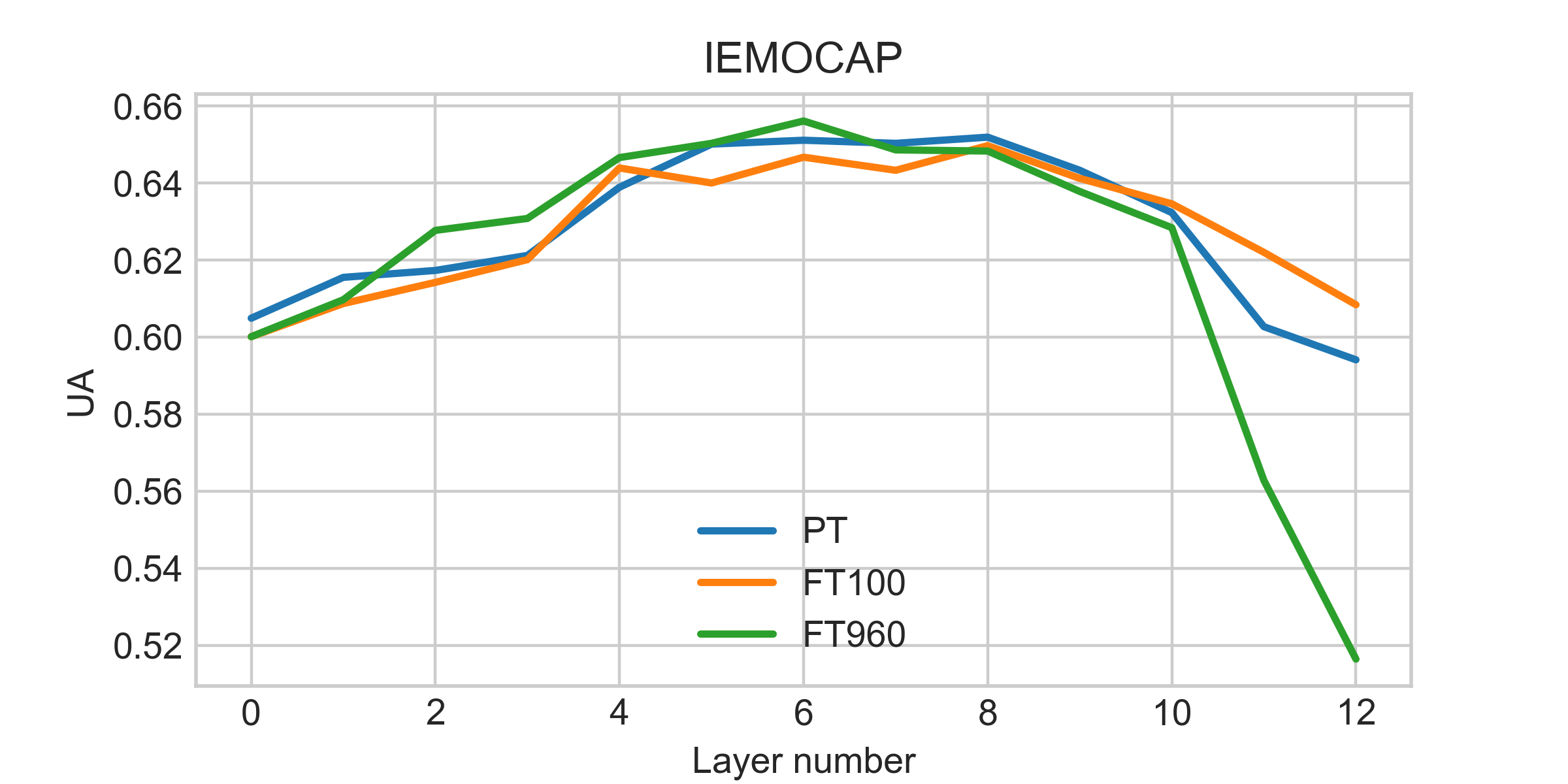}}
  \subfigure{\includegraphics[width=0.8\textwidth]{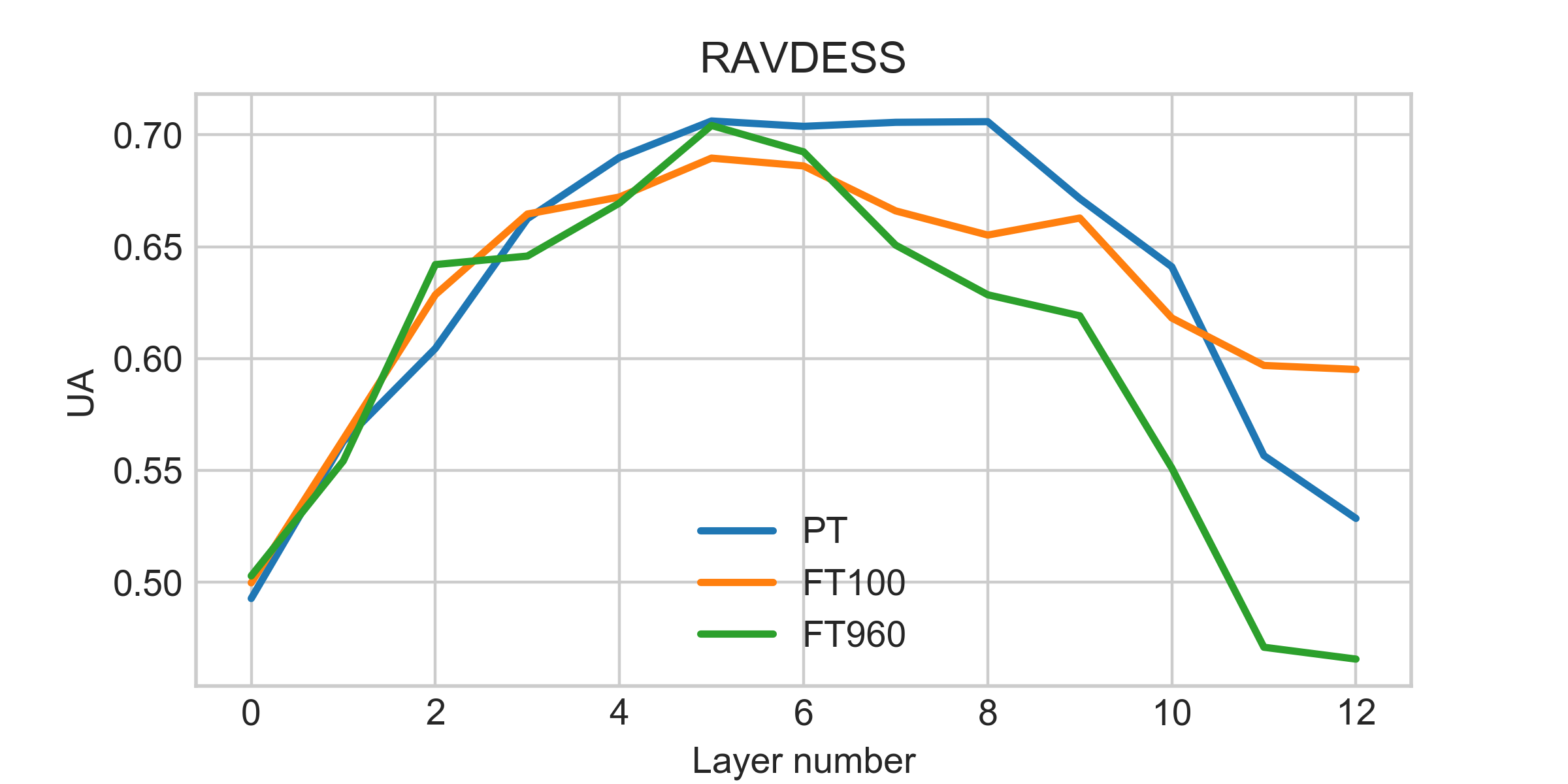}}
  \caption{SER accuracy comparison using models. PT: pre-trained; FT100: fine-tuned on 100h of Librispeech; FT960: fine-tuned on 960h of Librispeech.}
  \label{chap3/fig:acc}
\end{figure}

\begin{table}[t]
\centering
\caption{UA (\%) using different inputs and models.}
\label{chap3/tab:acc}
\begin{tabular}{llccc}
\hline
\multirow{2}{*}{\textbf{Input}} & \multirow{2}{*}{\textbf{Model}} & \multicolumn{2}{c}{\textbf{Corpus}} \\
 &  & IEM & RAV \\ \hline
\multirow{3}{*}{Best layer} & \textit{PT} & 65.19 & \colorbox{pink}{70.62} \\
 & \textit{FT100} & 64.97 & 68.96 \\
 & \textit{FT960} & \colorbox{lime}{65.61} & 70.42 \\ \hline
 \multirow{3}{*}{Last layer} & \textit{PT} & 59.41 & 52.85 \\
 & \textit{FT100} & \colorbox{lime}{60.84} & \colorbox{pink}{59.51} \\
 & \textit{FT960} & 51.64 & 46.56 \\ \hline
\multirow{3}{*}{Average} & \textit{PT} & 64.93 & 67.26 \\
 & \textit{FT100} & 64.72 & \colorbox{pink}{67.40} \\
 & \textit{FT960} & \colorbox{lime}{65.51} & 64.20 \\ \hline
\multirow{3}{*}{Average w/o last two} & \textit{PT} & 65.11 & 67.36 \\
 & \textit{FT100} & 64.90 & \colorbox{pink}{67.50} \\
 & \textit{FT960} & \colorbox{lime}{65.87} & 65.56 \\ \hline
\multirow{3}{*}{Weighted average} & \textit{PT} & 65.28 & 68.47 \\
 & \textit{FT100} & 64.94 & \colorbox{pink}{68.89} \\
 & \textit{FT960} & \colorbox{lime}{65.67} & 65.11 \\ \hline
\end{tabular}
\end{table}

Figure~\ref{chap3/fig:acc} depicts the trends of layer-wise UA on the two corpora. We include layer 0 (the output of the CNN encoder right before the Transformers) in accordance with prior works. We see that: 

\begin{enumerate}
    \item Before the best middle layer (layer 6 for IEM and layer 5 for RAV), all three models (\textit{PT, FT100, and FT960}) show the same trend: accuracies go up and are relatively close, but then start dropping after the middle layer. This upward-downward trend is possibly related to the acoustic-linguistic property of W2V2: frame-level inputs are encoded by the Transformers until the middle layer \citep{pasad2021layer}. At this point, the representations encode phonetic information but have not yet lost much of the original acoustic properties. This makes the middle layers contain the most useful information for SER. In subsequent layers, the representations gradually encode word identity and word meaning more strongly \citep{pasad2021layer}. At this stage, potential ASR errors together with the loss of the original acoustic information appear to lead to drops in SER accuracy.

    \item On IEM, there are barely any differences among the UAs until layer 11, while on RAV, the differences after the middle layer are more dramatic. This phenomenon is plausible as RAV only has two fixed statements, yet IEM contains various sentences, allowing fine-tuned W2V2 models to make more use of linguistic information, which makes up for the acoustic loss. In RAV, however, every sentence is repeated with each emotion, which means linguistic information has no contribution to emotion discrimination.

    \item In general, \textit{PT} \textgreater \ \textit{FT100} \textgreater \ \textit{FT960} from the middle layer but \textit{FT100} clearly outperforms the other two on the last two layers. We assume that moderate fine-tuning enables W2V2 to achieve a good acoustic-linguistic balance, as word information has been found encoded by the last two layers in fine-tuned models \citep{pasad2021layer}, and such a balance helps \textit{FT100} achieve better performance on the last two layers.
\end{enumerate}

Table~\ref{chap3/tab:acc} compares the SER accuracies using different inputs and models, and yields the following findings:

\begin{enumerate}
    \item UAs of the best layer of \textit{PT} and \textit{FT960} are close and higher than \textit{FT100}. So, although moderate fine-tuning enables the model to capture both acoustic and linguistic properties for \textit{FT100}, neither of them is fully encoded causing a decrease in accuracy for the best layer of \textit{FT100}, compared to \textit{PT} and \textit{FT960}.

    \item The situation is reversed on the last layer. Compared to \textit{FT100}, \textit{PT} lacks linguistic information and \textit{FT960} relies too much on imperfect linguistic information while losing acoustic information due to ``over'' fine-tuning.

    \item Word-level information does not help SER on RAV, as mentioned before, which makes the deeper layers of \textit{FT960} the worst. Hence, it is reasonable that \textit{FT960} generates better performance on IEM yet worse performance on RAV when taking the average on all layers, the average without the last two layers, or the weighted average on all layers as input.

    \item In corpora like RAV, where emotional cues are conveyed only through speech acoustics, the best middle layer may suffice as input for SER. However, for corpora like IEM, emotional expression is also influenced by linguistic content.

    \item Except for the ``best'' layer inputs, \textit{FT960} and \textit{FT100} produce the best results on IEM and RAV, respectively. This differs from the patterns in Figure~\ref{chap3/fig:acc} from which we would expect \textit{FT960} to perform the worst on IEM and \textit{PT} the best on RAV. It means that the performance obtained by averaging layer outputs does not equal the average of all layer performance, which demonstrates that representations of different layer contain different information contributing to SER.
\end{enumerate}

\subsection{Probing Paralinguistic Properties}
Compared to handcrafted features, Self-SL representations are hard to explain as they are not designed by human knowledge of speech signals. For example, paralinguistic features have proven useful and have long been used for emotion detection or clinical speech analysis \citep{haider2019assessment}. On the other hand, Self-SL representations are formed by reconstruction loss or contrastive loss from speech, but whether, why, and how these losses contribute to downstream tasks other than ASR are still unknown.

In this experiment, we measure the similarities between each layer's output and different types of paralinguistic features to see how W2V2 retains well-known acoustic correlates of speech perception. We evaluate the similarity using Canonical Correlation Analysis (CCA) \citep{hardoon2004canonical}. CCA finds linear projections of two multivariate feature sets such that the correlations between the projected variables are maximized. It is particularly suitable for our task because it can capture shared information between the model representations and paralinguistic features even if they exist in different representational spaces. By using CCA, we can quantitatively assess how closely the internal representations of W2V2 align with known acoustic correlates.

The selected paralinguistic features are listed in Table~\ref{chap3/tab:para}, which are mainly based on eGeMAPS \citep{eyben2015geneva}, commonly used as a minimal set of features for SER. The groupings are directly adopted from eGeMAPS, except that Shimmer and Jitter are grouped under voice quality, as they have proven especially useful yet are often overlooked in SER \citep{li2007stress,jacob2016speech}. We also extract 40 MFCCs as linguistic (phone) features for comparison. We downsample them to make their sequence lengths comparable to W2V2 representations as required by CCA.

\begin{sidewaysfigure*}
  \centering
  \includegraphics[width=\textwidth]{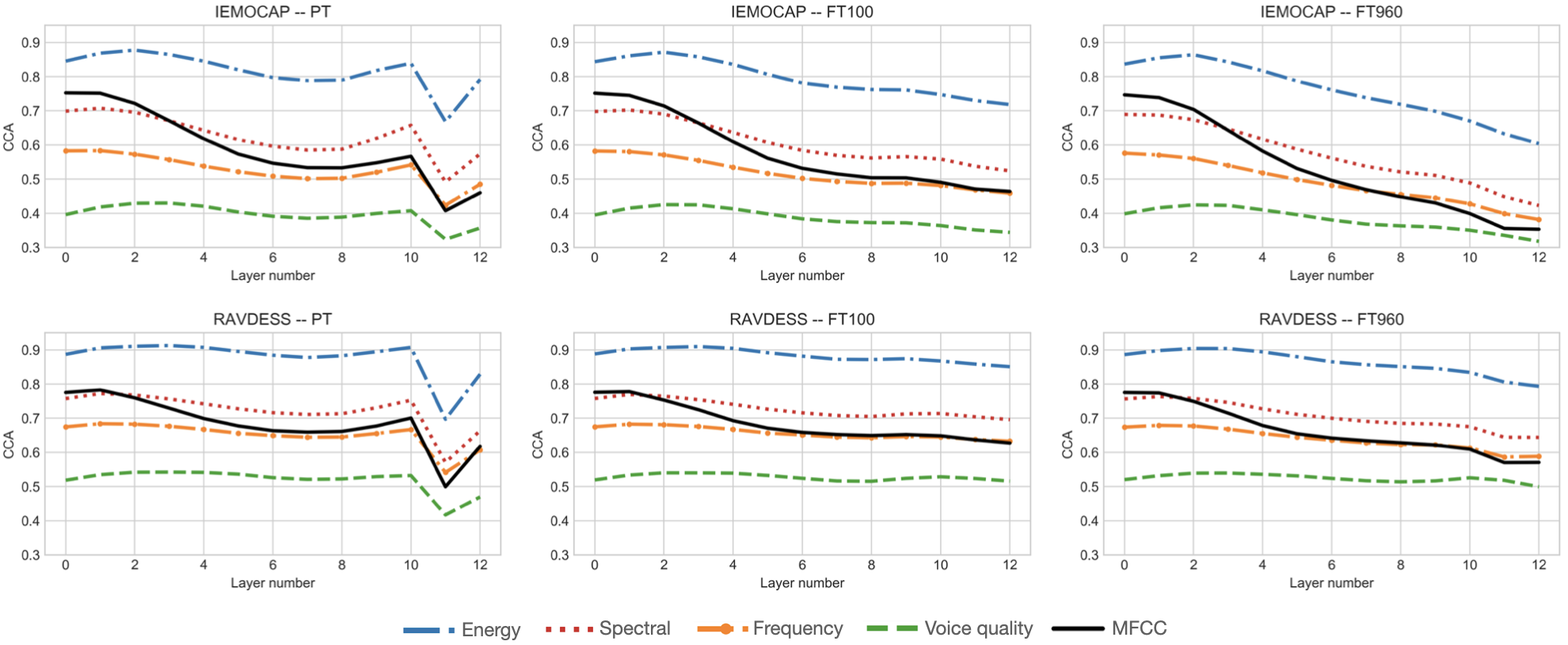}
  \caption{CCA similarity comparison for paralinguistic property.}
  \label{chap3/fig:cca}
\end{sidewaysfigure*}

Figure~\ref{chap3/fig:cca} shows the layer-wise CCA on IEM and RAV using three W2V2 models. It can be noted that:

\begin{enumerate}
    \item The curves of paralinguistic features are flatter than those of MFCCs, indicating that W2V2 does not focus on as much paralinguistic information. In particular, voice quality features seem barely taken into the encoding process.
    
    \item When looking at \textit{PT} models, we can see a reverse trend from the middle layer, which reinforces a view that the \textit{PT} model follows an autoencoder style behavior where deeper layers ``reconstruct'' the input \citep{pasad2021layer}. The reverse trend of similarity with MFCCs on IEM is weaker than that on RAV, possibly demonstrating that the linguistic complexity makes the reconstruction process harder and more error-prone. Hence, the deeper representations are more similar to MFCCs on RAV than on IEM. The peculiar pattern of the last two layers is due to the training objective of masked segment prediction.
    
    \item When looking at the fine-tuned models, we note that the similarities keep decreasing since the models have been fine-tuned towards ASR and learn to compute speech information from frame to phone, and then to word level with layer depth \citep{pasad2021layer}. This phenomenon reinforces our explanation of the accuracy drop in Figure~\ref{chap3/fig:acc} that acoustic properties are being replaced by linguistic ones that contain errors.

    \item The graphs indicate that the overall similarity variation on IEM is larger than on RAV. This is again, likely due to the fact that RAV has much less linguistic variation overall, which we in turn see as less change in CCA through the layers. Moreover, since how we say a sentence is affected by its content, the CCA variation of paralinguistic features is larger on IEM than on RAV. Finally, since the layer outputs contain more complex linguistic information, the overall CCA values for paralinguistic features on IEM are lower than those on RAV, no matter the starting values or overall values. However, the starting values of similarities with MFCCs on both corpora are almost the same, further suggesting that W2V2 focuses on learning linguistic information rather than paralinguistic.
\end{enumerate}

\begin{table}[t]
\centering
\caption{Extracted paralinguistic features.}
\label{chap3/tab:para}
\begin{tabular}{ll}
\hline
\textbf{Feature set} & \textbf{Low-level descriptors} \\ \hline
Energy & Loudness; Harmonics-to-noise ratio \\ \hline
Frequency & \begin{tabular}[c]{@{}l@{}}Pitch; Formant 1;\\ Formant 1, 2, 3 frequency\end{tabular} \\ \hline
Spectral & \begin{tabular}[c]{@{}l@{}}Alpha ratio; Hammarberg index;\\ Formant 1, 2, and 3 relative energy;\\ Spectral slope 0-500 Hz, 500-1500 Hz;\\ Harmonic difference H1-H2 and H1-A3\end{tabular} \\ \hline
Voice quality & Jitter; Shimmer \\ \hline
\end{tabular}
\end{table}

\subsection{Probing Layer Correlations}
To better understand how different layer outputs are correlated with each other before and after fine-tuning for ASR, and how W2V2 encodes information and contributes to SER, we calculate pair-wise CCA similarities of W2V2 representations from every layer and plot the similarities using heat maps to visualize the correlations. We only discuss IEM, as the same patterns are found on RAV.

\begin{figure}[t]
  \centering
  \includegraphics[width=0.98\textwidth]{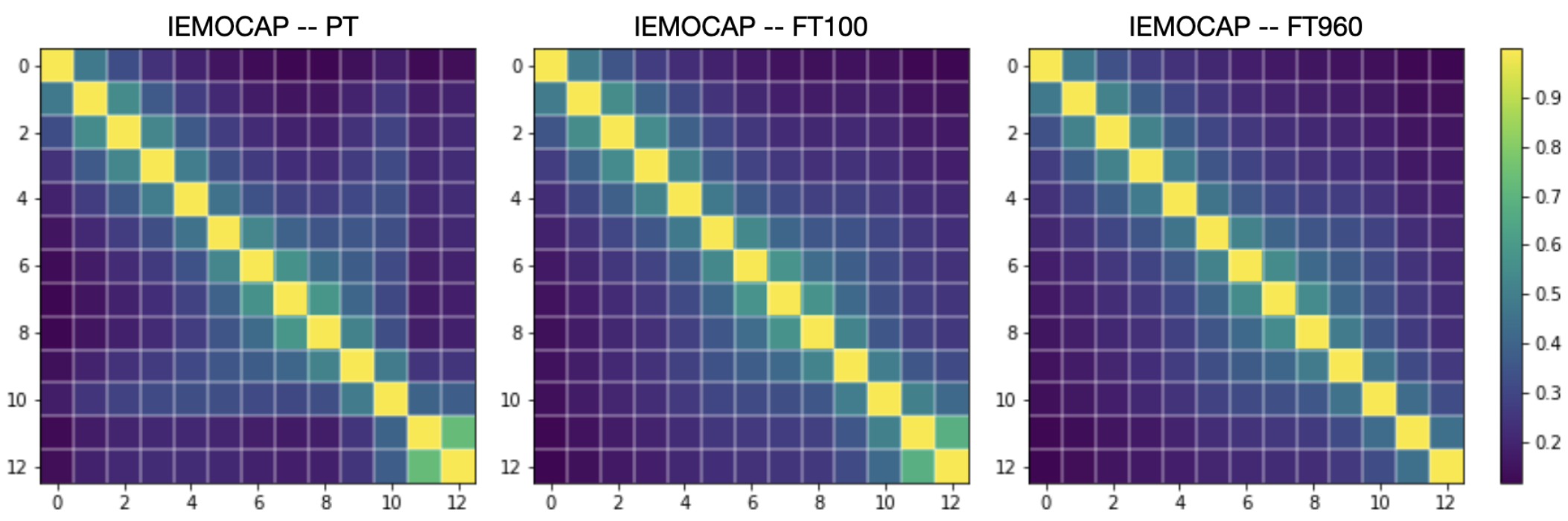}
  \caption{Pair-wise correlations of layer representations.}
  \label{fig:pair}
\end{figure}

\begin{table}[t]
\centering
\caption{Correlations between the last two and prior layers.}
\label{tab:pair}
\begin{tabular}{llccccc}
\hline
\textbf{Layer} & \textbf{Model} & \textbf{Layer 6} & \textbf{Layer 7} & \textbf{Layer 8} & \textbf{Layer 9} & \textbf{Layer 10} \\ \hline
11 & \textit{\begin{tabular}[c]{@{}l@{}}PT\\ FT100\\ FT960\end{tabular}} & \begin{tabular}[c]{@{}c@{}}0.25\\ \colorbox{lime}{0.29}\\ 0.24\end{tabular} & \begin{tabular}[c]{@{}c@{}}0.24\\ \colorbox{lime}{0.30}\\ 0.26\end{tabular} & \begin{tabular}[c]{@{}c@{}}0.24\\ \colorbox{lime}{0.31}\\ 0.27\end{tabular} & \begin{tabular}[c]{@{}c@{}}0.31\\ \colorbox{lime}{0.38}\\ 0.32\end{tabular} & \begin{tabular}[c]{@{}c@{}}0.45\\ \colorbox{lime}{0.53}\\ 0.43\end{tabular} \\ \hline
12 & \textit{\begin{tabular}[c]{@{}l@{}}PT\\ FT100\\ FT960\end{tabular}} & \begin{tabular}[c]{@{}c@{}}0.26\\ \colorbox{lime}{0.27}\\ 0.22\end{tabular} & \begin{tabular}[c]{@{}c@{}}0.25\\ \colorbox{lime}{0.28}\\ 0.22\end{tabular} & \begin{tabular}[c]{@{}c@{}}0.26\\ \colorbox{lime}{0.29}\\ 0.23\end{tabular} & \begin{tabular}[c]{@{}c@{}}0.31\\ \colorbox{lime}{0.34}\\ 0.26\end{tabular} & \begin{tabular}[c]{@{}c@{}}0.43\\ \colorbox{lime}{0.44}\\ 0.32\end{tabular} \\ \hline
\end{tabular}
\end{table}

Figure~\ref{fig:pair} shows the pair-wise correlations:

\begin{enumerate}
    \item In the \textit{PT} model, we can notice an arrow-like shape from layers 0--10. Specifically, layers 9 and 10 show higher correlations with shallow layers compared to those in \textit{FT100} and \textit{FT960}. This suggests that representations in these layers become more general, making certain shallow-layer information easier to extract, which also explains the reverse trend observed in the \textit{PT} model. However, such a pattern seems specific to these two layers as it is not obvious in prior deep layers, even layer 8, which also accounts for the reverse trend in Figure~\ref{chap3/fig:cca}. After fine-tuning, this phenomenon disappears. The two layers become the same as shallow layers that have high correlations with nearby layers, and the correlations get weaker with distance.

    \item In \textit{PT} model, the last two layers are highly correlated with each other, even more obvious than the prior adjacent layers. However, the correlation gets weaker as fine-tuning goes. The color becomes dimmer in \textit{FT100} and further dimmer in \textit{FT960}. Nevertheless, an interesting phenomenon appears in \textit{FT100}: the similarities between last two layers (especially layer 11) and the prior deep layers (layer 6 to 10) become higher. We present the CCA values in Table~\ref{tab:pair}. It is obvious that the correlations in \textit{FT100} are the highest, but they decrease in \textit{FT960}. Moreover, the lower right part is slightly brighter in \textit{FT100} than in \textit{FT960} (values are skipped as limited space) indicating the deeper layers are more correlated with each other, which means low-level linguistic information (e.g., phonetics) generally exists. These phenomena validate our assumption that moderate fine-tuning enables W2V2 to achieve a good acoustic-linguistic balance but over fine-tuning ``forces'' the model to concentrate on learning high-level linguistic properties (e.g., word meaning) towards ASR.
\end{enumerate}

\subsection{Probing Hierarchical Properties}
Speech generally follows the hierarchy of frame, phone, syllable, word and utterance from low to high level, that contribute differently to speech perception and understanding \citep{yenigalla2018speech,pascual2019learning}. Since Self-SL enables frames to capture context information, the representations are expected to contain higher-level meanings. To verify this, we prepare the extracted paralinguistic features at frame, phone, and word levels and measure their similarities with W2V2 representations using CCA, respectively. As our purpose is only to verify whether W2V2 features contain high-level speech information, we do not use forced alignment to determine the perfect boundaries. Instead, we adopt a less accurate yet efficient approach to compute hierarchical features: we approximate a phone by averaging five consecutive frames (we also tried to add overlap, but the results didn't make much difference): a word by averaging five consecutive phones, based on the fact that frame length is set as 25 ms when being extracted, and phone length varies from 50 ms to 200 ms (five frames on average) and word length from 250 ms to 1000 ms (five phones on average) \citep{chen2022speechformer}.

We use all the paralinguistic features provided by eGeMAPS and implement the composition of hierarchical features. Then we downsample the paralinguistic features or the W2V2 representations depending on their lengths to make them comparable. Finally, we compute the CCA differences (${CCA}_{phone}-{CCA}_{frame}$ and ${CCA}_{word}-{CCA}_{phone}$) which represent how similar the higher-level features with W2V2 representations are compared to the lower-level ones.

\begin{figure}[t]
  \centering
  \includegraphics[width=0.8\textwidth]{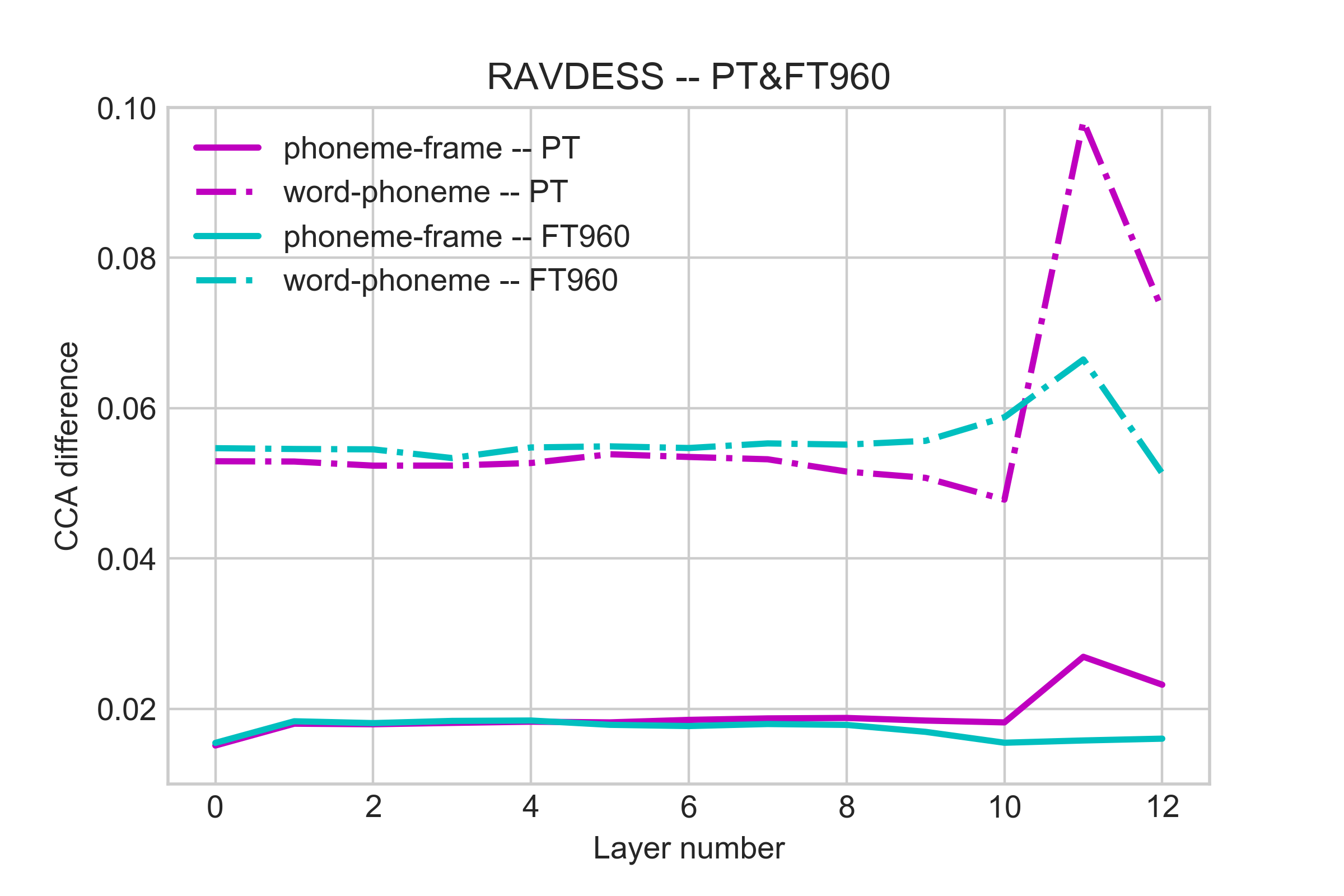}
  \caption{Hierarchical CCA similarity differences.}
  \label{chap3/fig:hie}
\end{figure}

From Fig~\ref{chap3/fig:hie}, we can note that:

\begin{enumerate}
    \item Higher-level paralinguistic features do have higher similarities with W2V2 representations as the difference values are all positive. Besides, the value of ${CCA}_{word}-{CCA}_{phone}$ is even higher than ${CCA}_{phone}-{CCA}_{frame}$, which means W2V2 representations are more similar to word-level paralinguistic information.

    \item The CCA differences barely change until layer 11 and become larger in the last two layers, which is due to the masked segment prediction enabling them to capture more context information (which is high level), especially on layer 11.

    \item The curves of fine-tuned models are flatter because the last two layers become more coherent with the other layers by fine-tuning. Note that since the paralinguistic property is affected by linguistic property in IEM, the patterns are not as clear as RAV. Therefore, yet we observed similar trend in IEM, we only illustrate it using RAV.
\end{enumerate}

\subsection{Probing Emotion Bias}
Different emotions have different paralinguistic patterns \citep{li2007stress,lugger2007relevance}. For example, hot angry and happy emotions usually have high intensity and pitch, while sad and calm emotions have low intensity \citep{luengo2010feature}, Hence, we calculate CCA similarities between paralinguistic features with W2V2 representations of every emotion for discriminative analysis. We also use all the paralinguistic features in eGeMAPS as in the previous task.

\begin{figure}[t]
  \centering
  \subfigure{\includegraphics[width=0.8\textwidth]{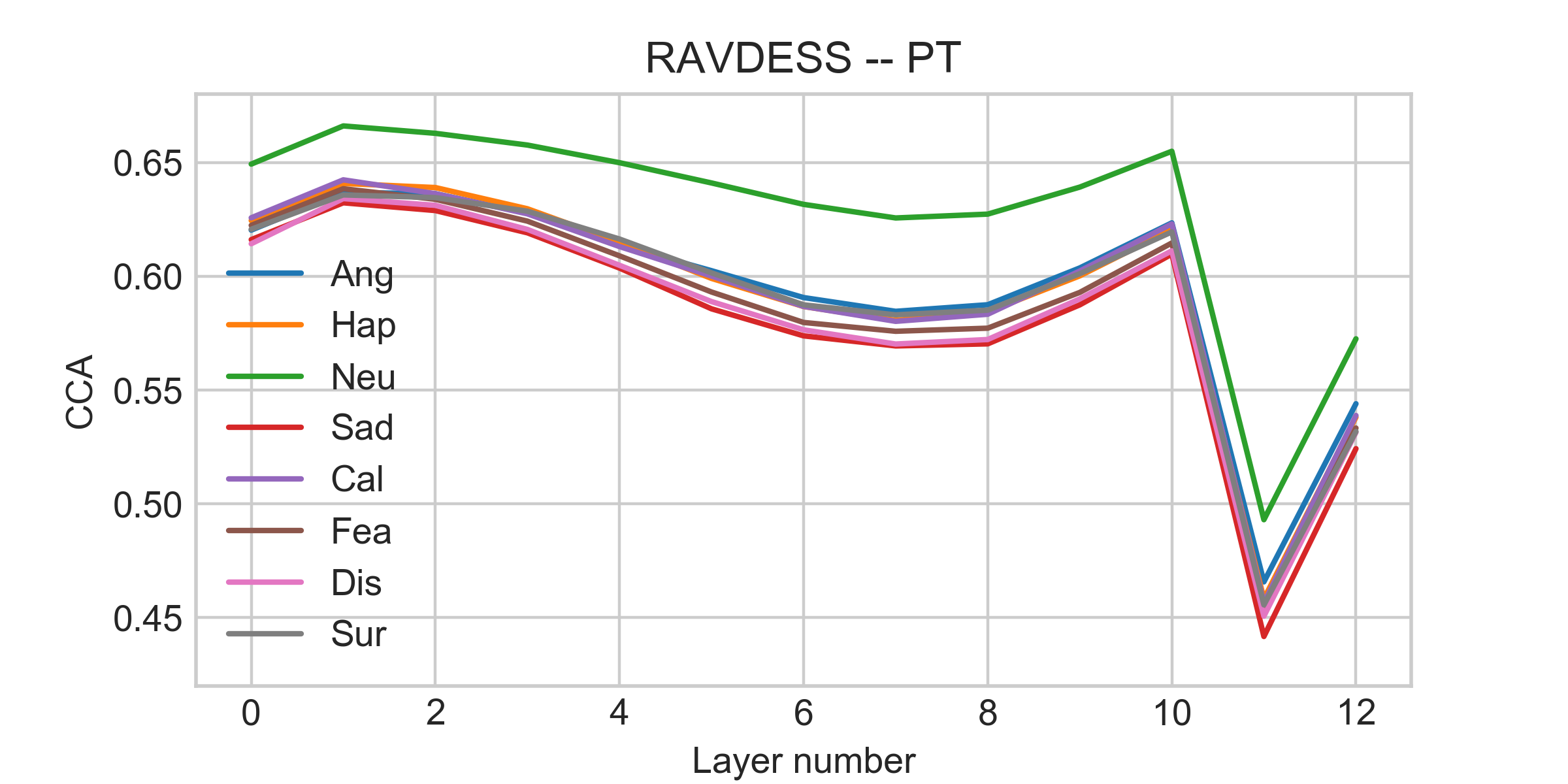}}
  \subfigure{\includegraphics[width=0.8\textwidth]{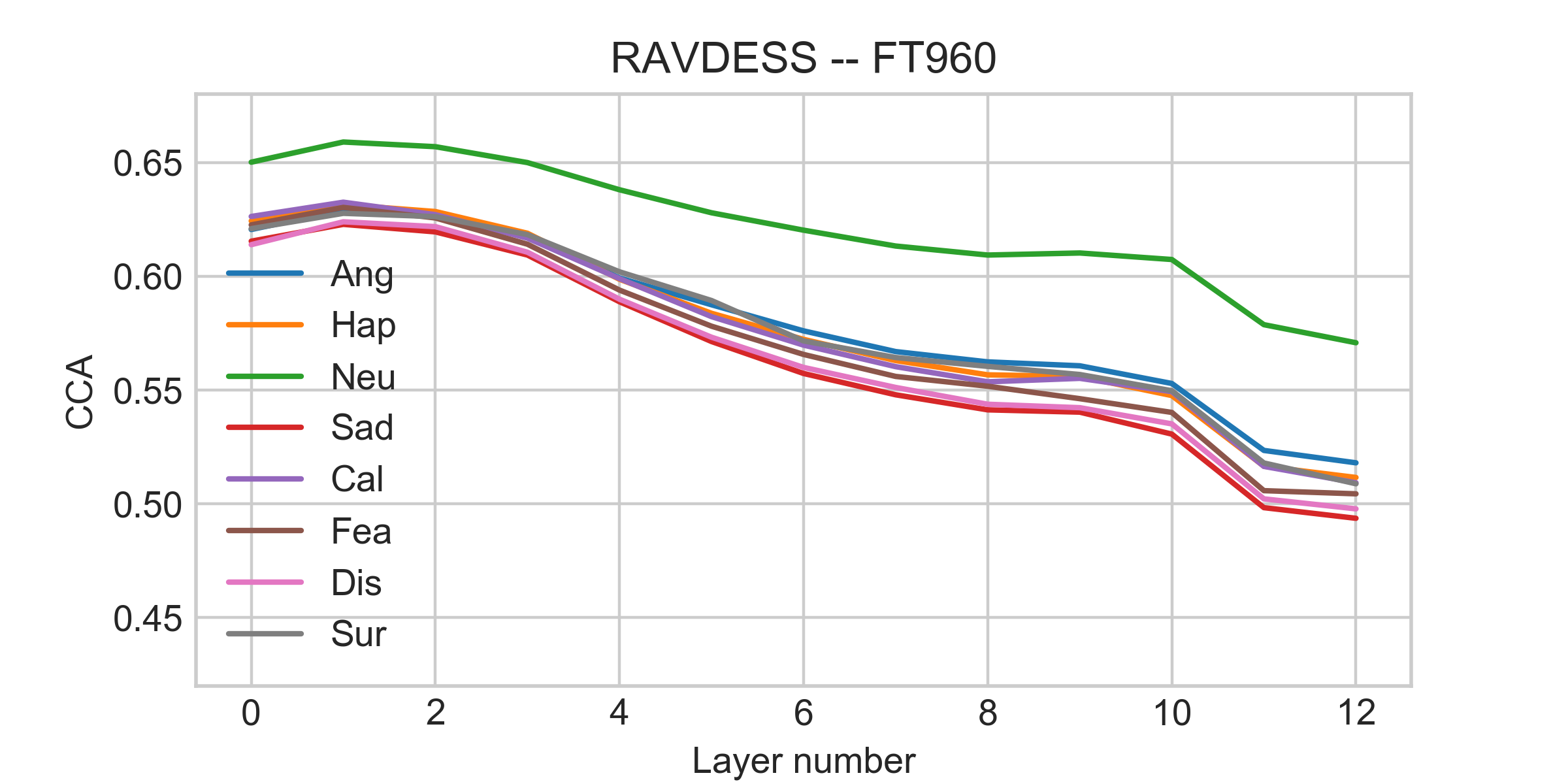}}  
  \caption{Discriminative analysis for emotion bias.}
  \label{chap3/fig:dis}
\end{figure}

As illustrated in Fig~\ref{chap3/fig:dis}:

\begin{enumerate}
    \item Higher similarities between paralinguistic features with W2V2 representations are found in \textit{Neutral} emotion for both the \textit{PT} model and the \textit{FT} models, pointing to interesting observations: a) \textit{Neutral} is likely more frequently represented within Librispeech, as it is a corpus of read audiobooks where most emotional cues arise only within speech of fictional characters, i.e. bias in the data during pre-training of W2V2 consequently results in learned representations which are emotion-agnostic; b) the pre-training pretext task in W2V2 (predicting masked segments) is not sufficient to learn a truly generalized representation in which different emotions are captured effectively. We also see that the curves converge after the middle layer on \textit{PT} model. This again indicates that the deeper layers (except the last two) of \textit{PT} model reconstruct the acoustic input.

    \item The curves become even less distinguishable at the last two layers, indicating again the autoencoder type of learning resulting from the masked segment prediction does not help distinguish emotions. This may be because the paralinguistic information of a masked segment is difficult to predict from unmasked segments, as paralinguistic information is more spontaneous and less contextual compared to linguistic information. The masked segment prediction discriminates linguistic information while blurring the paralinguistic difference among frame segments, which makes the paralinguistic properties of every emotion become similar, resulting in the close curves.

    \item For the \textit{FT960} model, the distances between the curves increase with depth. It seems that W2V2 not only avoids encoding paralinguistic information, but consistently discards some paralinguistic features as learning proceeds, especially in speech that contains rich paralinguistic information, e.g., voice quality, which leads to increasing differences in the similarities between W2V2 representations and paralinguistic features across emotions (otherwise, the distances should not change with layer depth).
\end{enumerate}

\subsection{Summary}
As speech foundation models are usually used simply as feature extractors for SER, there is a lack of understanding and approaches to analyze these models on emotional speech. To this end, we study W2V2, a representative speech foundation model, demonstrating how state-of-the-art ASR models should be explored for SER. We found that representations from W2V2 created for ASR lacks certain paralinguistic information. We also contribute to understanding the types of representations W2V2 learns by thoroughly comparing layer outputs in their correlations and SER. The hierarchy and bias analysis pave the path for better usage of W2V2 on downstream tasks. Our endeavor contributes to bridging the long-existing gap between ASR and SER research.

\section{Hierarchical Fusion with Joint ASR Fine-Tuning}
\label{chap3:hierarchical}
Considering the satisfactory SER accuracy achieved by W2V2 and its limited emphasis on paralinguistic features, as shown in the previous section, we propose using a single W2V2 model as the backbone, jointly fine-tuning it on emotional speech while simultaneously training a SER model.  

Although both ASR and SER use speech signals as input, ASR operates more at the frame level, whereas SER recognizes emotion on larger timescales. Previous work \citep{fayek2016correlation} has demonstrated that hidden features from the initial layers of both ASR and SER tasks are transferable, with their relevance gradually fading through deeper layers. \citet{yoon2018multimodal} used the Google Cloud Speech API to generate transcription that were then fused with MFCCs for SER. They achieved 69.1\% WA, likely due to their low WER (5.53\%). \citep{sahu19_interspeech} concatenated GloVe features obtained from Wit.ai API text transcription with acoustic features, obtaining a  62.9\% UA using an LSTM based model. \citet{feng2020end} jointly trained ASR-SER using log mel-scale filter bank outputs and decoder outputs from a pre-trained ASR model, and achieved 68.6\% WA. \citet{cai2021speech} used a single W2V2 as the training model for both ASR and SER rather than separating two tasks, and used 10-fold cross-validation, which usually yields better results than 5-fold; \citet{zhou2020transfer} fine-tuned the pre-trained ASR model on IEMOCAP before transferring the ASR features for SER.

However, it remains unclear which ASR features benefit SER and in what way. To address this, we compare layer-wise ASR hidden features and ASR text outputs, and propose a hierarchical fusion of both into the pipeline for joint SER training.

\subsection{Experimental Setup}
For the \textbf{dataset}, we use IEMOCAP and removed utterances that did not have transcription, bringing the total number of utterances used in this study to 5,500. For the \textbf{ASR model}, we use the “wav2vec2-base-960h” model. Performance is evaluated using Unweighted Accuracy (UA).

For our proposed model, the audio time-series is encoded by the W2V2 model for ASR, and separately as MFCCs or the paralinguistic feature set from eGeMAPS (Table~\ref{chap3/tab:para}) for SER. On the ASR path, the W2V2 representations are then decoded to text by a word-level Connectionist Temporal Classification (CTC) decoder. Both the hidden state output and the text are extracted. A RoBERTa model \citep{liu2019roberta} is then used to extract lexical features from text output. On the SER path, a max pooling with kernel size 2 is conducted on the acoustic features (MFCCs or eGeMAPS) to obtain an acoustic representation.

The ASR hidden output (short for hidden layer output), text output, and pooled acoustic features, are then encoded using 2-layer Bidirectional Long Short-Term Memory (Bi-LSTM) networks. Each layer contains 32 hidden units followed by a dropout with probability of 0.5. A self-attention layer with 16 heads and 64 nodes is applied on the output of the Bi-LSTM, which generates a fixed-length vector as the feature encoding. 

\subsection{Feature Fusion and Emotion Output}

\begin{figure}
    \centering
    \includegraphics[width=\textwidth]{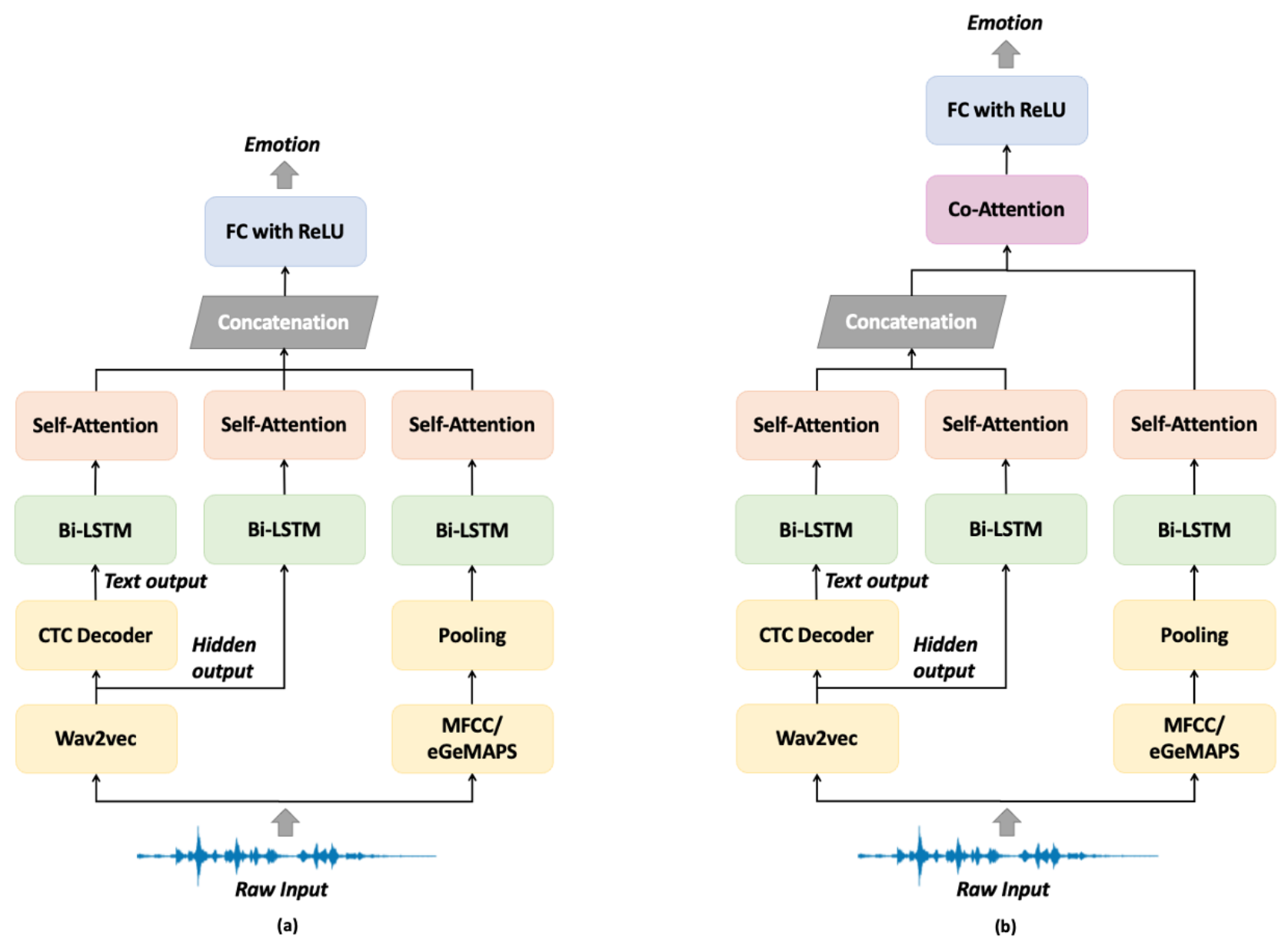}
    \caption{Baseline models. (a): concatenation; (b): concatenation with co-attention fusion.}
    \label{chap3/fig:model-fusion}
\end{figure}

\begin{figure}
    \centering
    \includegraphics[width=0.44\textwidth]{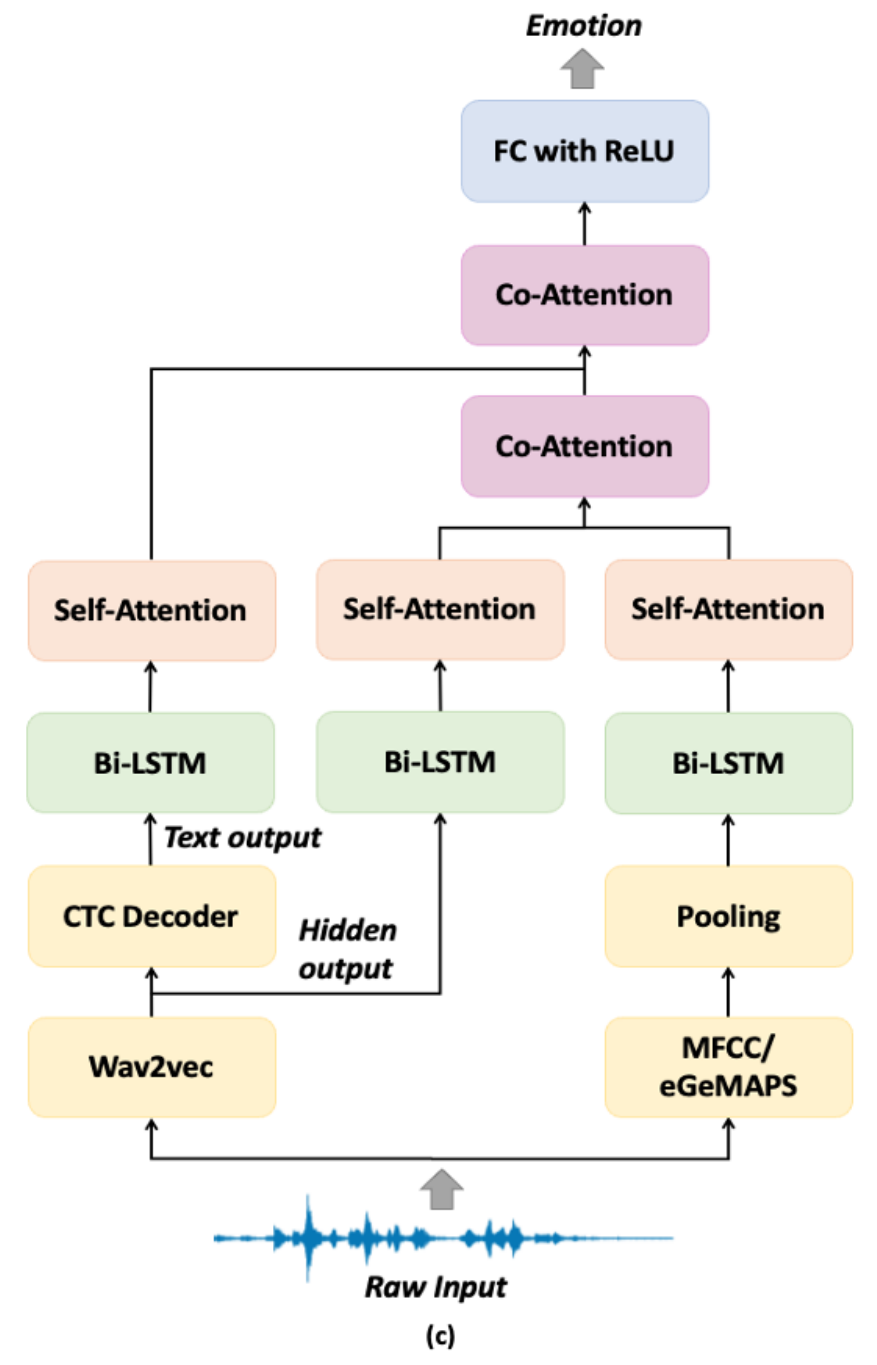}
    \caption{The proposed model. (c): hierarchical co-attention fusion with MFCC or eGeMAPS paralinguistic features as the acoustic input.}
    \label{chap3/fig:hier-model}
\end{figure}

We fuse the three encodings from the self-attention layer to produce a final vector for emotion classification. We compared our model with three baseline models: \textbf{(a) concatenation} and \textbf{(b) concatenation with co-attention fusion, as shown in Figure~\ref{chap3/fig:model-fusion}}. \textbf{Our proposed hierarchical co-attention fusion is shown in Figure~\ref{chap3/fig:hier-model}}. The co-attention mechanism \citep{lu2019vilbert} concatenates two hidden-state vectors which exchange key-value pairs in self-attention, allowing features from one input channel to be incorporated into the other:

\begin{align}
    H_C &= Concat(H_{C_1}, H_{C_2}) \\
 H_{C_i} &= MultiHead(Q_{A}, K_{B}, V_{B})W^O \\
        &= Concat(head_1, ..., head_n) \\
 head_i &= Attention(Q_{A}W_i^Q,K_{B}W_i^K,V_{B}W_i^V)
\end{align}
where $W^O$, $W_i^Q$, $W_i^K$, and $W_i^V$ are trainable parameters. $Q_{A}$ represents the query from one input channel, while $K_B$ and $V_B$ represent the key and value from the other. The value of $n$ is 16, and $H_C$ denotes the final concatenated hidden states of co-attention.

Hierarchical approaches for emotion recognition have proven useful in fusing different-level or different-type of features in previous works \citep{tian2016recognizing}. Inspired by these works and the hierarchical characteristics of speech \citep{pascual2019learning}, we propose to fuse the input features from low to high level in a hierarchical manner using co-attention. Because MFCC and eGeMAPS paralinguistic features are extracted within frames, W2V2 features contain within-utterance context, and RoBERTa features can learn cross-sentence context, we refer to them as frame-level, utterance-level, and corpus-level respectively, and hierarchically fuse them in this order to generate a fixed-length vector. This is passed to a fully connected output layer with \textit{Softmax} activation function to generate the probability distributions over emotion classes.

The model is optimized by the multi-task loss function:
\begin{align}
    \mathcal{L} &= \lambda \mathcal{L}_{\text{ASR}} + (1-\lambda) \mathcal{L}_{\text{SER}},
\end{align}
where $\mathcal{L}_{\text{ASR}}$ and $\mathcal{L}_{\text{SER}}$ are the losses for ASR and SER, respectively. We vary $\lambda$ from 0.1 to 0.9 in increments of 0.2 to examine whether ASR or SER should serve as the primary task in the joint training framework. We use cross-entropy as the loss function and Adam optimizer with a learning rate of $1.0 \times 10^{-4}$ and decay rate of $1.0 \times 10^{-4}$. The gradients are clipped with a threshold of 5.0. The batch size is set to 20, and the number of epochs is limited to 100. If the training of either task (ASR or SER) converges (i.e., its training loss does not decrease for three consecutive epochs), we terminate its training, allowing the joint model to focus on the other task. We perform 5-fold CV and use UA to assess the performance\footnote{Note that the experimental settings have been revised from those in our published work \citep{li2022fusing} and resulted in improved results.}.

\subsection{Results}

\begin{table}
    \centering
    \caption{Performance comparison of using different models, features and fusion methods. $\lambda$ = 0.1.}
    \label{chap3/tab:result}
    \resizebox{\textwidth}{!}{
    \begin{tabular}{lllc}
    \toprule
    \textbf{Model} & \textbf{Feature} & \textbf{Fusion approach} & \textbf{UA} \\ \midrule
    \textbf{1)} SER (\textit{baseline}) & MFCC + Ground-truth transcription & Concatenation & 66.5\% \\
     & & Co-attention & 67.3\% \\
    & eGeMAPS + Ground-truth transcription & Concatenation & 67.4\% \\
     & & Co-attention & \textbf{68.4}\% \\ \hdashline
    \textbf{2)} ASR-SER & MFCC + Middle layer reps. & Concatenation & 63.9\% \\
     & & Co-attention & 64.4\% \\
    & eGeMAPS + Middle layer reps. & Concatenation & 66.9\% \\
     & & Co-attention & 67.7\% \\ \hdashline
    \textbf{3)} ASR-SER & MFCC + ASR text output & Concatenation & 63.7\% \\
     & & Co-attention & 64.0\% \\
    & eGeMAPS + ASR text output & Concatenation & 66.7\% \\
     & & Co-attention & 67.0\% \\ \hdashline
    \textbf{4)} ASR-SER & MFCC + Middle layer reps. + ASR text output & Concatenation & 64.1\% \\
     & & Co-attention & 64.5\% \\
     & & Hierarchical co-attention & 66.1\% \\
    & eGeMAPS + Middle layer reps. + ASR text output & Concatenation & 67.1\% \\
     & & Co-attention & 67.5\% \\
     & & Hierarchical co-attention & \textbf{68.4\%} \\ \bottomrule
    \end{tabular}
    }
\end{table}

The experimental results are presented in Table~\ref{chap3/tab:result}. We set $\lambda$ = 0.1, as it provides the best performance. The results include: \textbf{1)} the baseline model, which combines acoustic features with human transcription in the SER model (i.e., without ASR); the joint ASR-SER model comprising \textbf{2)} acoustic features + ASR hidden output; \textbf{3)} acoustic features + text output; and \textbf{4)} our proposed full model. We utilized the middle-layer ASR output from the W2V2 model, as it achieves the best SER performance according to Table~\ref{chap3/fig:acc}.

We notice that:

\begin{enumerate}
    \item Both ASR hidden output and text output help improve the SER performance over the acoustic features, although the difference is small with ASR hidden output slightly better than ASR text output. This is consistent with a previous claim in the literature that ``ASR features can be more robust than the text output of ASR'' \citep{feng2020end}.
    \item In general, the performance of co-attention fusion outperforms that of concatenation. It appears that the relatedness of two input channels is learned by attention.
    \item The eGeMAPS paralinguistic features contributes slightly more than MFCCs when combining with ground-truth transcription. However, when combining with the middle layer representations, the difference becomes larger. This is plausible as W2V2 representations lack some paralinguistic information as shown in Figure~\ref{chap3/fig:cca} in the previous section. Thus, combining both makes the features more comprehensive for SER.
    \item Our proposed approach: the joint ASR-SER model incorporating both ASR hidden output and text output using hierarchical co-attention achieves the same UA as the eGeMAPS + ground-truth co-attention result. This indicates that our approach can help ameliorate ASR errors by combining paralinguistic features, hidden ASR features, and ASR text output.
\end{enumerate}

To better understand the internal representations of the W2V2 model and their relevance to SER, we conducted a detailed analysis focusing on the contribution of different W2V2 layers. While Chapter~\ref{chap3:acoustic} presented a layer-wise study on SER using W2V2 features, here we take a look into how it functions with joint ASR fine-tuning by comparing the hidden-state outputs from each layer. As shown in Table~\ref{chap3/tab:layer}, the results align with the findings in Figure~\ref{chap3/fig:acc} where the SER accuracy increases till the middle layer then drops.

Additionally, the overall SER performance in this joint training setup is noticeably lower than the results reported in Figure~\ref{chap3/fig:acc}, where SER was performed independently without simultaneous ASR tuning. We hypothesize that this degradation stems from the impact of the ASR training in the joint optimization process. In other words, since the model architecture is built upon W2V2, the emotional characteristics (such as voice quality features) in the representation space may be suppressed because it is primarily fine-tuned for speech recognition. This phenomenon highlights an inherent trade-off in multitask learning setups, where optimizing for one task, in this case, word recognition, may inadvertently diminish the model’s capacity to capture fine-grained emotional cues.

\begin{table}
\centering
\caption{Layer-wise analysis results. $\lambda$ = 0.1.}
\label{chap3/tab:layer}
\begin{tabular}{llc}
\toprule
\textbf{Model} & \textbf{Hidden Output} & \textbf{UA} \\ \midrule
ASR-SER & layer 0 reps. & 59.2\% \\
 & layer 1 & 60.4\% \\
 & layer 2 & 60.8\% \\
 & layer 3 & 61.5\% \\
 & layer 4 & 62.8\% \\
 & layer 5 & 63.5\% \\
 & layer 6 & \textbf{64.4}\% \\
 & layer 7 & 64.0\% \\
 & layer 8 & 63.0\% \\
 & layer 9 & 62.5\% \\
 & layer 10 & 62.1\% \\
 & layer 11 & 55.3\% \\
 & layer 12 & 51.9\% \\ \bottomrule
\end{tabular}
\end{table}

To verify our hypothesis, we investigate the impact of the weight of the ASR training $\lambda$ using the hierarchical co-attention model proposed, as it performs best. As seen in Table~\ref{chap3/tab:asrloss}, the SER accuracy (i.e., UA) drops with $\lambda$ increases, whereas WER barely changes. It is likely because the smaller the ASR training weight, the longer time ASR reaches convergence so that the W2V2 representations can better be adapted to SER task. On the other hand, as it takes only two to three epochs for ASR training to converge as we noted during the experiments, $\lambda$ barely impacts the ASR performance (i.e., WER). In contrast to \cite{feng2020end} which noted a large WER drop using a conventional encoder-decoder ASR model in joint ASR-SER training, wav2vec 2.0 demonstrated the potential for achieving satisfactory performance in both tasks through speech foundation models.

\begin{table}
\centering
\caption{The impact of ASR training weight $\lambda$.}
\label{chap3/tab:asrloss}
\begin{tabular}{llc}
\toprule
\textbf{$\lambda$} & \textbf{UA} & \textbf{WER} \\ \midrule
0.1 & \textbf{68.4\%} & 18.9\% \\
0.3 & 67.2\% & 18.8\% \\
0.5 & 66.6\% & 18.8\% \\
0.7 & 66.0\% & 18.9\% \\
0.9 & 64.7\% & 18.8\% \\ \bottomrule
\end{tabular}
\end{table}

\subsection{Summary}
Following the experiments in Section~\ref{chap3:acoustic}, this section extends the analysis of wav2vec 2.0 by integrating its ASR training purpose into a joint training with SER. We compared the joint ASR-SER performance using different acoustic inputs and ASR outputs, as well as investigated the effectiveness of layer-wise hidden output and the impact of ASR training. The findings contribute to better understanding of the relationship between ASR and SER through speech foundation models.

\section{Chapter Summary}
\label{chap3:summary}
This chapter presents a comprehensive investigation into the relationship between the speech foundation model wav2vec 2.0 and SER. Section~\ref{chap3:acoustic} conducts a series of probing experiments on emotional speech using W2V2. Building on the findings from Section~\ref{chap3:acoustic}, Section~\ref{chap3:hierarchical} introduces a joint ASR-SER training framework that incorporates ASR outputs into SER and integrates paralinguistic features to address the paralinguistic shortage of wav2vec 2.0. A limitation of this chapter is its exclusive focus on the wav2vec 2.0 series. Although other self-supervised models may exhibit similar behavior, further investigation is worthwhile for comparison. Additionally, the supervised ASR model Whisper is expected to demonstrate distinct patterns. Nevertheless, the work presented in this chapter lays the groundwork for exploring ASR models for SER, providing a foundation for leveraging the strengths of ASR while mitigating its weaknesses.

The next chapter builds upon this one by shifting focus from acoustic to linguistic aspects, emphasizing the analysis of ASR lexical output.

\chapter{Utilizing ASR Transcription for SER}
\label{chap4}

\section{Introduction}
Recent work on SER consistently shows the benefits of incorporating both textual and acoustic features on benchmark corpora \citep{sebastian2019fusion, yoon2018multimodal}. Although some researchers have also attempted to translate this approach into the wild, e.g., in-car voice systems \citep{li2021feeling}, it is still rare to see SER applications in our daily lives. One of the major reasons is that the majority of SER research uses human annotation, i.e., gold-standard manual transcriptions. In contrast, even for the `lab' emotion corpora, transcriptions from a SOTA ASR system can result in high WERs. This means that very few of the findings obtained in the lab can be replicated in the wild.

To bridge this gap, it is necessary to exploit the imperfect textual data generated by ASR for emotion recognition. While previous studies proposed using ASR transcriptions for SER \citep{sahu19_interspeech}, the effect of ASR on emotion and vice-versa is not always clear. For example, a high WER on emotional speech is generally assumed to be a result of the distortion relative to neutral speech \citep{fernandez2004computational}. Thus, even though ASR is a relatively well studied area, key questions about how it can be used in SER remain unresolved, such as 1) how confident can we be of ASR on different emotions? and 2) how is the SER performance affected by ASR errors? Thus, to make true progress in SER, we need to understand the interrelationship between ASR and SER.

In this section, we look at these fundamental yet long-standing issues by exploring the relationship between ASR results and emotional speech from the linguistic perspective, with the objective of pushing SER research closer to realistic use scenarios. Specifically, we first take a deep look into the word distribution of emotion corpora and how they are misrecognized in speech recognition. We analyze WER variation to see how ASR performance is affected by different emotions and how this relates to word-level confidence scores.

\section{Exploring ASR Transcription of Emotional Speech}
\label{chap4:asremotion}
\subsection{Experimental Setup}
\subsubsection{Datasets}
We use three corpora: IEMOCAP, CMU-MOSI, and MELD to cover as many speech conditions as possible for generalizability. For IEMOCAP (IEM), we use four emotion classes: \textit{angry, happy (+excited), neutral, and sad}. For IEM and MELD, we removed utterances whose transcription is blank or whose audio file is too long due to mis-processing in corpora construction, bringing the total utterance numbers to 5 500 and 13 689, respectively.

\subsubsection{ASR Models}
For the ASR models, we use the Kaldi Librispeech ASR model\footnote{\url{https://kaldi-asr.org/models/m13}} (KLA), a self-supervised model: wav2vec2-base-960h\footnote{\url{https://huggingface.co/facebook/wav2vec2-base-960h}} (W2V2), a Conformer model\footnote{\url{https://github.com/espnet/espnet/blob/master/egs/librispeech/asr1/RESULTS.md}} (CONF) from ESPnet \citep{watanabe2018espnet}, and the medium model of Whisper\footnote{\url{https://openai.com/research/whisper}} (WHIS). KLA, W2V2, and CONF are pre-trained on Librispeech960, and WHIS on 680\,000~h of multilingual and multitask data collected from the web. The ASR systems output whole words instead of characters. We use these four models for generalizability.

\subsection{Experimental Results}
In this section, we analyze the influence of emotion on ASR by conducting a series of experiments to identify sources of errors across corpora and word classes and quantifying the effect of WER on SER performance.

\subsection{WER on Emotional Speech}

\begin{table}[ht]
\centering
\caption{WER (\%) of the ASR models on the emotion corpora.}
\label{chap4:wer}
\begin{tabular}{lccc}
\hline
 & \textbf{IEM} & \textbf{MOSI} & \textbf{MELD} \\ \hline
KLA & 36.8 & 40.9 & 58.5 \\ 
W2V2 & 32.7 & 35.4 & 57.8 \\ 
CONF & 27.1 & 30.1 & 52.1 \\ 
WHIS & 12.3 & 17.3 & 34.8 \\ \hline
\end{tabular}
\end{table}

\begin{figure*}[ht]
  \centering
  \includegraphics[width=0.5\textwidth]{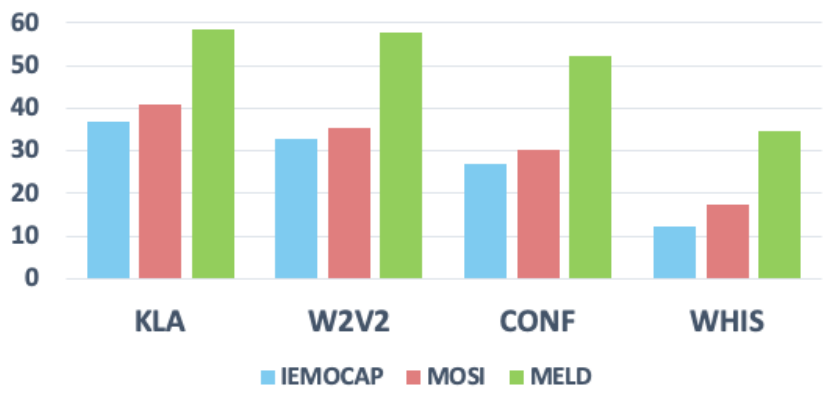}
  \caption{WER (\%) of the ASR models on the emotion corpora.}
  \label{chap4/fig:wer}
\end{figure*}

We compute the WERs of the four ASR models on the three emotion corpora as the basis for analysis, as shown in Table~\ref{chap4:wer} and Figure~\ref{chap4:wer}. Among the four models, WHIS shows the best performance on all the corpora, greatly outperforming the other three. We see large performance differences among corpora, even though they are all US or UK English emotional speech. This could be due to the different recording settings, the noise that occurred, or the vocabulary that was said.

\subsection{Exploring ASR Errors Based on Part-of-Speech}
\label{chap4/sec:pos}
In addition to the well-known acoustic factors that affect ASR (e.g., recording quality, noise), we would like to know if lexical factors, such as word distribution also affects WER. To investigate this, we analyze the words in each corpus based on two aspects: Part-of-Speech (PoS) tag and affective score. In particular, we explore whether words of a particular class or with strong affective coloring have a strong impact on ASR performance.

\textbf{PoS.} We select seven classes of PoS tags: \textit{Noun, Verb, Adjective, Adverb, Wh-word, Function-word} from the Penn Treebank \citep{marcus1993building} and \textit{Stop-word} from the stopwords corpus of NLTK \citep{bird2006nltk}. The tagging was also conducted using NLTK.

\textbf{Affective score.} We use an affective words database that has nearly 14 thousand English words \citep{warriner2013norms} to refer the affective scores. Each word is rated in three affective dimensions: Valence (V), Arousal (A), and Dominance (D), ranging from 1 (weak) to 9 (strong). We divide the scores into three classes: low (1-3), medium (4-6), and high (7-9) according to their overall mean value, respectively.

We define three metrics to see how different classes of words affect the ASR performance. The Word Ratio (WR) is used to see the distribution of different classes of words in a corpus. The Error Ratio (ER) indicates if misrecognized words largely come from a specific class. The Class Error Rate (CR) measures how difficult words in a class are to recognize. We define them as follows.

\begin{equation}
\text{word ratio} = \dfrac{\text{word count per class}}{\text{total word count}}
\end{equation}

\begin{equation}
\text{error ratio} = \dfrac{\text{word error count per class}}{\text{total word error count}}
\end{equation}

\begin{equation}
\text{class error rate} = \dfrac{\text{word error count per class}}{\text{total word count per class}}
\end{equation}

\begin{figure*}[ht]
  \centering
  \includegraphics[width=0.65\textwidth]{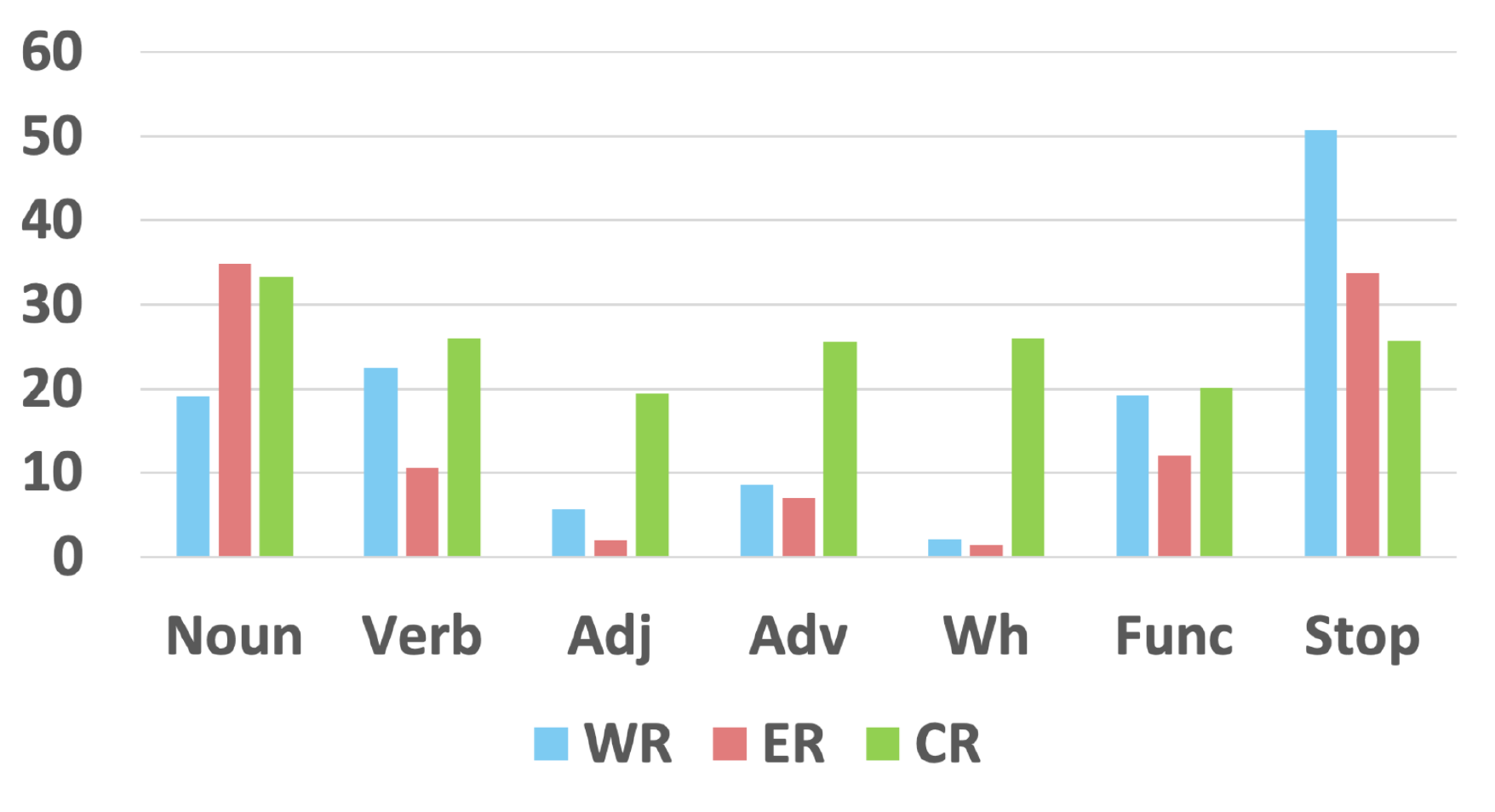}
  \caption{WR, ER, and CR (all in \%) based on PoS tag (IEMOCAP).}
  \label{chap4/fig:pos}
\end{figure*}

The WR\footnote{Note that \textit{Stop} words can have multiple PoS tags as \textit{Stop} is not in the Penn Treebank, making the summation of the word ratios greater than 100\%.}, ER, and CR based on PoS tag and affective score are shown in Table~\ref{chap4/tab:pos} and \ref{chap4/tab:affect}, respectively, with illustrative examples from IEMOCAP shown in Figure~\ref{chap4/fig:pos} and Figure~\ref{chap4/fig:affect}. For brevity, we averaged the values over the four ASR models. As the ER values from the four ASR models are close and the CR values are proportionally distributed, taking the average does not affect the general findings.

\begin{table}[ht]
\caption{WR, ER, and CR (all in \%) based on PoS tag.}
\centering
\label{chap4/tab:pos}
\begin{tabular}{lrrrrrrrrr}
\hline
 & \multicolumn{3}{c}{\textbf{IEM}} & \multicolumn{3}{c}{\textbf{MOSI}} & \multicolumn{3}{c}{\textbf{MELD}} \\
 & WR & ER & CR & WR & ER & CR & WR & ER & CR \\ \hline
\textit{Noun} & 19.1 & 34.8 & \multicolumn{1}{r|}{33.3} & 21.7 & 36.8 & \multicolumn{1}{r|}{45.3} & 25.3 & 44.0 & 58.0 \\
\textit{Verb} & 22.5 & 10.6 & \multicolumn{1}{r|}{26.0} & 19.3 & 8.4 & \multicolumn{1}{r|}{25.9} & 20.3 & 8.2 & 34.5 \\
\textit{Adj} & 5.7 & 2.0 & \multicolumn{1}{r|}{19.4} & 10.6 & 3.5 & \multicolumn{1}{r|}{27.5} & 5.7 & 2.0 & 32.1 \\
\textit{Adv} & 8.6 & 7.0 & \multicolumn{1}{r|}{25.6} & 8.6 & 5.1 & \multicolumn{1}{r|}{23.5} & 7.6 & 5.9 & 38.8 \\
\textit{Wh} & 2.1 & 1.4 & \multicolumn{1}{r|}{26.0} & 1.7 & 0.7 & \multicolumn{1}{r|}{21.0} & 2.4 & 1.8 & 44.1 \\
\textit{Func} & 19.2 & 12.1 & \multicolumn{1}{r|}{20.1} & 25.1 & 15.6 & \multicolumn{1}{r|}{24.1} & 17.5 & 10.2 & 30.3 \\
\textit{Stop} & 50.7 & 33.7 & \multicolumn{1}{r|}{25.7} & 50.0 & 30.0 & \multicolumn{1}{r|}{24.8} & 46.8 & 29.5 & 33.7 \\ \hline
\end{tabular}
\end{table}

Table~\ref{chap4/tab:pos} shows that ER generally increases with WR. This is reasonable, as the more words in one class, the more errors will be generated. However, this does not hold for all PoS classes: the WRs for \textit{Noun}, \textit{Verb} and \textit{Func} are similar, but the ER is much lower for \textit{Verbs}. That is, ASR performs better on verbs than expected based on verb frequency. On IEM, \textit{Verb} has a higher WR than \textit{Noun} and \textit{Func} but much lower ER, and similarly for \textit{Adj} and \textit{Adv} on MOSI. This indicates that in ASR models, some classes of words are more difficult to recognize than others. The \textit{Noun} class is the most difficult to recognize, showing the highest CRs across the corpora. However, other word classes do not have uniform patterns across the corpora. For instance, \textit{Wh} words are the easiest to recognize on MOSI yet the second hardest on MELD. 

The CRs of most word classes increase from IEM to MELD, which is in line with the change of WERs in Table~\ref{chap4:wer}. This is plausible as the word error count increases with the WER. Nevertheless, the rate of growth of CR varies from class to class. For example, \textit{Noun} saw a 74\% increase in CR (33.3 to 58.0) compared to 33\% for \textit{Verb} (26.0 to 34.5). Again, this reflects the fact that word classes have different ERs, perhaps because they have different tolerances for speech distortion.

\subsection{Exploring ASR Errors Based on Affective Scores}

\begin{table}[ht]
\caption{WR, ER, and CR (all in \%) based on affective score.}
\centering
\label{chap4/tab:affect}
\begin{tabular}{lrrrrrrrrr}
\hline
 & \multicolumn{3}{c}{\textbf{IEM}} & \multicolumn{3}{c}{\textbf{MOSI}} & \multicolumn{3}{c}{\textbf{MELD}} \\
 & WR & ER & CR & WR & ER & CR & WR & ER & CR \\ \hline
$V_{low}$ & 2.5 & 3.6 & \multicolumn{1}{r|}{36.1} & 2.3 & 3.6 & \multicolumn{1}{r|}{43.2} & 2.0 & 2.9 & 43.0 \\
$V_{mid}$ & 10.8 & 13.8 & \multicolumn{1}{r|}{29.6} & 9.3 & 11.7 & \multicolumn{1}{r|}{33.9} & 9.8 & 14.2 & 37.7 \\
$V_{high}$ & 13.9 & 16.2 & \multicolumn{1}{r|}{26.3} & 15.6 & 18.9 & \multicolumn{1}{r|}{34.3} & 12.2 & 16.2 & 35.5 \\
$A_{low}$ & 15.0 & 17.4 & \multicolumn{1}{r|}{24.8} & 12.3 & 13.8 & \multicolumn{1}{r|}{30.9} & 13.2 & 17.3 & 35.0 \\
$A_{mid}$ & 11.1 & 14.7 & \multicolumn{1}{r|}{30.7} & 13.6 & 17.5 & \multicolumn{1}{r|}{36.5} & 10.1 & 15.1 & 39.4 \\
$A_{high}$ & 1.0 & 1.4 & \multicolumn{1}{r|}{32.8} & 1.3 & 2.0 & \multicolumn{1}{r|}{43.6} & 0.7 & 1.0 & 40.0 \\
$D_{low}$ & 1.0 & 1.4 & \multicolumn{1}{r|}{28.1} & 1.1 & 1.8 & \multicolumn{1}{r|}{47.2} & 0.8 & 1.2 & 38.9 \\
$D_{mid}$ & 16.7 & 20.5 & \multicolumn{1}{r|}{27.7} & 15.6 & 20.0 & \multicolumn{1}{r|}{36.2} & 15.0 & 20.9 & 36.9 \\
$D_{high}$ & 9.4 & 11.5 & \multicolumn{1}{r|}{27.4} & 10.6 & 11.5 & \multicolumn{1}{r|}{29.2} & 8.1 & 11.2 & 36.7 \\ \hline
\end{tabular}
\end{table}

\begin{figure*}[ht]
  \centering
  \includegraphics[width=0.8\textwidth]{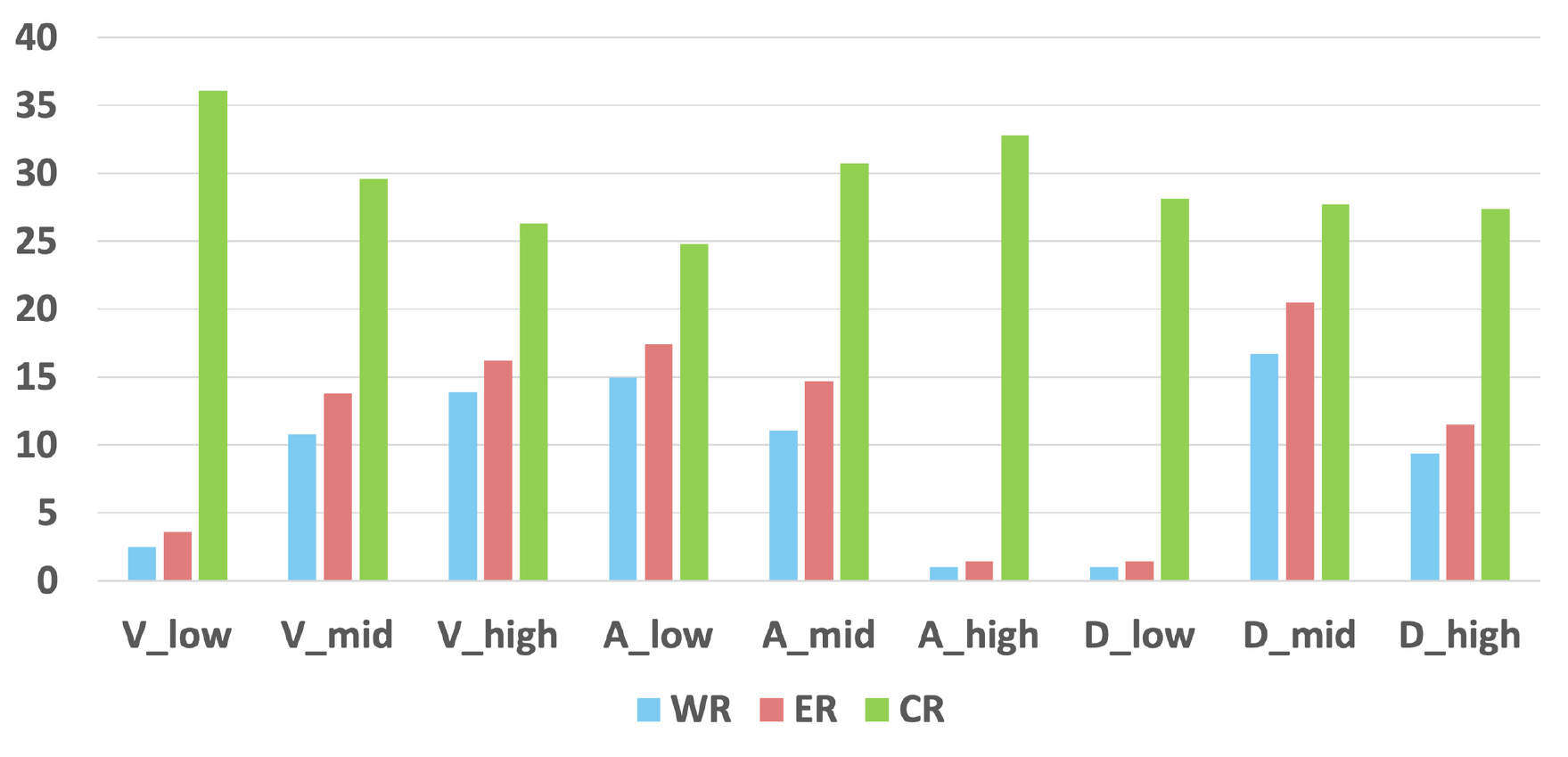}
  \caption{WR, ER, and CR (all in \%) based on affective score (IEMOCAP).}
  \label{chap4/fig:affect}
\end{figure*}

Looking at the affective score-based distribution in Table~\ref{chap4/tab:affect} and Figure~\ref{chap4/fig:affect} and again see that, in general, the higher the WR, the higher the ER. However, in this case, we do not see classes with similar WRs but very different ERs. Nevertheless, differences still remain: CR decreases as valence and dominance scores increase, i.e., high valence and dominance words are better recognized. In contrast, we see that higher-arousal words are more difficult to recognize. Even so, unlike the classes based on PoS tags, the CRs have a relatively balanced distribution here, and their rates of increase with the WER are at the same level.

We see similar but not identical patterns when we look at the results for individual ASR models. We omit detailed results for brevity but note per-class differences in performance, e.g., \textit{Nouns} are better recognized by KLA (32\% ER) compared to CONF (35\% ER) even though the latter has a lower overall WER. In general, no one model performed the best across all PoS or affective score classes.

\subsection{Exploring WER with Utterance Length}
One of the major factors that is likely to have an effect on ASR performance is utterance length. Generally, longer utterances are more likely to have more word errors compared to shorter ones because there are more opportunities for the ASR model to make errors \citep{sahu19_interspeech}. However, we hypothesize that longer utterances contain more contextual information that certain ASR models (e.g., W2V2) can use to compensate for the speech distortion.

\begin{figure*}[ht]
  \centering
  \includegraphics[width=0.65\textwidth]{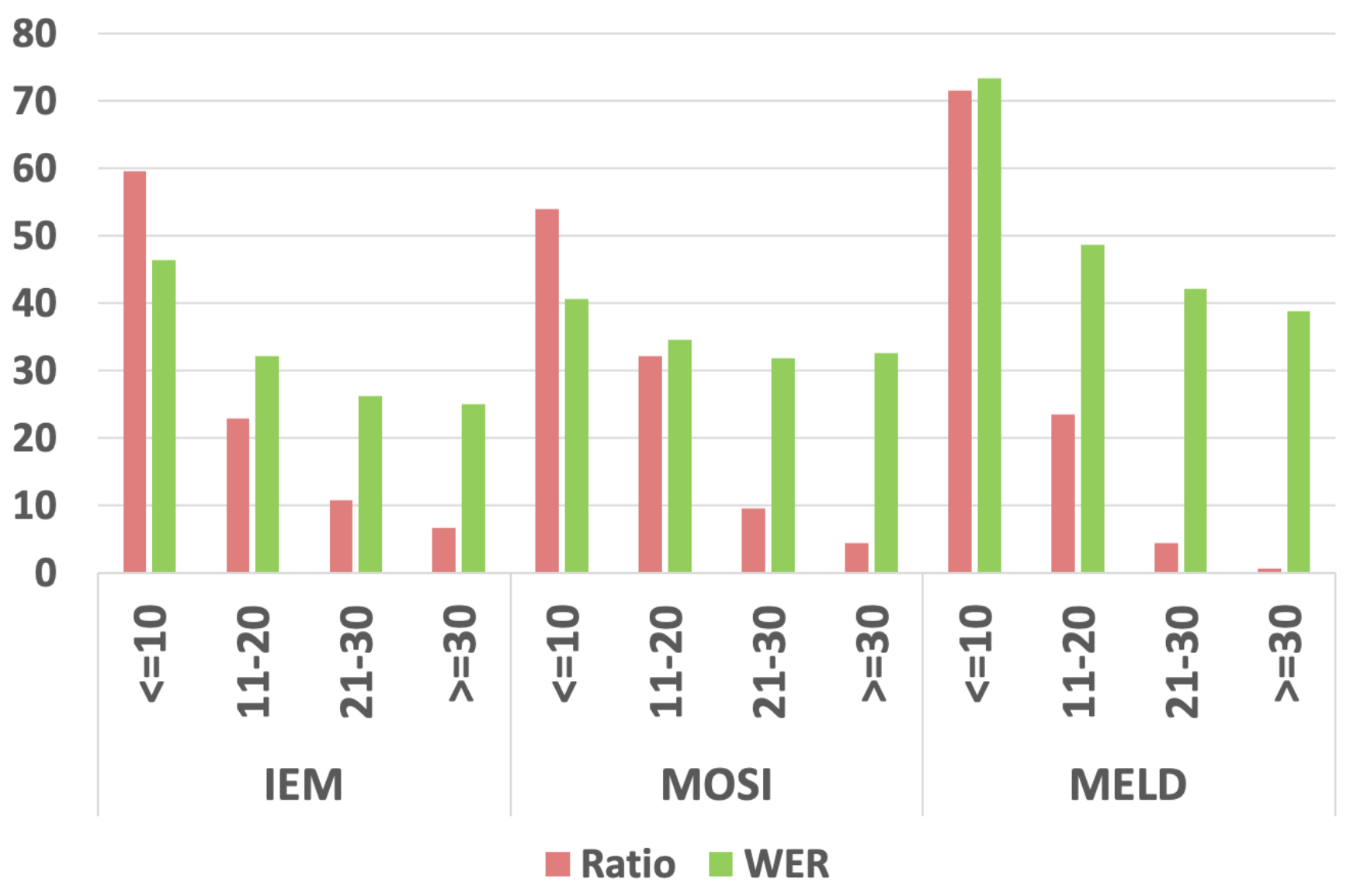}
  \caption{WER and ratio (both in \%) according to different utterance lengths (number of words N).}
  \label{chap4/fig:wordlength}
\end{figure*}

To investigate this, we analyze how WER and WR change with utterance length. The four ASR models displayed the same trends, so Table~\ref{chap4/tab:wordlength} and Figure~\ref{chap4/fig:wordlength} shows results for W2V2 for brevity. We see that the shorter the utterances, the higher the WER, regardless of corpus. These support our hypothesis and additionally show that if a corpus contains many short utterances, ASR may not work well.

\begin{table}[ht]
\centering
\caption{WER and ratio (both in \%) according to different utterance lengths (number of words N).}
\label{chap4/tab:wordlength}
\begin{tabular}{lrrrrrr}
\hline
 & \multicolumn{2}{c}{\textbf{IEM}} & \multicolumn{2}{c}{\textbf{MOSI}} & \multicolumn{2}{c}{\textbf{MELD}} \\
N & Ratio & WER & Ratio & WER & Ratio & WER \\ \hline
$\le$10 & 59.6 & 46.4 & 53.9 & 40.6 & 71.5 & 73.3 \\
11-20 & 22.9 & 32.2 & 32.2 & 34.5 & 23.5 & 48.6 \\
21-30 & 10.8 & 26.3 & 9.5 & 31.9 & 4.4 & 42.1 \\
$\ge$30 & 6.7 & 25.0 & 4.4 & 32.6 & 0.6 & 38.8 \\ \hline
\end{tabular}
\end{table}

\subsection{Exploring ASR Performance on Different Emotions}
It has long been argued that the acoustic characteristics of emotional speech, e.g., prosody variation, can deteriorate ASR performance \citep{goldwater2010words}. Word-level factors, however, are rarely studied. Based on the findings from the above experiments, we conducted a word-level analysis based on different emotions. Specifically, we look at the WER alongside the ratio of \textit{Noun} and short utterance (word count less than 10), which we found to be major factors affecting ASR performance above.

As MOSI does not contain discrete emotion labels, we omit it in this exploration task. From Table~\ref{chap4/tab:asremotions}, we can see the WER on Neutral is not the best but instead worse than most of the other emotions in both corpora. As neutral speech is the least emotionally ``distorted'', the reason could be that the ratio of short utterances in Neutral is very high, which offsets the benefit of less emotional distortion. We did not see a large discrepancy in the ratio of \textit{Noun} between emotions, although we found that a high ratio harms ASR performance in Section~\ref{chap4/sec:pos}.

\begin{table}[ht]
\centering
\caption{WER, ratio of Noun words, and short utterances (all in \%) according to different emotions on IEM and MELD.}
\label{chap4/tab:asremotions}
\begin{tabular}{lcccc}
\hline
\textit{IEM} & \textbf{Ang} & \textbf{Hap} & \textbf{Neu} & \textbf{Sad} \\ \hline
WER & 22.8 & 38.9 & 36.3 & 29.5 \\ 
Noun & 20.0 & 21.3 & 20.4 & 19.5 \\ 
$\le$10 & 44.8 & 55.7 & 60.5 & 52.7 \\ \hline
\end{tabular}
\end{table}
\begin{table}[ht]
\centering
\begin{tabular}{lccccccc}
\hline
\textit{MELD} & \textbf{Ang} & \textbf{Dis} & \textbf{Fea} & \textbf{Joy} & \textbf{Neu} & \textbf{Sad} & \textbf{Sur} \\ \hline
WER & 52.7 & 53.4 & 58.6 & 59.9 & 58.3 & 52.9 & 65.3  \\ 
Noun & 26.2 & 27.6 & 26.4 & 29.7 & 26.9 & 25.2 & 24.5 \\ 
$\le$10 & 64.1 & 60.1 & 62.1 & 72.4 & 72.8 & 62.0 & 82.8 \\ \hline
\end{tabular}
\end{table}

\begin{figure*}[ht]
  \centering
  \includegraphics[width=0.95\textwidth]{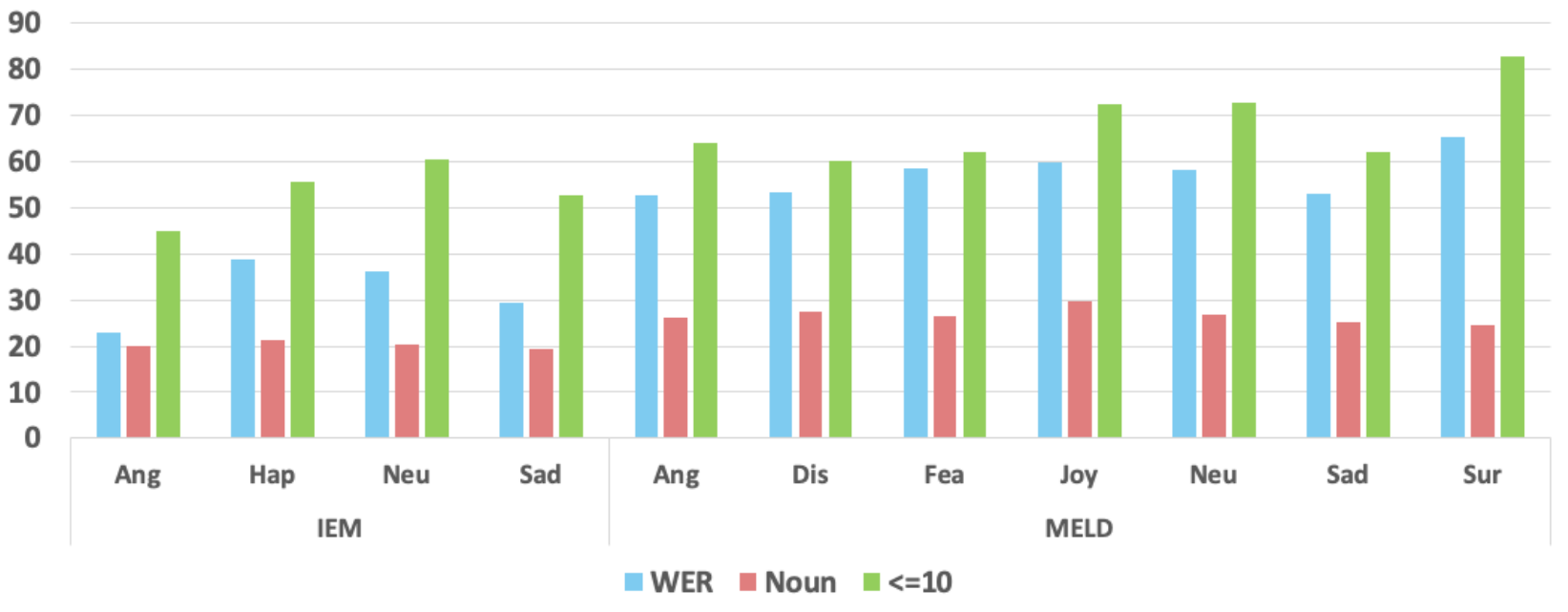}
  \caption{WER, ratio of Noun words, and short utterances (all in \%) according to different emotions on IEM and MELD.}
  \label{chap4/fig:asremotions}
\end{figure*}

\subsection{Exploring ASR Confidence Scores Across Emotions}
Confidence scores can help identify words of high recognition quality for better use of ASR transcriptions. \cite{santoso2021speech} and \cite{santoso2022speech} adjusted the attention weights for words using their confidence scores. \cite{pan2020improving} proposed removing words with low confidence and selecting words with the highest confidence from multiple ASR hypotheses \citep{pan2021using}. However, the measurement and use of confidence scores is a complicated task due to many reasons such as out-of-domain words and overconfidence \citep{qiu2021learning,li2022improving}. Hence, we analyze how confidence varies with emotion by exploring the word confidence scores. We compared two ASR models: CONF and KLA (as the confidence scores produced by the two models are more accessible), and also compared the confidence scores generated by KLA with and without calibration. The calibration was applied by using the official procedure in Kaldi\footnote{\url{https://github.com/kaldi-asr/kaldi/blob/master/egs/tedlium/s5/local/confidence_calibration.sh}}. Since we noticed similar patterns of confidence in all three corpora, we only show the results of IEM for brevity.

In Table~\ref{chap4/tab:confidence}, we observe a great difference in confidence scores between ASR models. The scores for correctly recognized words are similar between CONF and KLA, but for incorrectly recognized words, the score is 0.70 in KLA compared to 0.87 in CONF. Overall, KLA scores are generally lower than those of CONF. Additionally, in the emotions Angry, Neutral, and Sad, confidence scores increase with utterance length: the longer the utterance, the higher the confidence, which aligns with the WER trend shown in Table~\ref{chap4/tab:wordlength}. However, this pattern does not appear in Angry.

\begin{table}[ht!]
\centering
\caption{Distribution of confidence scores of the ASR models. (The CONF column is illustrated in Figure~\ref{chap4/fig:CONF}. The last two columns are illustrated in Figure~\ref{chap4/fig:KLA}).}
\label{chap4/tab:confidence}
\begin{tabular}{lcccc}
\hline
\multicolumn{1}{l}{} & \textbf{N} & \textbf{CONF} & \textbf{KLA} & \textbf{KLA (calibrated)} \\ \hline
Correct &  & 0.97 & 0.95 & 0.62 \\
Incorrect &  & 0.87 & 0.70 & 0.53 \\ \hline
Ang & \begin{tabular}[c]{@{}c@{}}$\le$10\\ 11-20\\ 21-30\\ $\ge$30\end{tabular} & \begin{tabular}[c]{@{}c@{}}0.96\\ 0.96\\ 0.96\\ 0.96\end{tabular} & \begin{tabular}[c]{@{}c@{}}0.91\\ 0.91\\ 0.90\\ 0.90\end{tabular} & \begin{tabular}[c]{@{}c@{}}0.62\\ 0.61\\ 0.61\\ 0.61\end{tabular} \\ \hline
Hap & \begin{tabular}[c]{@{}c@{}}$\le$10\\ 11-20\\ 21-30\\ $\ge$30\end{tabular} & \begin{tabular}[c]{@{}c@{}}0.92\\ 0.94\\ 0.95\\ 0.94\end{tabular} & \begin{tabular}[c]{@{}c@{}}0.80\\ 0.83\\ 0.86\\ 0.85\end{tabular} & \begin{tabular}[c]{@{}c@{}}0.57\\ 0.58\\ 0.59\\ 0.58\end{tabular} \\ \hline
Neu & \begin{tabular}[c]{@{}c@{}}$\le$10\\ 11-20\\ 21-30\\ $\ge$30\end{tabular} & \begin{tabular}[c]{@{}c@{}}0.91\\ 0.94\\ 0.95\\ 0.96\end{tabular} & \begin{tabular}[c]{@{}c@{}}0.81\\ 0.86\\ 0.88\\ 0.89\end{tabular} & \begin{tabular}[c]{@{}c@{}}0.58\\ 0.59\\ 0.59\\ 0.60\end{tabular} \\ \hline
Sad & \begin{tabular}[c]{@{}c@{}}$\le$10\\ 11-20\\ 21-30\\ $\ge$30\end{tabular} & \begin{tabular}[c]{@{}c@{}}0.92\\ 0.95\\ 0.96\\ 0.97\end{tabular} & \begin{tabular}[c]{@{}c@{}}0.81\\ 0.87\\ 0.89\\ 0.92\end{tabular} & \begin{tabular}[c]{@{}c@{}}0.58\\ 0.59\\ 0.60\\ 0.61\end{tabular} \\ \hline
\end{tabular}
\end{table}

\begin{figure*}[ht!]
  \centering
  \includegraphics[width=0.75\textwidth]{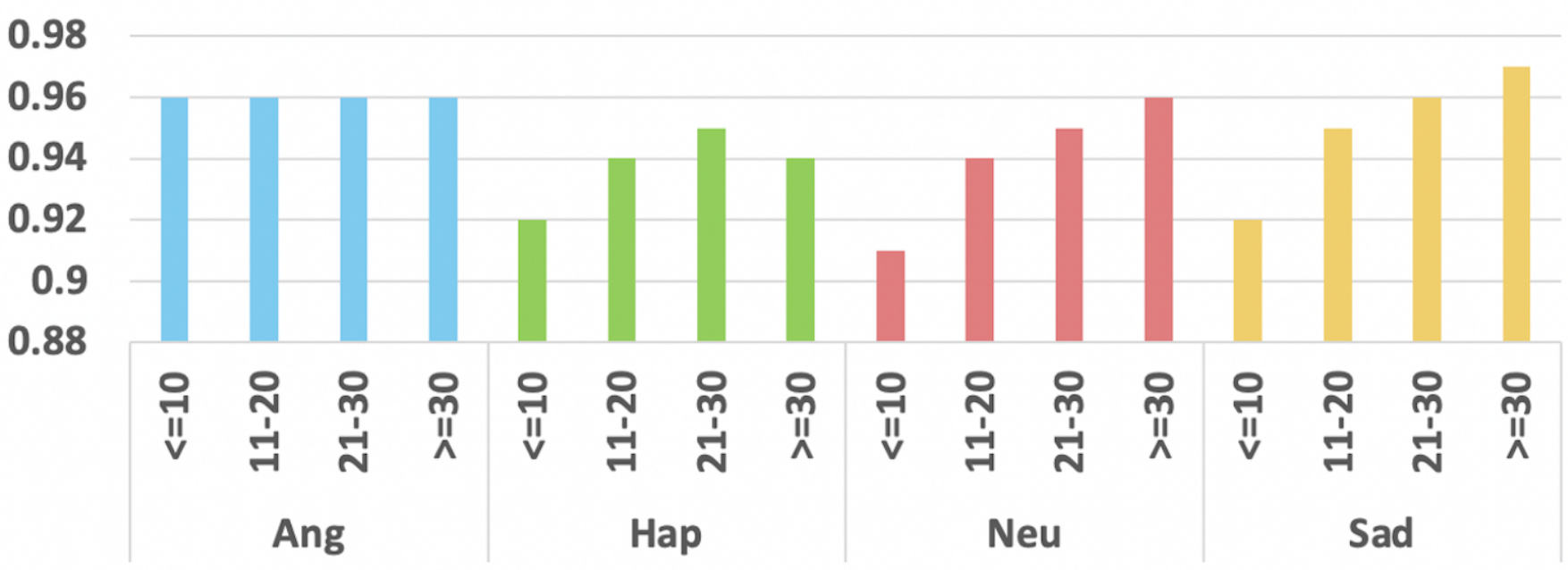}
  \caption{Distribution of confidence scores of CONF model.}
  \label{chap4/fig:CONF}
\end{figure*}

\begin{figure*}[ht!]
  \centering
  \includegraphics[width=0.8\textwidth]{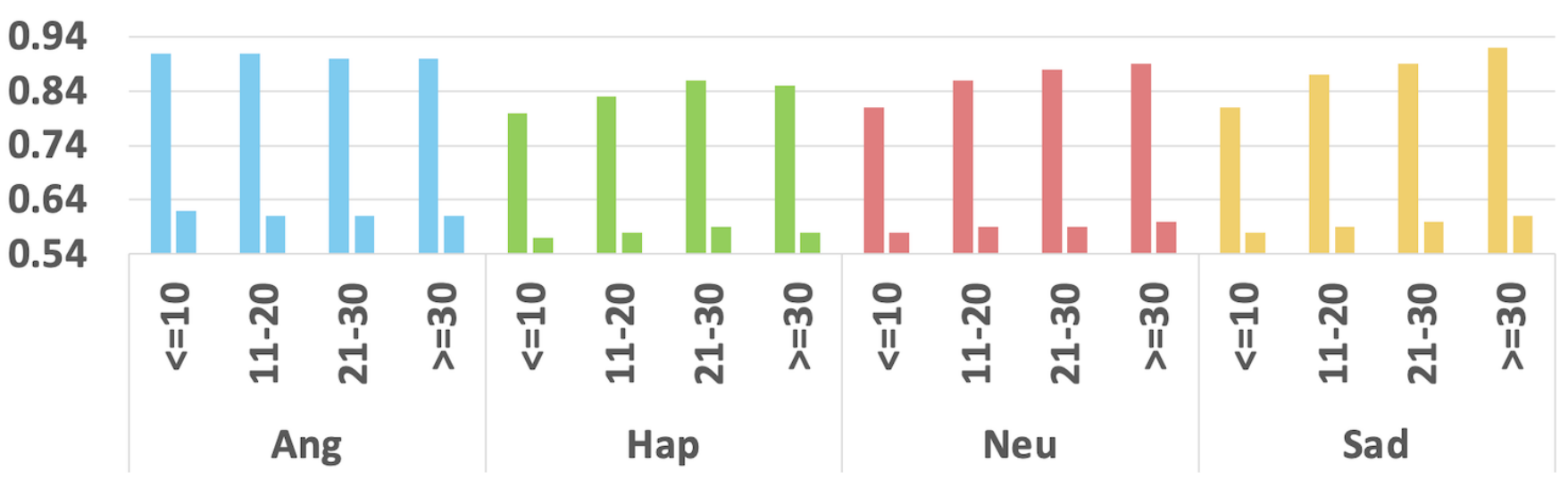}
  \caption{Distribution of confidence scores of KLA model w/o and w/ calibration (left and right per utterance length).}
  \label{chap4/fig:KLA}
\end{figure*}

By comparing the confidence scores with and without calibration for KLA, we observe that the gap in confidence between correctly and incorrectly recognized words becomes smaller. Similarly, the influence of utterance length nearly disappears, indicating that emotions are treated more equally after calibration. Calibration also mitigates the overconfidence commonly exhibited by end-to-end ASR models \citep{li2021confidence}. The overall confidence for each emotion corresponds to their respective WERs in Table~\ref{chap4/tab:asremotions}, with ASR performance being the best on Angry and the worst on Happy. These findings highlight that confidence scores vary with emotion, utterance length, and ASR model. Therefore, these variations need to be carefully considered and confidence scores properly calibrated when they are used to adjust word contributions in SER.

\subsection{Summary}
In this section, we addressed a long-existing yet understudied issue: the relationship between ASR lexical output and emotion. We demonstrated that word distribution in a corpus is a key factor that affects ASR performance and that differences exist between emotions. The frequencies of different PoS tags, affective scores, and utterance lengths may have a large impact on ASR performance. For example, the more the Noun words and short utterances, the higher the WER, regardless of the speaking style (speech collected in the lab, from monologues or TV shows). Moreover, different ASR models behave similarly in these trends, but with minor differences: some classes of words are easier to recognize for one model but maybe harder for another. Moreover, the confidence scores generated by different models and with calibration may show large differences, so the use of confidence scores should be carefully examined. We expect these findings to provide a valuable foundation for future SER research using ASR transcription, such as knowledge integration for LLM prompting.

\section{Impact of ASR Transcription Quality on SER}
\label{chap4:asrquality}
After examining ASR transcription of emotional speech, we now investigate how the quality of ASR transcription affects SER. Over the past several years, significant progress has been achieved in SER through the utilization of ASR transcriptions, specifically, in chronological order:

\cite{yoon2018multimodal} proposed a deep dual recurrent encoder model that simultaneously utilizes speech signal and transcriptions from the Google Cloud Speech API. \cite{sahu19_interspeech} utilized two commercial ASR systems to generate transcriptions for bimodal SER (speech + text), resulting in a relative loss of unweighted accuracy compared to ground-truth transcription. \cite{li2020learning} introduced a temporal alignment mean-max pooling mechanism to capture subtle and fine-grained emotions in utterances, alongside a cross-modality excitement module for sample-specific adjustments on embeddings.

\cite{santoso2021speech} proposed using a confidence measure to adjust the importance weights in ASR transcription based on the likelihood of a speech recognition error in each word, effectively mitigating the effects of ASR errors on SER performance. \cite{wu2021emotion} introduced a dual-branches model for ASR error robustness, with a time-synchronous branch combining speech and text modalities and a time-asynchronous branch integrating sentence text embeddings from context utterances. \cite{shon2021leveraging} generated pseudo labels on ASR transcription for semi-supervised speech sentiment analysis.

While these studies have highlighted the effectiveness of integrating text features from ASR transcription, there is still a lack of understanding regarding how WER and fusion techniques impact SER. Therefore, we undertake a benchmark study utilizing diverse ASR models and fusion techniques, conducting SER on various emotion corpora to get a clearer picture of the effect of transcription errors on SER.

\subsection{Experimental Setup}
\subsubsection{ASR models}
We adopt the following 11 models as they are widely used in the speech area and can provide varying WERs. \\
\textbullet\ \textit{wav2vec2-base-\{100h,960h\}} \\
\textbullet\ \textit{wav2vec2-large-960h} \\
\textbullet\ \textit{wav2vec2-large-960h-lv60-self} \\
\textbullet\ \textit{HuBERT-large-ls960-ft} \\
\textbullet\ \textit{WavLM-libri-clean-100h-base-plus} \\
\textbullet\ \textit{Whisper-\{tiny, base, small, medium, large-v2\}.en}

\subsubsection{Datasets}
To ensure generalizability, we utilize three corpora: IEMOCAP, CMU-MOSI, and MSP-Podcast, to include diverse speech conditions and various evaluation metrics for both discrete and continuous emotions.

For IEMOCAP, we focus on four emotion classes: \textit{angry, happy (including excited), neutral, and sad}. We exclude utterances with blank transcription from this corpus, resulting in a total of 5 500 utterances.
CMU-MOSI comprises 2 199 monologue video clips sourced from YouTube, each annotated with sentiment scores ranging from -3 to 3.
For MSP-Podcast, we use its Release 1.11 version and evaluate its performance on the Test1 set. We exclude utterances without an emotion label, resulting in a total of 104 663 utterances.

Following the literature \citep{tsai2019multimodal,hazarika2020misa}, we compute \textbf{Acc4} (unweighted four-class accuracy) for IEMOCAP, Concordance Correlation Coefficient (\textbf{CCC}) for MSP-Podcast, and \textbf{Acc2} (binary accuracy), \textbf{Acc7} (seven-class accuracy), and Mean Absolute Error (\textbf{MAE}) for CMU-MOSI.

\subsubsection{SER model}
We employ RoBERTa-base as the text encoder, as it is widely used for text-based SER \citep{siriwardhana2020jointly,luo2023fine}. Given that transcription generated by models other than \textit{Whisper} lack punctuation, we remove punctuation from the transcription of \textit{Whisper} models to ensure a fair comparison. All letters are lowercased for consistency. A backbone SER model is built for all corpora. Since our goal is not to achieve state-of-the-art performance, the model simply comprises two dense layers: the first encodes RoBERTa output of dimension 768 into hidden states of dimension 128, and the second further encodes it into a dimension of 16. We use ReLU as the activation function between the dense layers. In the case of IEMOCAP, we apply one output layer with Softmax activation for classification. For MSP-Podcast and CMU-MOSI, which involve regression tasks, no final output activation is applied. We set the learning rate as $5.0 \times 10^{-4}$ for IEMOCAP and CMU-MOSI, and $1.0 \times 10^{-4}$ for MSP-Podcast, using the AdamW optimizer. The weight decay is set as $1.0 \times 10^{-5}$, and the batch size is 64. For training, we employ five-fold cross-validation on IEMOCAP (100 epochs) and follow the official training/validation/testing sets on CMU-MOSI (100 epochs) and MSP-Podcast (30 epochs). The random seeds are kept consistent across all experiments. Note that the SER model is trained and tested on the same transcription source, rather than being trained on ground truth and then tested on ASR transcription. 

\subsection{Benchmarking SER Across WERs with Text Only}
We present the WERs and corresponding SER performance based on transcription from different ASR models. From Table~\ref{chap4/tab:wer_ser}, we observe that:

\begin{table}
    \centering
    \caption{Benchmarking SER performance based on ASR transcription across corpora. $\uparrow$: higher the better. $\downarrow$: lower the better. \textbf{\textcolor{red}{Red bold}}: better performance than the ground truth.}
    \includegraphics[width=0.52\textwidth]{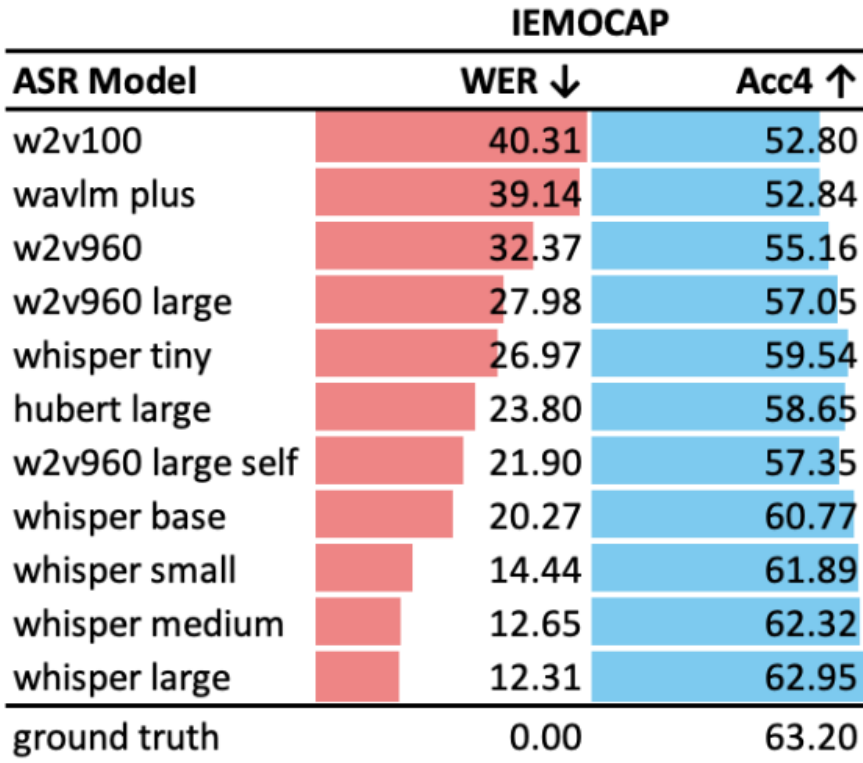}
    \includegraphics[width=0.85\textwidth]{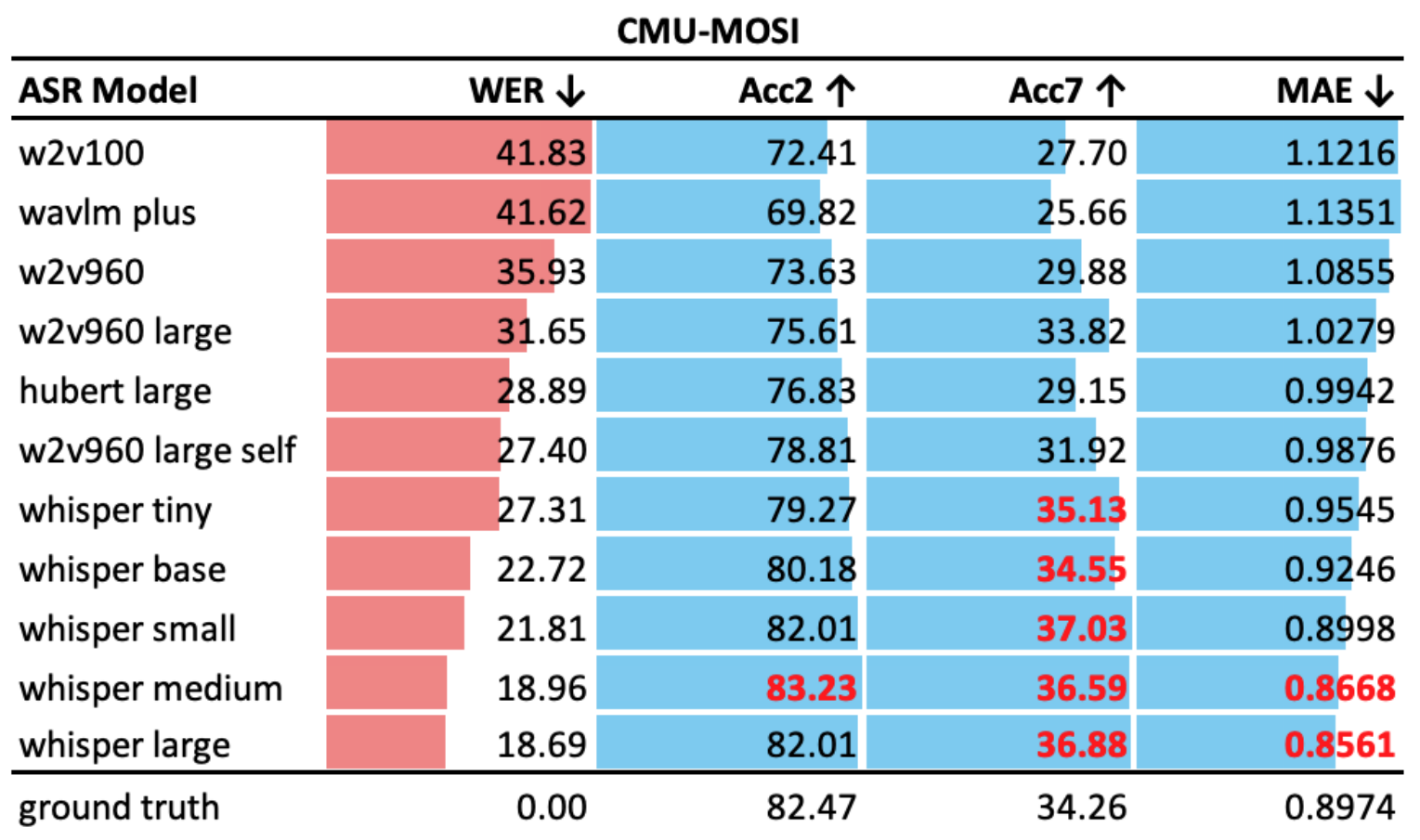}
    \includegraphics[width=0.85\textwidth]{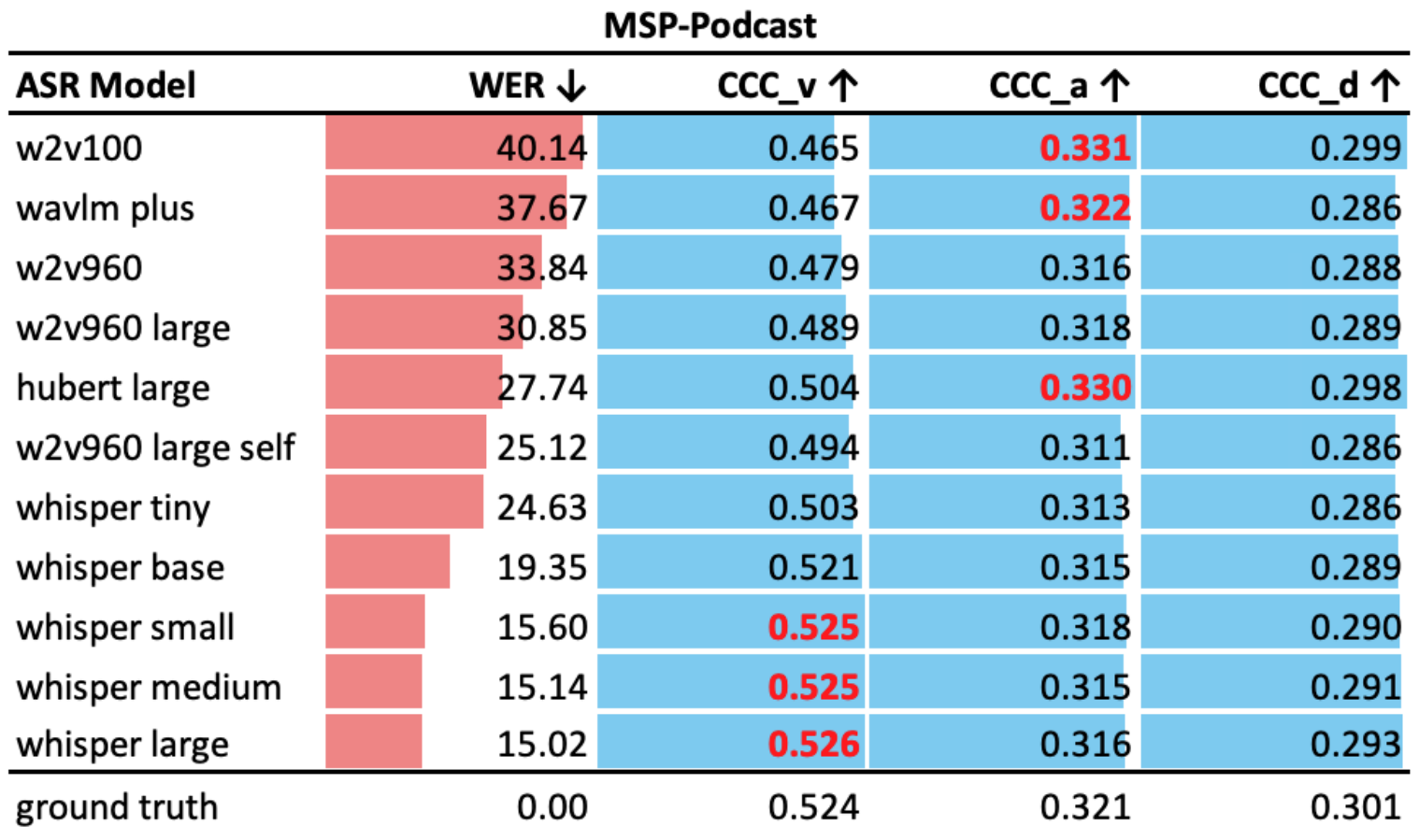}
    \label{chap4/tab:wer_ser}
\end{table}

\begin{enumerate}
    \item SER performance generally decreases as WER increases. On IEMOCAP and MOSI, there is nearly a 10\% accuracy decrease with WERs around 40\%, regardless of Acc2, Acc4, or Acc7. However, exceptions exist. For example, on IEMOCAP, \textit{Whisper-tiny} produces higher accuracy than its neighboring models, even though the WERs are similar. Additionally, \textit{W2V960-large-self} yields worse accuracy compared to its neighboring models, despite having a relatively lower WER compared to others. The same phenomenon can also be found with CMU-MOSI: the Acc2 of \textit{W2V100} is higher than that of \textit{WavLM plus}, and the Acc2 of \textit{Whisper-large} is lower than that of \textit{Whisper-medium}. This might be due to certain words being misrecognized as words that have little effect on or even positively contribute to their ground-truth emotion labels.

    \item SER is robust to relatively low WER, and in some cases, it is even better with ASR errors. From IEMOCAP, it is observed that a WER of approximately 12\% has minimal impact on SER performance compared to ground-truth transcription. Moreover, in CMU-MOSI, ASR errors can potentially enhance SER: transcription generated by specific \textit{Whisper} models outperform those based on ground truth. Prior research has shown that sentiment analysis remains robust against ASR errors \citep{tokuhisa2008emotion}, as both positive and negative sentiments can consist of a range of emotional states and therefore require less granularity. Additionally, we find that in certain cases (e.g., relatively low WER), ASR errors do not diminish SER performance. Recent research on speech-based dementia detection also indicates that ASR errors can offer valuable insights for dementia classification \citep{li2024useful}, possibly due to specific types of errors (e.g., repetitions, disfluencies) that may reveal indicators of dementia.

    \item Different metrics have different sensitivities to WER. On CMU-MOSI, Acc7 shows a distinct pattern compared to Acc2 and MAE. Acc2 and MAE demonstrate consistent and smooth variations, whereas Acc7 appears random and lacks a discernible pattern. Interestingly, in nearly half of the cases (all from \textit{Whisper}), automatic transcription outperforms the ground-truth transcripts in terms of Acc7. This discrepancy may stem from the mismatch between the regression model used during training (since MOSI is labeled with continuous values) and the classification metric applied to Acc7, where predicted and ground-truth values are grounded (i.e, the Acc7 is based on grounded regression score). This mismatch could potentially blur accuracy and render SER insensitive to WER.

    \item Different emotion dimensions exhibit varying degrees of robustness to ASR errors. From MSP-Podcast, it is evident that valence, arousal, and dominance exhibit distinct patterns. Firstly, the CCC of valence mirrors the pattern observed in Acc4 for IEMOCAP and Acc2 for CMU-MOSI, suggesting that valence shares similarities with categorical emotion in terms of robustness to ASR errors. Given that valence is conceptually similar to sentiment (indicating positivity or negativity), this alignment is plausible. Secondly, arousal and dominance do not exhibit a clear correlation with WERs. Our observations are consistent with the conventional understanding that valence is more influenced by textual content, whereas arousal and dominance are more influenced by audio cues \citep{schuller2009acoustic, wagner2023dawn, li2019expressing}.
\end{enumerate}
    
Indeed, when we replaced text features with middle-layer speech features from \textit{W2V960-base} in the SER model, we obtained CCC\_v, CCC\_a, and CCC\_d values of 0.531, 0.635, and 0.558, respectively, further validating our observations.

\subsection{Benchmarking SER Across WERs with Bimodal Fusion}
\label{chap4/sec:bimodal}
We integrate speech features to investigate the robustness of existing fusion techniques in handling ASR errors in real-world scenarios. Since speech is not the primary focus of this study, we simply use the middle-layer speech features from \textit{W2V960-base}. For fusion techniques, we employ the following approaches: \\
\textbullet\ \textit{\textbf{Early fusion}}: text and speech features are concatenated at the embedding level. \\
\textbullet\ \textit{\textbf{Late fusion}}: text and speech features are learned independently by their respective models, and the final decision is determined based on their outputs. \\
\textbullet\ \textit{\textbf{Cross-attention fusion}}: text and speech features are attended to each other via attention mechanism and then concatenated. \\
\textbullet\ \textit{\textbf{Tensor fusion}} \citep{zadeh2017tensor}: unimodal information and bimodal interactions are learned explicitly and then aggregated. \\
\textbullet\ \textit{\textbf{Non-local gate-based (NL-gate) fusion}} \citep{wang2018non}: NL-gate is incorporated with multiple directions and at multiple positions with query-key-value within the attention mechanism. \\
\textbullet\ \textit{\textbf{Modality-invariant and -specific fusion (MISA)}} \citep{hazarika2020misa}: combining both modality-invariant and modality-specific features.

For fairness, we keep the backbone SER model used in the previous section unchanged, modifying only the input dimension of the first dense layer to match the output dimension of the hidden states from each fusion model. Additionally, we modify \textit{Tensor Fusion} and \textit{MISA} to receive bimodal inputs, as they were originally designed for trimodal inputs. We present results for IEMOCAP and MSP-Podcast to cover both categorical and dimensional emotions. Regarding CMU-MOSI, integrating speech did not largely improve performance, as text alone already yields strong results. Results are presented in Table~\ref{chap4/tab:fusion}.

\begin{table*}
    \centering
    \caption{Benchmarking SER performance based on ASR transcription with fusion techniques. Maximum diff: maximum difference between the performance of ground truth and that of ASR transcription. \textbf{\textcolor{red}{Red bold}}: better performance than the ground truth.}
    \includegraphics[width=\textwidth]{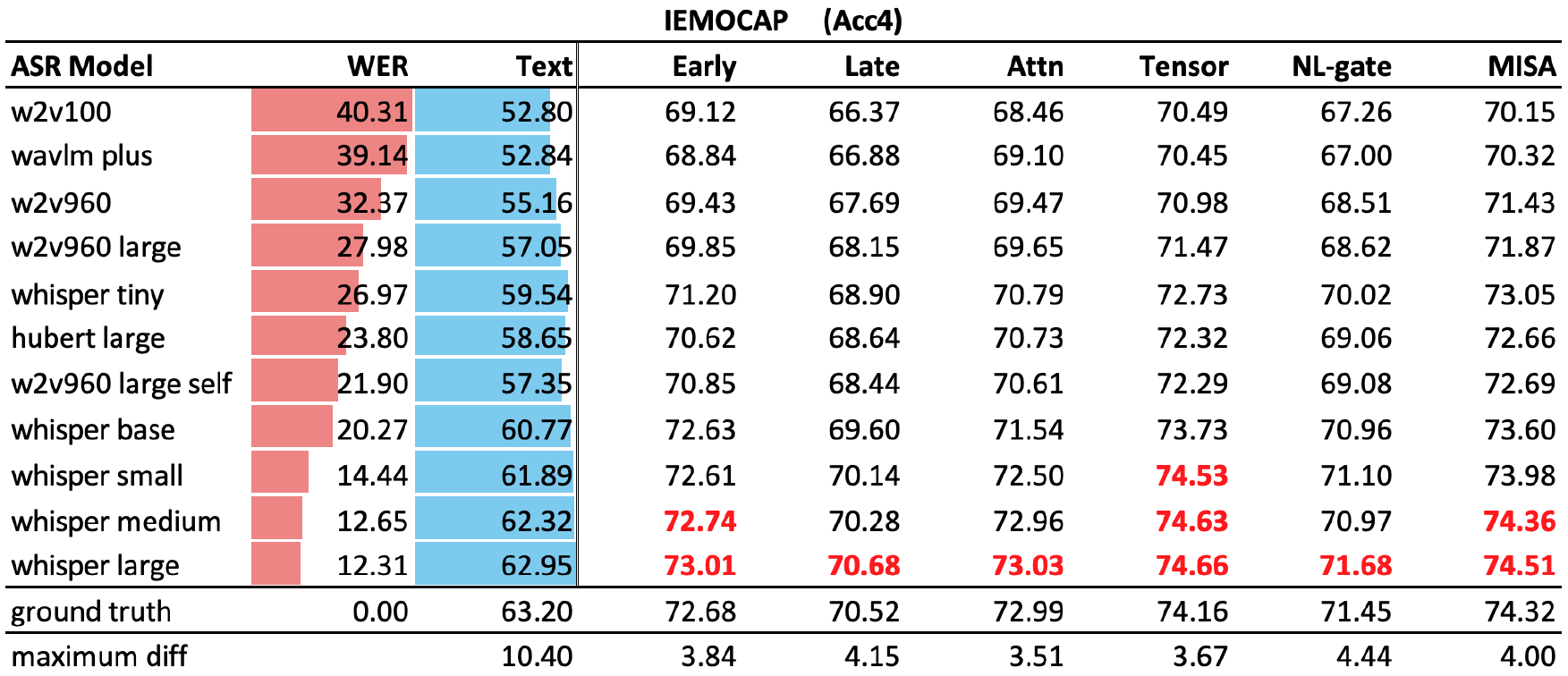}
    \includegraphics[width=\textwidth]{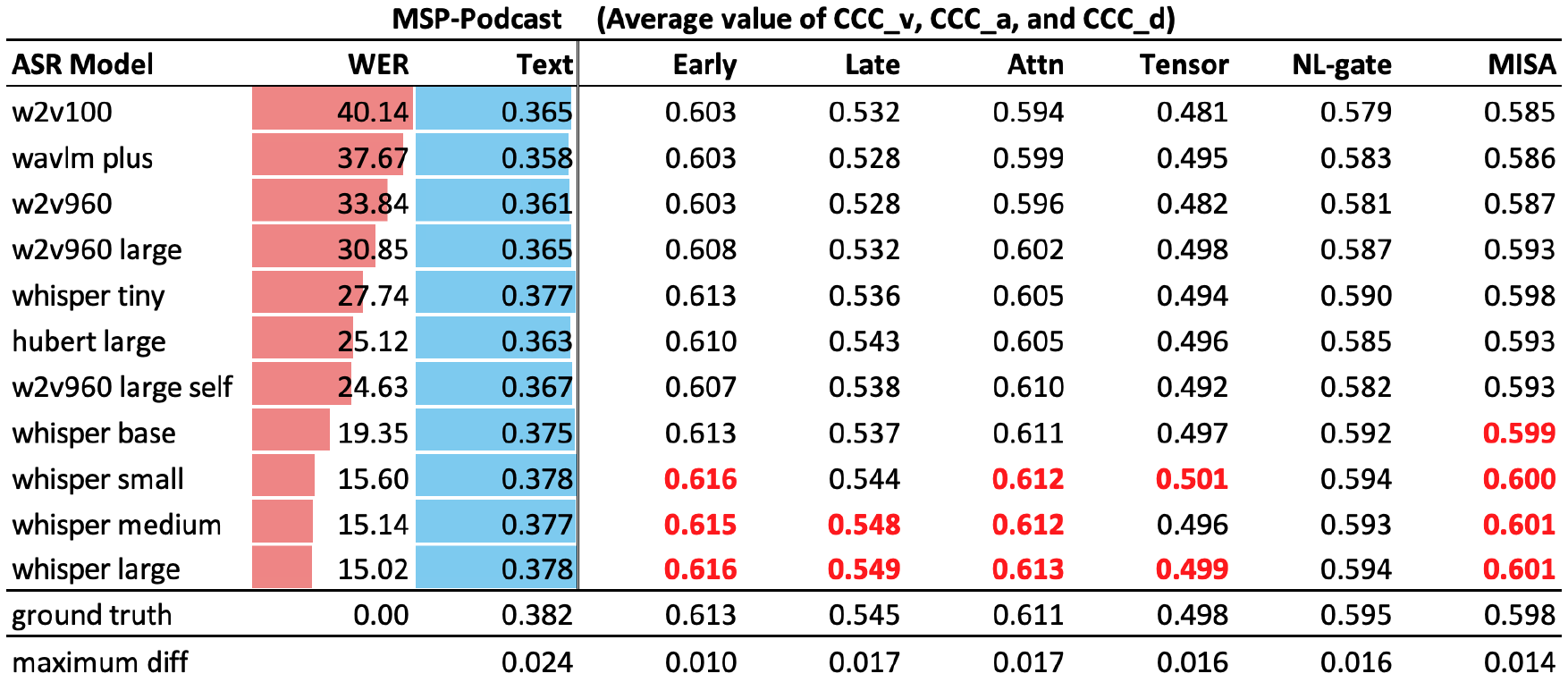}
    \label{chap4/tab:fusion}
  \end{table*}

It can be seen that:
\begin{enumerate}
    \item Fusing speech features largely mitigates the negative impact of increasing WER. The decrease in Acc4 based on WER reaches 10\% without fusion on IEMOCAP, but only 4\% with fusion. Moreover, with most fusion techniques, the SER performance is even better on transcription with relatively low WER than on ground truth. These findings confirm the benefits of bimodal SER in real applications, which can be achieved by using a pre-trained ASR model as a base for jointly generating transcription and speech features \citep{li2022fusing, cai2021speech}.

    \item There is no optimal WER-robust fusion approach. Due to different amounts of encoded information, the performances of fusion techniques are not directly comparable. For instance, \textit{Tensor Fusion} and \textit{MISA} encode more information by incorporating both unimodal and bimodal features, or both modality-invariant and -specific features. However, their effectiveness varies across different corpora. For example, they perform well on IEMOCAP (categorical) but poorly on MSP-Podcast (dimensional), indicating large inconsistency. Thus, it is difficult to identify a single best-performing technique that is universally effective in addressing ASR errors. Nevertheless, the maximum difference between the performance of ground truth and that of ASR transcription is still indicative of the robustness of the fusion techniques. We observe that although \textit{cross-attention fusion} may not produce the best performance, it yields a relatively small maximum difference, suggesting that it is less affected by WER. This observation is reasonable since cross-attention captures the relatedness between bimodal inputs, which changes dynamically once the words in text have changed, ensuring that the relatedness is least affected by ASR errors.
\end{enumerate}

\subsection{Summary}
\label{chap4:discussion}
We conducted a benchmark of SER performance using ASR transcripts with varying WERs and explored mainstream fusion techniques to assess the impact of ASR performance on SER. Our findings revealed several novel insights: \textbf{1)} SER can tolerate relatively low WERs, especially in real-life speech scenarios. \textbf{2)} Bimodal SER with transcripts containing approximately 10\% errors may not perform worse than those with ground-truth text, particularly with powerful Whisper models. However, further analysis is necessary to understand the nature and locations of ASR errors, as well as the mechanisms underlying fusion techniques. Based on this work, we have launched a challenge aimed at addressing ASR errors for text-based SER \citep{yang2024large}.

\section{Chapter Summary}
\label{chap4:summary}
This chapter presents a comprehensive study of ASR transcription of emotional speech. Section~\ref{chap4:asremotion} investigates the ASR transcription through word-level analysis to understand how different emotions affect specific words. Section~\ref{chap4:asrquality} conducts a benchmark study on the impact of WER in SER, examining how WER and fusion techniques influence both classification and regression performance. This chapter contributes to a deeper understanding of the relationship between ASR's lexical output and emotional speech, aiming to leverage ASR transcription in SER.

The next chapter builds on this foundation by introducing novel ASR error correction techniques to improve ASR transcription quality for enhanced SER performance.

\chapter{Correcting ASR Errors for SER}
\label{chap5}

\section{Introduction}
Building on the findings from the previous chapter, we now focus on addressing the challenges posed by ASR transcription errors in the context of SER. In this chapter, we explore ASR Error Correction (AEC), a post-processing approach, to enable more reliable use of ASR transcription for SER. Specifically, we explore emerging Large Language Models (LLMs) in Section~\ref{chap5:LLM} and traditional Sequence-to-Sequence (S2S) in Section~\ref{chap5:S2S} for the best application, hoping for joint AEC and SER.

\subsection{LLM-Based Approaches}
Recent studies have suggested that LLMs have the ability to reason about emotional content \citep{sap2019social}. This finding has encouraged researchers to further explore the ``emotional intelligence'' (e.g., emotion recognition, interpretation, and understanding) of LLMs. For example, \citep{wang2023emotional} developed a psychometric assessment focusing on emotion understanding to compare the emotional intelligence of LLMs and humans. They found that most LLMs achieved above-average Emotional Quotient (EQ) scores, with GPT-4 surpassing 89\% of human participants with an EQ of 117.

Therefore, the use of LLMs in text-based emotion recognition has emerged as a resource-efficient and effort-saving alternative to human annotators and traditional emotion classifiers for two main reasons: \textbf{1)} Emotion recognition requires substantial human effort. Typically, multiple annotators are needed for each sample to reach a majority vote, ensuring accurate assessment. Although platforms like Amazon Mechanical Turk provide a relatively efficient solution, concerns persist regarding privacy leaks, subjective bias, and the reliability of annotations \citep{feng2024foundation}. \textbf{2)} Despite the advancements of state-of-the-art deep learning technologies, training an emotion classifier involves multiple steps, including feature extraction and model training, which require careful consideration of which features, models, and algorithms to use.

To this end, researchers have recently started exploring and utilizing LLMs for emotion annotation and recognition. These works include various approaches, such as using multiple-step prompting with LLMs \citep{hama2024emotion}, examining prompt sensitivity \citep{amin2024prompt}, integrating outputs from multiple LLMs \citep{feng2024foundation}, and incorporating acoustic information \citep{santoso2024large}. Despite these efforts, understanding of the usage, efficacy, and reliability of LLM-based approaches remains limited, especially on ASR transcription. Therefore, building upon existing literature, we conduct various experiments on LLM-based emotion recognition to investigate effective prompting practices.

\subsection{S2S-Based Approaches}
While S2S models have been established for AEC \citep{jiang2021sequence,hrinchuk2020correction,dutta2022error}, research on Low-Resource Out-of-Domain (LROOD) scenarios is limited. Many previous studies were performed on the same large corpus without considering the LROOD problem \citep{zhang2021end,tanaka2021cross}, leaving challenges remain such as determining effective Pre-Training (PT) and Fine-Tuning (FT) strategies with LROOD data. None of the prior work has considered the characteristics of the ASR models that are the source of transcript generation. Moreover, although some studies have used data augmentation to produce more erroneous sentences for LROOD downstream corpora for AEC training \citep{dutta2022error,ghosh2024failing}, we argue that such arbitrary augmentation is unreliable because the error patterns of the augmented data differ from the original ASR errors. We hypothesize that different ASR models may produce distinct patterns of ASR errors due to their respective training process (e.g., training data, algorithms, targets), which requires that the AEC model be trained for corresponding ASR domain errors.

Additionally, acoustic information has proven useful for crossmodal AEC \citep{lin2023multi,radhakrishnan2023whispering}, but it is not always possible to acquire audio sources for the PT stage (e.g., due to privacy or other ethical issues). Therefore, determining how to better incorporate audio features and which acoustic features are useful remains an open question, considering high-WER speech usually contains low-quality audio that can introduce distortions into the crossmodal training. Finally, very few studies have applied corrected ASR transcripts to downstream tasks to evaluate AEC extrinsically.

To improve ASR transcriptions for SER, we explore both LLM-based and S2S approaches in this chapter. For the LLM-based approach, we first investigate the use of LLMs for emotion recognition on ASR transcriptions, examining their effectiveness across varying levels of WER. We then introduce a novel prompting framework, Revise-Reason-Recognize (\textsc{R3}), which integrates AEC and reasoning with emotion-specific prompts to identify emotions from ASR transcriptions.

For the S2S-based approach, we propose a crossmodal method for AEC and conduct a thorough analysis comparing AEC performance with and without PT or FT on LROOD scenarios. Additionally, we compare transcriptions from different ASR models with similar WERs, enhance crossmodal AEC by incorporating speech representations only during the FT stage, and perform SER using transcripts corrected by our proposed AEC method.

\section{LLM-Based Emotion Recognition on ASR Transcription}
\label{chap5:LLM}

Interest in LLM-based emotion recognition has surged recently, with the availability of pre-trained models. Studies in this area have explored a variety of approaches: \cite{gong2023lanser} inferred emotion labels using three different prompting approaches: text generation, mask filling, and textual entailment, employing a fine-grained emotion taxonomy. \cite{feng2024foundation} proposed an ensemble approach that integrates outputs from multiple LLMs, leveraging a Mixture of Experts (MoE) reasoning model. They trained emotion classifiers using MoE-generated emotion labels from both ground-truth and ASR transcriptions, and tested these classifiers on ground-truth labels, demonstrating comparable performance in emotion classification. \cite{hama2024emotion} employed a multi-step prompting technique with few training samples for text emotion recognition. \cite{santoso2024large} and \cite{latif2023can} incorporated textual acoustic feature descriptors into prompts. \cite{zhang2024refashioning} investigated several approaches, including in-context learning, few-shot learning, accuracy, generalization, and explanation. \cite{amin2024prompt} examined various prompting techniques, including chain-of-thought, role-play, and their variations.

Despite these advancements, we argue that each approach has its limitations, leaving several concerns unaddressed. For example, \cite{zhang2024refashioning} did not study prompting, \cite{hama2024emotion} only proposed a multi-step prompting without further exploration. The prompting techniques tested by \cite{amin2024prompt} were not specifically designed for emotion. \cite{santoso2024large} and \cite{latif2023can} only considered basic acoustic features, without incorporating more emotion-related properties. Furthermore, both \cite{feng2024foundation} and \cite{gong2023lanser} used ASR transcriptions generated by \textit{Whisper}, whose output has been shown to be robust in emotion recognition even with ASR errors, as demonstrated in Chapter~\ref{chap4:asrquality}. However, this setting is not ideal for LLMs handling challenging ASR transcriptions in real-world emotion applications.

\subsection{Methodology}
To address the issues above, we \textbf{1)} Develop prompts that incorporate emotion-specific knowledge; \textbf{2)} Incorporate AEC to refine transcriptions for robust emotion recognition; and \textbf{3)} Explore LLM training schemes to further improve the performance.

\subsubsection{Prompting with Emotion-Specific Knowledge}
\begin{figure}
    \centering
    \includegraphics[width=0.8\columnwidth]{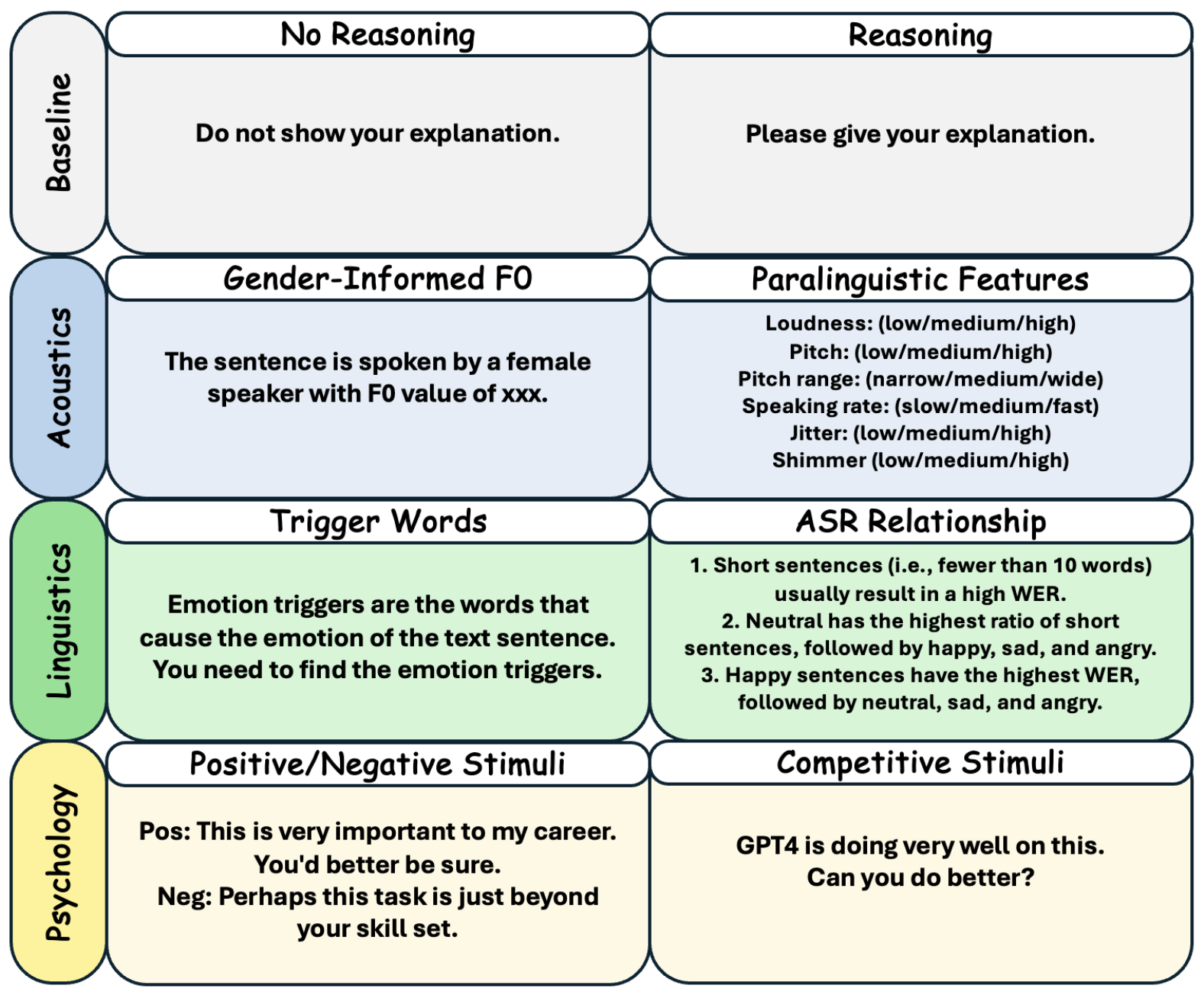}
    \caption{Emotion-specific prompts used in this work.}
    \label{chap5/fig:prompt}
\end{figure}

In light of the relationship between emotion and relevant disciplines, we extract useful knowledge from acoustics, linguistics, and psychology, to develop emotion-specific prompts for emotion recognition. The prompts are presented in Figure~\ref{chap5/fig:prompt}.

\textbf{Acoustic information} plays a crucial role in distinguishing speech emotions. Features like energy, pitch, and speaking rate have proven useful when been incorporated into prompts as textual descriptors \citep{santoso2024large}. Similarly, gender information, which is highly correlated with pitch, has also been shown to be useful \citep{latif2023can,li2019improved}. However, these features are insufficient to fully describe fine-grained differences in emotions beyond the Big Four. Hence, \textit{we hypothesize that additional acoustic features can enhance LLMs' emotion recognition ability. We propose including pitch range, jitter, and shimmer to incorporate mid-level prosody (between frame-level and utterance-level) and voice quality}. The acoustic features are extracted using openSMILE \footnote{https://audeering.github.io/opensmile-python/} and Mid-Level Prosodic Feature Toolkit \footnote{https://www.cs.utep.edu/nigel/midlevel/} calculated based on the threshold. The thresholds are taken from the lowest 30\% and the highest 30\% of the ordered numerical value of the acoustic feature. Each acoustic feature is classified as low, mid, or high.

\textbf{Linguistic structure} is essential for understanding emotion in text. For example, Section~\ref{chap4:asremotion} investigated the impact of part-of-speech, affective score, and utterance length on emotion. However, to our knowledge, only one work, \cite{singh2023language} has utilized linguistics via identifying emotion triggers (i.e., words that elicit the emotion) when prompting LLMs for emotion prediction. Thus, \textit{we hypothesize that LLM-based emotion recognition can benefit from more linguistic knowledge as LLM processing is inherently text-based. We propose including ASR-emotion relationships among emotion category, WER, and utterance length} as outlined in Section~\ref{chap4:asremotion}.

\textbf{Psychological theories}, such as self-monitoring, social cognitive theory, and cognitive emotion regulation have proven effective in improved LLMs' performance across various tasks \citep{li2023large,wang2024negativeprompt}. Therefore, \textit{we hypothesize that LLMs' emotion recognition ability can resonate with their emotional intelligence and thus be enhanced. We propose incorporating positive and negative stimuli from \citep{li2023large,wang2024negativeprompt}, as well as create our novel competitive stimuli.}

\subsubsection{Emotion Recognition with ASR Error Correction}
Traditional emotion recognition often struggles with imperfect text as we have shown in Section~\ref{chap4:asrquality}. \textit{We argue that it is more challenging to prompt LLMs for emotion recognition on ASR transcription compared to human transcription, due to the presence of word errors}. Therefore, we propose the \textsc{R3} prompting pipeline to perform emotion recognition with AEC and reasoning on ASR transcriptions. The R3 pipeline involves three steps: \textbf{Revise}, where ASR errors are corrected based on N-best hypotheses; \textbf{Reason}, where the LLMs self-explain based on the corrected transcriptions and emotion-specific knowledge; and \textbf{Recognize}, where the emotion is recognized. To incorporate AEC into our prompts, we follow an AEC-specific Alpaca prompt \citep{yang2023generative}, which uses the \textit{``You are an ASR error corrector''} instruction, guiding the LLMs to perform error correction. As LLMs have proven their ability in both AEC and emotion recognition \citep{yang2024large}, this format is expected to facilitate seamless integration with our emotion prompting, instructing the LLMs to function simultaneously as both an ASR error corrector and an emotion recognizer.

\subsubsection{Exploring LLM Training Schemes}
To understand how LLM training schemes contribute to emotion recognition, we explore context-aware learning, in-context learning, and instruction tuning. For \textbf{context-aware learning}, we organize the sentences in the conversation order and compare different context windows (i.e., the number of sentences preceding the sentence to be recognized). For \textbf{in-context learning}, we test and compare several few-shot cases. For \textbf{instruction tuning}, we apply Parameter-Efficient Fine-Tuning (PEFT) using LoRA \citep{hulora}. PEFT using LoRA is a method for adapting large pre-trained models to new tasks without updating all the model parameters. Instead, it introduces a small number of trainable parameters into the model, significantly reducing the computational cost and memory usage of fine-tuning. We set learning rate, weight decay, and epoch to $1.0 \times 10^{-4}$, $1.0 \times 10^{-5}$, and 5, respectively, and keep the LoRA configuration at its default settings, after several times of model tuning.

\subsection{Experimental Setup}
For the \textbf{datasets}, we use IEMOCAP and the Test1 set of MSP-Podcast. We combine \textit{excited} with \textit{happy} and use the Big Four classes for IEMOCAP. For the \textbf{ASR models}, we adopt the following ten to generate diverse transcripts to form 10-best ASR hypotheses, following the model choice in Section~\ref{chap4:asrquality} for consistency:

\textbullet\ \textit{wav2vec2-base-\{100h,960h\}} \\
\textbullet\ \textit{wav2vec2-large-960h} \\
\textbullet\ \textit{wav2vec2-large-960h-lv60-self} \\
\textbullet\ \textit{HuBERT-large-ls960-ft} \\
\textbullet\ \textit{WavLM-libri-clean-100h-base-plus} \\
\textbullet\ \textit{Whisper-\{tiny, base, small, large-v2\}.en}

For the \textbf{LLMs}, we use \textit{Llama-2-7b-chat-hf}, \textit{Llama-2-13b-chat-hf}, and \textit{Falcon-7b-instruct}, following a recent LLM-based emotion recogntiion study \citep{feng2024foundation} for comparison. The temperature and maximum token length are set to $1.0 \times 10^{-4}$ and 100, respectively. Our main experiments are conducted on the IEMOCAP dataset using Llama-2, following \cite{feng2024foundation}, partly because both IEMOCAP and Llama-2 are among the most commonly used datasets and LLMs. Results from \textit{Falcon} and the MSP-Podcast dataset are provided as supplementary experiments when appropriate.

\subsection{Experiments and Results}
We present the results and discussion based on multiple exploration tasks. We replace responses that fall outside the emotion classes with \textit{neutral} (roughly 5\%). Unweighted Accuracy (UA) is used to measure the results.

\begin{table}[th]
\centering
\large
\caption{Emotion recognition accuracy on transcriptions of increasing WER. $\uparrow$: higher the better. $\downarrow$: lower the better.}
\begin{tabular}{lcc}
\hline
\multirow{2}{*}{\textbf{WER\%$\downarrow$ (\textit{Transcription source})}} & \multicolumn{2}{c}{\textbf{UA\%$\uparrow$}} \\
    & \textbf{\textit{7b-chat-hf}} & \textbf{\textit{13b-chat-hf}} \\ \hline
0.00 (\textit{Ground-truth}) & 44.50 & 47.43 \\ \hdashline
12.3 (\textit{Whisper large}) & 41.77 & 44.27 \\
14.4 (\textit{Whisper small}) & 41.47 & 43.98 \\
20.2 (\textit{Whisper base}) & 41.16 & 43.70 \\
21.9 (\textit{W2V960 large self}) & 41.12 & 43.59 \\
23.8 (\textit{HuBERT large}) & 41.36 & 43.88 \\
26.9 (\textit{Whisper tiny}) & 40.80 & 43.14 \\
27.9 (\textit{W2V960 large}) & 40.49 & 43.10 \\
32.3 (\textit{W2V960}) & 40.00 & 43.01 \\
39.1 (\textit{Wavlm plus}) & 38.01 & 40.12 \\
40.3 (\textit{W2V100}) & 38.09 & 40.19 \\ \hline
\end{tabular}
\label{chap5/tab:acc-wer}
\begin{minipage}{8.5cm}
\centering
\footnotesize \textit{Data: IEMOCAP. LLM: Llama-2.}
\end{minipage}
\end{table}

\subsubsection{1. Do WERs have an impact on LLM prompting?}
In this task, we use the \textbf{baseline no reasoning} prompt: \textit{Predict the emotion from \{the emotion classes\}. Do not show your explanation.} From Table~\ref{chap5/tab:acc-wer}, we see that WERs do impact LLM prompting. Even the best-performing ASR transcription (i.e., from \textit{Whisper large}) shows more than a 4\% loss compared to ground-truth text. This finding contradicts a previous claim that LLM-based emotion recognition is robust to ASR errors \citep{feng2024foundation}. We believe this discrepancy arises from their \textbf{1)} use of \textit{Whisper large}, which provides relatively accurate transcriptions, and \textbf{2)} introduction of a fifth emotion class, `other', to filter out unconfident labels. Furthermore, LLM-based performance remains relatively stable within certain WER ranges. The accuracy decrease does not linearly correlate with the WER increase. Finally, LLMs benefit from more parameters as the 13b model consistently outperforms the 7b model.

\subsubsection{2. Does emotion-specific knowledge help?}
In this task, we use each of the \textbf{emotion-specific} prompts and their combinations for emotion recognition and compare their effectiveness on both ground-truth and ASR transcriptions. For brevity, we use one ASR transcription, whose WER ranked in the middle, as the representative (i.e., \textit{HuBERT large}). Results are presented in Table.~\ref{chap5/tab:prompt}. We see that: 

\begin{enumerate}
    \item All emotion-specific prompts improve the performance, demonstrating the efficacy of our proposed approach by incorporating emotion-specific knowledge. However, the improvement is less pronounced on ASR transcription, highlighting the necessity for AEC.

    \item Our proposed paralinguistic information improves on \citep{santoso2024large}, verifying our hypothesis that additional paralinguistic features are beneficial. Furthermore, the improvement in 8-class is more obvious, confirming that these additional features help in distinguishing finer-grained emotions (see Table~\ref{chap5/tab:para}).

    \item Linguistic knowledge generally contributes the most, even on ASR transcription. This means that LLMs benefit from identifying emotional trigger words and understanding the ASR-emotion relationship. This ASR-emotion does apply to ground-truth text, as it is specifically developed for ASR transcription.

    \item The steady improvement from psychological prompts confirms our hypothesis that LLMs' emotion recognition ability can be affected by psychological setting. Interestingly, among the psychological prompts, stimuli with negative affect perform the best, aligning with the finding from \cite{wang2024negativeprompt} (though it needs a thorough investigation as a future work).

    \item Surprisingly, the baseline reasoning prompt did not improve performance. By investigating the responses, however, we found this is likely due to the LLM hallucinations, where they often described the (acoustic) tone despite having only text input.

    \item Majority voting underperforms most single prompts, aligning with the finding of \citet{santoso2024large}. Finally, identifying the best prompt combination for both ground-truth and ASR transcriptions, we see that linguistics contributes the most to the latter by having both trigger words and ASR relationships.
\end{enumerate}

\begin{table}[ht]
\centering
\caption{Emotion recognition accuracy by using emotion-specific prompts. $\uparrow$: higher the better.}
\begin{tabular}{llcc}
\hline
\multicolumn{2}{c}{\multirow{2}{*}{\textbf{Prompt}}} & \multicolumn{2}{c}{\textbf{UA\%$\uparrow$}} \\
\multicolumn{2}{c}{} & \multicolumn{1}{l}{\textbf{\textit{\small{Ground-truth}}}} & \multicolumn{1}{l}{\textbf{\textit{\small{HuBERT large}}}} \\ \hline
Baseline & 1) No reasoning & 44.50 & 41.36 \\
 & 2) Reasoning & 43.83 \small{($-0.70$)} & 40.07 \small{($-1.29$)} \\
Acoustics & 3) Gender info & 45.59 \small{($+1.09$)} & 42.22 \small{($+0.86$)} \\
 & 4) Paralinguistic info & 46.25 \small{($+1.75$)} & 43.02 \small{($+1.66$)} \\
Linguistics & 5) Trigger words & 46.80 \small{($+2.30$)} & 43.45 \small{($+2.09$)} \\
 & 6) ASR relationship & / & 44.10 \small{($+2.74$)} \\
Psychology & 7) Positive stimuli & 45.90 \small{($+1.40$)} & 42.35 \small{($+0.99$)} \\
 & 8) Negative stimuli & 47.43 \small{($+2.93$)} & 42.76 \small{($+1.40$)} \\
 & 9) Competitive stimuli & 45.81 \small{($+1.31$)} & 41.98 \small{($+0.65$)} \\ \hdashline
\multicolumn{2}{c}{Majority voting} & 45.72 \small{($+1.22$)} & 42.94 \small{($+1.58$)} \\
\multicolumn{2}{c}{4 + 5 + 8} & \textbf{48.96 \small{($+4.46$)}} & 44.30 \small{($+2.94$)} \\
\multicolumn{2}{c}{4 + 5 + 6 + 8} & / & \textbf{44.47 \small{($+3.11$)}} \\ \hline
\end{tabular}
\label{chap5/tab:prompt}
\begin{minipage}{8.5cm}
\centering
\small \textit{Data: IEMOCAP. LLM: Llama-2-7b-chat-hf.}
\end{minipage}
\end{table}

\begin{table}[ht]
\centering
\caption{Accuracy comparison with and without Pitch Range (PR), Jitter (Ji), and Shimmer (Sh). $\uparrow$: higher the better.}
\begin{tabular}{lcc}
\hline
\multirow{2}{*}{\textbf{Prompt}} & \multicolumn{2}{c}{\textbf{UA\%$\uparrow$}} \\
 & \textit{\textbf{IEMOCAP}} & \textit{\textbf{MSP-Podcast}} \\ \hline
Baseline no explanation & 44.50 & 35.70 \\
Paralinguistic info & \textbf{46.25} \small{($+1.75$)} & \textbf{37.37} \small{($+1.67$)} \\
\ \ \ -- w/o PR, Ji, Sh & 45.89 \small{($+1.39$)} & 36.73 \small{($+1.03$)} \\ \hline
\end{tabular}
\label{chap5/tab:para}
\begin{minipage}{8.5cm}
\centering
\small \textit{LLM: Llama-2-7b-chat-hf.}
\end{minipage}
\end{table}

\subsubsection{3. Does the proposed R3 prompt help?}
In this task, we use the \textbf{R3} prompt: \textit{You are an ASR error corrector and emotion recognizer. Generate the most likely transcript from \{the 10-best ASR hypotheses\} and predict the emotion from \{the emotion classes\} with reasoning based on the provided knowledge.} For comparison, we conduct an ablation study, removing AEC or reasoning. We use \textit{4+5+6+8} as the emotion knowledge since it has proven the best.

\begin{table}[ht]
\centering
\caption{Performance comparison. $\uparrow$: higher the better.}
\begin{tabular}{lccc}
\hline
\multirow{2}{*}{\textbf{Prompt}} & \multicolumn{2}{c}{\textbf{UA\%$\uparrow$}} \\
 & \small{\textbf{\textit{Llama-2-7b}}} & \small{\textbf{\textit{Llama-2-13b}}} & \small{\textbf{\textit{Falcon-7b}}} \\ \hline
R3 & \textbf{49.72} & \textbf{52.27} & \textbf{47.24}  \\
\ \ \ -- w/o AEC & 43.29 & 47.20 & 43.00  \\
\ \ \ -- w/o reasoning & 47.48 & 50.01 & 45.49 \\ \hline
\end{tabular}
\label{chap5/tab:R3}
\begin{minipage}{8.5cm}
\centering
\small \textit{Data: 10-best of IEMOCAP. Models: -chat-hf \& -instruct.}
\end{minipage}
\end{table}

\begin{figure}
    \centering
    \includegraphics[width=\textwidth]
    {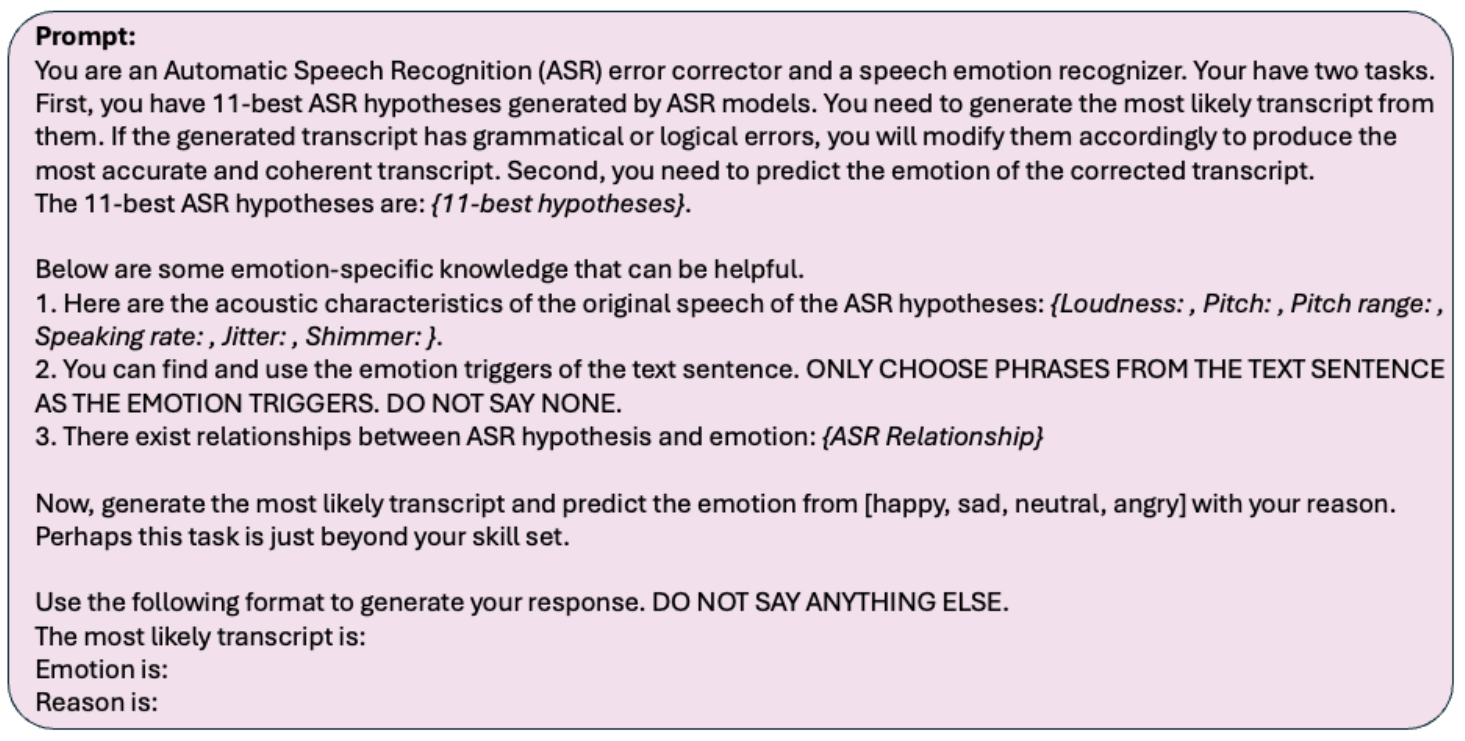}
    \includegraphics[width=\textwidth]
    {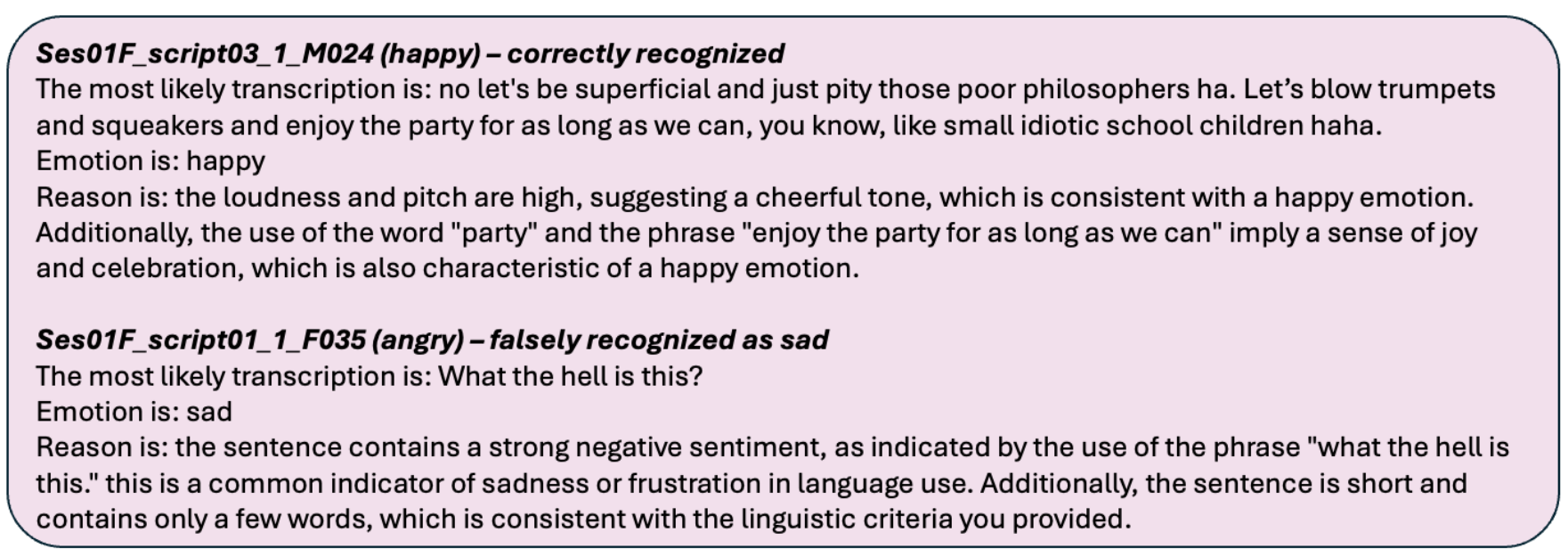}
    \caption{Examples of LLM reasoning with emotion-specific knowledge of the R3 prompt.}
    \label{chap5/fig:r3}
\end{figure}

As shown in Table~\ref{chap5/tab:R3}, both AEC and reasoning contribute to the effectiveness of our R3 prompt. Moreover, when incorporating our proposed emotion-specific knowledge, reasoning improves the performance, in contrast to the decrease observed when emotion-specific knowledge was not provided (see Table~\ref{chap5/tab:prompt}). This suggests that emotion recognition is particularly challenging for LLMs to reason without relevant information. The examples in Figure~\ref{chap5/fig:r3} illustrate how the R3 prompt helps LLMs in reasoning with emotion-specific knowledge, regardless of whether the recognition is correct.

\subsubsection{4. Do LLM training schemes help?}
In this task, we apply the \textbf{R3} prompt with context-aware learning (windows of 5 and 25), in-context learning (5-shot and 10-shot of the training data), and instruction tuning. For instruction tuning, we perform cross-validation by applying PEFT on every four sessions of IEMOCAP, testing on the remaining session, and then averaging the results, which is the same as typical SER model training. We do not compare performance across these three approaches due to their different settings.

\begin{table}[ht]
\centering
\caption{Performance via LLM training. $\uparrow$: higher the better.}
\begin{tabular}{lcccccc}
\hline
 & \multicolumn{2}{c}{\textbf{\textit{Context-aware}}} & \multicolumn{2}{c}{\textbf{\textit{In-context}}} & \textbf{\textit{Tuning}} & \multicolumn{1}{c}{\multirow{2}{*}{\textbf{\textit{Altogether}}}} \\
 & \textbf{\textit{5}} & \textbf{\textit{25}} & \textbf{\textit{5}} & \textbf{\textit{10}} & \textbf{\textit{PEFT}} \\ \hline
\textbf{UA\%$\uparrow$} & 54.35 & 62.46 & 50.74 & 54.36 & 64.67 & 70.35 \\ \hline
\end{tabular}
\label{chap5/tab:training}
\begin{minipage}{8.5cm}
\centering
\small \textit{Data: 10-best of IEMOCAP. Model: Llama-2-7b-chat-hf.}
\end{minipage}
\end{table}

From Table~\ref{chap5/tab:training}, it is evident that each LLM training scheme improves the baseline performance of using R3 on 10-best ASR hypotheses (49.72 in Table~\ref{chap5/tab:R3}). For context-aware learning and in-context few-shot learning, longer context windows and more samples yield higher accuracy. Instruction tuning leads to the highest performance, which is reasonable as the LLM has learned from the training data. Notably, a long context window also results in UA greater than 60\%, indicating the potential to utilize conversational knowledge in real-world LLM-based emotion recognition without tuning the models. Altogether (context-aware: 25 $+$ in-context: 10 $+$ PEFT tuning), they further improved the performance to 70.35\%.

\subsubsection{5. Are LLMs sensitive to minor prompt variations?}
In this task, we investigate whether LLM-based emotion recognition is sensitive to minor prompt variations. During our experiments, we observed that prompts with slight differences but the same meaning, such as variations in word choice or the order of provided emotion classes, can largely impact task performance. In Table~\ref{chap5/tab:sensitivity}, we modify the \textbf{baseline no reasoning prompt} by changing either the word \textit{Predict} to \textit{Select} or the order of the emotion classes.

\begin{table}[ht]
\centering
\large
\caption{Performance comparison of prompt variations. (A: Angry, H:Happy, N: Neutral, S: Sad). $\uparrow$: higher the better.}
\begin{tabular}{lcccc}
\hline
 & \multicolumn{2}{c}{\textbf{\textit{Word usage}}} & \multicolumn{2}{c}{\textbf{\textit{Emotion order}}} \\
 & \textbf{\textit{Predict}} & \textbf{\textit{Select}} & \textbf{\textit{A, H, N, S}} & \textbf{\textit{H, N, A, S}} \\ \hline
\textbf{UA\%$\uparrow$} & 44.50 & 41.12 & 44.50 & 40.87 \\ \hline
\end{tabular}
\label{chap5/tab:sensitivity}
\begin{minipage}{8.5cm}
\centering
\footnotesize \textit{Data: IEMOCAP. LLM: Llama-2-7b-chat-hf.}
\end{minipage}
\end{table}

This aligns with recent findings that LLMs can behave differently due to subtle changes in prompt formatting, such as separators and case, regardless of model size, number of few-shot examples, or instruction tuning \citep{sclarquantifying}. We believe this issue is a major factor hindering the widespread use of LLMs for emotion recognition and similar tasks, thus suggest that future studies evaluating LLMs with prompts should report the performance across plausible prompt variations.

\subsection{Summary}
In this section, we proposed emotion-specific prompts by incorporating relevant knowledge from acoustics, linguistics, and psychology. We also compared LLM-based emotion recognition on both ground-truth and ASR transcriptions, confirming the necessity of AEC. Consequently, we developed the \textsc{Revise-Reason-Recognize} prompting pipeline that integrates AEC, reasoning, and emotion recognition, which proves effective. Additionally, by investigating several LLM training schemes, we confirmed the value of longer context windows, more few-shot samples, and instruction tuning. Finally, we uncovered the sensitivity of LLMs to minor prompt variations. This research is expected to bridge the gap between existing studies on LLMs and emotion recognition.

\section{S2S-Based ASR Error Correction and Emotion Recognition}
\label{chap5:S2S}
After studying the LLM-based approach and understanding its advantages and limitations (e.g., unreliability due to prompt variations and limited explainability), we investigate the S2S-based approach, which does not share these limitations, for comparison.

End-to-end AEC, which maps erroneous transcripts to ground-truth text using an S2S approach, has become prevalent in scenarios where ASR is treated as a black box \citep{mani2020asr,liao2023improving}. Furthermore, some work has used both acoustic information and ASR hypotheses as input instead of text-only data, achieving crossmodal AEC \citep{lin2023multi,du2022cross,radhakrishnan2023whispering}. 

Despite these advances, S2S AEC is still a challenging task, especially for \textit{Low-Resource Out-of-Domain (LROOD) data}. Therefore, we explore this relatively unexplored aspect, aiming to provide a comprehensive analysis, which in turn provides a better understanding of AEC. The definition of LROOD in this context involves two conditions:

\begin{itemize}
    \item The data used for the downstream AEC task and the data used to pre-train the AEC model come from different datasets. These datasets are constructed for different speech tasks and purposes, with the downstream dataset being much smaller in scale than the pre-training dataset.
    \item The audio source from the downstream AEC task can be used to fine-tune the AEC model, but it is substantially different from that used in pre-training. Alternatively, the pre-trained AEC model may have been trained without using any audio source at all.
\end{itemize}

These two conditions are not met in prior S2S AEC studies. For example, previous work often used the same dataset for model pre-training, fine-tuning, and testing \citep{zhang2021end,tanaka2018neural}, or consistently relied on either text source alone or a combination of text and audio sources throughout the entire process \citep{lin2023multi,radhakrishnan2023whispering}. Additionally, the amount of fine-tuning data was large enough to allow the pre-trained AEC model to be fully adapted to the downstream datasets \citep{mani2020asr,liao2023improving}. In contrast, our work is the first to explore AEC under challenging LROOD settings.

The exploration steps with respective research problems and hypotheses are as follows.

\begin{enumerate}
    \item While S2S models have been established for AEC, research on LROOD scenarios is limited. \textit{Many previous studies were performed on the same large corpus without considering the LROOD problem \citep{zhang2021end,tanaka2018neural}, leaving challenges remain such as determining effective Pre-Training (PT) and Fine-Tuning (FT) strategies with LROOD data}. Therefore, we compare AEC performance with and without PT or FT on LROOD data using an S2S model.

    \item None of the prior works has considered the characteristics of the ASR models that are the source of transcript generation. Moreover, although some studies have used data augmentation to produce more erroneous sentences for LROOD downstream corpora for AEC training, we argue that such arbitrary augmentation is unreliable because the error patterns of the augmented data differ from the original ASR errors. \textit{We hypothesize that different ASR models may produce distinct patterns of ASR errors (e.g., some may have more insertions, substitutions, or deletions, and some may remove or retain disfluencies), which requires that the AEC model be trained for corresponding ASR domain errors}. Thus, through a comparative analysis using transcripts obtained from different ASR models with nearly the same WER, we investigate this issue and refer to it as \textsc{asr domain discrepancy}.

    \item Acoustic information has proven useful for crossmodal AEC \citep{lin2023multi,radhakrishnan2023whispering}, but \textit{it is not always possible to acquire audio sources for the PT stage (e.g., due to privacy or other ethical issues)}. Therefore, determining how to better incorporate audio features and which acoustic features are useful remains an open question, considering high-WER speech usually contains low-quality audio that can introduce distortions into the crossmodal training. To address this, we improve crossmodal AEC by incorporating Discrete Speech Units (DSUs) from Self-SSL represetations only in the FT stage, representing a resource-efficient and effort-saving approach.

    \item \textit{Very few studies have applied corrected ASR transcripts to downstream tasks to evaluate AEC extrinsically}. Hence, we conduct SER using the transcripts corrected by our proposed AEC approach, validating its potential for downstream applications.
\end{enumerate}

To our knowledge, we are the first to address the LROOD problem in AEC. The closest works are \citet{lin2023multi, chen2024s, zhang2021end, tanaka2021cross}, which fused audio into the language model as crossmodal AEC. They either fine-tuned models on large downstream data or used the same corpus with audio in all phases of model training. In this section, however, we propose DSU fusion in FT only and provides insights for tasks that require high-quality transcripts yet are constrained by limited resources.

\subsection{Proposed Architecture}
\begin{figure}[ht]
\centering
\includegraphics[width=\textwidth]{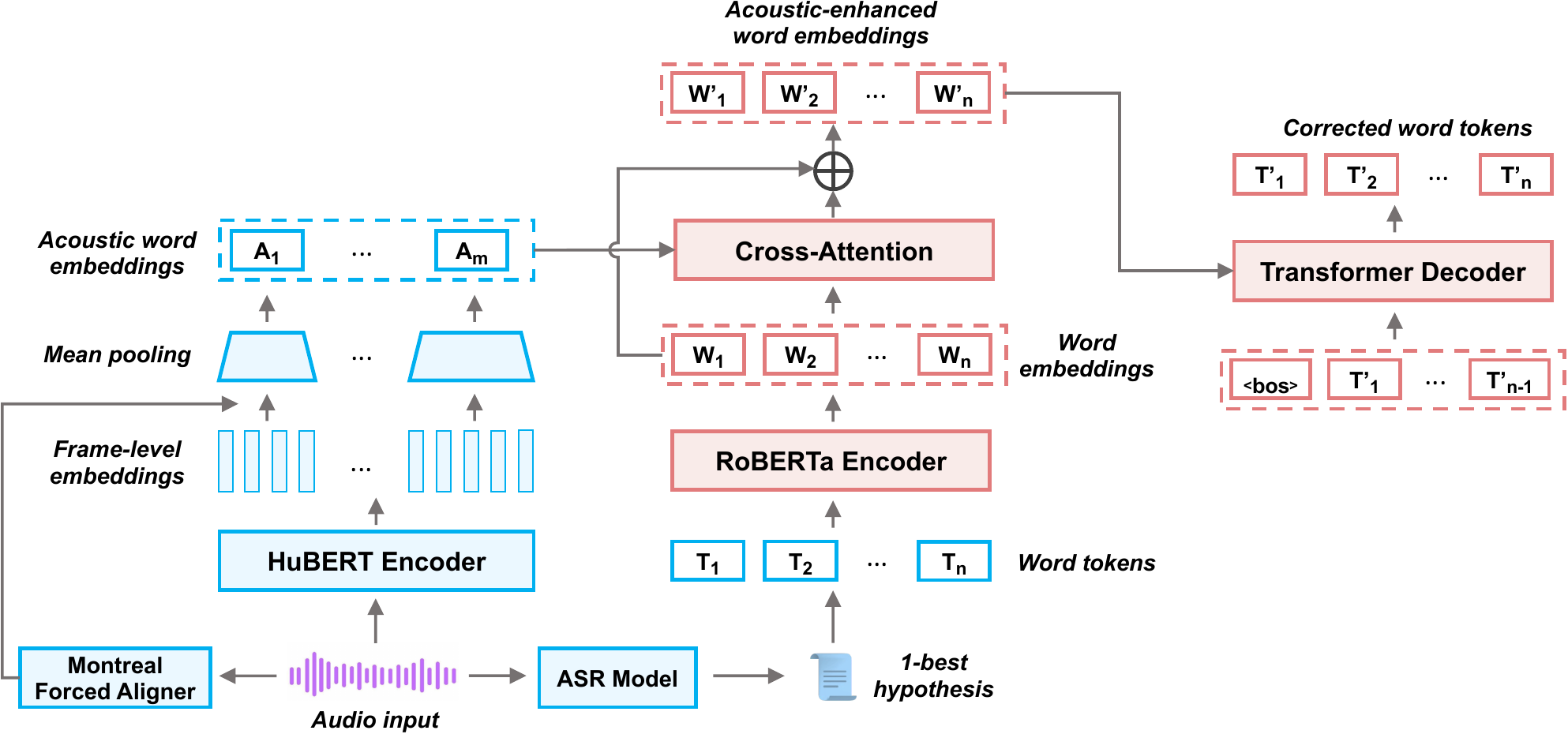}
\caption{Architecture of our crossmodal AEC with discrete speech units. (\textcolor{pink}{Pink}: trainable; \textcolor{cyan}{Blue}: frozen).}
\label{chap5/fig:model}
\end{figure}

The architecture of our proposed crossmodal S2S-based approach is shown in Figure~\ref{chap5/fig:model}. On one path, the audio input is transcribed by the ASR model into a textual transcript, which is then tokenized for the \textit{RoBERTa-base} encoder to generate word embeddings. On the other path, the \textit{HuBERT} encoder produces Self-Supervised Representations (SSR) from the audio input, followed by mean pooling to generate SSR-based Acoustic Word Embeddings (AWEs) as DSUs. Note that the Montreal Forced Aligner \citep{mcauliffe2017montreal} was used on the ASR transcripts beforehand to determine the word boundaries for mean pooling. Next, cross-attention aligns the AWEs and word embeddings to obtain the acoustic-enhanced word embeddings for the Transformer decoder to produce corrected word tokens. Our motivation for using this architecture is to highlight our effectiveness even with the most basic components.

\subsection{Experimental Setup}
Two \textbf{datasets} are primarily used: Common Voice and IEMOCAP. Common Voice is one of the largest publicly available multilingual and diverse speech datasets. We adopt its 13.0 English version, using $1.5 \times 10^{5}$ utterances from the training set and the entire test set, bringing the total number to $1.66 \times 10^{5}$. Since IEMOCAP consists of emotional speech with a large portion of scripted sessions, its wording is expected to differ from that of Common Voice. We follow our prior setting, using four basic emotions with 5,500 utterances whose transcripts are not empty.

Besides this, CMU-MOSI and MSP-Podcast are used for further validation. CMU-MOSI contains only 2,199 samples, which is approximately half the size of IEMOCAP. We randomly select 1 800 samples to fine-tune our AEC model and the remaining samples for testing. For MSP-Podcast, we adopt its Odyssey 2024 emotion challenge version, which contains a training set of 68,119 samples and a test set of 19,815 samples, making it approximately 16 times larger than IEMOCAP. These two corpora are used for: \textbf{1)} confirming the generalizability of our approach and further proving its effectiveness; and \textbf{2)} investigating the impact of data size (i.e., how our performance varies with different amounts of FT data).

For the \textbf{ASR models}, we use \textit{wav2vec 2.0} (\textit{W2V2}) in its \textit{base-960h} version, a \textit{Conformer} model (\textit{CONF}) from \textit{ESPnet}, and the \textit{Whisper} model in its \textit{tiny.en} version, because their distinct training processes are expected to produce distinct and variable ASR transcriptions. We combine the transcripts of \textit{W2V2}, \textit{CONF}, and the ground truth, resulting in a mixture of different system error types that mimic the transcript of a random ASR system (\textit{Random}). This mixture is then compared with the \textit{Whisper}-based transcript with a comparable WER to investigate the ASR domain discrepancy problem in AEC (see Section~\ref{chap5/sec:domain}--\textit{ASR Domain Discrepancy Problem}).

\textit{Whisper} is run on all four corpora, yielding satisfactory WERs (see Table~\ref{chap5/tab:wer}) for the following reason: the WER of Common Voice (for PT) is close to the others (for FT), ensuring the error ratios are consistent in PT and FT. Otherwise in PT, too many errors can result in a serious over-correction problem in subsequent FT and inference phases (which generally occurs in OOD scenarios, as we have observed in our experiments), while too few errors may lead to insufficient learning of error-gold pairs. This setting was usually ignored in the literature, where many studies trained the AEC model on Librispeech with less than 10\% WER \citep{guo2019spelling}, thereby hindering their generalizability. Subsequently, to discover the ASR domain discrepancy problem, \textit{which is a new concept presented by this work}, we create the transcript of IEMOCAP from the \textit{Random} model for comparison, which has almost the same WER as that from \textit{Whisper}.

\begin{table}[ht]
\centering
\caption{WERs (\%) of the ASR transcripts.}
\begin{tabular}{lll}
\hline
\textbf{ASR Model} & \textbf{Corpus} & \textbf{WER} \\ \hline
\multirow{2}{*}{\textit{Whisper}} & Common Voice & 19.11 \\
    & IEMOCAP & 17.18 \\
    & CMU-MOSI & 17.84 \\
    & MSP-Podcast & 17.65 \\ \hline
\textit{Random} & IEMOCAP & 17.12 \\ \hline
\end{tabular}
\label{chap5/tab:wer}
\end{table}

In the following experiments, we employ three metrics to evaluate AEC performance: WER, BLEU, and GLEU, which have been described in Chapter~\ref{chap2}. Utilizing all three metrics, we aim to comprehensively assess AEC quality from different perspectives.

\subsection{Experiments and Results}
We first investigate PT and FT without incorporating audio information, revealing the ASR domain discrepancy problem mentioned earlier. Subsequently, we incorporate different acoustic features and propose the use of DSUs for better audio-text alignment to generate corrected word tokens.

\subsubsection{Pre-Training \& Fine-Tuning}
To pre-train the AEC model, $1.66 \times 10^{5}$ samples from Common Voice were recognized by \textit{Whisper}, with 1 000 random samples held out as the test set, and the rest for training and validation with an 80\%-20\% split. The training aims to recover the gold transcripts from the ASR output. With a batch size of 256, an initial learning rate of $1.0 \times 10^{-5}$, and the Adam optimizer, we train the model for 30 epochs using cross-entropy loss and select the best checkpoint based on WER as the evaluation metric. Decoding is performed using beam search with a size of 5. The training framework is adopted from \cite{chen2023exploring}. Performance on the test set is shown in Table~\ref{chap5/tab:encoder}.

\begin{table}[ht]
\centering
\caption{AEC Performance on the test set of Common Voice.}
\scalebox{0.93}{
\begin{tabular}{lcccc}
\hline
\textbf{Model} & \textbf{WER} & \textbf{BLEU} & \textbf{GLEU} \\
\hline
\textit{Original ASR transcript} & 19.30 & 70.56 & 71.24 \\ 
\textit{Best checkpoint} & 18.19 & 72.14 & 72.45 \\
\hline
\end{tabular}
}
\label{chap5/tab:encoder}
\end{table}

Furthermore, to prevent the over-correction problem, we continue training this saved checkpoint on TED transcriptions \citep{cettolo-etal-2012-wit3} to learn to copy the gold transcripts (i.e., ground truth $\rightarrow$ ground truth) and potentially enhance its domain robustness. This continue-training lasts for two epochs, ensuring it does not overfit while maintaining correction stability. We save the checkpoint as the base model for subsequent experiments.

Next, we fine-tune this model on the training set of IEMOCAP for 40 epochs with a batch size of 64, an initial learning rate of $2.0 \times 10^{-5}$ (excluding the parameters of ``bias'' and ``LayerNorm.weight''), an epsilon of $1.0 \times 10^{-8}$, and the Adam optimizer. Following the standard five-fold split of IEMOCAP, the FT is performed five times (each time using four folds for FT and one for testing), and the final performance is reported based on the transcript composed of the corrected results obtained from five instances.

\begin{table}[ht]
\centering
\caption{Comparison results on IEMOCAP of w/ and w/o pre-training or fine-tuning.}
\begin{tabular}{ccccc}
\hline
\textbf{PT} & \textbf{FT} &  \textbf{WER$\downarrow$} & \textbf{BLEU$\uparrow$} & \textbf{GLEU$\uparrow$}  \\ \hline
\multicolumn{2}{c}{\textit{Original ASR transcript}} & 17.18 & 76.56 & 75.29 \\ \hdashline
\cmark & \xmark & 17.14 & 76.61 & 75.34 \\
\xmark & \cmark & 17.08 & 77.01 & 75.52 \\
\cmark & \cmark & \textbf{16.40} & \textbf{78.00} & \textbf{76.58} \\ \hline
\end{tabular}
\label{chap5/tab:pre_con}
\end{table}

Table~\ref{chap5/tab:pre_con} shows that without FT, a pre-trained model cannot perform well. The improvement of PT only is hardly noticeable on IEMOCAP, whereas the improvement is large on the test set of Common Voice (Table~\ref{chap5/tab:encoder}), despite their original ASR transcripts being of similar quality (Table~\ref{chap5/tab:wer}). This is likely due to the domain discrepancy between Common Voice and IEMOCAP, which results in the model pre-trained on the former being unable to recognize some erroneous OOD words in the latter. However, even without PT, the model can still improve transcript quality to some extent after FT\footnote{Technically, since there is no PT on Common Voice, it is not appropriate to use the term ``FT'' as the model is directly trained on IEMOCAP. However, we keep ``FT'' here for consistency.} on LROOD data (i.e., IEMOCAP).

Our best result comes from using both PT and FT, which indicates that the capacity learned during PT is activated and enhanced by FT. This combination well alleviates the LROOD problem.

\subsubsection{ASR Domain Discrepancy Problem}
\label{chap5/sec:domain}
To study the impact of ASR domain discrepancy, we conduct experiments by FT the AEC model on the output from another ASR system. Specifically, we use the transcript generated by \textit{Random} and compare it to the transcript generated by \textit{Whisper}. The results are presented in Table~\ref{chap5/tab:dis}.

We can observe that without PT on the transcript of Common Voice (which is generated by \textit{Whisper}), the difference in the metric values remains small after FT, compared to their original ASR transcripts. However, this pattern disappears with PT, as the transcript quality from \textit{Whisper} becomes better than that from \textit{Random}, highlighting the detrimental impact of ASR domain discrepancy. This phenomenon suggests that to correct transcripts from an ASR model, it is crucial to use the same ASR model as that used in PT (i.e., to continue using the same ASR model in both PT and FT). Nevertheless, the transcript quality from \textit{Random} still improves, indicating that PT on a large corpus, even if its transcript is from a different ASR model, is still indispensable in ASR error correction, particularly the LROOD scenarios.

\begin{table}
\centering
\caption{Comparison results on IEMOCAP of fine-tuning on transcript generated by different ASR models.}
\begin{tabular}{lccc}
\hline
\textbf{ASR Model} &  \textbf{WER$\downarrow$} & \textbf{BLEU$\uparrow$} & \textbf{GLEU$\uparrow$}  \\ \hline
\multicolumn{4}{l}{Original ASR transcript} \\
\textit{Whisper} & 17.18 & 76.56 & 75.29 \\
\textit{Random} & 17.12 & 76.64 & 75.38 \\ \hdashline
\multicolumn{4}{l}{\xmark pre-training} \\
\textit{Whisper} & 17.08 & 77.01 & 75.52  \\
\textit{Random} & 17.03 & 77.08 & 75.61 \\ \hdashline
\multicolumn{4}{l}{\cmark pre-training} \\
\textit{Whisper} & 16.40 & 78.00 & 76.58 \\
\textit{Random} & 16.54 & 77.57 & 76.42 \\ \hline
\end{tabular}
\label{chap5/tab:dis}
\end{table}

\subsubsection{Incorporation of Discrete Speech Units}
\label{chap5/sec:dsu}
So far, we have investigated how PT and FT contribute to text-only S2S AEC. To further improve the quality of error correction, we study the incorporation of acoustic information. Previous studies usually incorporated acoustic information in all stages: PT, FT, and testing, and only utilized continuous features such as mel-spectrogram or SSR \citep{lin2023multi,du2022cross}. However, we argue that these practices do not apply to LROOD scenarios for the following reasons:

\textbf{1)} The audio source of large-scale PT data is not always accessible due to privacy or other ethical concerns. \textbf{2)} The high-WER OOD speech usually contains distinct audio that can introduce acoustic distortions (e.g., prosodic variation or noise) into crossmodal training. \textbf{3)} It is challenging to align discrete word embeddings with continuous audio features. To this end, we propose to discretize the audio features from Self-SL representations to create Discrete Speech Units (DSUs) and avoid incorporating such acoustic information in PT, making it a resource-efficient and effort-saving approach.

We utilize Acoustic Word Embeddings (AWEs), which are fixed-dimensional vectors representing variable-length spoken word segments as DSUs. These vectors map acoustic features extracted from speech signals to vectors, where similar words or linguistic units have similar embeddings in the vector space \citep{maas2012word,levin2013fixed}. AWEs can capture information about phonetics and other acoustic aspects of speech, offering promising potential for word discrimination \citep{matusevych2020analyzing}.

Following recent studies on the analysis of AWEs from self-supervised speech models, we use SSR from \textit{HuBERT} with mean pooling followed by forced alignment to find the word boundary, as this practice is straightforward to implement and has been shown competitive performance in word discrimination tasks with the state of the art on English AWEs \citep{sanabria2023analyzing,saliba2024layer}. On the other hand, we also use mel-spectrogram and continuous SSR as the audio input for comparison. After a layer-wise analysis (as shown in Table~\ref{chap5/tab:layer-wise}), we use AWEs from \textit{HuBERT} layer 7 and SSR from \textit{HuBERT} layer 8 as they performed the best among all layers, respectively. This aligns with a previous finding that \textit{HuBERT} encodes the most word information between the middle layer and the last layer \citep{pasad2023comparative}.

\begin{table}[ht]
\centering
\caption{Layer-wise performance analysis using different acoustic features.}
\begin{tabular}{lccc}
\hline
\textbf{Acoustic Feature} &  \textbf{WER$\downarrow$} & \textbf{BLEU$\uparrow$} & \textbf{GLEU$\uparrow$}  \\ \hline
HuBERT SSR &  &  &  \\ \hdashline
\textit{layer 0} & 16.33 & 77.93 & 76.68 \\
\textit{layer 1} & 16.34 & 77.98 & 76.62 \\
\textit{layer 2} & 16.24 & 78.01 & 76.71 \\
\textit{layer 3} & 16.30 & 78.01 & 76.70 \\
\textit{layer 4} & 16.27 & 77.97 & 76.69 \\
\textit{layer 5} & 16.27 & 77.95 & 76.66 \\
\textit{layer 6} & 16.29 & 77.98 & 76.65 \\
\textit{layer 7} & 16.34 & 77.93 & 76.59 \\
\textit{layer 8} & \textbf{16.20} & \textbf{78.01} & \textbf{76.71} \\
\textit{layer 9} & 16.26 & 77.92 & 76.65 \\
\textit{layer 10} & 16.24 & 77.98 & 76.71 \\
\textit{layer 11} & 16.33 & 77.83 & 76.60 \\
\textit{layer 12} & 16.35 & 77.92 & 76.68 \\
\hline
HuBERT AWEs &  &  &  \\ \hdashline
\textit{layer 0} & 16.16 & 77.92 & 76.72 \\
\textit{layer 1} & 16.33 & 77.80 & 76.60 \\
\textit{layer 2} & 16.27 & 78.06 & 76.76 \\
\textit{layer 3} & 16.26 & 78.05 & 76.67 \\
\textit{layer 4} & 16.21 & 78.05 & 76.70 \\
\textit{layer 5} & 16.23 & 78.13 & 76.74 \\
\textit{layer 6} & 16.31 & 78.07 & 76.63 \\
\textit{layer 7} & \textbf{16.07} & \textbf{78.22} & \textbf{76.96} \\
\textit{layer 8} & 16.25 & 78.01 & 76.68 \\
\textit{layer 9} & 16.30 & 78.10 & 76.70 \\
\textit{layer 10} & 16.30 & 78.00 & 76.64 \\
\textit{layer 11} & 16.24 & 78.01 & 76.70 \\
\textit{layer 12} & 16.24 & 78.05 & 76.65 \\
\hline
\end{tabular}
\label{chap5/tab:layer-wise}
\end{table}

To incorporate the DSUs, we set the maximum sequence length as that of corresponding word embeddings and 0-pad the short sequence. To incorporate continuous features for comparison, we first downsample them to the same sequence length as the word embeddings using a fast Fourier transform. Unlike \textit{HuBERT}, which has the same feature dimension of 768 as \textit{RoBERTa}, we use a feed-forward layer for mel-spectrogram to expand its dimension to this size. After such pre-processing, we implement cross-attention to align acoustic features with word embeddings:
\begin{align}
A' = Attn(Q_{w},K_{a},V_{a}) = softmax(\frac{Q_{w}K_{a}^T}{\sqrt{d_k}})V_{a}
\end{align}
where $Q_{t}$, $K_{a}$, and $V_{a}$ represent the respective matrix for query (word embeddings), key (acoustic features), and value (acoustic features), $d_{k}$ is the size of a key vector, and $A'$ is the word-aligned acoustic features. Next, we sum $A'$ and $W$ for the Transformer decoder with optimizable parameters $\theta_{T}$ to generate a corrected version $W'$:
\begin{align}
W'=\arg\!\max_{W}P(W| addition(A', W);\theta_{T})
\end{align}
The results of fusing acoustic features are shown in Table~\ref{chap5/tab:summary} with previous experimental results included for comparison.

\begin{table}[ht]
\centering
\caption{Result summary on IEMOCAP.}
\begin{tabular}{lccc}
\hline
\textbf{Model} & \textbf{WER$\downarrow$} & \textbf{BLEU$\uparrow$} & \textbf{GLEU$\uparrow$}  \\ \hline
\textit{Original ASR transcript} & 17.18 & 76.56 & 75.29 \\ \hdashline
\textit{PT} & 17.14 & 76.61 & 75.34 \\
\textit{FT} & 17.08 & 77.01 & 75.52 \\
\textit{PT+FT} & 16.40 & 78.00 & 76.58 \\
\textit{PT+FT+mel-spec} & 17.36 & 76.82 & 75.48 \\
\textit{PT+FT+HuBERT SSR} & 16.20 & 78.01 & 76.71 \\
\textit{PT+FT+HuBERT AWEs} & \textbf{16.07} & \textbf{78.22} & \textbf{76.96} \\ \hline
\end{tabular}
\label{chap5/tab:summary}
\end{table}

We note that:

\begin{enumerate}
    \item Compared with other acoustic features, \textit{HuBERT} AWEs provide the best results across all metrics. This verifies our hypothesis that \textit{DSUs align more easily with word embeddings than continuous acoustic features}.

    \item The inclusion of mel-spectrogram worsens WER rather than improves it, which contrasts with findings in \cite{lin2023multi,zhang2021end,tanaka2021cross}. This phenomenon is reasonable and consistent with discussion in Section~\ref{chap5/sec:dsu}--\textit{Incorporation of Discrete Speech Units}: \textit{i)} IEMOCAP being emotional speech, contains intense prosody variation, making it challenging to encode phonetic information from mel-spectrogram; \textit{ii)} the small-size data for FT (4 400 samples with an average duration of 5 seconds) hinders the model from sufficiently learning linguistic information from mel-spectrogram, representing a low-resource scenario; \textit{iii)} our incorporation of audio features only happens during FT and testing, causing mel-spectrogram to struggle to provide sufficient information to word embeddings. In contrast, \cite{lin2023multi,zhang2021end,tanaka2021cross} conducted all training phases using the same large corpus, making their findings inapplicable to LROOD scenarios. This is possibly because the SSR DSUs contain semantic information and have been shown to be effective for word discrimination.

    \item Interestingly, despite mel-spectrogram worsening WER compared to the original ASR transcript and PT, BLEU and GLEU show improvement. This is likely because the corrected texts are more fluent and structurally correct with respect to the reference (favourable for BLEU and GLEU), while still containing word-level mistakes captured by WER. This evidences the importance of using multiple metrics to evaluate the quality of ASR transcription. Furthermore, it demonstrates the contribution of audio to high-level linguistic information, which corroborates our later finding in SER (Section~\ref{chap5/sec:ser}--\textit{Incorporation of Discrete Speech Units}).
\end{enumerate}

\subsubsection{Evaluation on Additional Corpora}
\label{chap5/sec:general}
As mentioned before, we test the performance of our proposed approach on two more corpora: CMU-MOSI and MSP-Podcast, to verify its generalizability. The results are shown in Table~\ref{chap5/tab:mosi} and \ref{chap5/tab:msp}. All experimental settings remain the same.

It can be noted that:

\begin{enumerate}
    \item PT fails to provide better results than the original ASR transcript on CMU-MOSI, whereas the performance improvement is large on MSP-Podcast. This phenomenon is likely due to the OOD problem: CMU-MOSI consists of monologue speech with opinions on specific topics (mainly about movies), containing a high proportion of OOD words, making the PT model trained on Common Voice less effective. In contrast, MSP-Podcast consists of natural, real-life speech recorded in podcast settings, sharing more linguistic similarities with Common Voice. This needs further corpora analysis for verification in future work.

    \item Both FT and the incorporation of AWEs bring performance improvements on CMU-MOSI, despite PT not being effective and the FT data being extremely limited at only 1 800 samples. Since the data size and domain similarity of IEMOCAP are between those of CMU-MOSI and MSP-Podcast, its performance improvement also falls in between (Table~\ref{chap5/tab:summary}). Furthermore, the performance improvement is even larger on MSP-Podcast, indicating that the more data available for FT, the better the performance. These findings demonstrate the efficacy of our approach in LROOD scenarios and also highlight its generalizability and potential across various scenarios.
\end{enumerate}

\begin{table}[ht]
\centering
\caption{Result summary on CMU-MOSI.}
\begin{tabular}{lccc}
\hline
\textbf{Model} & \textbf{WER$\downarrow$} & \textbf{BLEU$\uparrow$} & \textbf{GLEU$\uparrow$}  \\ \hline
\textit{Original ASR transcript} & 17.84 & 72.82 & 72.17 \\ \hdashline
\textit{PT} & 17.88 & 72.80 & 72.16 \\
\textit{PT+FT} & 17.65 & 73.31 & 72.63 \\
\textit{PT+FT+HuBERT AWEs} & \textbf{17.22} & \textbf{73.98} & \textbf{73.01} \\ \hline
\end{tabular}
\label{chap5/tab:mosi}
\end{table}

\begin{table}[ht]
\centering
\caption{Result summary on MSP-Podcast.}
\begin{tabular}{lccc}
\hline
\textbf{Model} & \textbf{WER$\downarrow$} & \textbf{BLEU$\uparrow$} & \textbf{GLEU$\uparrow$}  \\ \hline
\textit{Original ASR transcript} & 17.65 & 81.32 & 78.02 \\ \hdashline
\textit{PT} & 16.23 & 82.59 & 79.14 \\
\textit{PT+FT} & 14.73 & 83.16 & 80.84 \\
\textit{PT+FT+HuBERT AWEs} & \textbf{13.89} & \textbf{83.64} & \textbf{81.80} \\ \hline
\end{tabular}
\label{chap5/tab:msp}
\end{table}

\subsubsection{Performance Comparison with Literature}
To confirm the effectiveness of our S2S-based approach, we compare it with the following baselines:

\textbf{1)} Crossmodal AEC using continuous acoustic information: mel-spectrogram \citep{zhang2021end}

\textbf{2)} Crossmodal AEC using continuous acoustic information: self-supervised representations \citep{lin2023multi}.

\textbf{3)} Generative AEC using an LLM with 1-best ASR hypothesis and Alpaca prompt \citep{radhakrishnan2023whispering}.

\textbf{4)} Generative AEC using an LLM with N-best ASR hypothesis and Alpaca prompt \citep{radhakrishnan2023whispering}.

\textbf{5)} Generative AEC using an LLM with 1-best ASR hypothesis and Task-Activating prompt \citep{yang2023generative}.

\textbf{6)} Generative AEC using an LLM with N-best ASR hypothesis and Task-Activating prompt \citep{yang2023generative}.

\begin{figure}[ht!]
  \centering
  \includegraphics[width=\linewidth]{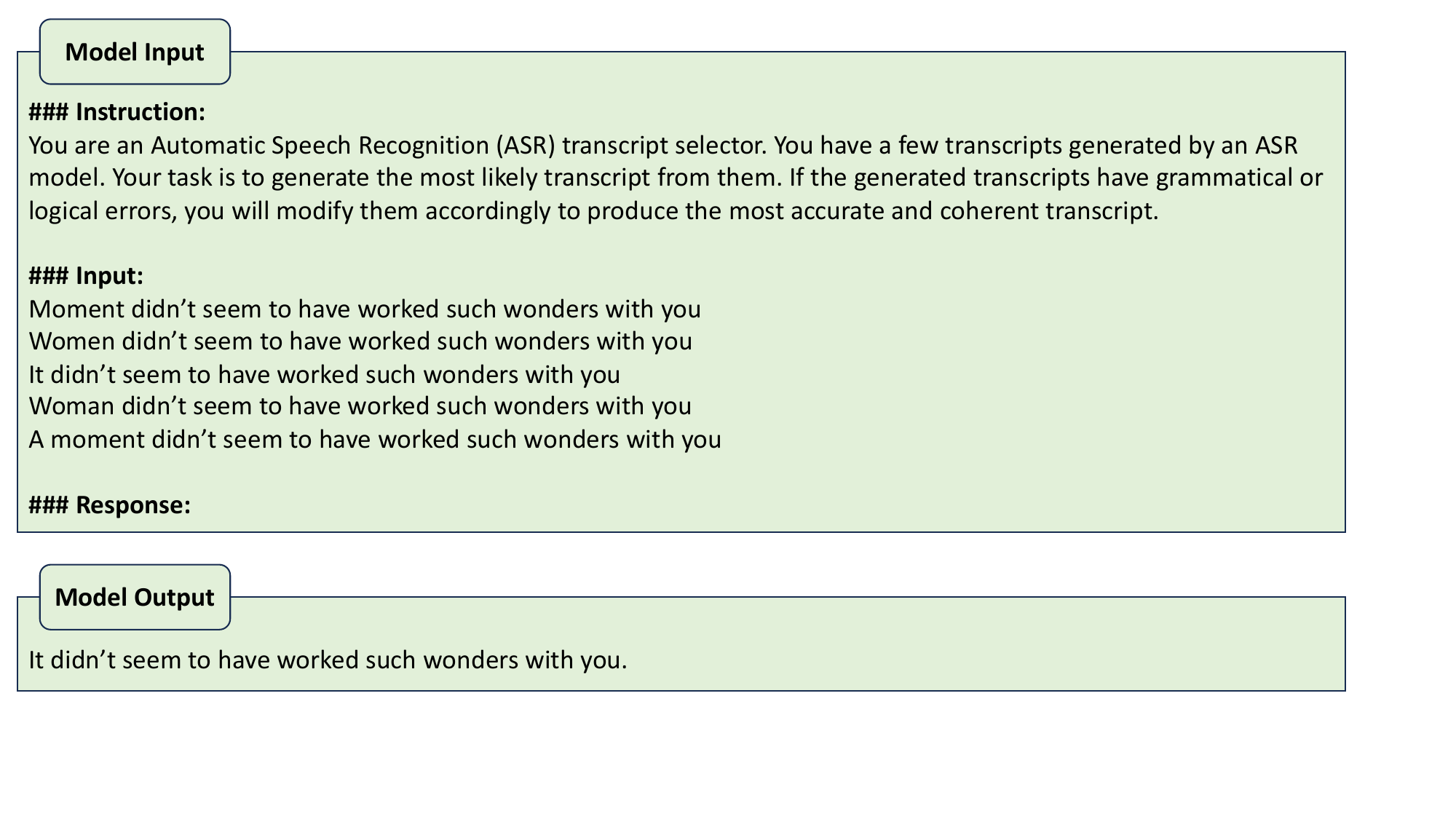}  
  \includegraphics[width=\linewidth]{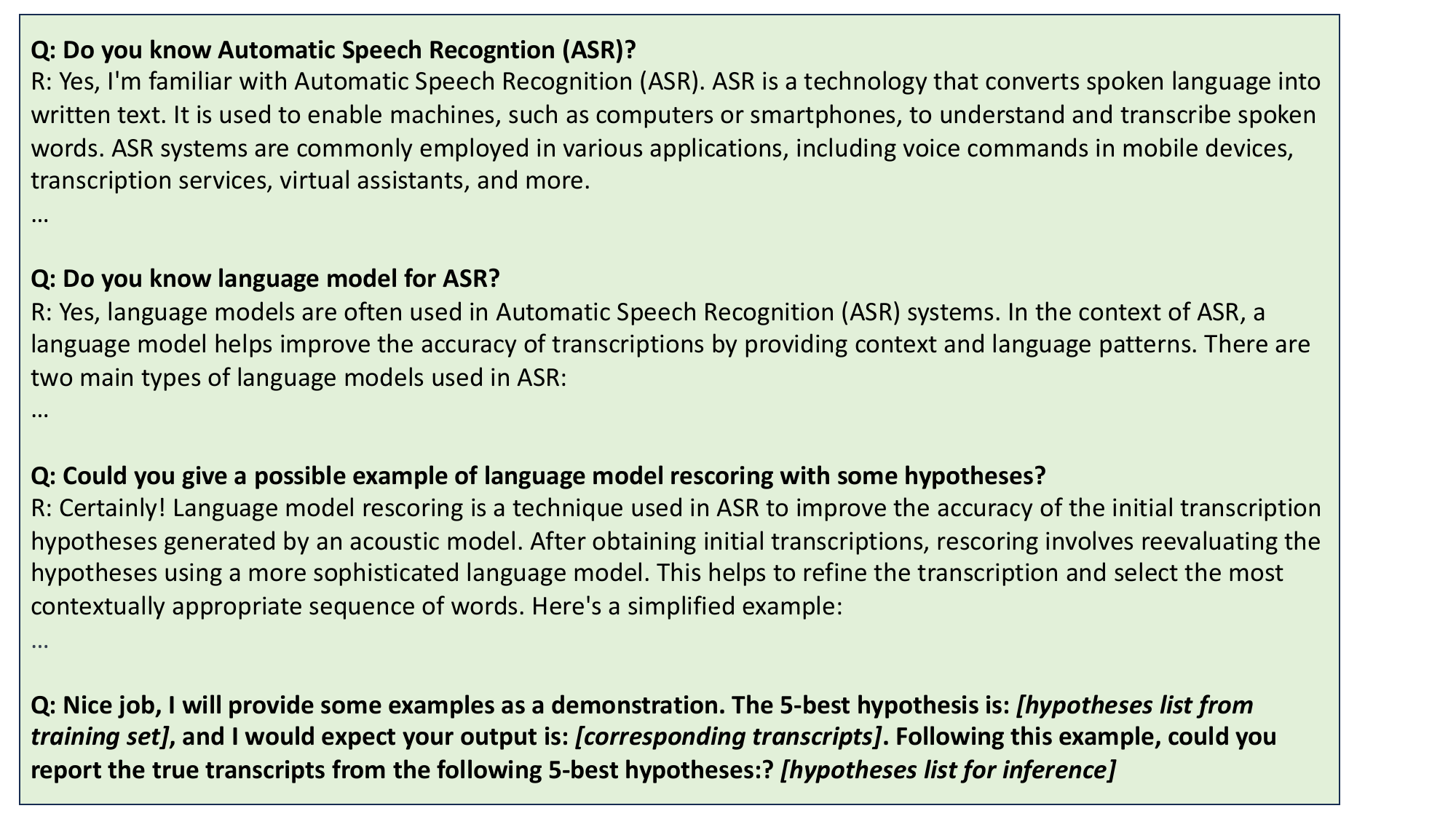}
  \caption{An illustration of the Alpaca prompt (upper) and Task-Activating prompt (below) used in this work.}
  \label{chap5/fig:aec-prompt}
\end{figure}

Since the comparisons with \textbf{1)} and \textbf{2)} have already been presented in Table~\ref{chap5/tab:summary} and discussed, we omit them here. For the remaining comparisons, we attempt the Alpaca prompt \citep{taori2023stanford} and Task-Activating (TA) prompt \citep{yang2023generative} using \textit{InstructGPT} on both 1-best and 5-best hypotheses. Alpaca prompt refers to a structured input format used in fine-tuning language models, particularly inspired by the Alpaca dataset created by Stanford. It typically follows an instruction-response format, where a model is given a natural language instruction and expected to generate a relevant response. This approach enhances the LLM's ability to follow complex instructions with minimal supervision. On the other hand, the TA prompt is designed to explicitly direct an LLM toward a specific task by embedding contextual cues and task-relevant information within the prompt. This method enhances the LLM's ability to generate accurate and task-specific responses, particularly in structured problem-solving scenarios. Figure~\ref{chap5/fig:aec-prompt} illustrates how the Alpaca prompt and TA prompt are used. The results are presented in Table~\ref{chap5/tab:comparison}.

\begin{table}[ht]
\centering
\caption{Performance comparison with generative AEC approaches.}
\begin{tabular}{lccc}
\hline
\textbf{Model} & \textbf{WER$\downarrow$} & \textbf{BLEU$\uparrow$} & \textbf{GLEU$\uparrow$}  \\ \hline
\textit{Original ASR transcript} & 17.18 & 76.56 & 75.29 \\
\textit{Our full model} & \textbf{16.07} & \textbf{78.22} & \textbf{76.96} \\ \hdashline
\textit{\textbf{3)} Alpaca prompt\textsubscript{1-best}} & 17.18 & 76.56 & 75.29 \\
\textit{\textbf{4)} Alpaca prompt\textsubscript{5-best}} & 17.01 & 76.97 & 75.44 \\
\textit{\textbf{5)} TA prompt\textsubscript{1-best}} & 17.18 & 76.57 & 75.30 \\
\textit{\textbf{6)} TA prompt\textsubscript{5-best}} & 16.62 & 77.99 & 75.98 \\ \hline
\end{tabular}
\label{chap5/tab:comparison}
\end{table}

From the comparison results, it can be observed that: the generative AEC approaches underperform our S2S crossmodal AEC approach, particularly as the 1-best hypothesis shows hardly any difference compared to the original ASR transcript, confirming our effectiveness for scenarios where only the 1-best hypothesis is available. However, compared to the generative LLM-based approach, the promising results of the S2S crossmodal approach depend on relatively high-quality ASR transcriptions and audio input. The two approaches are complementary rather than contradictory.

\subsubsection{S2S-Based ASR Error Correction for SER}
\label{chap5/sec:ser}
To verify the quality and usability of our AEC approaches in downstream applications, we compare SER performances using the corrected transcript and the original ASR transcript.

We train the SER model on the ground-truth transcript of the IEMOCAP training set and evaluate its performance on the ASR transcript of the test set, employing five-fold cross-validation. This is to validate the effectiveness of our ASR error correction method. Textual features are extracted using \textit{RoBERTa}. The SER model consists of two bidirectional LSTM layers (hidden state: 32), a self-attention layer (hidden state: 64, heads: 16), a dense layer (hidden state: 64) with ReLU activation, and an output layer with Softmax activation. We use the AdamW optimizer with a learning rate of $1.0 \times 10^{-4}$ and weight decay of $1.0 \times 10^{-5}$ and a batch size of 64. Training is performed for 150 epochs, and the reported results are the best Unweighted Accuracy (UA) achieved.

\begin{table*}[ht]
\centering
\caption{Comparison results of SER performance.}
\begin{tabular}{lccc|c}
\hline
\textbf{Transcript} & \textbf{WER$\downarrow$} & \textbf{BLEU$\uparrow$} & \textbf{GLEU$\uparrow$} & \textbf{UA$\uparrow$}  \\ \hline
\textit{Original} & 17.18 & 76.56 & 75.29 & 60.92 \\
\textit{Corrected} & 16.07 & 78.22 & 76.96 & \textbf{61.82} \\ \hline
\end{tabular}
\label{tab:ser}
\end{table*}

As expected, SER performance can be improved by using the corrected transcript. The UA increased from 60.92 to 61.82 (\textit{+0.90}) with WER decreasing from 17.18 to 16.07 (\textit{-1.11}), which represents a larger improvement, compared to the benchmark results in Table~\ref{chap4/tab:wer_ser}. This observation can be attributed to the fact that AEC with DSUs not only reduces WER but potentially does so via preserving syntax and semantics better, leading to higher usability in downstream tasks (resonates with the last finding in Section~\ref{chap5/sec:dsu}--\textit{Incorporation of Discrete Speech Units}). However, further analysis is needed to understand the nature of ASR errors: where they occur and how they are corrected.

\subsubsection{Joint ASR Error Correction with SER}
As the above experiments have proven that the corrected transcriptions are more accurate in SER, we hypothesize that the tasks of AEC and SER can benefit from each other. To verify this, we conduct a further experiment on joint AEC and SER, i.e., training a multi-task model that corrects the ASR transcriptions and utilizes the corrected transcriptions for SER simultaneously. Specifically, we keep the PT stage the same as in previous experiments but integrate the SER task during the FT stage. We only use \textit{PT+FT+HuBERT AWEs}, as this strategy performed best in previous experiments.

We use the same architecture as shown in Figure~\ref{chap5/fig:model}, with the only difference being the integration of an SER module. The SER module is consistent with that used in Section~\ref{chap5/sec:ser}--\textit{S2S-Based ASR Error Correction for SER}, taking the corrected ASR transcription as input, encoded by \textit{RoBERTa}, followed by two bidirectional LSTM layers (hidden state: 32), a self-attention layer (hidden state: 64, heads: 16), a dense layer (hidden state: 64) with ReLU activation, and an output layer with Softmax activation. The AdamW optimizer with a learning rate of $1.0 \times 10^{-4}$, weight decay of $1.0 \times 10^{-5}$, and a batch size of 64 is applied. The AEC loss and SER loss are equally weighted as 0.5. Experimental evaluation is also conducted on IEMOCAP for comparison, with results presented in Table~\ref{chap5/tab:joint}.

\begin{table}[ht]
\centering
\caption{Result comparison between separate training and joint training.}
\begin{tabular}{lcccc}
\hline
\textbf{Model} & \textbf{WER$\downarrow$} & \textbf{BLEU$\uparrow$} & \textbf{GLEU$\uparrow$} & \textbf{UA$\uparrow$} \\ \hline
\textit{Separate training} & 16.07 & 78.22 & 76.96 & 61.82 \\
\textit{Joint training} & \textbf{16.08} & \textbf{78.23} & \textbf{76.98} & \textbf{62.25} \\ \hline
\end{tabular}
\label{chap5/tab:joint}
\end{table}

It can be observed that jointly training SER with the FT AEC model results in better performance in both tasks compared to separate training. In particular, the SER accuracy increases more clearly, indicating that SER benefits more from joint training with AEC.

\subsection{Summary}
In this section, we pre-trained an S2S AEC model on large corpora and fine-tuned it on an LROOD corpus with the assistance of DSUs. The results indicate that for AEC on LROOD data, PT, FT, and DSUs are all important. Moreover, the ASR domain discrepancy problem requires attention and should be mitigated by using the same ASR model to generate transcripts in all phases of AEC applications. We compared different acoustic features and verified the superiority of DSUs over continuous features in aligning with word embeddings. A downstream SER task further demonstrated the improved quality of the corrected transcript, highlighting the applicability of our approach. Additionally, SER benefits from joint training with AEC, consistent with the findings of the previous section of LLM-based approach.

\section{Chapter Summary}
This chapter presents two approaches for ASR error correction for the use in SER. Section~\ref{chap5:LLM} introduces an LLM-based method that integrates emotion-specific prompts and ASR error correction to refine N-best ASR hypotheses and recognize emotion from the corrected transcription. Section~\ref{chap5:S2S} describes an S2S-based method that combines discrete acoustic units with word units for cross-modal ASR error correction, achieving better performance than the LLM-based method on the 1-best ASR hypothesis. The LLM-based approach demonstrates its effectiveness by incorporating acoustic information, linguistic structure, and psychological theories into prompt templates, leveraging N-best ASR hypotheses. Performance can be further enhanced by applying LLM training strategies such as in-context learning and context-aware training. It is flexible in terms of various prompting paradigms and LLM training schemes. However, a key limitation is its lack of reliability and stability due to sensitivity to prompt variations.

In contrast, the S2S-based approach takes the 1-best hypothesis as input and incorporates discrete speech units derived from HuBERT self-supervised representations, presenting a different setup compared to the LLM-based approach. Training the S2S model is flexible during the training process, as pre-training and fine-tuning are separable. However, it is limited by its reliance on the 1-best hypothesis, whose quality largely impacts performance. In comparison, the LLM-based method leverages N-best hypotheses, allowing for effective re-ranking and yielding semantically richer transcriptions. Whereas, the S2S model provides stable improvements and has no prompt concerns.

In future work, we aim to mitigate their limitations by combining both approaches into a two-step joint ASR error correction and emotion recognition framework, leveraging their respective strengths.

So far, we have laid the groundwork for combining audio and text input in multimodal SER. In the next chapter, we examine a specific issue: cross-modal incongruity.

\chapter{Addressing Cross-Modal Incongruity in SER}
\label{chap6}
\section{Introduction}
After examining and benchmarking the influence of ASR errors on SER performance, we turn our attention to a more subtle but critical challenge: cross-modal incongruity. Specifically, we investigate the discrepancies in emotional tendencies between different modalities, that arise due to either modality misalignment or ASR imperfections. Such incongruity can lead to conflicting emotional cues, ultimately hindering the effectiveness of multimodal SER systems.

In this chapter, we conduct an investigation into the origins and effects of the Inter-Modal Incongruity (IMI) issue through a structured, step-by-step analysis in Section~\ref{chap6/sec:exploration}. Building on the concept of hierarchical attention fusion introduced in Chapter~\ref{chap3:hierarchical}, we then present a hierarchical cross-modal Transformer in Section~\ref{chap6/sec:incongruity-fusion}. This model incorporates an incongruity-aware, modality-gated fusion mechanism to effectively mitigate the IMI problem in multimodal SER, promoting more consistent and reliable emotion recognition across different modalities. In addition, Section~\ref{chap6/sec:asr-fusion} describes the integration of both LLM-based and S2S-based ASR error correction methods, previously introduced in Chapter~\ref{chap5}, combined with the proposed modality-gated fusion. This integrated approach is specifically designed to reduce the cross-modal inconsistencies introduced by ASR errors for real-world applications.

\section{Exploration of Incongruity in Multimodal Inputs}
\label{chap6/sec:exploration}
Emotions are expressed in complex ways in human communication (e.g., via face, voice, and language). As such, multimodal fusion has become a hot topic in the past decade. Previous studies \citep{Xu2018,li2020attention} have shown that by taking advantage of complementary information from multiple modalities, affect\footnote{'affect' refers to the broader emotional experiences or internal emotional expressions, such as the underlying emotional states or moods, whereas 'emotion' typically refers to specific, recognized feelings like happiness or anger.} recognition can become more robust and accurate. However, several major issues remain unsolved, impeding the progress of Multimodal Information Processing (MIP). First, multimodal signals are not always strictly synchronous. For example, the visual signal usually precedes the audio by around 120 ms when people express emotion \citep{grant2001speech}. Second, different modalities may have different or even opposite affective tendencies, which makes affective states difficult to recognize. For instance, people can say negative content with a positive voice to express politeness \citep{laplante2003things} or with a smile to express sarcasm \citep{caucci2012social}.

Approaches tackling these issues have been proposed in prior work. For example, \citet{tsai2019multimodal} introduced the Multimodal Transformer (MulT) model to learn a pair-wise latent alignment with the Transformer structure, which directly attends to low-level features in multiple modalities to solve the asynchrony problem. \citet{Wu2021} proposed an incongruity-aware attention network that focuses on the word-level incongruity between modalities by assigning larger weights to words with incongruent modalities. Nevertheless, to capture as much information as possible for better performance, recent models usually repeatedly fuse specific or all modalities \citep{liang2018multimodal,tsai2019multimodal,li2022cross}, resulting in not only redundant features but also large model sizes that hinder their real-world use.

Before presenting our model, we first conduct an incongruity analysis to clarify the IMI problem and why it is necessary to address it.

\subsection{Datasets}
We use \textbf{CMU-MOSI}, \textbf{CMU-MOSEI}, and \textbf{IEMOCAP} for emotion recognition. Following prior work, we use four emotions (\textit{happy, sad, angry}, and \textit{neutral}) for the experimental evaluation in IEMOCAP, bringing in 4,453 samples. Moreover, to verify our approach in incongruity-rich scenarios for generalizability, we also use \textbf{UR-FUNNY} and \textbf{MUStARD} for further evaluation. UR-FUNNY is collected from TED talk videos, incorporating language, acoustic, and visual modalities, as well as the context preceding the punchline, for humor detection. The dataset consists of 5,000 humor and 5,000 non-humor instances. MUStARD is sourced from popular TV shows such as Friends, Big Bang Theory, Golden Girls, and Sarcasmaholics. This dataset comprises 690 video segments with labels indicating sarcasm or non-sarcasm. Humor and sarcasm often arise from multimodal incongruity, where conflicting cues from different communicative modes, such as language, visuals, and prosody, create a cognitive dissonance that prompts reinterpretation. This mismatch enhances the communicative complexity and engages the audience in understanding the incongruity, often leading to humorous or ironic effects \citep{attardo1994linguistic,yus2016humour}.

\subsection{Analysis of Inter-Modal Incongruity}
To address the IMI problem, there are typically two approaches: \textbf{1)} Always selecting one modality as the primary input. For instance, sentiment analysis, as well as humor and sarcasm detection have demonstrated the best performance when language is chosen as the primary input \citep{rahman2020integrating,ma2023multimodal,hasan2021humor}. However, \textit{it cannot be guaranteed that one modality consistently and predominantly contributes to the final task}. Even though language is of utmost importance, it does not imply that every sample is dominated by language. \textbf{2)} Employing a weighted sum to adjust the input from each modality, thereby avoiding their equal contributions \citep{rahman2020integrating,yang2020cm}. Yet, \textit{this approach overlooks the fact that some modalities can be more closely related or synchronized}. Fusing all modalities simultaneously without establishing a hierarchy may diminish the efficacy of fusion. Additionally, the practice of encoding the same input at both the word- and utterance-level for fusion has also been adopted to address incongruity \citep{Wu2021}. However, \textit{such an approach may introduce redundancy due to the repeated encoding of the same modality input}.

Furthermore, most research typically provides high-level examples to demonstrate how modalities may be misaligned. For instance, in a video clip from MUStARD for sarcasm detection, a positive sentence--``You're right, the party is fantastic,'' is presented with a facial expression of ``eye-rolling up'' and drawn-out syllables on the word ``fantastic'' \citep{Wu2021}. However, no evidence has been presented to illustrate how different modalities may mismatch at the latent level. Therefore, we perform an analysis using heatmaps to reveal how cross-attention highlights mismatched latent information. We conduct the following experiments on CMU-MOSEI. Note that approximately twice the number of examples presented were examined. We select the most representative cases for illustration, following previous work \citet{tsai2019multimodal,rahman2020integrating}.

\textbf{Exp 1}. Investigating how source modality enhances target modality via cross-attention. We use the example of $V\rightarrow T$ (text attended by vision). Next, we aim to see how the combination of two modalities collectively affects the third:

\textbf{Exp 2}. Investigating how the salient parts of target modality are represented by self-attention with and without the combination of two source modalities. We use the example of $(A+V)\rightarrow T$ (text attended by cross-attention-fused audio-vision). Further, we hope to know how different source modalities individually affect the target:

\textbf{Exp 3}. Investigating how the salient parts of target modality are represented by cross-attention when using different source modalities. We use the examples of $V\rightarrow T$ (text attended by vision) and $A\rightarrow T$ (text attended by audio).

The experimental setting is shown in Table~\ref{chap6:tab:setup}, and the visualization is shown in Figure~\ref{chap6:fig:exp1}, ~\ref{chap6:fig:exp2}, and ~\ref{chap6:fig:exp3}.

\begin{table}[ht]
\centering
\caption{Experimental setting for multimodal incongruity analysis.}
\begin{tabular}{r|cccc}
\hline
\textbf{Exp.} & \textbf{Target modal} & \textbf{Source modal} & \textbf{Cross-attention} & \textbf{Self-attention}\\ \hline
1 & $T$ & $V$ & $V\rightarrow T$ & / \\
2 & $T$ & $A + V$ & $(A + V)\rightarrow T$ & $T$\\
3 & $T$ & $A$ or $V$ & $A\rightarrow T$, $V\rightarrow T$ & / \\ \hline
\end{tabular}
\label{chap6:tab:setup}
\end{table}

\subsection{Results and Discussion}

\begin{figure*}[!ht]
    \center{\includegraphics[width=\textwidth]{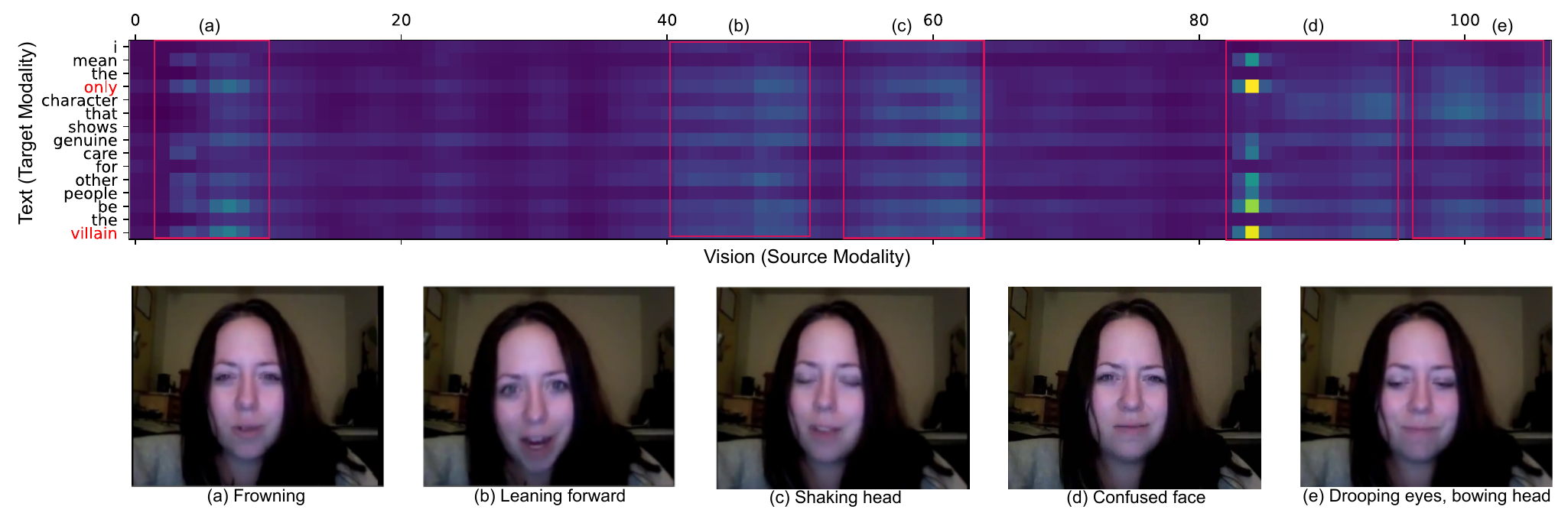}}
    \caption{Exp 1 -- Heatmap of highlighted hidden states using cross-attention on $V$ and $T$.}
    \label{chap6:fig:exp1}
\end{figure*}
    
\begin{figure*}[!ht]
    \center{\includegraphics[width=\textwidth]{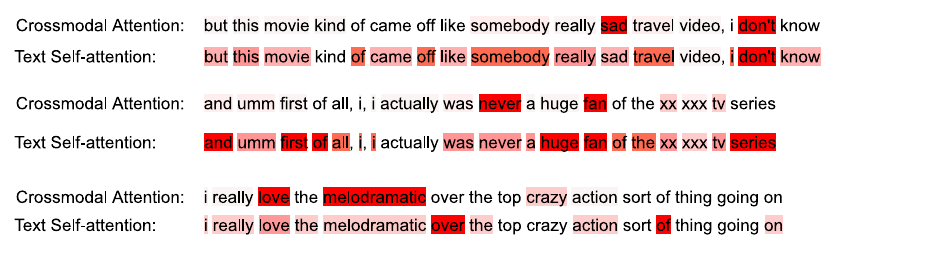}}
    \caption{Exp 2 -- Heatmap of highlighted words using self-attention w/ and w/o cross-attention.}
    \label{chap6:fig:exp2}
\end{figure*}

\begin{figure*}[!ht]
    \center{\includegraphics[width=\textwidth]{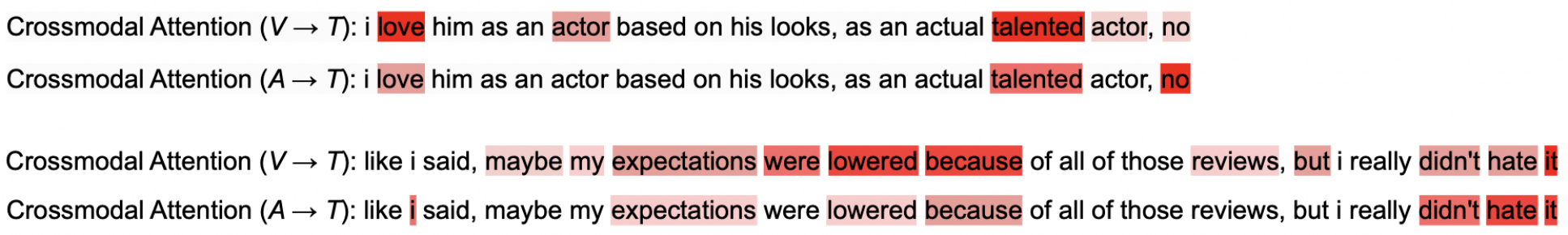}}
    \caption{Exp 3 -- Heatmap of highlighted words using different modalities in cross-attention.}
    \label{chap6:fig:exp3}
\end{figure*}

Figure~\ref{chap6:fig:exp1} shows the video frame (x-axis) and text words (y-axis). The salient affective information captured by cross-attention is highlighted in red boxes. It can be observed that the highlighted parts are due to obvious facial or behavior changes of the character in the video, such as frowning or shaking head. The cross-attention successfully highlights the meaningful words associated with a facial expression (e.g., ``only'', ``villain'').

In Figure \ref{chap6:fig:exp2}, it can be noted that when fused with the combination of source modalities, $T$ focuses more on the words related to emotion with less noise from other words. For example, when with cross-attention, the word ``sad'' is the most salient in the first sentence, yet much less important with self-attention. The same is true for the word ``never'' in the second sentence and the words ``love'' and ``melodramatic'' in the third sentence.

In Figure~\ref{chap6:fig:exp3}, we observe that when fused with different individual source modalities, the target modality $T$ can be enhanced with disparate affective tendencies. When using $V$ as the source modality, the words ``love'' and ``talented'' are the most highlighted in the first sentence, representing a positive meaning. When using $A$, however, ``no'' is the most focused word, showing negation is important. Similarly, ``but'' showing the turnaround in the second sentence is captured by $V$ yet ignored by $A$, and the two draw attention to different parts. These phenomena demonstrate that different modalities may contain mismatched affective tendencies. The existence of IMI has been found by high-level inter-modal comparison \citep{Desai2022}, especially in sentiment analysis \citep{li2019expressing} and sarcasm detection \citep{Wu2021}. This analysis indicates that incongruity also exists latently, resulting in hidden states in one modality being affected by the other.

Based on the above findings, we find that cross-attention does help multimodal fusion by aligning two modalities to highlight the salient affective information in the target modality with complementary information from the source. According to the attention mechanism \citep{vaswani2017attention}, this process can be described as mapping the Query (from the target) to the Key (from the source) and obtain scores for the Value (from the source). However, such a process could malfunction if the modalities have mismatched affective tendencies, which leaves the IMI difficult to resolve at the latent level.

\section{Incongruity-Aware Modality-Gated Fusion}
\label{chap6/sec:incongruity-fusion}
To exploit the advantages of cross-attention while solving the above problems, we propose a new multimodal fusion approach: Hierarchical Crossmodal Transformer with Dynamic Modality Gating (HCT-DMG), which improves on existing methods in two aspects: \textbf{1)} Several previous approaches treated all modalities equally and fused them at every step, leaving incongruity in the fusion \citep{tsai2019multimodal,sahay2020low}, while our HCT-DMG approach initially fuses the auxiliary modalities before integrating the primary modality in the final step. This strategy avoids excessive influence on the effect of the primary modality. \textbf{2)} Some prior work determined a primary modality based on the hierarchy of modalities used before modeling \citep{rahman2020integrating,hazarika2020misa}. Such a practice is empirical and results in a fixed hierarchy. Thereby, the weighting pattern (e.g., $T\oplus W_{1}A\oplus W_{2}V$) cannot be changed during model training even though other hierarchies may become better suited to the task. In contrast, HCT-DMG automatically selects and dynamically changes the primary modality in each training batch and constructs the hierarchy accordingly. Therefore, our proposed approach can \textit{reduce redundancy}, \textit{improve model reliability in incongruent scenarios} and enable the model to be \textit{modality-agnostic}.

\begin{figure}
    \centering
    \includegraphics[width=\textwidth]{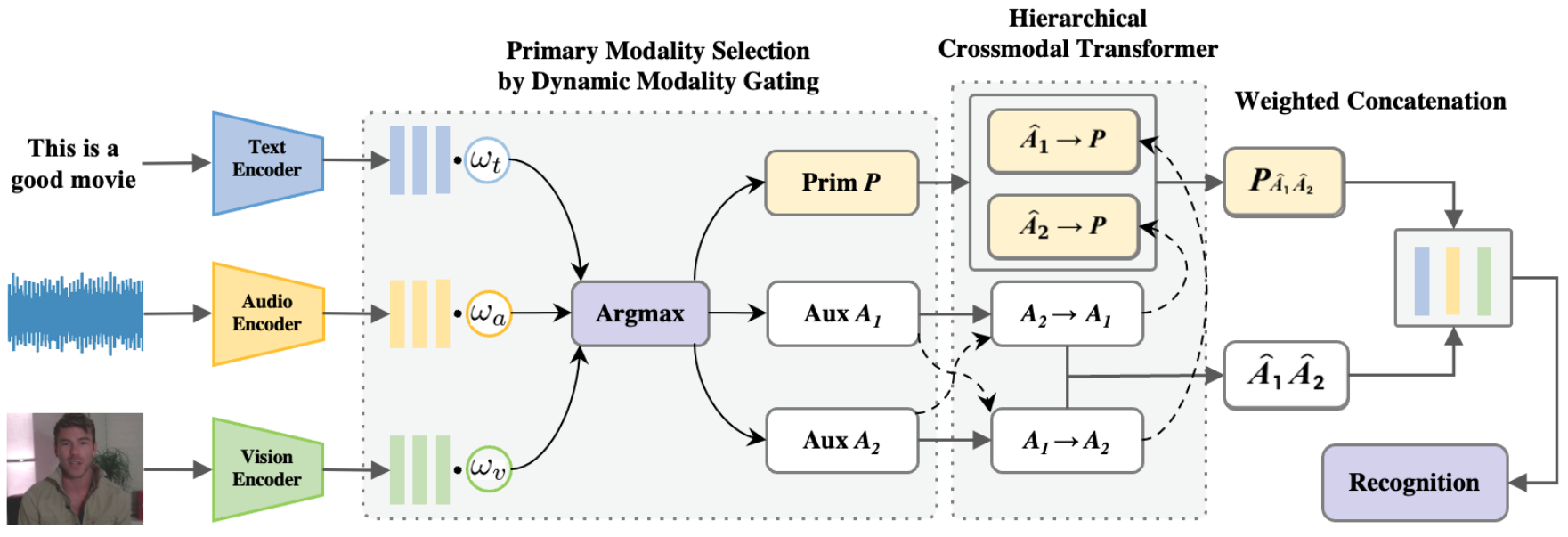}
    \caption{Architecture of HCT-DMG. Prim and Aux are short for primary and auxiliary. Dot lines denote crossmodal attention.}
    \label{chap6:fig:HCT-DMG}
\end{figure}

The architecture is shown as Figure~\ref{chap6:fig:HCT-DMG}. HCT-DMG is constructed based on three modalities: Text ($T$), Audio ($A$), and Vision ($V$), and consists of four components: feature encoder, dynamic modality gating, hierarchical crossmodal Transformer, and weighted concatenation. Note that HCT-DMG supports modalities not limited to $T$, $A$, and $V$, as the dynamic modality gating enables to construct the best hierarchy for any three inputs.

\noindent{\textbf{Feature Encoder.}}
\label{chap6/sec:feature}
The input features are first fed into 1D Convolutional (Conv1D) networks to integrate local contexts and project the features into the same hidden dimension. Then the features are passed to the Gated Recurrent Unit (GRU) networks, which encode global contexts by updating their hidden states recurrently and model the sequential structure. We use two sets of input features: one uses the same conventional feature extractors as \citet{Tsai2018Learning,tsai2019multimodal} and \citet{sahay2020low} for comparison, while the other uses Large Pre-trained Models (LPM) for performance improvement, which will be described in Section~\ref{chap6/sec:setup}.

\noindent{\textbf{Dynamic Modality Gating.}}
DMG determines which modality should be the primary one by the trainable weight for each modality during training, rather than by manual selection. Specifically, each modality is assigned a trainable weight whose value is based on its contribution to the final task. The larger the contribution of a modality, the larger its weight value. The sum of all trainable weights equals to 1, and we allow the weights to be updated in every training batch to ensure that DMG can be well adapted to any input size by working dynamically. We will present the process of how to select the primary modality with DMG in Section~\ref{chap6/sec:mg}.

\begin{algorithm}[!ht]
    \small
    \SetKwFunction{isOddNumber}{isOddNumber}
    \SetKwInOut{KwIn}{Input}

    \KwIn{Primary modality $P$;
    Auxiliary modalities $A_1$ and $A_2$;
    Fixed modality index $Idx_P, Idx_{A_1}, Idx_{A_2} = 0, 1, 2$;
    Dynamic modality gating weight $W$ initialized with \texttt{nn.Param(softmax([1, 1, 1]))}}
   
    \textbf{repeat} \\
    \For{model input}
    {
        
        Sample $Label$, $Text$, $Audio$, $Vision$;
        
        $W' \leftarrow \texttt{softmax(}W\texttt{)}$;

        $P \leftarrow P * W'[Idx_P]$;
        
        $A_1, A_2 \leftarrow A_1 * W'[Idx_{A_1}], A_2 * W'[Idx_{A_2}]$;

        $Idx_{P'} \leftarrow \texttt{argmax(}W'\texttt{)}$;

        \If{$Idx_{P'} == Idx_{A_1}$}
        {
            $P, A_1 \leftarrow A_1, P$
        }
        
        \If{$Idx_{P'} == Idx_{A_2}$}
        {
            $P, A_2 \leftarrow A_2, P$
        }
        
        $Pred \leftarrow \texttt{HCT(}P, A_1, A_2\texttt{)}$;
        $\texttt{update(}Pred, Label\texttt{)}$;
    }
    \textbf{until} no \textit{model input}
    \caption{Principle of HCT-DMG.}
    \label{chap6:alg-dmg}
\end{algorithm}

\noindent{\textbf{Hierarchical Crossmodal Transformer.}}
As a variant of self-attention, cross-attention \citep{lu2019vilbert} transforms the signals from the source modality into a different set of Key-Value pairs to interact with the target modality (in multimodal fusion, cross-attention is usually referred to as crossmodal attention), which has proven useful in various domains \citep{zhang2022cross,rashed2022context}. The crossmodal Transformer used here is the same as that of MulT \citep{tsai2019multimodal}, which is a deep stacking of several cross-attention blocks with layer normalization and positional embeddings. Unlike MulT, which has six crossmodal Transformers in the same step, we use two in the first step $A_1$ and $A_2$ to obtain enhanced auxiliary modalities $\hat{A}_1$ and $\hat{A}_2$:
\begin{equation}
    \hat{A}_1=CMT(A_2\rightarrow{A_1})
\end{equation}
\begin{equation}
    \hat{A}_2=CMT(A_1\rightarrow{A_2})
\end{equation}

Then in the second step, another two crossmodal Transformers are used to yield the enhanced primary modality $\hat{P}_{\hat{A}_1}$ and $\hat{P}_{\hat{A}_2}$:
\begin{equation}
\hat{P}_{\hat{A}_1}=CMT(\hat{A}_1\rightarrow{P})
\end{equation}
\begin{equation}    \hat{P}_{\hat{A}_2}=CMT(\hat{A}_2\rightarrow{P})
\end{equation}

where $P$ denotes the primary modality. The pseudo code of the working principle of HCT-DMG is shown in Algorithm~\ref{chap6:alg-dmg}.

\noindent{\textbf{Weighted Concatenation.}}
After obtaining the enhanced $\hat{P}_{\hat{A}_1}$ and $\hat{P}_{\hat{A}_2}$, we concatenate them and use the self-attention to find its salient parts as the final primary representation $\hat{P}_{\hat{A}_1\hat{A}_2}$:

\begin{equation}
\hat{P}_{\hat{A}_1\hat{A}_2}=SA(Concat\left[\hat{P}_{\hat{A}_1};\ \hat{P}_{\hat{A}_2}\right])
\end{equation}

At this point, crossmodal representations for every modality have been generated: $\hat{A}_1$, $\hat{A}_2$, and $\hat{P}_{\hat{A}_1\hat{A}_2}$. We concatenate them as the final multimodal representation $Z$:

\begin{equation}
Z=Concat\left[W_{1}\hat{A}_1;\ W_{2}\hat{A}_2;\ \hat{P}_{\hat{A}_1\hat{A}_2}\right]
\end{equation}

where $W_{1}$ and $W_{2}$ are the weight matrices, which are learned by the model itself to control how much auxiliary information to extract.

\subsection{Experimental Setup}
\label{chap6/sec:setup}
Same as the analysis experiments, we use \textbf{CMU-MOSI}, \textbf{CMU-MOSEI}, \textbf{IEMOCAP}, \textbf{UR-FUNNY} and \textbf{MUStARD}. On MOSI and MOSEI, we evaluate the performances using the following metrics: 7-class accuracy (Acc7: sentiment score in the same scale as the labeled scores); binary accuracy (Acc2: positive/negative sentiment polarity); F1 score; Mean Absolute Error (MAE); and the correlation of the recognition results with ground truth (Corr). On IEMOCAP, we report the binary classification accuracy (one versus the others) and F1 score. On UR-FUNNY and MUStARD, we report the performance on binary classification accuracy (Acc2: humor/non-humor or sarcasm/non-sarcasm).

For MOSI, MOSEI, and IEMOCAP, we use the CMU-SDK \citep{zadeh2018multi}, which splits the datasets into training/validation/testing folds. For UR-FUNNY and MUStARD, the training, validation, and testing sets have already been split and provided. As described in Section \ref{chap6/sec:feature}, we use both conventional features and LPM features. The conventional features are obtained by using GloVe \citep{pennington2014glove}, FACET \citep{facet}, and COVAREP \citep{p56} for \textit{T}, \textit{V}, and \textit{A}, respectively. For the LPM, we use BERT \citep{devlin2018bert} and WavLM \citep{chen2022wavlm} for \textit{T} and \textit{A}, respectively. We do not use LPM for \textit{V} as they (e.g., CLIP \citep{radford2021learning}) did not show steady improvement as that for \textit{A} and \textit{T}.

\subsubsection{Baselines}
\label{sec:5.2.1}
We perform a comparative study against our approach, considering four aspects: \textbf{1)} models using conventional features; \textbf{2)} models using LPM features; \textbf{3)} models using the same crossmodal Transformer as ours; \textbf{4)} models with similar sizes to ours. The baselines are as below:

Early Fusion LSTM (\textbf{EF-LSTM}) and Late Fusion LSTM (\textbf{LF-LSTM}) \citep{Tsai2018Learning}. Attention or Transformer-based fusion: \textbf{RAVEN} \citep{Wang2019}, \textbf{MulT} \citep{tsai2019multimodal}. Graph-based fusion: \textbf{Graph-MFN} \citep{Zadeh2018}. Low-rank-based fusion: \textbf{LMF} \citep{Liu2018}. Cyclic translations-based fusion: \textbf{MCTN} \citep{pham2019found}. Context-aware attention-based fusion: \textbf{CIA} \citep{chauhan2019context}. Multi-attention Recurrent-based fusion: \textbf{MARN} \citep{zadeh2018multi}. Temporal memory-based fusion: \textbf{MFN} \citep{zadeh2018memory}. Memory fusion by incorporating the information from the preceding context: \textbf{C-MFN}. Recurrent multiple stages-based fusion: \textbf{RMFN} \citep{liang2018multimodal}. Low-rank Transformer-based fusion: \textbf{LMF-MulT} \citep{sahay2020low}. Modality-invariant and -specific fusion using LPM: \textbf{MISA} \citep{hazarika2020misa}. Integration of the preceding context and external knowledge: \textbf{HKT} \citep{hasan2021humor}. Cascade residual autoencoder-based: \textbf{MMIN} \citep{zhao2021mmin}. Masked multihead attention-based: \textbf{TATE} \citep{zeng2022tate}. Multimodal reconstruction-based: \textbf{MRAN} \citep{luo2023mran}. Modality translation-based: \textbf{TransM} \citep{wang2020transm} and \textbf{MTMSA} \citep{liu2024mtmsa}. Graph network-based: \textbf{GCNet} \citep{lian2023gcnet}. Knowledge distillation-based: \textbf{CorrKD} \citep{li2024corrkd}. Adversarial learning-based: \textbf{HRLF} \citep{li2024hrlf}. We also include several of the above-mentioned models enhanced by Connectionist Temporal Classification (CTC) \citep{tsai2019multimodal}.

Note that not all of the above baselines are employed for comparison across all five datasets for the following reasons:

\begin{enumerate}
    \item As the recognition of fine-grained emotions can be largely improved by word-aligned data, we do not use the baselines with such alignment for a fair comparison on IEMOCAP.

    \item On MOSI, MOSEI, and IEMOCAP, the current State-Of-The-Art (SOTA) results are achieved by sophisticated approaches such as Self-MM \citep{yu2021learning}, DMD \citep{li2023decoupled}, CorrKD \citep{li2024corrkd}, and MAG-XLNet \citep{rahman2020integrating}. However, their approaches fundamentally differ from the aforementioned baselines as they either generated additional unimodal labels using multimodal information and labels (i.e., ground-truth) or integrated multimodal information into the Transformer by modifying its structure. Therefore, we exclude them from the comparison since all the models listed in the table concentrate exclusively on fusion methods without introducing extra training tasks or making modifications to the Transformers or pre-trained encoders.

    \item Since humor and sarcasm detection are relatively new tasks in affective computing compared to sentiment analysis and emotion recognition, there are fewer baselines available. Therefore, we also include their variants that used different language models for comparison. We omit the size comparison as there are no comparable small models in the literature. For the HKT baseline, we use the performance without its proposed humor centric features because these features represent an additional modality, resulting in an unfair comparison with the others based on \textit{T}, \textit{V}, \textit{A}. For the same reason in 2), we also exclude MAG-XLNet for comparison on UR-FUNNY and MUStARD.
\end{enumerate}

\subsection{Results and Discussion}
\begin{table}[!ht]
    \centering
    \caption{Comparison results of sentiment analysis on MOSI.}
    \begin{tabular}{l|ccccc}
    \hline
    \multirow{2}{*}{Model} & \multicolumn{5}{c}{CMU-MOSI} \\
     & Acc7$\uparrow$ & Acc2$\uparrow$ & F1$\uparrow$ & Corr$\uparrow$ & MAE$\downarrow$ \\ \hline
    EF-LSTM & 33.7 & 75.3 & 75.2 & 0.608 & 1.023 \\
    RAVEN & 33.2 & 78.0 & 76.6 & 0.691 & 0.915 \\
    MCTN & 35.6 & 79.3 & 79.1 & 0.676 & 0.909 \\
    CTC+EF-LSTM & 31.0 & 73.6 & 74.5 & 0.542 & 1.078 \\
    CTC+RAVEN & 31.7 & 72.7 & 73.1 & 0.544 & 1.076 \\
    CTC+MCTN & 32.7 & 75.9 & 76.4 & 0.613 & 0.991 \\
    MARN & 34.7 & 77.1 & 77.0 & 0.625 & 0.968 \\
    MFN & 34.1 & 77.4 & 77.3 & 0.632 & 0.965 \\
    RMFN & 38.3 & 78.4 & 78.0 & 0.681 & 0.922 \\
    LMF & 32.8 & 76.4 & 75.7 & 0.668 & 0.912 \\
    CIA & 38.9 & 79.8 & 79.5 & 0.689 & 0.914 \\
    MTMSA & - & 84.9 & - & - & -  \\
    MMIN & - & 82.3 & - & - & -  \\
    TATE & - & 84.9 & - & - & -  \\
    MRAN & - & 83.9 & - & - & - \\
    TransM & - & - & 82.6 & - & - \\
    GCNet & - & - & 83.2 & - & -  \\
    CorrKD & - & - & 83.9 & - & -  \\
    HRLF & - & - & 84.2 & - & -  \\
    MISA$^{\dag}$ & 41.4 & 81.9 & 81.8 & \textbf{0.762} & \textbf{0.810} \\
    MulT$^{\diamond}$ \small{(1.07M)} & 34.3 & 80.3 & 80.4 & 0.645 & 1.008 \\
    LMF-MulT$^{\diamond}$ \small{(0.86M)} & 34.0 & 78.5 & 78.5 & 0.681 & 0.957 \\
    LF-LSTM \small{(1.24M)} & 33.7 & 77.6 & 77.8 & 0.624 & 0.988 \\ \hdashline
    HCT-DMG &  &  &  & \\
    \ \ \ \textit{Conven.} \small{(0.78M)} & 39.4 & 82.5 & 82.5 & 0.710 & 0.881 \\
    \ \ \ \textit{LPM}$^{\dag}$ \small{(0.83M)} & \textbf{41.8} & \textbf{85.1} & \textbf{84.8} & 0.732 & 0.855 \\ \hline
    \end{tabular}
    \label{chap6:tab:mosi}
    \end{table}
    
    \begin{table}[!ht]
    \caption{Comparison results of emotion recognition on MOSEI.}
    \centering
    \begin{tabular}{l|ccccc}
    \hline
    \multirow{2}{*}{Model} & \multicolumn{5}{c}{CMU-MOSEI} \\
     & Acc7$\uparrow$ & Acc2$\uparrow$ & F1$\uparrow$ & Corr$\uparrow$ & MAE$\downarrow$ \\ \hline
    EF-LSTM & 47.4 & 78.2 & 77.9 & 0.642 & 0.616 \\
    RAVEN & 50.0 & 79.1 & 79.5 & 0.662 & 0.614 \\
    MCTN & 49.6 & 79.8 & 80.6 & 0.670 & 0.609 \\
    CTC+EF-LSTM & 46.3 & 76.1 & 75.9 & 0.585 & 0.680 \\
    CTC+RAVEN & 45.5 & 75.4 & 75.7 & 0.599 & 0.664 \\
    CTC+MCTN & 48.2 & 79.3 & 79.7 & 0.645 & 0.631 \\
    LMF & 48.0 & 82.0 & 82.1 & 0.677 & 0.623 \\
    Graph-MFN & 45.0 & 76.9 & 77.0 & 0.540 & 0.710 \\
    CIA & 50.1 & 80.4 & 78.2 & 0.590 & 0.680 \\
    TransM & - & - & 81.5 & - & - \\
    GCNet & - & - & 82.4 & - & -  \\
    CorrKD & - & - & 82.2 & - & -  \\
    HRLF & - & - & 82.9 & - & -  \\
    MISA$^{\dag}$ & 51.8 & \textbf{84.2} & \textbf{84.0} & 0.724 & 0.568 \\
    MulT$^{\diamond}$ \small{(1.07M)} & 50.4 & 80.7 & 80.6 & 0.677 & 0.617 \\
    LMF-MulT$^{\diamond}$ \small{(0.86M)} & 49.3 & 80.8 & 81.3 & 0.668 & 0.620 \\
    LF-LSTM \small{(1.24M)} & 48.8 & 77.5 & 78.2 & 0.656 & 0.624 \\ \hdashline
    HCT-DMG &  &  &  &  & \\
    \ \ \ \textit{Conven.} \small{(0.78M)} & 50.6 & 81.6 & 81.9 & 0.691 & 0.593 \\
    \ \ \ \textit{LPM}$^{\dag}$ \small{(0.83M)} & \textbf{53.2} & \textbf{84.2} & \textbf{84.0} & \textbf{0.752} & \textbf{0.535}\\ \hline
    \end{tabular}
    \label{chap6:tab:mosei}
    \end{table}
    
    \begin{table}[!ht]
    \caption{Comparison results of emotion recognition on IEMOCAP.}
    \centering
    \begin{tabular}{l|cccccccc}
    \hline
    \multirow{3}{*}{Model} & \multicolumn{8}{c}{IEMOCAP} \\
     & \multicolumn{2}{c}{Happy} & \multicolumn{2}{c}{Sad} & \multicolumn{2}{c}{Angry} & \multicolumn{2}{c}{Neutral} \\
     & Acc & F1  & Acc & F1 & Acc & F1 & Acc & F1 \\ \hline
    CTC+EF-LSTM & 76.2 & 75.7 & 70.2 & 70.5 & 72.7 & 67.1 & 58.1 & 57.4 \\
    CTC+RAVEN & 77.0 & 76.8 & 67.6 & 65.6 & 65.0 & 64.1 & 62.0 & 59.5 \\
    CTC+MCTN & 80.5 & 77.5 & 72.0 & 71.7 & 64.9 & 65.6 & 49.4 & 49.3 \\
    MulT$^{\diamond}$ \small{(1.07M)} & 85.6 & 79.0 & 79.4 & 70.3 & 75.8 & 65.4 & 59.5 & 44.7 \\
    LMF-MulT$^{\diamond}$ \small{(0.86M)} & 85.6 & 79.0 & 79.4 & 70.3 & 75.8 & 65.4 & 59.2 & 44.0 \\
    LF-LSTM \small{(1.24M)} & 72.5 & 71.8 & 72.9 & 70.4 & 68.6 & 67.9 & 59.6 & 56.2 \\ \hdashline
    HCT-DMG  &  &  &  & \\
    \ \ \ \textit{Conven.} \small{(0.78M)} & 85.6 & 79.0 & 79.4 & 70.3 & 75.8 & 65.4 & 61.0 & 50.5 \\
    \ \ \ \textit{LPM}$^{\dag}$ \small{(0.83M)} & \textbf{87.1} & \textbf{81.6} & \textbf{82.4} & \textbf{73.2} & \textbf{79.0} & \textbf{68.8} & \textbf{63.2} & \textbf{60.3} \\ \hline
    \end{tabular}
    \label{chap6:tab:iem}
    \end{table}
    
    \begin{table}[!ht]
    \caption{Comparison results of humor and sarcasm detection on UR-FUNNY and MUStARD.}
    \centering
    \begin{tabular}{l|cc}
    \hline
    \multirow{2}{*}{Model} & \multicolumn{1}{c}{UR-FUNNY} & \multicolumn{1}{c}{MUStARD} \\
     & Acc2$\uparrow$ & Acc2$\uparrow$ \\ \hline
    C-MFN (GloVe) & 65.23 & - \\
    C-MFN (ALBERT)$^{\dag}$ & 61.72 & - \\
    MISA (GloVe) & 68.60 & - \\
    MISA (BERT)$^{\dag}$ & 69.62 & 66.18 \\
    MISA (BERT)$^{\dag}$ [punchline only] & 70.61 & - \\
    MISA (ALBERT)$^{\dag}$ & 69.82 & 66.18 \\
    HKT$^{\dag}$ & \textbf{76.36} & 75.00 \\
    \hdashline
    HCT-DMG &  &   \\
    \ \ \ \textit{Conven.} & 68.54 & 73.29 \\
    \ \ \ \textit{LPM}$^{\dag}$ & 75.09 & \textbf{76.62} \\ \hline
    \end{tabular}
    \label{chap6:tab:urfunny}
    \end{table}

    The comparison results are shown in Tables~\ref{chap6:tab:mosi}, \ref{chap6:tab:mosei}, \ref{chap6:tab:iem}, and \ref{chap6:tab:urfunny}. Note that MulT and MISA are our reproduced results using their official code. ${\dag}$ denotes models using LPM features. ${\diamond}$ denotes models using the same crossmodal Transformer as ours. On all five datasets, it can be seen that HCT-DMG achieves better results on almost every metric than the baselines when using LPM features, showing the effectiveness of our proposed approach. 

Moreover, when using conventional features, HCT-DMG improves every metric on MOSI and MOSEI, and almost every metric on IEMOCAP compared to the models of similar size. It also achieves highly competitive performance even compared to some LPM baselines on MOSI, UR-FUNNY, and MUStARD. In addition, compared to LMF-MulT and MulT, which use the same crossmodal Transformer as ours, HCT-DMG still outperforms them, especially on MOSI and MOSEI by a large margin. These demonstrate the superiority of our proposed approach. 

To the best of our knowledge, HKT is the SOTA model on humor and sarcasm detection in the literature thus far. Our result on UR-FUNNY does not surpass theirs, but we achieve higher result on MUStARD, demonstrating the competitiveness of our model compared to the SOTA. Note that HKT was developed specifically for experimentation on these two datasets with a punchline detection module, its effectiveness is unlikely to generalize to MOSI, MOSEI, and IEMOCAP for sentiment analysis and emotion recognition. On the contrary, HCT-DMG performs well on all five datasets, demonstrating that our proposed approach can address the IMI problem not only in general incongruous scenarios (humor and sarcasm) but also in complex ones (sentiment and emotion). Therefore, the generalizability of HCT-DMG is well-established, allowing for its extension to various MIP tasks.

Furthermore, as the cross-attention results in Exp. 2 in Section~\ref{chap6/sec:exploration} are produced by using HCT-DMG, Figure~\ref{chap6:fig:exp2} clearly shows that our approach strengthened the salient affective parts.

Finally, we present some hard samples in Table~\ref{chap6:tab:example}, where incongruity exists. It can be observed that MulT (using the same crossmodal Transformer as ours yet repeatedly fusing every modalities) fails to handle these hard samples (videos available\footnote{\url{https://sites.google.com/view/taclsubmission}}) producing results that contradict ground truths. In contrast, our approach can recognize true sentiments with very close scores.

In summary, the effectiveness of our proposed HCT-DMG has been demonstrated from different angles by the experimental results, the heatmap visualization, and the recognition scores of several hard samples.

\begin{table}
\centering
\caption{Examples containing incongruity from CMU-MOSI.}
\resizebox{\textwidth}{!}{
\begin{tabular}{llccc}
\hline
\# & \multicolumn{1}{c}{Spoken words + acoustic and visual behaviors}                                & Ground truth & MulT  & Ours   \\ \hline
1 &
  \begin{tabular}[c]{@{}l@{}}``And that's why I was not excited about the fourth one.''\\ + Uninterested tone and facial expression\end{tabular} &
  \colorbox{lime}{-1.4} &
  \colorbox{pink}{1.185} &
  \colorbox{lime}{-1.416} \\ \hline
2 &
  \begin{tabular}[c]{@{}l@{}}``I give Shrek Forever After directed by Mike Mitchell a grade of B minus.''\\ + Smile face\end{tabular} &
  \colorbox{pink}{1.0} &
  \colorbox{lime}{-0.576} &
  \colorbox{pink}{0.959} \\ \hline
  3 &
  \begin{tabular}[c]{@{}l@{}}``Um in general um, the little kids seemed to like it that were in there.''\\ + Skeptical tone and facial expression\end{tabular} &
  \colorbox{pink}{0.8} &
  \colorbox{lime}{-1.151} &
  \colorbox{pink}{0.700} \\ \hline
4  & \begin{tabular}[c]{@{}l@{}}``I honestly want the aliens to win.''\\ + Negative tone (somewhat disdainful) \end{tabular} & 
  \colorbox{lime}{-1.6} & 
  \colorbox{pink}{0.995} & 
  \colorbox{lime}{-1.906}  \\ \hline
\end{tabular}
}
\label{chap6:tab:example}
\end{table}

\subsection{Analysis and Discussion}
To verify that our approach alleviates the IMI issue as well as to demonstrate how DMG dynamically changes the primary modality, we conduct some further studies. To showcase the advantages of our approach and for brevity, we present the analysis results using only conventional features.

\subsubsection{Alleviation of the IMI Problem}

\begin{figure}[!ht]
\centering
\includegraphics[scale=0.22]{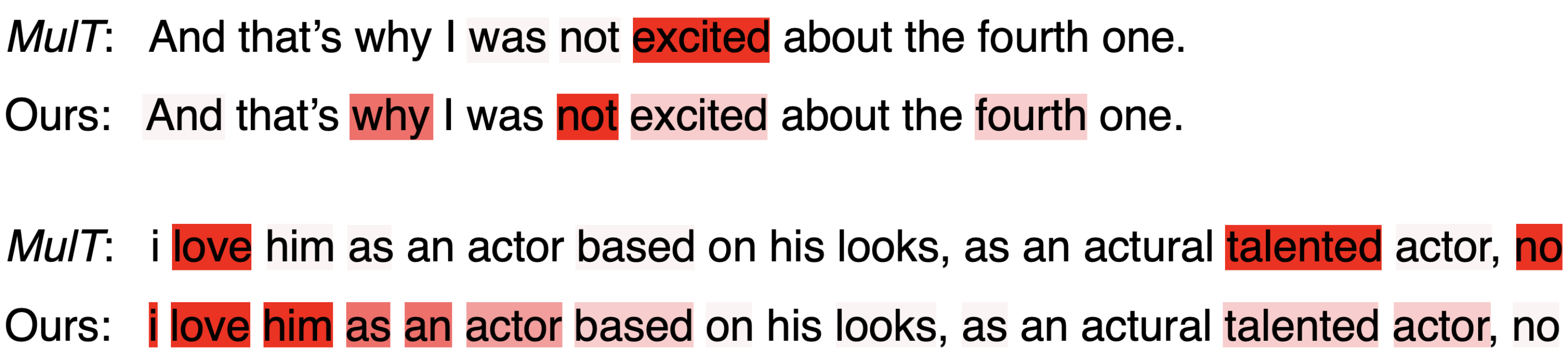}
\caption{Comparison examples of IMI.}
\label{chap6:fig:incongruity}
\end{figure}

First, as shown in Figure~\ref{chap6:fig:exp3} in Section~\ref{chap6/sec:exploration}, the affective words are enhanced by the auxiliary modalities. However, $V$ focuses more on positive words while $A$ highlights negation the most, which likely changes affective tendency. We re-implemented MulT and extracted the attention of its enhanced $T$ in the final stage to compare with ours. In the first example of Figure \ref{chap6:fig:incongruity}, it can be seen that MulT is confused by the IMI, resulting in the negative word being barely focused on, which yields a completely opposite affective tendency to the ground truth. In the second example, the positive and negative words are treated equally by MulT, leaving the IMI unsolved. These occur because MulT fuses $V$ and $A$ with $T$ at the same level, and simply concatenates two enhanced $T$. In contrast, our approach gives little attention to the word ``no'', showing that the IMI is resolved at the latent level as the hidden states of ``no'' are barely encoded by the cross-attention. Together with the examples in Table \ref{chap6:tab:example}, it is observed that our approach can greatly alleviate the IMI problem for these examples.

\subsubsection{Automatic Modality Selection by DMG}
\label{chap6/sec:mg}
\begin{figure}[ht]
    \centering
    \includegraphics[width=0.7\textwidth]{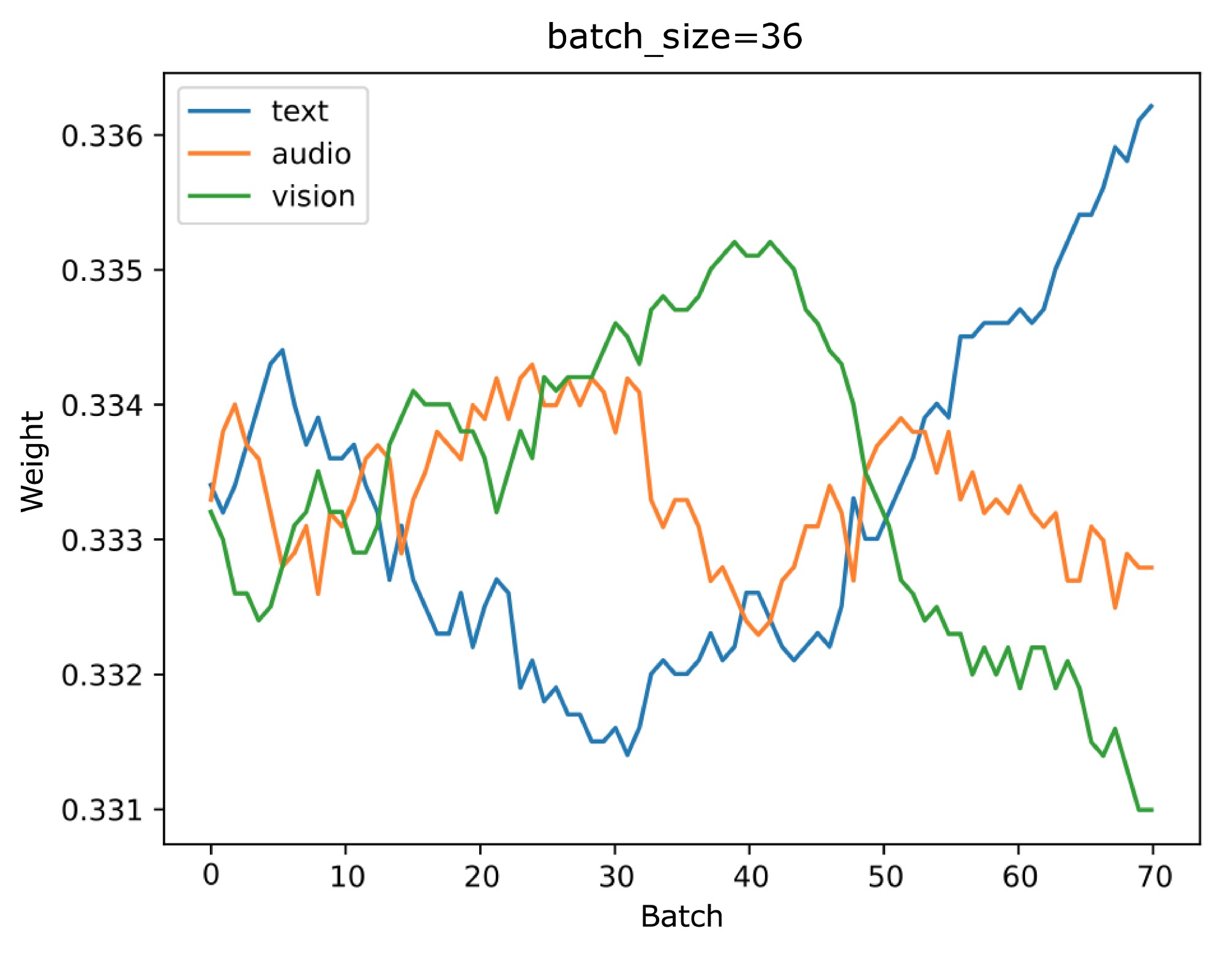}
    \caption{Weight variation in the first epoch on MOSI dataset.}
    \label{chap6:fig:fig.16}
\end{figure}

\begin{figure}[ht]
    \centering
    \includegraphics[width=0.7\textwidth]{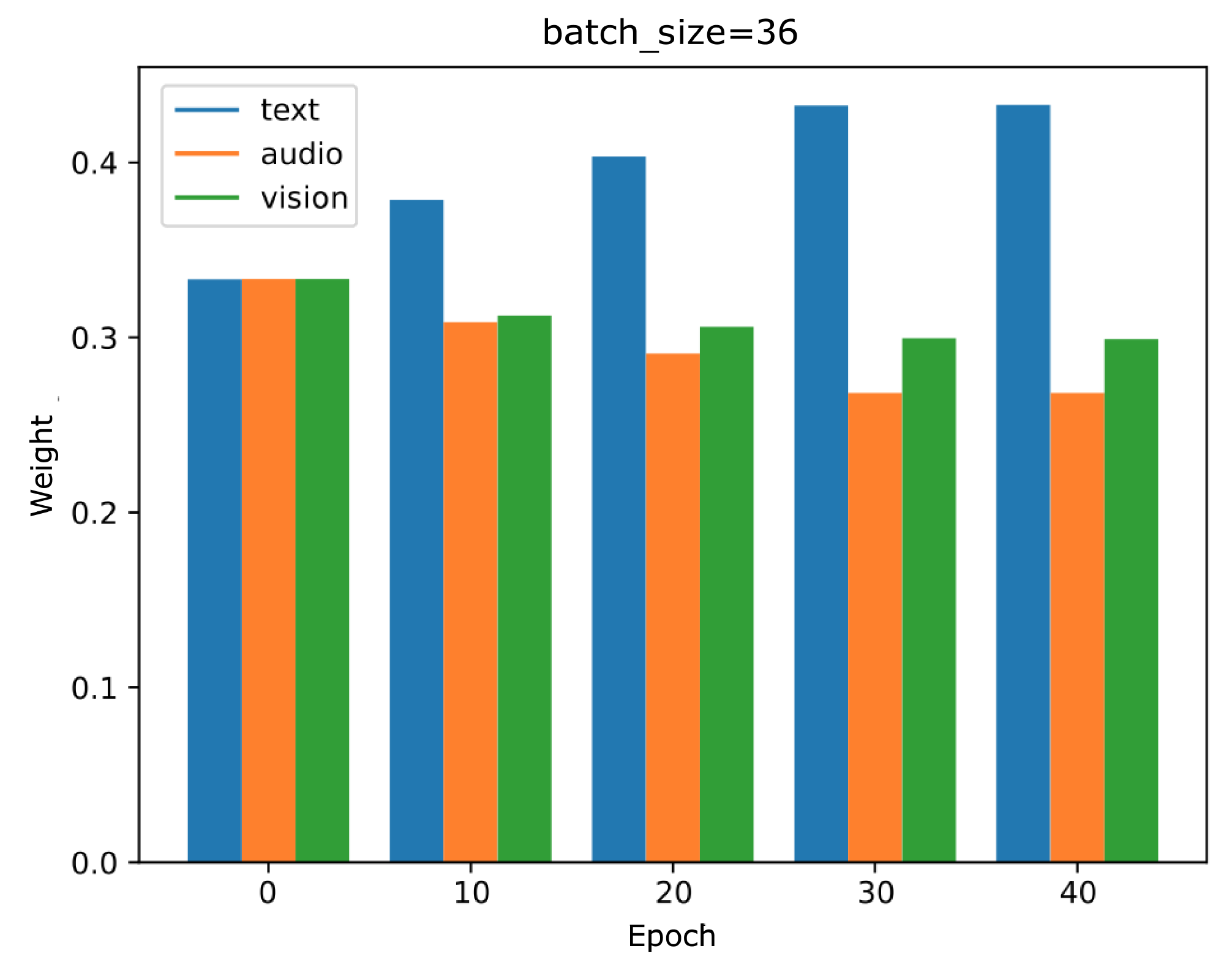}
    \caption{Variation of average weight with epoch on MOSI dataset.}
    \label{chap6:fig:fig.17}
\end{figure}

As the DMG automatically selects the primary modality by adjusting the weight for each modality, we show how the weights vary during training using CMU-MOSI. The weight of a modality denotes the confidence that this modality is selected as the primary one. Figure~\ref{chap6:fig:fig.16} shows how weights of the modalities vary in the first epoch. It can be noted that $T$ is not the primary one at the beginning but gradually dominates after batch 60. Figure~\ref{chap6:fig:fig.17} shows the variation in the average weight of each modality with epoch. It can be seen that $T$ indeed dominates, and the weight distribution starts to converge at around epoch 40. Meanwhile, $V$ gradually surpasses $A$ with epoch although the opposite is true in the first epoch in Figure~\ref{chap6:fig:fig.16}. It can be inferred that DMG can dynamically change the weights as the training goes on, demonstrating the efficacy of our approach.

Furthermore, since HCT-DMG works without the need to understand which modalities are used, its utility can be expected to expand beyond more general scenarios with different inputs. 

\subsubsection{Ablation Study on Removal of DMG}
\label{sec:ablation}

Although DMG automatically selects $T$ as the primary modality, we hope to see if this phenomenon really brings the best performance compared to prior work whose hierarchies are fixed. Thus, we remove DMG and construct an HCT model with three hierarchies, in each of which either $T$, $A$, or $V$ was manually selected as the primary modality respectively for performance comparison. As shown in Table~\ref{chap6:tab:ablation}, selecting $T$ as the primary modality achieves the best performance on every metric on MOSI and MOSEI and on most metrics on IEMOCAP, verifying the rationality and efficacy of the automatic modality selection by DMG.

Moreover, although the results of selecting $T$ as the primary modality are the best among the three, overall scores on all three datasets are slightly lower than HCT-DMG's in Tables~\ref{chap6:tab:mosi}, \ref{chap6:tab:mosei}, \ref{chap6:tab:iem}, and \ref{chap6:tab:urfunny}. This phenomenon is reasonable because the primary modality keeps changing during training, even though $T$ is primary overall. Such a dynamic property cannot be encoded by the manual selection of the primary modality whereas it can be encoded by HCT-DMG and thus yields better results. This is consistent with the weight variation observed in Figure~\ref{chap6:fig:fig.16} and \ref{chap6:fig:fig.17}, which reaffirms the validity of our approach. In addition, the results on IEMOCAP, where no modality consistently dominates, align with their close weights (all slightly above 0.33 by the experiment in Section~\ref{chap6/sec:mg} using IEMOCAP).

\begin{table}[!ht]
\centering
\caption{Performance comparison by manually selecting different primary modalities with convention features.}
\label{chap6:tab:ablation}
\begin{tabular}{c|ccccc}
\hline
Primary & \multicolumn{5}{c}{CMU-MOSI} \\ 
modality & Acc-7$\uparrow$ & Acc-2$\uparrow$ & F1$\uparrow$ & Corr$\uparrow$ & MAE$\downarrow$ \\ \hline
$T$ & \textbf{38.9} & \textbf{82.5} & \textbf{82.6} & \textbf{0.717} & \textbf{0.859}\\
$A$ & 37.5 & 81.3 & 81.3 & 0.705 & 0.883 \\
$V$ & 38.3 & 80.9 & 81.0 &  0.679 & 0.909 \\ \hline
\end{tabular}
\label{tab:4}
\end{table}

\begin{table}[!ht]
\centering
\begin{tabular}{c|ccccc}
\hline
Primary & \multicolumn{5}{c}{CMU-MOSEI} \\ 
modality & Acc-7$\uparrow$ & Acc-2$\uparrow$ & F1$\uparrow$ & Corr$\uparrow$ & MAE$\downarrow$ \\ \hline
$T$ & \textbf{50.1} & \textbf{81.8} & \textbf{81.9} & \textbf{0.685} & \textbf{0.601}\\
$A$ & 47.5 & 79.6 & 80.3 & 0.650 & 0.644 \\
$V$ & 48.7 & 80.8 & 81.0 &  0.659 & 0.633 \\ \hline
\end{tabular}
\end{table}

\begin{table}[!ht]
\centering
\begin{tabular}{c|cccccccc}
\hline
Primary & \multicolumn{8}{c}{IEMOCAP} \\
modality & \multicolumn{2}{c}{Happy} & \multicolumn{2}{c}{Sad} & \multicolumn{2}{c}{Angry} & \multicolumn{2}{c}{Neutral} \\
 & Acc & F1 & Acc & F1 & Acc & F1 & Acc & F1 \\ \hline
$T$ & \textbf{85.6} & \textbf{79.4} & \textbf{79.5} & \textbf{70.6} & 75.8 & 65.4 & 59.6 & \textbf{55.6} \\
$A$ & 85.4 & 79.4 & 78.8 & 70.4 & 75.4 & 65.5 & 59.3 & 52.4 \\
$V$ & 85.6 & 79.2 & 79.4 & 70.3 & \textbf{75.9} & \textbf{65.6} & \textbf{60.4} & 53.3 \\ \hline
\end{tabular}
\end{table}

We omit UR-FUNNY and MUStARD for brevity as the prior work by HKT has explored the contribution, where each modality was individually used, and the other two were removed \citep{hasan2021humor}. They observed that in humor detection, the contribution of each modality followed the order of $T$ > $A$ > $V$, while in sarcasm detection, the order is $T$ = $A$ > $V$.

\subsection{Integration with LLMs}
\label{chap6/sec:llm}
To further validate our approach, we explore the performance using LLMs. Given the necessity of a sufficient amount of data to fine-tune an LLM into a multimodal LLM, we combine MOSI, MOSEI, and IEMOCAP. To align these datasets, we unify their emotion labels into three categories: \textit{positive, negative, and neutral}. These datasets are trained together, with separate evaluations conducted for each.

\begin{figure}[ht]
    \centering
    \includegraphics[width=0.8\textwidth]{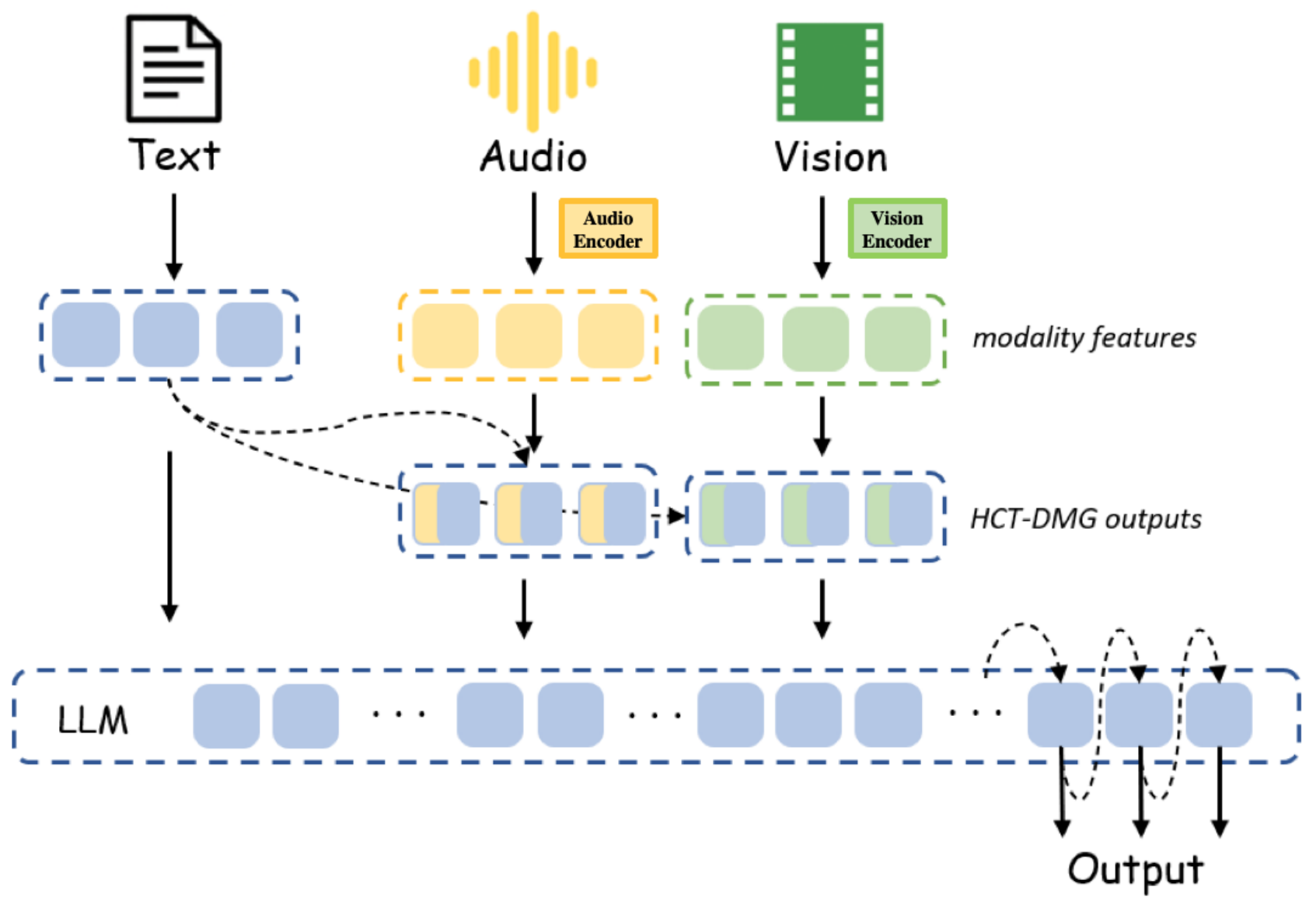}
    \caption{LLM integrated with HCT-DMG (LLM-HCT).}
    \label{chap6:fig:llm-hct}
\end{figure}

We use \textit{LlaMA-3.1-8b} \citep{dubey2024llama} as the LLM and employ a single-layer linear projection to align the multimodal inputs with the text feature space, following the setup of MiniGPT-4 \citep{zhu2023minigpt}. Furthermore, to perform end-to-end emotion-aware feature extraction for multimodal alignment, we use \textit{emotion2vec plus large} as the speech encoder and HSemotion as the visual encoder. Emotion2vec \citep{ma2023emotion2vec} is a self-supervised speech emotion representation model designed to extract universal emotional features from speech signals. It utilizes self-supervised online distillation, combining utterance-level and frame-level losses during pre-training on open-source unlabeled emotion data. HSemotion \citep{savchenko2022hsemotion} is a high-speed and lightweight facial emotion recognition model. It employs multi-task learning to simultaneously predict basic emotions, valence-arousal values, and expression intensity from images or video frames. We employ Emotion2vec and HSemotion to extract emotion-aware speech and visual representations, aiming to enhance the emotional capabilities of the LLM. Due to resource constraints, the model is trained with a batch size of 2 for 5 epochs. The architecture is shown in Figure~\ref{chap6:fig:llm-hct}. The task command, emotional sentence, and available emotion options are integrated into a single text input to prompt the LLM: ``\textit{Please analyze the emotion of the given sentence and choose the emotion from \{Positive, Neutral, Negative\}}''. Since LLMs are primarily trained on textual data, we designate text as the primary modality to leverage their strong reasoning abilities in this domain (in other words, assigning audio or visual input as the primary modality severely impairs their reasoning performance).

Considering that fine-tuning a pure text-based LLM model to a multimodal model requires a large amount of data, we concatenate three datasets for joint training, but evaluate them separately. Specifically, we standardize the labels of the MOSI, MOSEI, and IEMOCAP datasets into three categories: positive, neutral, and negative, as shown in Table~\ref{chap6:tab:llm_data}. Audio and visual inputs are encoded into 1024 and 1280 dimensions, respectively, and then projected into 4096 dimensions to align with the text feature dimension of \textit{LlaMA-3.1-8b}.

\begin{table}[ht]
\caption{The data distribution of the dataset used for training the LLM.}
\label{chap6:tab:llm_data}
\resizebox{\columnwidth}{!}{%
\begin{tabular}{l|ccc|ccc|ccc|c}
\hline
\multirow{2}{*}{Sentiment} & \multicolumn{3}{c|}{MOSI} & \multicolumn{3}{c|}{MOSEI} & \multicolumn{3}{c|}{IEMOCAP} & \multirow{2}{*}{Total} \\
 & Train & Valid & Test & Train & Valid & Test & Train & Valid & Test &  \\ \hline
Positive & 680 & 124 & 277 & 8,062 & 932 & 2,289 & 1,393 & 180 & 176 & 14,113 \\
Neutral & 54 & 13 & 30 & 3,544 & 433 & 1,026 & 1,331 & 161 & 161 & 6,753 \\
Negative & 552 & 92 & 379 & 4,572 & 507 & 1,353 & 3,194 & 400 & 400 & 11,449 \\
Total & 1,286 & 229 & 686 & 16,178 & 1,872 & 4,668 & 5,918 & 741 & 737 & 32,315 \\ \hline
\end{tabular}%
}
\end{table}

We evaluate our LLM-HCT in comparison with two baseline strategies:

\begin{itemize}
    \item \textit{\textbf{LLM-Concat}}: Features from the audio and visual modalities are concatenated to those from the text modality.
    \item \textit{\textbf{LLM-MulT}}: Each pair among the three modalities undergoes cross-modal attention, and the resulting outputs are then concatenated together (same as the fusion strategy of MulT \citep{tsai2019multimodal}).
    \item \textit{\textbf{LLM-HCT}}: The three modalities are combined using our proposed HCT-DMG (as in Figure~\ref{chap6:fig:llm-hct}).
\end{itemize}

As shown in Table~\ref{tab:llm_exps}, our proposed scheme still outperforms MulT within the LLM framework and achieves the best performance among the three approaches. Interestingly, MulT leads to a performance drop, even compared to simple concatenation, possibly due to the added complexity of aligning modalities through cross-modal attention. The repeated fusion of visual and audio information may hinder the LLM from effectively learning relevant information from the textual modality. In contrast, LLM-HCT designates text as the primary modality and hierarchically integrates information from the other two modalities into it, minimizing divergence from the original text feature space and enabling more effective adaptation of visual and audio features to the text modality. Finally, it is noticed that with the visual modality, the results are worse than those without visual modality on MOSI. We believe that it is likely due to two reasons: \textbf{1)} Audio and text typically align well, particularly in tasks involving emotion, sentiment, or intent. Since the LLM is pre-trained solely on text, aligning visual features with text is more challenging. \textbf{2)} Introducing a third modality increases model complexity (particularly because visual features often lack consistency with textual and audio features, as noted in Reason 1) and increases data requirements. Due to the limited size of MOSI, the model is susceptible to overfitting on the training data.

\begin{table}
\centering
\caption{Comparison results using LLMs.}
\resizebox{\columnwidth}{!}{%
\begin{tabular}{l|cc|cccccc}
\hline
\multirow{2}{*}{Setting} & \multirow{2}{*}{Audio} & \multirow{2}{*}{Visual} & \multicolumn{2}{c}{MOSI} & \multicolumn{2}{c}{MOSEI} & \multicolumn{2}{c}{IEMOCAP} \\ 
 &  &  & Acc & F1 & Acc & F1 & Acc & F1 \\ \hline
LLM-Concat & \ding{51} & \ding{55} & 78.51 & 78.51 & 72.17 & 72.17 & 76.64 & 76.64 \\
LLM-Concat & \ding{51} & \ding{51} & 77.97 & 77.97 & 72.66 & 72.66 & 78.56 & 78.56 \\
LLM-Cross & \ding{51} & \ding{55} & 78.26 & 78.26 & 71.94 & 71.94 & 76.60 & 76.60 \\
LLM-Cross & \ding{51} & \ding{51} & 77.43 & 77.43 & 72.06 & 72.06 & 77.91 & 77.91 \\
LLM-HCT & \ding{51} & \ding{55} & \textbf{80.17} & \textbf{80.17} & 72.49 & 72.49 & 77.42 & 77.42 \\
LLM-HCT & \ding{51} & \ding{51} & 79.70 & 79.70 & \textbf{73.05} & \textbf{73.05} & \textbf{78.93} & \textbf{78.93} \\ \hline
\end{tabular}%
}
\label{tab:llm_exps}
\end{table}

\subsection{Summary}
In this section, we analyzed cross-attention-based multimodal fusion and propose a Hierarchical Crossmodal Transformer with Dynamic Modality Gating for incongruity-aware multimodal affect recognition. \textbf{1)} We demonstrate the existence of inter-modal incongruity at the latent level due to cross-attention. Specifically, we show that cross-attention can help to capture affective information across modalities and enhance salient parts in the target modality, but it can also induce mismatched affective tendencies from different modalities. \textbf{2)} We proposed HCT-DMG, which dynamically selects the primary modality during training. This model requires fewer fusion operations and does not repeatedly fuse a single modality, reducing the parameters to approximately 0.8M while largely outperforming existing models of similar size. \textbf{3)} We further analyzed the mechanism and feasibility of automatic modality selection by DMG and show that the selection process supports the primacy of text in prior MIP studies, adding new insights to the literature. \textbf{4)} We tested the performance of HCT-DMG on five datasets for sentiment analysis, emotion recognition, humor and sarcasm detection. HCT-DMG achieves remarkable results, consistently outperforming almost all of the baseline models across all tasks. It highlights the versatility and effectiveness of our proposed approach in handling various multimodal affect recognition tasks. \textbf{5)} We expanded the experiments using LLMs, integrating multimodal features into \textit{LlaMA-3.1-8b}. The results further demonstrates the generalizability and effectiveness of our proposed method.

\section{ASR Error-Robust Modality-Gated Fusion}
\label{chap6/sec:asr-fusion}
Since ASR errors bring noise to the text modality, it is expected that they could result in incongruity between the text and other modalities. In this section, we propose an ASR Error-Robust Modality-Gated Fusion that combines our ASR error correction techniques from Chapter~\ref{chap5} and the incongruity-aware modality-gated fusion from Section~\ref{chap6/sec:incongruity-fusion}.

\subsection{Experimental Setup}
\subsubsection{Settings}
We experimented on both bimodal SER (text + audio) and trimodal (text + audio + vision) for a comprehensive SER evaluation. For the bimodal scenario, we use the same datasets and experimental settings as the benchmarking experiment in Section~\ref{chap4/sec:bimodal}. For the trimodal scenario, we follow the same settings as in the previous section (Section~\ref{chap6/sec:incongruity-fusion}).

\subsubsection{Incongruity from ASR Errors}

\begin{table}[h]
    \centering
    \includegraphics[width=\textwidth]{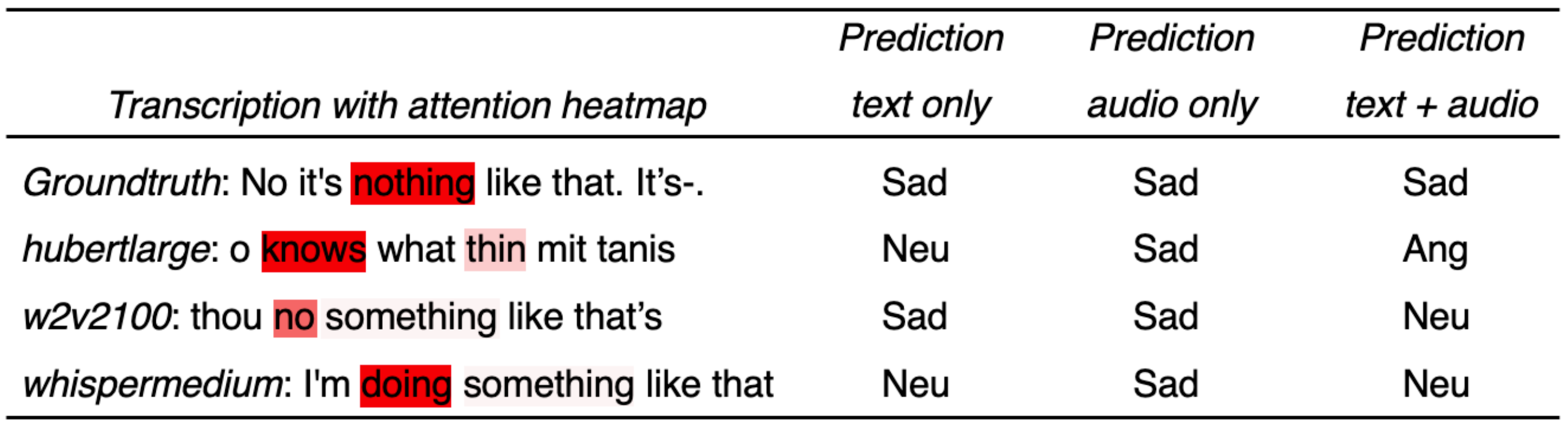}
    \caption{Incongruity brought by ASR errors. The groundtruth label is \textit{Sad}.}
    \label{chap6:tab:asrincongruity}
\end{table}

First of all, we present a case study to show the incongruity brought by ASR errors. As shown in Table~\ref{chap6:tab:asrincongruity}, the ASR error in the text modality causes incongruity with the audio modality. We compute the attention score and create the attention heatmap during the attention fusion of text and audio for SER. The attention heatmap successfully captures the salient word ``nothing'' in the ground truth transcription but fails to do so in the ASR transcriptions due to recognition errors. Even when either the text or audio modality alone correctly identifies the emotion from the ASR output, their fusion results in incorrect predictions. This case study underscores the misalignment between the text and audio modalities introduced by ASR errors, which can degrade performance in multimodal SER tasks.

\subsection{Proposed Framework}
\begin{figure}[ht]
  \centering
  \includegraphics[width=\textwidth]{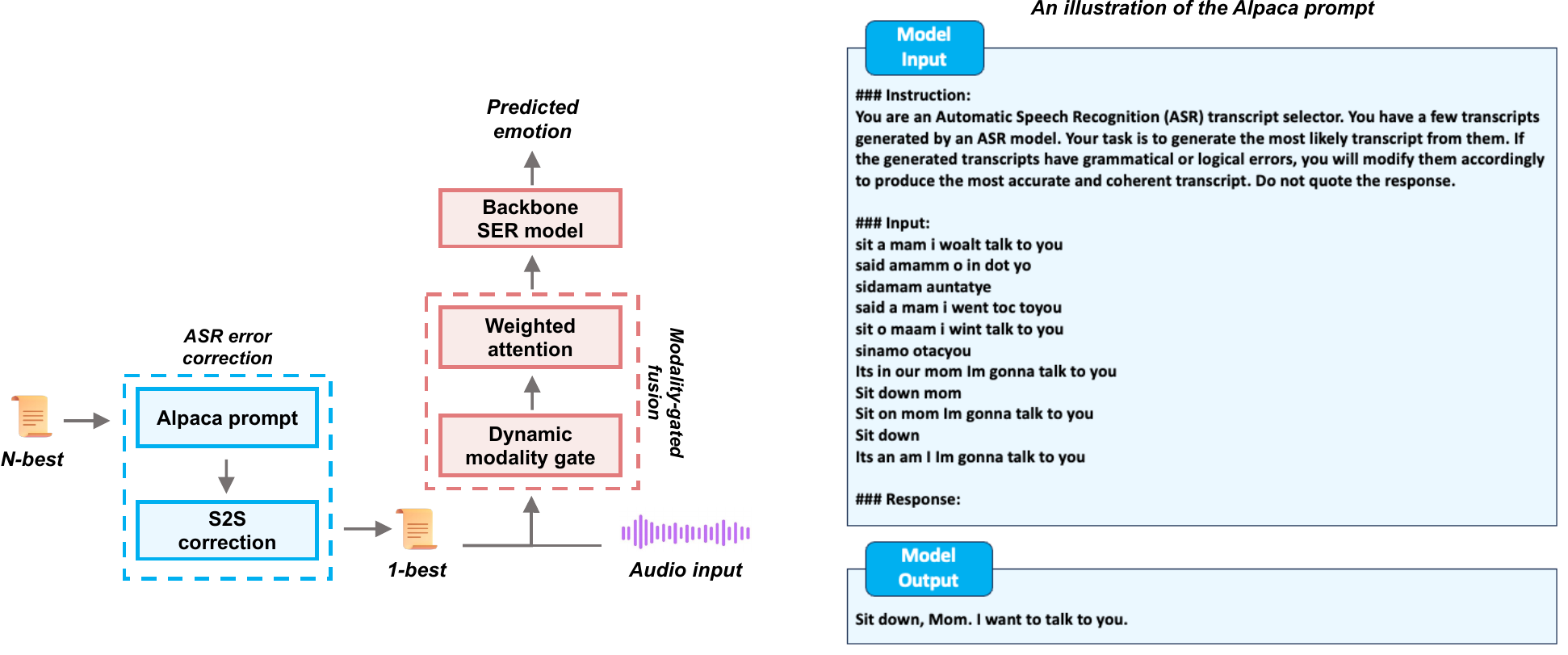}
  \caption{Proposed ASR error-robust modality-Gated fusion framework. \textcolor{cyan}{\textbf{Blue}}: ASR error correction; frozen. \textcolor{pink}{\textbf{Pink}}: Modality-gated fusion; trainable.}
  \label{chap6:fig:proposal}
\end{figure}

To address both the WER and incongruity issues, we propose an ASR Error-Robust Modality-Gated Fusion framework involving ASR error correction and modality-gated fusion, as illustrated in Figure~\ref{chap6:fig:proposal} (left side). This framework integrates insights from previous sections, specifically the LLM-based ASR error correction (Alpaca prompt module) and the S2S-based ASR error correction from Chapter~\ref{chap5}, as well as the modality-gated fusion from Section~\ref{chap6/sec:incongruity-fusion}.

\subsubsection{ASR Error Correction}
We combine the transcriptions from all used ASR models, forming N-best (N=11) hypotheses. Compared to using N-best hypotheses directly from a single ASR model, which offer limited diversity, such a combination offers a more diverse range of transcription results. Furthermore, this approach facilitates the comparison of the SER performance with each of the ten ASR models, as depicted in Table~\ref{chap4/tab:fusion}.

Then, the ASR error correction process consists of a two-stage process. First, we utilize the Alpaca prompt \citep{taori2023stanford}, which guides an LLM to generate the most likely transcription (i.e., 1-best hypothesis) from the N-best hypotheses and correct any grammatical or logical errors. An illustration of the Alpaca prompt is depicted on the right of Figure~\ref{chap6:fig:proposal}. We employ InstructGPT, which has demonstrated effectiveness in LLM-based ASR error correction \citep{radhakrishnan2023whispering}.

In the second stage, we further refine the 1-best transcription by utilizing our proposed S2S ASR error correction model in Chapter~\ref{chap5}, which was pre-trained on the \textit{Whisper-tiny} output $y'$ (i.e., 1-best hypothesis) of Common Voice to recover the corrected sequence $y$ (i.e., gold annotation), with an optimizable parameter $\theta$:
\begin{align}
y=\arg\!\max_{y}P(y| y',\theta)
\end{align}

\subsubsection{Modality-Gated Fusion}
Based on the quality-enhanced transcription, we adapt the trimodal incongruity-aware fusion method proposed in Section~\ref{chap6/sec:incongruity-fusion} to our bimodal inputs, considering potential distortion of original emotion in the spoken content due to ASR errors, which may introduce incongruence between emotions in audio and text.

Specifically, each modality is initially assigned an equal trainable weight. These weights dynamically adjust during training based on the respective contribution of each task, with larger contributions resulting in higher values. The sum of the weights always equals 1, and they are updated in every training batch to ensure dynamic adaptation to any input batch size. Subsequently, the weights are multiplied by corresponding modality features to perform weighted cross-attention fusion.
The pseudo-code is shown in Algorithm~\ref{chap6:alg}, where $H$ represents the hidden representation as input to the backbone SER model. $CrossAttn$ and $SelfAttn$ denote multihead attention, the former with the first argument as query and the second as key and value, and the latter with the argument as query, key, and value. The head number is set to eight.

\begin{algorithm}[!ht]
    \small
    \SetKwFunction{isOddNumber}{isOddNumber}
    \SetKwInOut{KwIn}{Input}

    \KwIn{
    Audio modality $A$; Text modality $T$;
    Dynamic modality gating weight $W^{1}$ initialized with \texttt{nn.Param(torch.tensor([1, 1]))};
    Feature concatenation weight $W^{2}$ initialized with \texttt{nn.Param(torch.tensor([1, 1, 1]))}}  
   
    \For{model input}
    {
        
        Sample $A$, $T$;
        
        $W^{1} \leftarrow \texttt{Softmax(}W^{1}\texttt{)}$;

        $W^{2} \leftarrow \texttt{Softmax(}W^{2}\texttt{)}$;

        $W^{1}_{A} \leftarrow W^{1}[0]$, $W^{1}_{T} \leftarrow W^{1}[1]$;

        $W^{2}_{A} \leftarrow W^{2}[0]$, $W^{2}_{T} \leftarrow W^{2}[1]$, $W^{2}_{AT} \leftarrow W^{2}[2]$,
                
        \eIf{$\texttt{argmax(}W'\texttt{)}$ == 0}
        {
            $A' = W^{1}_{A} * A$;

            $T' = W^{1}_{T} * CrossAttn(A, T)$;
        }{        
            $A' = W^{1}_{A} * CrossAttn(T, A)$;

            $T' = W^{1}_{T} * T$;
        }

        $H = SelfAttn(Concat\left[A'; T'\right])$;
                
        $H' = Concat\left[W^{2}_{A} * A, W^{2}_{T} * T, W^{2}_{AT} * H\right]$;
        
    }
    \caption{Modality-gated fusion.}
    \label{chap6:alg}
\end{algorithm}

Finally, we apply the same backbone SER model, keeping all other settings unchanged (e.g., the text and audio encoders) as in Section~\ref{chap4/sec:bimodal}. We believe that this combination of dynamic modality gate and weighted attention can effectively mitigate the negative impact of ASR errors by consistently focusing on the most salient parts in both inputs for SER.

\subsection{Results}
The results of the bimodal scenario are presented in Table~\ref{chap6:tab:result}. We include only the MAE on CMU-MOSI and the average CCC on MSP-Podcast for brevity. Our approach further reduces WER and improves SER performance, demonstrating its effectiveness. While it does not surpass the best transcription in terms of MAE on CMU-MOSI, the values are very close. Moreover, in line with the findings in \cite{he2024mf}, our results confirm that employing proper ASR error-robust approaches can exceed SER performance based on ground-truth text using ASR transcription. This phenomenon likely occurs because the LLM/S2S models use context to rewrite the transcript in a way that better aligns with the projected emotion, causing certain misrecognized words to be interpreted as contributing positively to the ground-truth emotions, rather than indicating that ASR errors are preferable in SER.

\begin{table}[ht]
\centering
\caption{Performance comparison. $\uparrow$: higher the better. $\downarrow$: lower the better.}
\begin{tabular}{lrr|rr|rr}
\multicolumn{1}{c}{\textbf{}} & \multicolumn{2}{c|}{\textbf{IEMOCAP}} & \multicolumn{2}{c|}{\textbf{CMU-MOSI}} & \multicolumn{2}{c}{\textbf{MSP-Podcast}} \\ \hline
 & \textbf{WER$\downarrow$} & \textbf{Acc4$\uparrow$} & \textbf{WER$\downarrow$} & \textbf{MAE$\downarrow$} & \textbf{WER$\downarrow$} & \textbf{CCC$\uparrow$} \\ \hline
best trans & 12.31 & 74.66 & 18.69 & 0.8558 & 15.02 & 0.616 \\
ground truth & 0.00 & 74.32 & 0.00 & 0.8902 & 0.00 & 0.613 \\ \hline
ours & \textbf{10.12} & \textbf{76.66} & \textbf{17.00} & \textbf{0.8557} & \textbf{12.85} & \textbf{0.618}
\end{tabular}
\label{chap6:tab:result}
\end{table}

\subsection{Summary}
In this section, we proposed an ASR Error-Robust Modality-Gated Fusion that integrates ASR error correction and modality-gated fusion. This framework demonstrated superior performance compared to baseline models in our experimental results and holds potential for various tasks where ASR transcription serves as the text source.

\section{Chapter Summary}
\label{chap6/sec:summary}
Building on the foundations laid in earlier chapters, this chapter examines cross-modal incongruity in SER, arising from either modality misalignment or imperfections in ASR. Such incongruity may introduce conflicting emotional signals and undermine SER accuracy. To tackle this, we incorporate our proposed ASR error correction methods and present novel incongruity-aware fusion models to enhance the robustness of multimodal SER systems.

We begin by analyzing cross-modal incongruity in multimodal SER inputs, as detailed in Section~\ref{chap6/sec:exploration}. To mitigate this issue, Section~\ref{chap6/sec:incongruity-fusion} proposes a hierarchical attention-based fusion framework that employs an incongruity-aware modality-gated mechanism to reduce the effects of conflicting emotional cues. Additionally, in Section~\ref{chap6/sec:asr-fusion}, we augment this framework by integrating ASR error correction into the modality-gated fusion process, resulting in an ASR error-robust fusion approach aimed at minimizing incongruity caused by ASR errors and improving overall SER outcomes.

\chapter{Semi-Supervised Labeling for Effort-Saving SER}
\label{chap7}
\section{Introduction}
After exploring ASR models, correcting ASR errors, and integrating ASR outputs, we aim to address another issue in SER: significant human efforts in emotion labeling, through the insights about ASR we have obtained so far. Furthermore, we also aim to evaluate our approach in another speech-based cognitive task: Alzheimer's dementia detection, for generalizability.

Speech classification tasks, particularly those involving cognitive states, are valuable for a wide range of applications, including human-computer interaction \citep{deng2004speech}, health monitoring \citep{fleury2008sound}, and clinical diagnosis \citep{moell2025order}. However, human annotation for these tasks is often expensive, time-consuming, and requires extensive subjective assessment. This data scarcity issue hinders the progress of these classifications in real-world applications. To address this challenge, Semi-Supervised Learning (Semi-SL) offers a promising approach by leveraging both limited labeled data and a larger amount of unlabeled data for various classification tasks, such as hate speech detection \citep{d2020label}, emotion recognition and sound event detection \citep{zhang2012semi}, as well as sleepiness detection and gender classification \citep{zhang2013co}.

Generally, Semi-SL methods can be categorized into two main types: generating \textit{reliable pseudo-labels} and building \textit{reliable models} using limited data. For example, pseudo-labeling involves creating labels for unlabeled data based on predictions from an iteratively trained model \citep{zhu2021speech}. Consistency regularization ensures that the model produces consistent predictions for augmented versions of the same input (e.g., with added noise) \citep{lu2019semi}. These two types are often complementary and are commonly used together in existing Semi-SL frameworks \citep{feng2022semi}. Moreover, the advancements in self-supervised learning have further improved Semi-SL by utilizing large amounts of unlabeled data to pre-train Semi-SL models \citep{zhang2022censer,lai2021semi}.

Among the literature, the most relevant studies to our work focus on pseudo-label generation. \cite{d2020label} used label propagation, transducing labels from labeled data to unlabeled data with probabilistic transitions for hate speech classification. They found that semi-supervised learning based on label propagation improves hate speech classification in extremely low-resource scenarios, but the performance gains diminish as the amount of labeled data increases. \cite{zhang2012semi} incorporated unlabeled data with high confidence levels (i.e., classification probabilities generated by the trained classifier) into the training set and resampled the original labeled data for sound event classification. They also proposed dividing acoustic features into two views, selecting high-confidence instances in each view, and aggregating them with their predictions into the initial training sets per iteration \citep{zhang2013co}. \cite{zhu2021speech} applied noisy student training \citep{xie2020self} to emotion recognition, using a teacher model trained on labeled data to infer soft labels for unlabeled data. \cite{feng2022semi} proposed incorporating federated learning, utilizing both labeled and unlabeled data at each local client in a multi-view pseudo-labeling approach.

Despite these advances, selecting high-confidence pseudo-labeled data remains challenging. In this chapter:

$\bullet$ We propose a novel Semi-SL framework, integrating multi-view pseudo-labeling that leverages both acoustic (Chapter~\ref{chap3}) and linguistic characteristics (Chapter~\ref{chap4}) to select the most confident data for model training.

$\bullet$ We employ Fréchet Audio Distance (FAD) as a reference-free method to cluster unlabeled data based on acoustic similarity.

$\bullet$ We use task-specific prompts from Chapter~\ref{chap5} to predict labels from ASR transcripts, leveraging insights from acoustics, linguistics, and psychology.

$\bullet$ We examine multiple fusion methods, including the modality-gated fusion from Chapter~\ref{chap6} in the context of Semi-SL to build the bimodal classifier.

\section{Proposed Model}
\begin{figure}[ht]
    \centering
    \includegraphics[width=\textwidth]{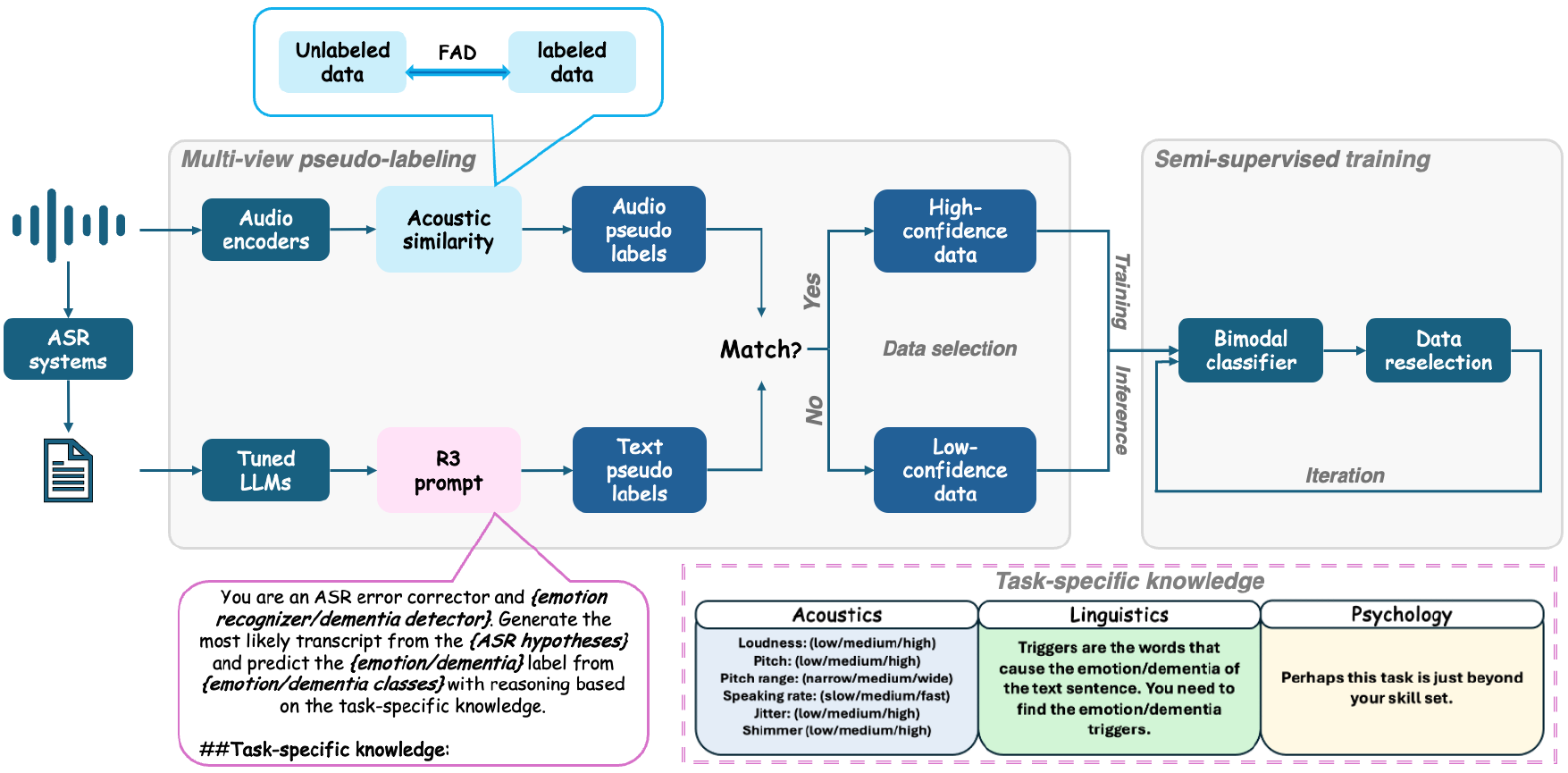}
    \caption{Framework of our proposed semi-supervised speech classification with pseudo-labeling and task-specific knowledge.}
    \label{chap7/fig:framework}
\end{figure}

As illustrated in Figure~\ref{chap7/fig:framework}, the Semi-SL framework with our proposed multi-view pseudo-labeling consists of two paths: acoustic and linguistic. The acoustic path utilizes the similarity between labeled and unlabeled data based on diverse audio embeddings, while the linguistic path employs LLMs to predict class labels from ASR transcriptions using task-specific knowledge. If the generated pseudo-labels from both paths align, we consider them as high-confidence data for training a bimodal classifier. Otherwise, the data are treated as low-confidence and will be further predicted using the trained bimodal classifier. The semi-supervised training of the bimodal classifier will iterate until a predefined criterion is met.

\subsection{Multi-View Pseudo-Labeling}
\subsubsection{Acoustic Path}
We extract acoustic features from multiple audio encoders trained with different objectives to reduce the bias of relying on a single one, inspired by a previous study in the music field \citep{li2024rethinking}. We use the following four audio encoders, resulting in four sets of embeddings for calculating the respective FAD scores:

$\bullet$ \textbf{\textit{VGGish}}: convolutional embeddings \citep{hershey2017cnn}

$\bullet$ \textbf{\textit{EnCodec}}: low-rate audio codecs \citep{defossezhigh}

$\bullet$ \textbf{\textit{wav2vec 2.0}}: self-supervised acoustic embeddings \citep{baevski2020wav2vec}

$\bullet$ \textbf{\textit{CLAP}}: contrastive audio-text embeddings \citep{elizalde2024natural}

Since these models are pre-trained for different purposes, they are used to capture distinct acoustic characteristics, providing complementary embeddings that help reduce the bias in speech classification associated with relying on a single audio encoder.

Given the embeddings of labeled data $\mathbf{X}^{l}$ and unlabeled data $\mathbf{X}^{u}$, the FAD score is calculated using multivariate Gaussians from two embedding sets $X^{l}(\mu_l, \Sigma_l)$ and $X^{u}(\mu_u, \Sigma_u)$ as follows:
\begin{align}
F(X^{l}, X^{u}) = ||\mu_l - \mu_u|| ^2 + tr(\Sigma_l + \Sigma_u - 2\sqrt{\Sigma_l\Sigma_u})
\end{align}
where $tr$ is the trace of a matrix.

Compared to traditional similarity metrics, such as cosine similarity and Euclidean distance, FAD is specifically designed for audio assessment, reflecting the perceptual similarity between two audio embedding distributions \citep{roblek2019fr}. It has proven effective in distinguishing real and synthetic audio \citep{gui2024adapting}, classifying audio with different emotions \citep{li2024rethinking}, and measuring acoustic similarity between emotional music and speech \citep{sun2025exploring}. Therefore, we use FAD to measure the acoustic similarity between labeled and unlabeled data. The four FAD scores, calculated based on four sets of embeddings, are averaged to obtain the final FAD score. Finally, the unlabeled data with the smallest FAD score relative to the labeled class will be assigned that class label as the \textit{acoustic pseudo-labels}. An example of this process for emotion pseudo-labeling is shown in Table~\ref{chap7/tab:audio}. From the VGGish embedding, although the lowest individual FAD score corresponds to \textit{Happy}, the lowest average score is for \textit{Angry}. Thus, the unlabeled data is labeled as \textit{Angry}. Note that the scores are not comparable across different audio encoders, as they are calculated based on different embeddings.

\begin{table}[ht]
\centering
\caption{An example of emotion pseudo-labeling using FAD. \textbf{Bold}: the smallest average FAD score, indicating its pseudo-label (i.e., Angry in this case).}
\label{chap7/tab:audio}
\begin{tabular}{c|c|c|c|c}
\hline
 & \textbf{Angry} & \textbf{Happy} & \textbf{Neutral} & \textbf{Sad} \\ \hline
\textbf{VGGish} & 4.12 & 3.98 & 6.87 & 12.20 \\
\textbf{EnCodec} & 35.33 & 42.56 & 57.24 & 89.65 \\
\textbf{wav2vec 2.0} & 54.66 & 58.49 & 88.78 & 109.02 \\
\textbf{CLAP} & 45.46 & 182.65 & 141.75 & 230.39 \\ \hdashline
\textbf{Average} & \textbf{34.64} & 71.42 & 73.41 & 110.57 \\ \hline
\end{tabular}
\end{table}

\subsubsection{Linguistic Path}
Previous speech classification tasks that used textual information typically relied on ground-truth text. However, in real-world applications, ASR is the only text source, and its transcriptions are usually noisy and contain errors, which can lead to incorrect classifications or pseudo-labels. We argue that \textit{it is more challenging to prompt LLMs for classification tasks based on ASR transcriptions compared to human transcriptions due to the presence of word errors.}

To address this, we use the \textsc{Revise-Reason-Recognize} (R3) prompting pipeline introduced in Section~\ref{chap5:LLM} to perform speech classification with ASR Error Correction (AEC) and reasoning on ASR transcriptions. As described, the R3 pipeline involves three steps: \textsc{Revise}, where ASR errors are corrected based on N-best hypotheses; \textsc{Reason}, where the LLMs self-explain based on the corrected transcriptions and task-specific knowledge; and \textsc{Recognize}, where the label is identified.

For the ASR systems, we adopt the following ten models, the same as those used in Section~\ref{chap6/sec:exploration}, to generate diverse transcriptions and form 10-best ASR hypotheses:

$\bullet$ \textit{wav2vec2-base-\{100h,960h\}}

$\bullet$ \textit{wav2vec2-large-960h}

$\bullet$ \textit{wav2vec2-large-960h-lv60-self}

$\bullet$ \textit{HuBERT-large-ls960-ft}

$\bullet$ \textit{WavLM-libri-clean-100h-base-plus}

$\bullet$ \textit{Whisper-\{tiny, base, small, large-v2\}.en}

To perform AEC, we follow an AEC-specific Alpaca template \citep{yang2023generative}, which uses the \textit{``You are an ASR error corrector''} instruction, guiding the LLMs to perform error correction. As LLMs have demonstrated their ability in both AEC and emotion recognition \citep{yang2024large}, we expect that this capability can be extended to dementia detection from ASR transcriptions as well. The revised ASR transcriptions will be used for subsequent text feature extraction to train the bimodal classifier. For reasoning, we design task-specific knowledge that incorporates acoustics, linguistics, and psychology as in Figure~\ref{chap7/fig:framework}.

We first apply Parameter-Efficient Fine-Tune (PEFT) on the following three LLMs with the LoRA adapter \citep{hulora} using the labeled data with the R3 prompt:

$\bullet$ \textit{Llama2-7b-chat-hf}

$\bullet$ \textit{Llama2-13b-chat-hf}

$\bullet$ \textit{Falcon-7b-instruct}

The objective of PEFT is to adapt a pretrained language model to either emotion or AD detection using a small number of task-specific parameters, while leveraging the knowledge from the pre-trained model. The optimization focuses on improving the model’s performance on the task at hand, typically by minimizing a task-specific loss function. By utilizing LoRA, PEFT achieves efficient fine-tuning by updating only a subset of the model’s parameters, which reduces computational cost and training time. The goal is to maximize task performance with minimal resource usage, especially when limited labeled data is available in this semi-supervised setting.

The learning rate, weight decay, and number of epochs are set to $1.0 \times 10^{-4}$, $1.0 \times 10^{-5}$, and 5, respectively, with AdamW optimizer used. The three fine-tuned LLMs are then prompted with the R3 prompt to predict class labels from ASR transcriptions of the unlabeled data. Finally, majority voting is applied to the predicted labels of the three LLMs to generate the \textit{linguistic pseudo-labels}.

\subsubsection{Data Selection}
The acoustic and linguistic pseudo-labels generated from the two paths are combined to select the most confident data for semi-supervised training. Data with matching pseudo-labels from both paths are selected as high-confidence, while data with differing pseudo-labels are considered low-confidence. Together with the labeled data, the high-confidence data will be used to train the bimodal classifier in the first iteration, ensuring robust initial training.

\subsection{Semi-Supervised Training}
\subsubsection{Bimodal Classifier}

The bimodal classifier consists of pre-trained feature encoders: \textit{HuBERT} \citep{hsu2021hubert} (we use HuBERT here, a distinct model from those used in the audio path, to provide a new perspective, trying to avoid redundant information in the pseudo-labels generated by the classifier and the audio path) and \textit{RoBERTa} \citep{liu2019roberta} that extract audio and text features, respectively, and a classification model that uses these features to generate a prediction label. For PEFT the encoders and training the classification model, the learning rate, weight decay, number of epochs, and batch size are set as $1.0 \times 10^{-4}$, $1.0 \times 10^{-5}$, 30, and 64, respectively. The AdamW optimizer is used. For the classification model, we examine the following four fusion methods:

\begin{itemize}
    \item \textbf{\textit{Early fusion}}: text and audio features are concatenated at the embedding level
    \item \textbf{\textit{Cross-attention fusion}}: text and audio features are attended to each other via attention and then concatenated \citep{li2022fusing}
    \item \textbf{\textit{Tensor fusion}}: unimodal information and bimodal interactions are learned explicitly and then aggregated \citep{zadeh2017tensor}
    \item \textbf{\textit{Modality-gated fusion}}: primary modality is dynamically adjusted in each training step, as proposed in Section~\ref{chap6/sec:incongruity-fusion}. As we only consider audio and text modalities, it aligns with that used in Section~\ref{chap6/sec:asr-fusion} for bimodal fusion
\end{itemize}

\subsubsection{Iteration}
After training the bimodal classifier, low-confidence data are predicted and labeled using the trained classifier\footnote{Here, 'predicted' refers to the model generating a class prediction (i.e., assigning a probability distribution over emotion classes) for the unlabeled data, while 'labeled' refers to assigning the most probable class as a pseudo-label to that data point based on the model’s prediction.}. In most previous Semi-SL studies, model-labeled data are fully trusted and incorporated into the training set in the next iteration. However, as training progresses, mislabeled data (noise) may accumulate, leading to a cycle of erroneous learning \citep{zhu2005semi}. To address this issue, we choose not to fully trust the model-labeled data. Instead, the pseudo-label generated by the bimodal classifier is compared with pseudo-labels from multi-view pseudo-labeling. If the model pseudo-label matches either the acoustic or linguistic pseudo-label, the data are added to the training set for the next iteration. Otherwise, they remain low-confidence and will be predicted in the next iteration.

In each iteration, we update the training set by adding model-labeled data and randomly removing 20\% of the initial high-confidence data to avoid over-reliance on multi-view pseudo-labeling. The model is reinitialized in every iteration to reduce overfitting and bias. The maximum number of iterations is set to 40. However, if there is no performance improvement on the validation set for two consecutive iterations, the iteration will be terminated. The process is summarized in Algorithm~\ref{alg:iteration}.

\begin{algorithm}
    \small
    \caption{Iteration Process}
    \label{alg:iteration}
    \SetKwInOut{Input}{Input}
    \Input{
      $H$: Bimodal classifier; $S_v$: Validation set; $D_c$: Confident (high-confidence) data; $D_u$: Unconfident (low-confidence) data; $L^a$: Acoustic pseudo-labels; $L^l$: Linguistic pseudo-labels; $L^h$: Model pseudo-labels; $I$: Maximum number of iterations
    }
    \For{$i = 1, \dots, I$}
    {
        Train classifier $H^i \gets f(D^i_c)$\;
        Evaluate $L_e \gets H^i(S_v)$\;
        \If{performance on $D_e$ does not improve}{
            \textbf{break}\;
        }
        Generate pseudo-label $L^i_h \gets h^i(D_u)$\;
        \If{$L^i_h$ equals $L^i_a$ \textbf{or} $L^i_l$}{
            $L^i_u \gets L^i_h$\;
        }
        Update $D^{i}_c$ and $D^{i}_u$\;
        Reinitialize classifier $H^i$\;
    }
    \textbf{end}
\end{algorithm}

\section{Experimental Setup}
\subsubsection{Datasets}
We use IEMOCAP for emotion recognition and ADReSSo \citep{luz2021detecting} for dementia detection. For IEMOCAP, we focus on the Big Four emotion classes and exclude utterances with blank transcriptions, resulting in 5,500 utterances (\textit{1,103 angry, 1,615 happy+excited, 1,704 neutral, 1,078 sad}). For ADReSSo, since there are no human labels in the test set to verify our approach, we use only the training set and focus on binary classes: Alzheimer's Dementia (AD) and Cognitively Normal (CN). Due to the long duration of each audio file and the presence of interviewer's speech, we segment all files using the official segmentation information, extracting participants' speech, which results in 2,268 utterances (\textit{1,200 AD, 1,068 CN}). Punctuation and extra whitespace are removed, and all text is converted to lowercase. The revised ASR transcriptions by R3 prompt yield word error rates of 11.48\% and 30.25\% for IEMOCAP and ADReSSo, respectively (the groundtruth transcription of ADReSSo are created by Whisper automatic transcription, followed by a native speaker's correction, as there is no ground truth provided in the dataset).

\subsubsection{Settings}
For both tasks, we use an 80/10/10 split for training, validation, and testing, applying the ratio equally to each class to ensure a balanced distribution. Additionally, the ground-truth labeling rates for the training data are compared at 20\%, 25\%, and 30\%, with the remaining data labeled using our method. All results are measured using Unweighted Accuracy (UA). Random seeds are kept consistent across all experiments. Other settings have been detailed in the previous section.

\subsection{Results and Discussion}

Four baselines are used for comparison:

\begin{itemize}
    \item \textbf{\textit{Supervised\_full}}: the classification model is trained on the entire training data (i.e., the 80\% split)
    \item \textbf{\textit{Supervised\_limited}}: the classification model is trained on the limited labeled data (20\%, 25\%, and 30\%) without pseudo-labeling the unlabeled data
    \item \textbf{\textit{Decision merging}}: two classification models are trained using audio and text, respectively, and their probability distribution are merged to select high-confidence data for the next iteration
    \item \textbf{\textit{Co-training}}: two classification models are trained using audio and text, respectively. High-confidence data selected by each model are added to the training set for the other model in the next iteration \citep{blum1998combining,zhang2018leveraging}
\end{itemize}

\begin{table}[ht]
    \centering
    \caption{Performance comparison with \textit{supervised\_full} and \textit{supervised\_limited} (results in UA\%). $N$: ground-truth labeling rate of training data. \textbf{BOLD}: best performance in each ground-truth labeling rate.}
    \label{chap7/tab:results}
    \resizebox{\textwidth}{!}{
    \begin{tabular}{llccccccc}
        \hline
         & \multirow{2}{*}{\textbf{Fusion}} & \multirow{2}{*}{\textbf{S\_full}} & \multicolumn{2}{c}{\textbf{$N=30\%$}} & \multicolumn{2}{c}{\textbf{$N=25\%$}} & \multicolumn{2}{c}{\textbf{$N=20\%$}} \\
         &  &  & {\textbf{S\_limited}} & {\textbf{Ours}} & {\textbf{S\_limited}} & {\textbf{Ours}} & {\textbf{S\_limited}} & {\textbf{Ours}} \\ \hline
        \multirow{4}{*}{\textbf{\begin{tabular}[c]{@{}l@{}}Emotion\\ recognition\end{tabular}}} & \textit{Early} & 73.67 & 70.01 & \textbf{72.52} (+2.51) & 69.76 & \textbf{72.04} (+2.28) & 68.89 & \textbf{71.20} (+2.31) \\
         & \textit{C-attn} & 74.02 & 71.20 & \textbf{73.79} (+2.59) & 69.97 & \textbf{72.87} (+2.90) & 69.03 & \textbf{71.85} (+2.82) \\
         & \textit{Tensor} & 75.18 & 72.00 & \textbf{74.79} (+2.79) & 70.42 & \textbf{73.25} (+2.83) & 69.52 & \textbf{72.34} (+2.82) \\
         & \textit{M-gated} & 75.53 & 72.03 & \textbf{75.10} (+3.07) & 71.19 & \textbf{73.90} (+2.71) & 70.88 & \textbf{73.61} (+2.73) \\ \hdashline
        \multirow{4}{*}{\textbf{\begin{tabular}[c]{@{}l@{}}Dementia\\ detection\end{tabular}}} & \textit{Early} & 80.03 & 75.93 & \textbf{79.04} (+3.11) & 73.55 & \textbf{78.75} (+5.20) & 71.87 & \textbf{78.10} (+6.23) \\
         & \textit{C-attn} & 80.32 & 76.77 & \textbf{79.99} (+3.22) & 73.76 & \textbf{78.89} (+5.13) & 71.18 & \textbf{78.25} (+7.07) \\
         & \textit{Tensor} & 80.71 & 76.62 & \textbf{80.12} (+3.50) & 74.82 & \textbf{79.48} (+4.66) & 72.01 & \textbf{78.83} (+6.82) \\
         & \textit{M-gated} & 81.23 & 77.14 & \textbf{80.87} (+3.73) & 74.78 & \textbf{79.71} (+4.93) & 72.14 & \textbf{79.11} (+6.97) \\ \hline
        \end{tabular}}
\end{table}

The comparison of our method with \textit{supervised\_full} and \textit{supervised\_limited} are shown in Table~\ref{chap7/tab:results}. It can be observed that

\begin{enumerate}
    \item With only 30\% labeled data, our method achieves performance competitive with the \textit{supervised\_full} baseline.
    \item Our proposed method outperforms the \textit{supervised\_limited} baseline, which lacks multi-view pseudo-labeling and semi-supervised training to augment the labeled data, particularly in dementia detection.
    \item The less ground-truth labeled data available, the more effective our method is in dementia detection, likely because binary classes are easier for pseudo-labeling.
    \item When ground-truth labels are limited, classification performance of the \textit{supervised\_limited} baseline for both tasks drops largely compared to the \textit{supervised\_full} training. Our proposed method, however, mitigates this drop by more than 2\% in emotion recognition and 3\%-7\% in dementia detection.
    \item Modality-gated fusion performs best among the four fusion methods. This is reasonable as it dynamically selects the primary modality contributing most to the classification tasks, thereby reducing the impact of ASR errors in the text modality.
\end{enumerate}

As bimodal fusion does not apply to the \textit{decision merging} and \textit{co-training} baseline settings, we select our best-performing results for comparison. Note that the principle of \textit{decision merging} is the same as that of a fusion method: late fusion (or decision-level fusion) \citep{snoek2005early}. The difference is that we further use the merged probability to determine the high-confidence data for iteration, whereas late fusion directly outputs results based on the highest probability. Here, we refer to it as \textit{decision merging} to avoid confusion with late fusion, as we have used fusion techniques in the previous experiment.

For both \textit{decision merging} and \textit{co-training}, we set the threshold at 0.5 for emotion recognition and 0.7 for dementia detection. For example, the fourth emotion class will be selected as the pseudo-label for an unlabeled data with a probability distribution of \{0.1, 0.1, 0.3, 0.5\}, and the unlabeled data will be added to the training set as high-confidence data. In general, a lower threshold is more suitable for multi-class tasks, while a higher threshold is preferred in binary tasks with higher risk of misclassification. Therefore, using 0.5 for emotion recognition and 0.7 for dementia detection is reasonable and aligns with confidence-based rejection methods in previous studies \citep{chow1970optimum,geifman2017selective,jiang2018trust}, where a reasonably high threshold is typically selected to balance accuracy and rejection rate.

\begin{table}[ht]
\centering
    \caption{Performance comparison with \textit{decision merging} and \textit{co-training} (results in UA\%). $N$: ground-truth labeling rate of training data. \textbf{BOLD}: best performance in each rate.}
    \label{chap7/tab:result2}
\begin{tabular}{llcc}
\hline
 &  & \textbf{\begin{tabular}[c]{@{}c@{}}Emotion\\ recognition\end{tabular}} & \textbf{\begin{tabular}[c]{@{}c@{}}Dementia\\ detection\end{tabular}} \\ \hline
\multirow{3}{*}{\textbf{$N=30\%$}} & \textit{Decision merging} & 67.60 & 74.55 \\
 & \textit{Co-training} & 69.99 & 74.73 \\
 & \textit{Ours} & \textbf{75.10} & \textbf{80.87} \\ \hdashline
\multirow{3}{*}{\textbf{$N=25\%$}} & \textit{Decision merging} & 66.34 & 72.70 \\
 & \textit{Co-training} & 69.04 & 72.87 \\
 & \textit{Ours} & \textbf{73.90} & \textbf{79.71} \\ \hdashline
\multirow{3}{*}{\textbf{$N=20\%$}} & \textit{Decision merging} & 66.12 & 71.07 \\
 & \textit{Co-training} & 69.00 & 72.10 \\
 & \textit{Ours} & \textbf{73.61} & \textbf{79.11} \\ \hline
\end{tabular}
\end{table}

From Table~\ref{chap7/tab:result2}, we can observe that:

\begin{enumerate}
    \item Our method largely outperforms the two baselines, likely for two reasons: It generates higher-confidence unlabeled data (since the pseudo labels come not only from the classifier but also from the audio and text pseudo labels, enhancing reliability), which iteratively trains the classifier to improve performance; second, our bimodal classifier takes both modalities as input during training. In contrast, although the two baselines also consider two modalities, their classifiers are separate, each relying on a single modality and ignoring the interrelatedness between them.
    \item Although both \textit{decision merging} and \textit{co-training} consist of two classifiers that output respective labels, the latter performs better than the former, especially in emotion recognition. This result is plausible since \textit{decision merging} could potentially weaken the better prediction if the other probability is greatly incorrect. On the contrary, by incorporating high-confidence data judged by the other view into training set, \textit{co-training} enables the two models to learn from each other indirectly. Its relatively lower effectiveness in dementia detection is likely due to the low-quality audio, which makes one of the models less powerful.
\end{enumerate}

\subsubsection{Effect of Multi-View Pseudo-Labeling Components}
We explore the contributions of each audio encoder and LLM in multi-view pseudo-labeling by keeping either the acoustic or linguistic path unchanged (i.e., full path) while adding an individual encoder from the other path. For brevity, Table~\ref{chap7/tab:ablation} presents the results of early fusion with 30\% labeled data omitting the other conditions. The results show that \textbf{1)} The acoustic path contributes more than the linguistic path. \textbf{2)} \textit{CLAP} and \textit{Falcon} perform best among the acoustic and linguistic encoders, respectively.

\begin{table}[ht]
    \centering
    \caption{Contribution of each component of the multi-view pseudo-labeling under the scenario of 30\% labeled data (results in UA\%).}
    \begin{tabular}{lcc}
        \hline
         & \textbf{Emotion recognition} & \textbf{Dementia detection} \\ \hline
        \textbf{Two paths} & 72.52 & 79.04 \\ \hdashline
        \multicolumn{3}{l}{\textbf{Full linguistic path}} \\
        \textit{  + VGGish} & 68.21 & 72.04 \\
        \textit{  + EnCodec} & 65.75 & 70.76 \\
        \textit{  + wav2vec 2.0} & 68.87 & 72.00 \\
        \textit{  + CLAP} & 69.41 & 72.62 \\ \hdashline
        \multicolumn{3}{l}{\textbf{Full acoustic path}} \\
        \textit{  + Llama2-7b} & 70.53 & 76.40 \\
        \textit{  + Llama2-13b} & 71.31 & 77.91 \\
        \textit{  + Falcon} & 72.00 & 78.88 \\ \hline
        \end{tabular}
    \label{chap7/tab:ablation}
\end{table}

\section{Chapter Summary}
In this chapter, we proposed a novel semi-supervised learning framework that introduces a multi-view pseudo-labeling method leveraging both acoustic and linguistic characteristics. This method utilized Fréchet audio distance and large language models to select the most reliable unlabeled data for augmenting the training set. Multiple fusion techniques have been compared to utilize multi-view knowledge for further enhancement of the framework. We evaluated our method on emotion recognition and dementia detection tasks, demonstrating that it outperforms fully-supervised, limited-supervised, and two SSL baselines. Our method achieved competitive performance compared to fully supervised learning while using less than 30\% of human-labeled data.

Compared to existing literature on semi-supervised training for speech classification tasks, we introduced a novel pseudo-labeling framework that leverages audio distance and outputs from LLMs. This approach improves the reliability of pseudo labels, as their confidence is informed by sources beyond the emotion classifier itself. As a result, the issue of error accumulation common in traditional semi-supervised learning algorithms is largely mitigated.

Several limitations remain, which we plan to address in future work: 1) exploring a wider range of audio encoders and LLMs to identify more reliable combinations; 2) refining the confidence threshold to better balance accuracy and the amount of high-confidence data for iteration; and 3) extending the method to additional speech classification tasks, especially those involving low-quality audio.

\chapter{Conclusions and Outlook}
\label{chap8}

\section{Review of Thesis Achievements}
Speech emotion is a natural, complex, and central component of human verbal communication. Analyzing and understanding the emotion in human speech is a crucial function for Artificial General Intelligence (AGI). The implementation of SER in real-world applications has gained traction across various industries \citep{li2023information,li2022information}. This thesis aims to accelerate this by integrating ASR functions.

With the evolution of speech technologies, the way of using speech models in SER has progressed with time. The premise of utilizing state-of-the-art speech foundation models in SER is understanding whether and how they function on emotional speech.

To address the first research question outlined in Chapter~\ref{chap1}, ``\textit{What is the interplay between ASR models and SER?}'', this thesis examines the behavior of a state-of-the-art speech foundation model, wav2vec 2.0, on emotional speech through a series of comprehensive experiments presented in Section~\ref{chap3:acoustic}. The results indicate that the representations generated by wav2vec 2.0, which are optimized for ASR tasks, often lack critical paralinguistic cues and exhibit distinct hierarchical structures and emotion biases. In Section~\ref{chap4:asremotion}, we further investigate how word distribution within a corpus influences ASR performance, uncovering notable emotion-specific variations. Factors such as PoS tag frequency, affective ratings, and utterance length were shown to greatly affect ASR accuracy. Certain word categories may be more easily recognized by one model but pose challenges for another. Additionally, Section~\ref{chap4:asrquality} presents an evaluation of SER systems using ASR-generated transcripts with varying WERs. We also examine common fusion strategies to understand how ASR quality impacts SER outcomes. To this end, we achieved the two objectives listed under this research question: 1) Analyzing how ASR systems process emotionally expressive speech and quantify their limitations; 2) Examining the impact of ASR errors on downstream SER models, particularly in real-world, spontaneous speech settings.

To address the second research question ``\textit{How to improve ASR transcription quality of small-scale emotional speech for SER?}'', Chapter~\ref{chap5} introduces both LLM-based and S2S-based approaches tailored to different ASR hypothesis scenarios, namely, N-best and 1-best outputs. Section~\ref{chap5:LLM} presents a method based on LLMs, which refines N-best ASR hypotheses through emotion-aware prompts and ASR error correction, ultimately enhancing emotion recognition from the improved transcriptions. In Section~\ref{chap5:S2S}, a S2S model is proposed that integrates discrete acoustic units with textual word units to perform cross-modal ASR error correction, outperforming the LLM-based method when applied to 1-best ASR outputs. The transcriptions enhanced by both methods have been utilized to address the third question, demonstrating their effectiveness in delivering higher-quality textual input for SER. To this end, the two objectives listed under this research question have been realized: 1) Adapting existing ASR error correction methods to low-resource emotional speech domains, with a focus on learning to correct emotion-specific transcription errors; 2) Improving SER performance by utilizing corrected ASR transcriptions in both text-only and speech-text fusion approaches.

In addressing the third research question ``\textit{Can we develop robust acoustic-lexical fusion frameworks for ASR-enhanced SER to mitigate error propagation?}'', Section~\ref{chap3:hierarchical} presents a joint ASR-SER training framework that hierarchically incorporates eGeMAPS paralinguistic features, intermediate representations from wav2vec 2.0, and ASR-generated transcriptions into the SER process. This design leverages the strengths of wav2vec 2.0 while addressing its limitations, resulting in performance comparable to that achieved using ground-truth transcripts. Furthermore, Section~\ref{chap6/sec:exploration} examines a key challenge in multimodal fusion: cross-modal incongruity. To address this, Section~\ref{chap6/sec:incongruity-fusion} introduces an incongruity-aware fusion strategy that helps the model discern the dominant affective tendency across modalities. Section~\ref{chap6/sec:asr-fusion} then presents an ASR error-robust framework combining ASR error correction with modality-gated fusion. This approach outperforms baseline models in our experiments and shows promise for text-only and bimodal fusion SER tasks where ASR transcriptions provide the textual input. Collectively, these contributions fulfill the two objectives: 1) Designing robust ASR-enhanced SER frameworks that effectively integrate speech and text cues while mitigating the impact of ASR errors; 2) Investigating novel fusion techniques, such as error-aware fusion and adaptive attention mechanisms to enhance SER robustness in real-life scenarios.

In response to the final research question ``\textit{Can effort-saving strategies be developed for scalable SER to reduce annotation costs?}'', Chapter~\ref{chap7} introduces an innovative semi-supervised learning framework based on a multi-view pseudo-labeling approach that incorporates both acoustic and linguistic cues. This framework employs Fréchet audio distance and LLMs to identify and select the most trustworthy unlabeled samples for training augmentation. Several fusion strategies are explored to integrate multi-view knowledge and further enhance model performance. Experiments on emotion recognition and dementia detection tasks demonstrate that the proposed approach surpasses baseline methods. Distinct from prior semi-supervised efforts in speech classification, our pseudo-labeling strategy combines audio similarity metrics and LLM outputs, yielding more reliable pseudo labels by incorporating external indicators of confidence beyond the emotion classifier itself. Consequently, the framework fulfills three objectives: 1) Developing novel SemiSL strategies for SER, incorporating both confidence-based and reliability-aware sample selection; 2) Exploring co-training methodologies that exploit multi-view learning across various feature encoders; 3) Reducing dependence on large-scale annotated emotion datasets and leverage ASR transcription, making SER models more cost-efficient and effort-saving.

\section{Toward Reliable and Applicable SER}
Despite the achievements presented in this thesis, several limitations remain and warrant further investigation.

First, as discussed in Section~\ref{chap3:acoustic}, speech foundation models such as wav2vec 2.0, although highly effective in ASR and other speech-related tasks, exhibit limited sensitivity to paralinguistic cues. These models are primarily optimized for phonetic accuracy and thus tend to abstract away emotion-relevant acoustic characteristics. This inherent bias poses a challenge for their direct application in emotion-sensitive domains such as mental health monitoring, customer service, and social robotics, where nuanced emotional perception is critical.

Second, while the ASR error correction strategies proposed in Chapter~\ref{chap5} improve transcription quality, their effectiveness remains highly dependent on the domain similarity between training data and deployment scenarios. Additionally, many emotionally salient acoustic events, such as vocal bursts, laughter, or sighs, are typically excluded from ASR transcriptions and cannot be adequately recovered through text-based correction methods. These omissions limit the emotional fidelity of the textual modality and may introduce blind spots in SER systems.

Third, although the fusion strategies developed in this thesis successfully integrate speech and text modalities, emotional signals across modalities are not always congruent, particularly in instances of sarcasm, irony, or emotionally masked expressions. Such cases demand a deeper level of contextual and pragmatic understanding. For instance, while Chapter~\ref{chap6} explores incongruity-aware fusion, the interpretability of the model remains limited. It is still unclear whether the model genuinely understands cross-modal incongruity or merely exploits superficial statistical patterns.

Finally, scalability poses an ongoing challenge. Although the LLM-based techniques introduced in Chapter~\ref{chap5} and the semi-supervised learning framework in Chapter~\ref{chap7} offer promising avenues for reducing manual annotation costs, their success relies on the capabilities of the underlying LLMs and the reliability of the emotion labels they produce. The mechanisms by which LLMs infer emotions from textual input remain questionable, and their failures are often difficult to diagnose. In high-stakes applications such as healthcare or security, even small errors can have serious consequences, highlighting the urgent need for more robust confidence estimation and model interpretability.

We are currently exploring several of these challenges in ongoing collaborations, with the aim of advancing toward more reliable and applicable SER systems. Future research should prioritize the following directions:

\begin{itemize}
\item Developing emotion-centric pretraining objectives for speech foundation models that better preserve affective and paralinguistic information.
\item Advancing multilingual and cross-cultural emotion modeling, including robust handling of code-switching and culturally specific affective expressions.
\item Designing context-aware fusion architectures capable of dynamically assessing and adapting to the reliability of different modalities.
\item Implementing uncertainty quantification and calibration techniques to improve robustness and trustworthiness in SER and relevant applications.
\item Conducting ethical and fairness assessments to ensure that SER systems do not propagate bias or misrepresent underrepresented demographic groups.
\end{itemize}

By addressing these limitations, we can pave the way for SER technologies that are not only technically robust but also socially responsible and truly applicable in real-world contexts.

\appendix

\noappendicestocpagenum
\addappheadtotoc

\renewcommand{\theequation}{\Alph{chapter}.\arabic{equation}}

\chapter{Research Outputs Related to This Thesis}
\label{append:appendix}

In addition to the papers directly stemming from this thesis as described in Section~\ref{chap1:outline}, there are also several papers related to this work. Below describes how they contribute to the topics relevant to this thesis.

1. Semi-Supervised Learning for Multimodal Speech and Emotion Recognition \citep{li2021semi}

This paper is the first submission of the PhD work, intended for the Doctoral Consortium at ICMI 2021. It outlines the planned direction of the thesis, proposing the ideas of hierarchical fusion and multi-view semi-supervised training, which lay the foundation for a significant portion of the thesis.

2. Alzheimer's Dementia Detection through Spontaneous Dialogue with Proactive Robotic Listeners \citep{li2022alzheimer}

This paper marks our initial exploration into speech-based detection of Alzheimer's disease, accepted to IEEE HRI 2022 late-breaking session. We proposed a method that integrates both speech and text information by utilizing the feature of ``focus words.'' The concept of focus words later inspired our use of trigger words in LLMs for emotion recognition and Alzheimer's dementia detection in the thesis work.

3. Robotic Speech Synthesis: Perspectives on Interactions, Scenarios, and Ethics \citep{li2022robotic}

This paper discusses the role of prosody in emotional speech, accepted to IEEE HRI Robo-identity 2 Workshop\footnote{https://sites.google.com/view/robo-identity2}, which provides a foundation for our subsequent exploration of whether embeddings from speech foundation models can encode emotion-related prosodic characteristics in the thesis work. Additionally, the paper addresses ethical issues surrounding synthesized speech, which inspired the special session ``Responsible Speech Foundation Models'' that we later organized at Interspeech \citep{liresponsible}.

4. A Cross-Domain Approach for Continuous Impression Recognition from Dyadic Audio-Visual-Physio Signals \citep{li2022cross}

This paper is our solution submitted to IUI 2022-Dyadic IMPRESSION Recognition Challenge\footnote{https://competitions.codalab.org/competitions/34190}. We integrated multimodal signals from both the speaker and the listener, including speech, eye gaze, facial expressions, and various physiological signals. We proposed two novel loss functions, Knowledge Distillation Loss and Similarity Enhancement Loss, to recognize the listener’s impression when receiving information from the speaker, outperforming the challenge baseline. It serves as an extension of the multimodal fusion work presented in this thesis.

5. Multimodal Dyadic Impression Recognition via Listener Adaptive Cross-Domain Fusion \citep{li2023multimodal}

This paper extends the above work by introducing a novel listener adaptation module, which measures and leverages the causal relationship between the speaker's and listener's signals as weights for multimodal fusion. This method further improved the performance over the prior approach and has been accepted to IEEE ICASSP 2023. The weight-adaptation multimodal fusion partly inspired our modality-gated fusion in the thesis work.

6. Transfer Learning for Personality Perception via Speech Emotion Recognition \citep{li2023transfer}

This paper explores transferring knowledge from an emotion recognition model based on wav2vec 2.0 to the task of personality detection, aiming to address the challenge of limited data in speech-based tasks. This work extends our thesis research on the wav2vec 2.0 model, further validating its applicability to emotion-related tasks. It has been accepted to Interspeech 2023.

7. I Know Your Feelings Before You Do: Predicting Future Affective Reactions in Human-Computer Dialogue \citep{li2023know}

In this paper, we proposed an emotion prediction framework that operates prior to emotion recognition in dialogue systems. By leveraging the speaker’s emotional and linguistic features (such as dialog acts), the system can predict the listener’s forthcoming emotional state. This approach contributes to smoother emotional dialogues and reduces computational resource usage. We analyzed real human emotional conversations to validate the feasibility of this approach. This work represents an attempt to apply our PhD research to real-world scenarios, and has accepted to CHI 2023 late-breaking work.

8. Empowering Dialogue Systems with Affective and Adaptive Interaction: Integrating Social Intelligence \citep{li2023empowering}

This paper builds upon the foundational work of this thesis and explores how to incorporate emotional intelligence and social intelligence into dialogue systems, enabling SER systems to be used in truly applicable ways. We proposed that dialogue systems should either simulate the user's emotions or maintain their own style, depending on the scenarios. This paper has been accepted to ACII 2023 Social and Affective Intelligence Workshop and won the outstanding paper award.

9. Enhancing Speech Emotion Recognition for Real-World Applications via ASR Integration \citep{li2023enhancing}

This paper is another submission to a Doctoral Consortium (at ACII 2023), outlining ongoing work and emphasizing the importance of integrating ASR into SER. It is a revision of our initial doctoral consortium paper presented at ICMI 2021 and lays the foundation for our later work on semi-supervised training.

10. Layer-Wise Analysis of Self-Supervised Acoustic Word Embeddings: A Study on Speech Emotion Recognition \citep{saliba2024layer}

This work is a collaborative study between me and the master's student Alexandra Saliba that I supervised in 2023. The work builds on our exploration of wav2vec 2.0 in the thesis. Instead of using continuous representations, we opted for Discrete Speech Units (DSUs), considering two key advantages they offer: First, a significant reduction in computational resources. Unlike the continuous representations with dimensions of (frame number, embedding), DSUs only have dimensions of (1, embedding). Second, DSUs have been shown to capture certain lexical information. We conducted a layer-wise analysis and applied DSUs to SER, which confirmed their effectiveness. It has been accepted to ICASSP 2024 Self-supervision in Audio, Speech and Beyond Workshop.

11. Beyond Voice Assistants: Exploring Advantages and Risks of an In-Car Social Robot in Real Driving Scenarios \citep{li2024beyond}

This paper represents another exploration of practical applications for speech emotion dialogue systems. We utilized the in-car voice assistant of a mass-produced vehicle to investigate how the WER of ASR impacts user perception, as well as how the speech personality and emotional dimensions of the in-car assistant influence the driving experience. This work was partially carried out prior to my PhD, during my time at Honda Innovation Lab. It connects my earlier industry research with my current academic pursuits, expanding our thesis work on ASR and SER into the domain of human-computer interaction.

12. Chance-Constrained Abnormal Data Cleaning for Robust Classification With Noisy Labels \citep{shen2024chance}

This is a collaborative work on the reliability of emotion labels, exploring data-cleaning methods to achieve robust classification despite noisy labels. Rather than using all the labels from a corpus, our approach focuses on filtering out abnormal data to improve classification tasks. We tested our method on IEMOCAP and enhanced the performance of supervised classification methods. It partly inspires the multi-view pseudo-labeling approach in this thesis to improve data reliability. This paper has been accepted to IEEE Transactions on Emerging Topics in Computational Intelligence.

13. 1st Place Solution to Odyssey Emotion Recognition Challenge Task1: Tackling Class Imbalance Problem \citep{chen20241st}

This paper is our solution submitted to Odyssey Emotion Recognition Challenge\footnote{https://www.odyssey2024.org/emotion-recognition-challenge}. We introduced an ensemble system that includes seven multi-modal models, which were trained independently with different loss functions and class weights. Specifically, the cross entropy loss and the focal loss were used. The system showed the best performance among a total of 68 submissions to the challenge, in all metrics. The S2S-based ASR error correction function and several fusion approaches in this thesis were used in this paper.

14. Rethinking Emotion Bias in Music via Frechet Audio Distance \citep{li2024rethinking}

This paper presents the work I conducted during my internship at Microsoft Research, where I extended part of my thesis research from speech to the domain of music. In this project, we established a benchmark for music emotion recognition by evaluating multiple music datasets and various audio encoders. Additionally, we improved two music generation models by integrating our proposed emotion generation module, which resulted in the production of more emotionally realistic music. We explored the biases introduced by different metrics, recognition tasks, and models, and proposed using Frechet Audio Distance as a solution. This method is applicable not only in the domain of music emotion but also in speech emotion and dementia analysis later studied in this thesis work. This paper has been accepted to IEEE ICME 2025.

15. Exploring Acoustic Similarity in Emotional Speech and Music via Self-Supervised Representations \citep{sun2025exploring}

This work is a collaborative study between me and the master's student Yujia Sun that I supervised in 2024. It is an extension of the wav2vec 2.0 exploration in this thesis. We compared two speech foundation models (wav2vec 2.0 and HuBERT) and one music foundation model (MERT) in terms of their performance on emotional speech and emotional music. We found that while speech and music SSL models do capture shared acoustic features, their behaviors can vary depending on different emotions. This paper has been accepted to IEEE ICASSP 2025. 

16. Cross-Lingual Speech Emotion Recognition: Humans vs. Self-Supervised Models \citep{han2025cross}

This work is a collaborative study between me and the master's student Zhichen Han that I supervised in 2024. It extends this thesis research by exploring the comparison between self-supervised speech models and human performance in cross-lingual SER. We used Mandarin Chinese, English, German, and Tianjin dialect (a Chinese regional dialect), along with three speech foundation models, to investigate how the same model performs in SER across different languages, and how it can be fine-tuned on the target language. The study found that a model trained on language A can, with simple fine-tuning on language B, achieve performance comparable to that of humans. We also discovered that speech foundation models are capable of identifying emotional segments within speech in cross-lingual tasks. This paper has also been accepted to IEEE ICASSP 2025.

17. Segmentation-Variant Codebooks for Preservation of Paralinguistic and Prosodic Information \citep{sanders2025segmentation}

This work is a collaborative study between me and another PhD student Nicholas Sanders in 2024. It extends Chapter 5 by exploring the effectiveness of discrete speech units of preserving prosodic and paralinguistic information in emotion recognition. We proposed novel segmentation-variant codebooks, which quantize speech at distinct linguistic
units (frame, phone, word, utterance), factorizing it into multiple streams of segment-specific discrete features. The results show that the segmentation-variant codebooks are significantly more effective at preserving prosodic and paralinguistic information across probing tasks, including emotion recognition. This paper has been accepted to Interspeech 2025.

18. Leveraging Cascaded Binary Classification and Multimodal Fusion for Dementia Detection through Spontaneous Speech \citep{liu2025leveraging}

This paper is a collaborative study between me and the researchers at University of Science and Technology of China. It extends the multimodal fusion techniques discussed in Chapter 4 to Alzheimer’s disease detection, and we submitted to the PROCESS Challenge 2025. For the three-class classification task (healthy control, mild cognitive impairment, and dementia), we propose a cascaded binary classification framework that fine-tunes pre-trained language models and incorporates pause encoding to better capture disfluencies. Experimental results on the test set outperform the baselines provided by the challenge in both tasks, demonstrating the robustness and effectiveness of our approach. This paper has also been accepted to Interspeech 2025.

19. Exploring Gender Bias in Alzheimer's Disease Detection: Insights from Mandarin and Greek Speech Perception \citep{he2025exploring}

This paper is a collaborative study between me and the researchers at University of Science and Technology of China. It extends the cross-lingual Alzheimer's research in Chapter 7. In a perception experiment involving 16 Chinese listeners evaluating both Chinese and Greek speech, we identified that male speech was more frequently identified as AD, with this bias being particularly pronounced in Chinese speech. Acoustic analysis showed that shimmer values in male speech were significantly associated with AD perception, while speech portion exhibited a significant negative correlation with AD identification. This paper has been accepted to National Conference on Man-Machine Speech Communication 2025.

20. HRAI 2025: The 1st Workshop on Holistic and Responsible Affective Intelligence \citep{li2025hrai}

This paper summarizes the workshop I led at ICMI 2025\footnote{https://icmi.acm.org/2025/workshops/}. It builds upon my entire thesis work, aiming to foster discussion and contributions toward holistic and responsible affective intelligence through advanced foundation models and machine learning techniques. The paper has been accepted to ICMI 2025.

21. Large Language Model Based Generative Error Correction: A Challenge and Baselines for Speech Recognition, Speaker Tagging, and Emotion Recognition \citep{yang2024large}

This paper presents the challenge summary and baseline systems for our hosted competition at IEEE SLT 2024\footnote{https://2024.ieeeslt.org/challenges/\#1715507565729-916ec1d3-b60d}. It is partially derived from the ASR error correction research discussed in Chapter 5. We provide baseline methods using LLMs for ASR error correction in speech recognition, speaker tagging, and emotion recognition tasks.








\edbibliography{bib/edengref}
\bibliographystyle{plainnat}

\end{document}